\documentclass[aps,prx,twocolumn,superscriptaddress,showpacs,final,floatfix,longbibliography]{revtex4-1}

\AtBeginDocument{
    \newwrite\bibnotes
    \def\bibnotesext{Notes.bib}
    \immediate\openout\bibnotes=\jobname\bibnotesext
    \immediate\write\bibnotes{@CONTROL{REVTEX41Control}}
    \immediate\write\bibnotes{@CONTROL{%
    apsrev41Control,author="08",editor="1",pages="1",title="0",year="1"}}
     \if@filesw
     \immediate\write\@auxout{\string\citation{apsrev41Control}}%
    \fi
}%
\usepackage{amsmath,amsfonts,amsthm,amssymb}
\usepackage{graphicx}
\usepackage[utf8]{inputenc}
\usepackage{braket}
\usepackage{setspace}
\usepackage{bbm}
\usepackage{color}
\usepackage{todonotes}
\usepackage{times}
\usepackage{siunitx}
\usepackage{comment}
\usepackage{graphicx}

\newcommand{\be}{\begin{equation}}

\newcommand{\ee}{\end{equation}}
\newcommand{\eea}{\end{eqnarray}}

\newcommand{\kommentar}[1]{}

\newcommand{\trace}{{\rm Tr}}

\newcommand{\de}{\mathrm{d}}

\newcommand{\imply}{\ \Rightarrow \ }
\newcommand{\aver}[1]{\langle #1 \rangle}

\newcommand{\diag}{\mathrm{diag}}

\newcommand{\id}{\openone}
\newcommand{\je}[1]{{\color{black} #1}}

\newcommand{\R}{\mathbb{R}}

\def\p{{\hat \varphi}}
\def\d{{\delta\hat \varrho}}

\def\Tc{t_{\rm comp}}
\def\Tcy{t_{\rm cycle}}

\renewcommand{\t}{\top}
\def\pp{\hat \varphi}
\def\dd{\delta\hat\varrho}
\def\RD{L}
\def\Npxl{N}
\def\Ntro{N_{\rm t}}
\def\di{\mathrm{d}}
\def\nGP{\rho_\mathrm{0}}
\newcommand{\discq}{{\bf \hat X}}
\newcommand{\discpp}{\pp^{(\Npxl)}}
\newcommand{\discdd}{\dd^{(\Npxl)}}

 \usepackage{soul}
 
\newcommand{\mg}[1]{{{\color{black} #1}}}
\def\i{\mathrm{i}}
\newcommand{\gv}[1]{{{\color{black} #1}}}
\newcommand{\nn}[1]{{{\color{black} #1}}}
\newcommand{\jsab}[1]{{{\color{black} #1}}}

\newcommand{\mhu}[1]{{{\color{black} #1}}}

\begin{document}

\title{Quantum field thermal machines}

\author{Marek Gluza*}
\affiliation{Dahlem Center for Complex Quantum Systems, Freie Universit{\"a}t Berlin, 14195 Berlin, Germany}%
\thanks{M.G., J.S., N.H.Y.N. and G.V. contributed equally}

\author{Jo\~ao Sabino*}

\affiliation{Vienna Center for Quantum Science and Technology, Atominstitut, TU Wien,  1020 Vienna, Austria}

\affiliation{Instituto Superior T\'{e}cnico, Universidade de Lisboa, Portugal}
\affiliation{Instituto de Telecomunica\c{c}\~oes, Physics of Information and Quantum Technologies Group, Lisbon, Portugal}
\thanks{M.G., J.S., N.H.Y.N. and G.V. contributed equally}

\author{Nelly H.Y. Ng*}
\affiliation{Dahlem Center for Complex Quantum Systems, Freie Universit{\"a}t Berlin, 14195 Berlin, Germany}%
\affiliation{School of Physical and Mathematical Sciences, Nanyang Technological University, 639673, Singapore}
\thanks{M.G., J.S., N.H.Y.N. and G.V. contributed equally}

\author{Giuseppe Vitagliano*}
\affiliation{Institute for Quantum Optics and Quantum Information (IQOQI), Austrian Academy of Sciences, 1090 Vienna, Austria}
\thanks{M.G., J.S., N.H.Y.N. and G.V. contributed equally}

\author{Marco Pezzutto}
\affiliation{Instituto de Telecomunica\c{c}\~oes, Physics of Information and Quantum Technologies Group, Lisbon, Portugal}

\author{Yasser Omar}
\affiliation{Instituto Superior T\'{e}cnico, Universidade de Lisboa, Portugal}
\affiliation{Instituto de Telecomunica\c{c}\~oes, Physics of Information and Quantum Technologies Group, Lisbon, Portugal}

\author{Igor Mazets}
\affiliation{Vienna Center for Quantum Science and Technology, Atominstitut, TU Wien,  1020 Vienna, Austria}%
\affiliation{Fakult{\"a}t f{\"ur} Mathematik, 
Universit{\"a}t Wien, 1090 Vienna, Austria}
\author{Marcus Huber}
\email{marcus.huber@univie.ac.at}
\affiliation{Vienna Center for Quantum Science and Technology, Atominstitut, TU Wien,  1020 Vienna, Austria}%
\affiliation{Institute for Quantum Optics and Quantum Information (IQOQI), Austrian Academy of Sciences, 1090 Vienna, Austria}

\author{J\"org Schmiedmayer}
\email{schmiedmayerjoerg@me.com}
\affiliation{Vienna Center for Quantum Science and Technology, Atominstitut, TU Wien,  1020 Vienna, Austria}%

\author{Jens Eisert}
\email{jense@zedat.fu-berlin.de}
\affiliation{Dahlem Center for Complex Quantum Systems, Freie Universit{\"a}t Berlin, 14195 Berlin, Germany}%

\begin{abstract}
Recent years have enjoyed an overwhelming interest in quantum thermodynamics, a field of research aimed at understanding thermodynamic tasks performed in the quantum regime. Further progress, however, seems to be obstructed by the lack of experimental implementations of thermal machines in which quantum effects play a decisive role. In this work, we introduce a blueprint of quantum field machines, which - once experimentally realized - would fill this gap. Even though the concept of the QFM presented here is very general and can be implemented in any many body quantum system that can be described by a quantum field theory. We provide here a detailed proposal how to realize a quantum machine in one-dimensional ultra-cold atomic gases, which consists of a set of modular operations giving rise to a piston. These can then be coupled sequentially to thermal baths, with the innovation that a quantum field takes up the role of the working fluid. In particular, we propose models for compression on the system to use it as a piston, and coupling to a bath that gives rise to a valve controlling heat flow. These models are derived within Bogoliubov theory, which allows us to study the operational primitives numerically in an efficient way. By composing the numerically modelled operational primitives we design complete quantum thermodynamic cycles that are shown to enable cooling and hence giving rise to a quantum field refrigerator. The active cooling achieved in this way can operate in regimes where existing cooling methods become ineffective. We describe the consequences of operating the machine at the quantum level and give an outlook of how this work serves as a road map to explore open questions in quantum information, quantum thermodynamic and the study of non-Markovian quantum dynamics.
\end{abstract}
\date{\today}
\maketitle 


\section{Introduction}
As elevated and set in stone as the basic principles of thermodynamics may appear, there is a development emerging that could not have been anticipated when this theory was being conceived.
Indeed, the basic laws have been  formulated in an effort to understand the functioning of macroscopic machines that can be described by classical physics.
However, due to advances in quantum technologies the question that currently begs for an answer is what happens if we consider heat engines for which \emph{quantum laws} and \emph{effects} are expected to play an important role. 
Indeed, there has been a significantly increased recent interest in exploring 
\emph{thermodynamic notions} in the \emph{quantum regime} 
\cite{Topical,PerspectiveKurizki,christian_review,KosloffReview,MillenReview,Janet,Niedenzu2019quantized}. 

One of the most notable insights that has been achieved in this context is, on the 
one hand, the increased role of knowledge and control giving rise to potentially 
superior performance of quantum machines. 
On the other hand, inevitable fluctuations of energy pose novel conceptual challenges in defining thermodynamic quantities at the quantum scale.
Additionally, in the quantum regime thermal and quantum correlations may range over substantial portions of the elements of the machine, possibly influencing its dynamics. 
These fundamental questions have
stimulated interesting experimental developments, e.g, fully controlling a quantum system such as a trapped ion
\cite{Rossnagel14,PhysRevLett.123.080602,Haenggi}, 
a single impurity electron spin in a silicon tunnel field-effect transistor \cite{PhysRevLett.125.166802}
or an electronic circuit~\cite{NewReferenceKSPekola} to engineer behaviour reminiscent of thermal machines. In 
ensembles of nitrogen vacancy centers in diamond,
first quantum signatures have just been observed 
\cite{PhysRevLett.122.110601}.

There is a caveat, however, 
constituting a serious road block in this avenue of research.
It arguably turns out to be excessively difficult to 
experimentally realize a machine that works in the thermodynamics regime and at the same time shows \emph{genuinely quantum} effects: 
This would be a physical system for which 

\emph{(i)} quantum 
mechanics is \emph{required} to derive an 
appropriate effective physical model describing its dynamics, with genuine quantum correlations potentially playing a major role and

\emph{(ii)} it is \emph{infeasible} to control its every single degree of freedom.

Thus, ideally such a machine 
would \gv{consist of a \nn{quantum} many-body system}. 
\gv{The genuinely quantum behaviour of such machines can in principle be witnessed by \nn{irregularities} 
of the system going against the natural direction of entropy increase. \nn{Such irregularities are, however,} generally difficult to observe due to the time-scales of their occurrence being long, and therefore easily dampened by external dissipation. That this is nevertheless possible has been demonstrated in the recent observations of many-body recurrences~\cite{Rauer19, schweigler2020decay}}.

\nn{Despite having the potential to play} a similar role for the development of quantum thermodynamics as the steam engine did for the classical theory of thermodynamics, at the present stage, such machines have yet to be devised.
This state of affairs seems a grave omission in particular in the light of the observation that it 
\je{has been} the study of the performance of machines that led to the 
development of classical thermodynamics in the first place.

\begin{figure}[tb]
	\centering	\includegraphics[width=0.95\columnwidth]{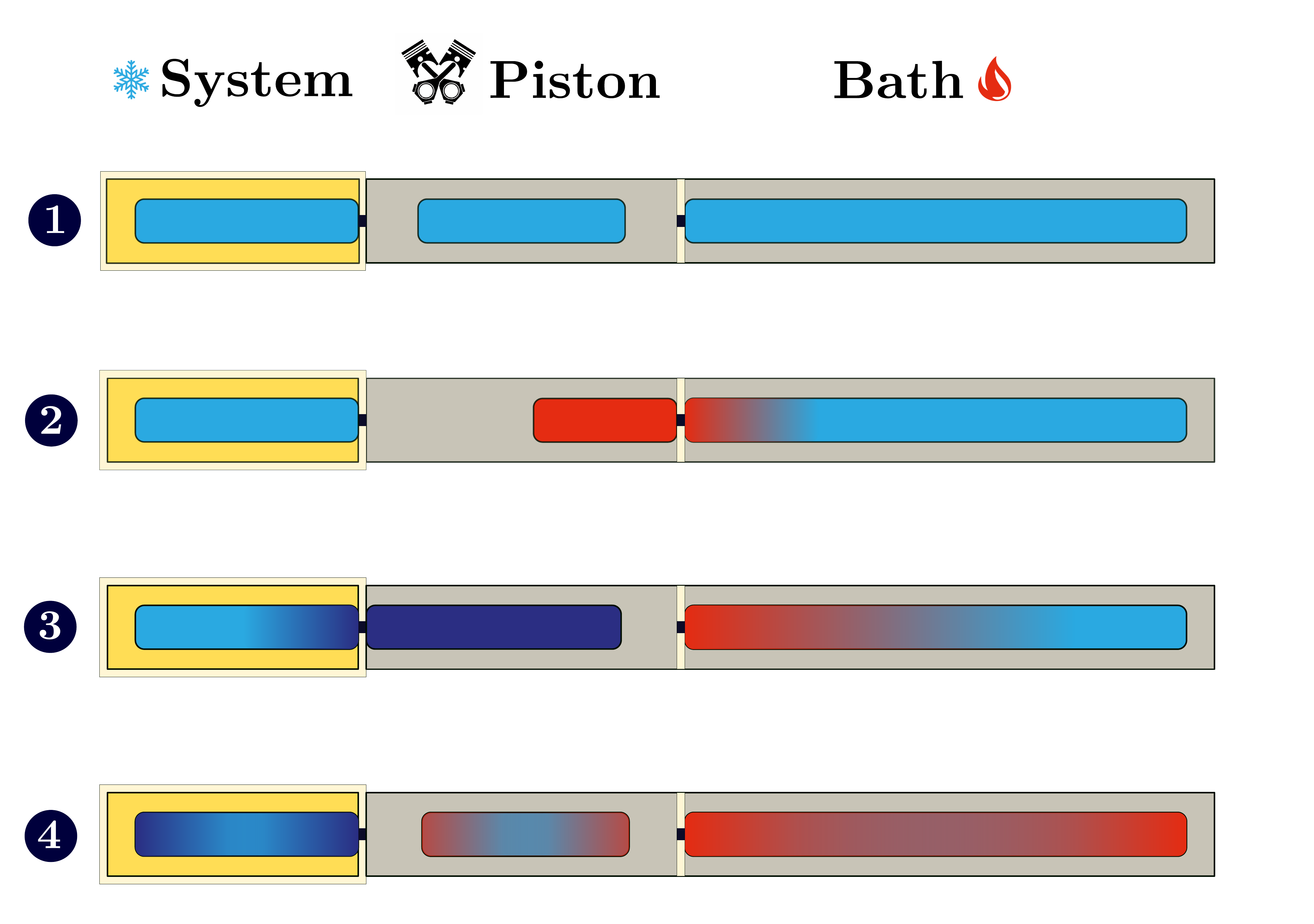}
\caption{\textbf{Quantum field refrigerator:}  
Similar to canonical ideas employed in ordinary thermal machines, we consider for a quantum many-body system a cycle consisting of a small set of control operations on quantum working fluids, concatenated in order to cool down a part of the machine (referred here as the ``system''). This is achieved through a protocol 
consisting of four steps:
1) Initialization of the system, the piston and the
bath at equal temperatures.
2) Compression of the piston and coupling to the bath which receives energy, and decoupling after the heat transfer.
3) Decompression of the piston, therefore decreasing its energy, then coupling to the system thus enabling heat transfer from the system to the piston.
4) Decoupling of the piston from the system and compression to initial size.  
 Through steps 1-4), we expect to achieve a decrease in  the system's energy, while the energy of the piston and bath should increase. This increase in energy happens in such a way that the piston and bath can be reused for multiple cycles before they saturate. All these operations can be implemented experimentally in an ultra-cold atomic gas, by shaping light fields that control the atoms.}
\label{fig:ThermalMachines}
\end{figure}

In this work, we propose a blueprint for a \emph{quantum field machine} (QFM) first conceived in Ref.~\cite{Schmiedmayer2018} that would, once experimentally realized, qualify as being a genuine quantum thermal machine in this sense. 
One of the central challenges here is a trade-off between a sufficient size of the machine to meaningfully allow for thermodynamic considerations -- after all, one has to make reference to thermal baths -- and sufficient control of the dynamics. Only if suitable levels of control can be reached, one can hope to transcend features of classical statistical mechanics and reveal genuine quantum behaviour of machines. Furthermore, elucidating quantum thermodynamic behaviour will be \gv{even more} important \gv{whenever} the envisioned machine actually manages to perform a task that would otherwise be impossible to achieve by other means.
\gv{A prominent example of such a potential task is {\it refrigeration}.

On the one hand, current cooling techniques applied to quantum systems (e.g., laser cooling, evaporative cooling) seem to have hit the ultimate, (semi-)classically possible limit; on the other hand, it is conceivable that quantum control over the cooling mechanism could serve to go beyond such limit.
In this sense, a genuine quantum machine could have revolutionary practical implications, very analogous to the steam engine example mentioned above.}
The QFM that we propose here intends \gv{precisely} to address all of the aforementioned challenges posed when building genuine quantum machines:

\textit{(i)} \gv{It is a} genuine complex quantum many-body system, \gv{describable by means of effective quantum field theories that capture emergent degrees of freedom using different scales \nn{ of refinement in the field theory model}. 
In this particular work} we focus on a QFM tuned  \gv{on a Gaussian} regime which is efficiently simulable numerically \cite{quantumreadout} \nn{and also a very good approximation for moderately short time scales}. \nn{We, however, note }\jsab{that it can be implemented in a strongly correlated regime where a Gaussian treatment or even a perturbative treatment is not possible}~\cite{Cazalilla,Giamarchi2004, Schweigler2017}.

\textit{(ii)} It offers potential new tools for quantum liquids and gases, e.g., by providing an additional stage of cooling which does not involve diluting the system and can be applied after the use of existing techniques

\textit{(iii)} The available degrees of controllability makes it possible to exploit strong correlations and coherences for probing quantum effects. This is achieved by steering \nn{the functioning of the machine by our understanding of} the physics of the system, instead of controlling individual degrees of freedom.

This anticipated device derives from ultra-cold atoms that in a tuneable fashion realize the full range from non-interacting to strongly correlated \nn{phononic} quantum fields~\cite{Mora03,Cazalilla,Giamarchi2004,popov2001functional,Gritsev07}, as can be implemented on an Atom Chip~\cite{Folman2000,FOLMAN2002263,reichel2011atom}. 
The feature that renders it a machine is the presence of programmable time-dependent potentials allowing to manipulate the quantum fields. Such time-dependent potentials have been implemented in a 1D experiment on an Atom Chip  by means of a \emph{digital micro-mirror device (DMD)}~\cite{Tajik2019}. 
That is to say, the DMD devices take the role of ``control knobs'' of the machine, \nn{in particular also being responsible for the input of work}. 
At the same time this field machine will operate at finite temperatures \nn{(in contrast to the majority of theoretical studies on quantum fields done with respect to the ground state)}, thus all these features come together when considering a QFM.

We shall start our investigation by laying out in Section~\ref{sec:QuntumFieldMachine} the concept of a QFM and describing its building blocks.
In Section~\ref{sec:Implementation} we give a detailed introduction on how to implement a quantum field machine using one-dimensional quasi-condensates manipulated on an Atom Chip with optical fields. 
In Section~\ref{Sec:three}, we present a numerical study of each primitive operation described in the introduction and in Section~\ref{sec:Otto} show how to compose them together to make a quantum field refrigerator and \jsab{compare how it performs compared state-of-the-art cooling techniques used in cold atoms experiments. Besides that, we discuss the phenomena of anomalous heat flow between two gases correlated by one of the primitives}
Finally, in Section \ref{sec:discussion} we complete the roadmap towards building a quantum field thermal machine by \gv{highlighting the near future directions of research that we will explore}.

\section{The quantum field machine} 
\label{sec:QuntumFieldMachine}

Thermodynamics is a versatile framework allowing to describe a large variety of machines. Any of these ordinary thermal machines can be explored in the quantum regime if one considers operating it under conditions where quantum effects prominently play a role.
This is the pathway we take in this work, by considering the working fluid to be a \emph{Bose-Einstein condensate (BEC)} and in the  one-dimensional regime more precisely we will consider quasi-condensates \cite{Petrov00}.
In order to investigate the influence of quantum effects on the machine, it is a necessity to consider an appropriate quantum model that describes the system. At the same time, it is also crucial to understand how the quantum evolution of a system can be used to implement certain abstract but well defined thermodynamic transformations, general enough to be independent of whether quantum effects are significantly involved or not.

A quantum thermal machine can be constructed by choosing few suitable building blocks and applying some operations on them in a cyclic fashion, forming a thermodynamic cycle.
For instance, as illustrated in Fig.~\ref{fig:ThermalMachines} it is instructive to consider a quantum thermal machine consisting of three elements, of which two are thermal baths, while the third is a piston shuttling between them. \nn{The relevant degrees of freedom in our machine are phonons, which we describe with an effective quantum field theory.}
With these ingredients it is, e.g., possible to run a
\emph{heat engine}, by allowing heat transfer from the hot bath to the cold one, while work can be extracted from the piston.
If quantum fluctuations play a significant role, their contribution would have to be taken into account for such a process.
Moreover, since the individual components of the machine are small and they feature relatively large energy fluctuations, the systems may exhibit complex out-of-equilibrium dynamics during the operation of the cycle. 
\nn{In this work, we demonstrate the reverse process: in particular, we operate the machine as a \emph{quantum field refrigerator}, using the piston to extract heat from one part of the machine and disposing it into another part. We show that with such an active cooling mechanism it is theoretically possible to cool down a system of ultra-cold atoms to a temperature regime in which other cooling methods are ineffective. }

In order to implement such quantum field machines, we identify \gv{two basic} operations which we call \emph{quantum thermodynamic primitives} (QTPs): \gv{a \emph{valve} and a \emph{piston}}. 
\gv{The first allows to control \emph{energy flow} between elements of the machine.
The second allows to \emph{control thermodynamic parameters} during a stroke: by changing the volume, we modify pressure or temperature via the equation of state\nn{~\cite{Mora03}}.
These basic ingredients of our thermodynamic protocols can be concatenated in a modular fashion to build up the complex range of potential applications \nn{for such a machine of interest}. 
In what follows, we will put particular
emphasis on providing details about the functioning of a quantum field refrigerator as illustrated in Fig.~\ref{fig:ThermalMachines}.
} 



\subsection*{Coupling and decoupling two quasi-condensates: A valve}
\label{sec:valve}

As depicted in Fig.~\ref{fig:ThermalMachines}, 
one of the essential ingredients for operating a quantum field machine is coupling its elements.
This will be in general realized by allowing excitations to tunnel through a barrier which controls energy flow between two parts, like in a valve.
When considering \gv{such a valve in the quantum regime,  we see some important differences compared to a similar operation in an ordinary thermal machine. Specifically:} 

\emph{(i)} In classical physics, merging of systems with identical density is largely featureless. \gv{In sharp contrast, with two quasi-condensates, even if initially uncorrelated, due to phase gradients at the interface of the two systems, \nn{\emph{excitations} of non-negligible magnitude are unavoidably created}, and this consequently leads to an overall energy and entropy increase.
Such quantum \emph{phase diffusion} effects \cite{lewenstein1996quantum,JavanainenWilkens,LeggettComment,JavanainenReply} can, however, be countered by enabling yet another quantum effect which is coherent tunneling through a barrier, leading to \emph{phase-locking}~\cite{Rauer2018,Schweigler_thesis,KaganeEtAl,Gritsev07}.}

\emph{(ii)} Conversely, splitting \gv{two quasi-condensates after they have established phase coherence} may introduce quantum noise \cite{MenottiAnglingCiracZollerSplitting,Gring2012} related to the \emph{dynamical Casimir effect} \cite{carusotto2010density,PhysRevA.99.053615}.
The production of excitations in this process, \nn{especially in} a finite system would add \nn{an even larger} amount of energy.

\emph{(iii)} The individual elements are systems which feature \emph{correlations} extending over sizeable lengths and times \nn{compared to the size and operation time scales of the machine}, unlike in ordinary thermal machines.
Notably, even at thermal equilibrium a single quasi-condensate has a finite thermal coherence length $\lambda_T\neq 0$ \cite{Schweigler2017,Schweigler_thesis} which would not be true if one were to simply set the reduced Planck constant to zero $\hbar \rightarrow 0$ entirely disregarding quantum effects.

\emph{(iv)} \gv{Operating a valve in the quantum regime features} \emph{recurrences} during the evolution, \gv{an effect that has been also experimentally} observed in Ref.~\cite{Rauer2018}, \gv{and is one}
of the signatures of \emph{non-Markovianity}.
\gv{This, among other consequences, implies that the concatenation of cycles of the QFM depends on \nn{very precise timing of the individual} elementary operations (i.e., the QTPs)}.

\subsection*{Compressing and decompressing: A piston}
\label{sec:piston}

The defining feature of a piston is that its size can be changed, which, via the equation of state~\cite{Mora03,chen2019interaction,Jaramillo_2016}, leads to a change of internal energy.
\gv{Because of this,} the main role of the piston is that even if all the parts of the quantum field machine are in thermal equilibrium, one can introduce temperature differences by performing work upon the piston.
This, in combination with the valve, enables heat flow in the desired direction.
Again, if the physics of the piston involves quantum effects one can expect certain differences to ordinary thermal machines.
For example:

\emph{(i)} While the energy is changing due to compression or decompression the piston may go out of thermal equilibrium, e.g., due to \emph{squeezing} of internal modes~\cite{quantumreadout,PhysRevA.99.053615}. 

\emph{(ii)} Internal dynamics in the QFM elements occur within time-scales comparable to timings of individual steps of the cycles considered.
In contrast, in classical thermal machines \gv{concrete time scales are not comparable and hence usually discarded.}
 
\emph{(iii)} The piston \gv{essentially consists of} a moving boundary which is closely related to the \emph{dynamical Casimir effect}~\cite{PhysRevA.99.053615}.

\gv{To conclude this section, let us emphasize that these effects are particularly relevant also for practical applications. For example,
while the amount of energy injected in a local operation is intensive, its effect are substantial. All of these effects jointly influence the quantitative performance of the quantum field machine, as we also observe in the numerical study that follows. In particular, regarding a general discussion about the efficiency of a quantum field machine see also Sec.~\ref{sec:Conclusions_efficiency}.}

\section{Implementing quantum field machines in 1D Bose-Einstein quasi-condensates}
\label{sec:Implementation}

This section discusses the \gv{basics for implementing a quantum field machine on ultra-cold one-dimensional gases}. In Sec.~\ref{sec:1dQFT} we describe the microscopic model and the related \emph{effective Hamiltonian} defining the energy of phononic fields.
Sec.~\ref{sec:longcontrol_1dQFT} describes concisely role of the DMD in \emph{engineering} the desired QTPs, closely matching the experimental state-of-the-art \cite{Tajik2019}.
Finally, 
we discuss various diagnostic methods in Sec.~\ref{sec:QuntumFieldMachine_readout}.

\subsection{Effective quantum field theory description of 1D cold atoms}
\label{sec:1dQFT}

Cold atomic gases at low temperatures and with a fixed average number of atoms are effectively one dimensional if the trap anisotropies are sufficiently large to constrain the dynamics in two (transversal) dimensions such that the dynamics effectively takes place in the remaining (longitudinal) direction \cite{Petrov00,Giamarchi2004}.
In this regime, 
the system is well described by the Lieb-Liniger Hamiltonian, which reads
\begin{align}\label{eq:HfullBEC}
\hat H_\text{LL} = \hspace{-3pt} \int \hspace{-2pt} \de z \hat \Psi^\dagger \biggl[ &\frac{-\hbar^2}{2m}\partial_z^2  + V(z,t) - \mu +\frac{g}2  \hat \Psi^\dagger \hat \Psi  \biggr] \hat \Psi \ .
\end{align}
Here $\hat \Psi(z)$ is the atomic annihilation operator at spatial position $z$ which satisfies bosonic exchange statistics $[\hat \Psi(z),\hat \Psi^\dagger(z^\prime)] = \delta(z-z^\prime)$.
The atomic mass is denoted by $m$ and $\hbar$ is the reduced Planck constant. 
The external potential  $V(z,t)$ is responsible for longitudinal trapping of the gas but \gv{can be also used} as a means 
of implementing the necessary control operations for the machine.
The quartic interaction has strength $g/2$ which is proportional to the scattering length of the atoms, and also depends on other characteristics of the trap, specific of the experimental implementation \cite{Rauer2018}.
Finally, $\mu$ is the chemical potential that can be fixed, e.g., by constraining the average number of atoms $N_\text{atoms}$.
\gv{In a semi-classical theory of such gases, the study of its evolution is constrained to the set of coherent states, thereby 
approximating the field operators by a classical wave function, that obeys the so-called {\it Gross-Pitaevskii (GP)} equation.

The variational ground state atomic density calculated from the GP equation, which we denote as $\nGP(z)$, has the interpretation of the mean-density profile that can be measured by in-situ density absorption \cite{Schweigler_thesis,Tajik2019} (see Eq.~\eqref{eq:rho_0z}, Appendix \ref{sec:app_readout}). 
By expressing the field operators in the polar decomposition,
\begin{equation}
\hat \Psi(z)= \sqrt{
{ \nGP(z) \hat \id +\dd(z)}}~
e^{\i\pp(z)},    
\end{equation}
the GP equation translates into a system of {\it hydrodynamic equations} of a superfluid in the density-phase variables~\cite{Cazalilla11}
\be
\begin{gathered}\label{eq:hydrogen2}
\partial_t \nGP + \partial_z (\nGP v) = 0 , \\
\partial_t v + v \partial_z v = - \tfrac{1}{m} \partial_z \left( V  -\mu + \tfrac 1 {\nGP} P  + Q \right),
\end{gathered}
\ee 
where we \je{have} defined $v(z,t)=\hbar \partial_z \varphi(z,t)/m$ as the fluid velocity, and the terms $P$ and $Q$ are referred to as the pressure and quantum pressure term respectively,
\begin{equation}
P=g\nGP^2/2, \quad Q=-\tfrac{\hbar^2}{2m\sqrt{\nGP}} \partial_z^2 \sqrt{\nGP}.
\end{equation}
Neglecting the last term $Q$, one obtains a set of Euler equations describing the flow of a non-viscous fluid with equation of state $P(\rho_0)=g\nGP^2/2$. This is a semi-classical approximation to our system.
In this approximation the gas has, at zero temperature, 
energy density $e(\nGP)=g\nGP^2/2$ and chemical potential $\mu(\nGP)=g\nGP$.
Such an approximation is nevertheless insufficient to capture all quantum effects we \je{aim at studying}.
Therefore, we employ a fully quantum treatment of the (linearised) evolution.

For} inhomogeneous systems, \nn{namely $\rho_0(z)\neq {\rm const.}$}, the \gv{model} cannot be solved exactly due to the quartic term.
However, it is well known that a quadratic approximation in the spirit of the \gv{Bogoliubov} theory captures low-energy excitations 
\cite{Mora03,Cazalilla} and works
very well for certain time-scales \cite{quantumreadout}.
The effective model is obtained by 
expanding the Hamiltonian up to second order in the density $\dd(z)$ and phase $\pp(z)$ fluctuation operators, which are again bosonic $[\dd(z),\pp(z^\prime)]= \i\delta(z-z^\prime) \hat \id$.
They represent phononic excitations of a cold atomic gas and their energy is given by the following 
\emph{effective phononic Hamiltonian}
\begin{equation}
\begin{split}
\hat H_\text{P}[\nGP]=\int
\di z \biggl[&\frac{\hbar^2 \nGP(z)}{2m}\left(\partial_z\pp(z)\right)^2+\frac g 2 \dd^2(z)  \biggr]\ \ ,\label{eq:H_quench}
\end{split}
\end{equation}
which can be decoupled in normal phononic modes.
An important feature of this model is that wave-packets travel with a speed of sound related to the mean density $c=\sqrt{g \nGP/m}$.

The model in Eq.~\eqref{eq:H_quench} provides a 
good effective description for experiments performed on an isolated quasi-condensate~\cite{Langen2013, Langen2015,Schweigler2017,Rauer2018,yangTLL}.
However, in our simulations the QFM couples its initially isolated elements.
Then, one has to additionally model what happens with the \emph{phase zero-modes} in the systems.
A phase zero-mode has the interpretation of the total momentum frame of the excitations. For an isolated system,
\gv{this mode allows for phase fluctuations without an energy cost}~\cite{lewenstein1996quantum,JavanainenWilkens,LeggettComment,JavanainenReply}.
However, when two thermal systems, each with their individual zero-mode, are coupled, the two zero-modes hybridize to form the joint zero-mode and one mode \gv{with fluctuations} that cost energy.
The energy cost can be large if the original phase
 zero-modes are non-trivially populated, since the phase difference
 of two independent systems is fully random. 
Nevertheless, this is different in the physical system where the energy \gv{changes continuously.} \nn{This can be described when considering a more refined modelling using the full Hamiltonian} \eqref{eq:HfullBEC}, which would dynamically induce \emph{phase-locking} between the two condensates during the process.
Via the large coupling expansion of $\hat H_\text{LL}$, or arguing phenomenologically, an effective model can be derived that reads
\begin{align}\label{eq:HamwithJmain}
    \hat H[\nGP]=\hat H_\text{P}[\nGP]+  2\pi \hbar \int\di z J(z)\nGP(z)  \pp^2(z) ,
\end{align}
where the additional term regularizes the zero-modes.
\gv{In our main simulations}, we make the modelling simplification $J= \text{const}$, effectively gapping-out the phase zero-modes \nn{across the condensate} at all times.
The presence of this additional term can be interpreted as the quasi-condensates being merged having been already phase-locked prior to the merging.
The phase-locking term effectively induces squeezing of the modes which can be analytically seen in the homogeneous case. \gv{Quantitatively, in the numerical study that follows we used a small value $J=\SI{20}{\milli\hertz}$. Meanwhile, Appendix~\ref{app:zero_mode} contains a further discussion on using a more generic $J(z)$.}

\subsection{Controlling the 1D quantum field simulator using a DMD}
\label{sec:longcontrol_1dQFT}

\jsab{To achieve the QTPs described in Section \ref{sec:QuntumFieldMachine}, the
longitudinal trapping potential $V(z,t)$ has to be precisely
manipulated. For that, \gv{it is possible} to create a dipole trap (which adds to the
magnetic chip trap) by shining blue-detuned light on the atoms,
which creates a conservative repulsive potential \cite{dipole_trap}. By spatially
manipulating this light, \gv{one would be able to} nearly arbitrarily shape the
trap or add features to the existing magnetic trap. 

Using a device such as the DMD for this purpose is a standard
technique for many cold atoms experiments \cite{Aidelsburger2017Relax2d, dmd_ha_roton, dmd_Zupancic,EckelCampbellPhysRevX.8.021021, dmd_boshier} (see also \je{Ref.}
\cite{atomtronics_review} for a review). 
In our specific platform, we use a device with 1920 $\times$ 1080 (full
HD) micro-mirrors that can be turned on (sending light to the atoms)
or off (sending light outside of the optical path). The whole 2D array
of mirrors spans a spatial region which is $\sim 10$ times the size of the
BEC and each mirror in the DMD contributes with a gaussian
distribution of light, with a width of 0.4 $\mu$m in the plane of the atoms. 
This is roughly twice the healing length and several orders of magnitude smaller than the phase coherence
length.  Moreover, \gv{the fact that} the DMD used \gv{in our platform} has a
refresh rate of 32 $\mu$s (3 orders of magnitude faster than the time
scale of the atoms), allows the 1D potentials \gv{to effectively vary continuously} in time. 
In fact, in \je{Ref.}\ \cite{Tajik2019}, it \je{has been} demonstrated that different 1D potential landscapes can be implemented with a very high degree of control in this experimental setup.}  

It is also worth stressing that 
optimal control techniques can be used for the realization of the valve and piston QTPs in the experiment in a way maximizing the stability of the system.
In Refs.~\cite{rohringer2015non,Gritsev_2010}, it has been demonstrated that, for the case of compressing the gas in a harmonic trap, it is possible to find short-cuts to adiabacity.
In this case, a single control parameter has been suitably optimized, which has been the frequency of the longitudinal harmonic trapping potential.
This has allowed to expand the gas without introducing longitudinal breathing of the mean density which hints that optimal control should also be important for implementing a piston using a DMD potential.
Similarly, for the valve it is important to switch on the coupling between the two systems, without introducing stray excitations into the system which again can be optimized by appropriately tailored time-dependent potentials using the DMD.
\mg{Performing optimal control of the elements of the QFM will have to take into account that a very fast manipulation of the cold-atomic gas can enter into a supersonic regime which leads to exciting physical effects that have been explored experimentally for the expansion of the gas \cite{EckelCampbellPhysRevX.8.021021} which is 
important for the piston and for local manipulation of the gas \cite{Wang_2015,dmd_kumar_2020,eckel2014hysteresis}
which is relevant for the valve.
}
 \subsection{Space and time resolved monitoring of thermodynamic transformations}
 \label{sec:QuntumFieldMachine_readout}

In order to monitor the operation of a quantum thermal machine, observables that reveal local and global information about the state of the system are needed.
Of special interest are for example atomic density, spectrum and occupation of excitations, or their coherences and correlations. 
These physical observables allow to monitor and understand the details of thermodynamic processes, such as heat or entropy flow during the operations and the global thermodynamic properties for the qualitative analysis.

There are several well established methods to probe 1D quantum systems.  
These range from in-situ measurements of density fluctuations \cite{PhysRevA.98.043604,PhysRevLett.116.050402,PhysRevLett.105.230402,PhysRevLett.96.130403,PhysRevLett.106.230405} to measuring phase fluctuations in time of flight by either ``density ripples'' \cite{Imambekov09, Manz11} or interference \cite{Schumm05,van2018projective}.  Information is extracted by analyzing the full distribution functions \cite{Hofferberth08} or correlation functions \cite{Langen2013, Langen2015,Schweigler2017,Schweigler19}.
It will be crucial to use these measurement methods to extract information about local properties of the system.
This detects the action of local control when implementing the envisioned operations and resolving the thermodynamic transformations occurring in the elements of the QFM.
Of specific interest, when probing the quantum thermodynamic processes, is the (local) occupations of excitations of the quantum fields, \gv{i.e., of the} phonons. 
We first observe that the energy of the phonons in the system is defined as the expectation value of the quadratic Hamiltonian \eqref{eq:H_quench}. \gv{Note that the coupling coefficient in the additional term in \eqref{eq:HamwithJmain} is chosen precisely such that its overall contribution to the energy is negligible and it renders negligible also the contribution of the zero modes while merging two systems. Thus, by integrating Eq.~\eqref{eq:H_quench} over the length of the condensate one would obtain the total energy of the system. On the other hand,}
access to the local phase-phase fluctuations
\begin{align}
C^{\phi\phi}(z,z^\prime) = \langle \pp(z)\pp(z^\prime) \rangle 
\end{align}
and to the second moments of local density fluctuations
\begin{align}
 C^{\rho\rho}(z,z^\prime) = \langle \dd(z)\dd(z^\prime)\rangle 
    \label{eq:dens_fluct}
\end{align}
also directly implies the knowledge of the local energy density, which is given by
\begin{align}
    \frac {\mathrm{d}E(z)}{\mathrm{d}z} = \frac{\hbar^2 \nGP(z)}{2m}\partial_{z_1}\partial_{z_2}{C^{\phi\phi}}\big|_{z_1=z_2=z}+\frac g 2 C^{\rho\rho}(z,z)\ .
    \label{eq:E_z}
\end{align}
Note that the cross-correlations between phase and density degrees of freedom 
  \begin{align}
    C^{\phi\rho}(z,z^\prime) = \langle \pp(z)\dd(z^\prime)\rangle 
    \label{eq:dens_fluct}
\end{align}
do not contribute to energy and vanish in thermal equilibrium, though may be non-zero during out-of-equilibrium dynamics.
At this point, two comments are in order.

\emph{(i)} The expression of the local energy \eqref{eq:E_z} needs to be regularized due to divergences at the point $z_1=z_2$. This is accounted for by considering a UV cut-off in the corresponding field theory, in order for the energy in the system to be finite.

\emph{(ii)} The UV cut-off emerges naturally in the experiment. This is due to its finite imaging resolution and effects of ``smearing'' in time of flight \cite{van2018projective}; therefore, one can only measure a coarse-grained expectation value of the fields averaged over a finite length scale $\sigma_{\rm res}$, and higher momentum modes cannot be detected.

The gradient of the phase operator $\hat v = \partial_z \pp$ can be interpreted as velocity of wave-packets traveling on top of the condensate \gv{(as per hydrodynamic description)}.
Thus, the first term in the Hamiltonian \eqref{eq:H_quench} can be thought of as the energy content related to the speed of wave-packets, while the other term to how much distortion to the local density they induce.
It is important to note that both contributions must be measured in order to have the complete information about the energy in the system.
As mentioned earlier, on the Atom Chip platform, it is possible to measure experimentally by observing the quasi-condensate \textit{in situ} transversely (from the side) by means of density absorption \cite{PhysRevA.98.043604,PhysRevLett.116.050402,PhysRevLett.105.230402,PhysRevLett.96.130403,PhysRevLett.106.230405}.
In Appendix~\ref{sec:app_readout}, we discuss how measurements of the local density fluctuations of the atomic gas 
gives access to direct measurement of the Gross-Pitaevskii profile $\nGP$ and the second moments of the density fluctuations $\Gamma^{\rho\rho}$. 
Additionally, in Appendix \ref{sec:app_readout} we describe a proposal for a tomographic reconstruction method similar to Ref.~\cite{quantumreadout}; based on out-of-equilibrium data of $\Gamma^{\rho\rho}(t)$ at different times $t$, one can recover $\Gamma^{\phi\phi}$.
This then provides access to the second moments of phase fluctuations and hence the energy in the phase sector can be extracted. 

Alternatively, one can envision interfering the system under study with a local oscillator (a large 3D BEC)  \cite{Aidelsburger2017Relax2d} or with an identical system \cite{Langen2013, Langen2015,van2018projective} to extract the local phase correlations $C^{\phi\phi}$. From them one can tomographically reconstruct correlations of density fluctuations $C^{\rho\rho}$~\cite{quantumreadout}.
If one can assume thermal equilibrium, then it is possible to extract the occupation numbers of phonons even from $C^{\phi\phi}$ alone \cite{Langen2015}.
Global parameters like temperature can then be obtained also by ``density ripples'' \cite{Imambekov09, Manz11}.
The temperature is typically extracted by means of an appropriate fit to the correlations of the fluctuations of the atoms after a time-of-flight expansion.

It is important to understand which thermodynamic transformations have a substantial effect that is clearly detectable in the experiment.
The precision for measuring the (changes) in temperature or energy in the system will depend on reliability of the state preparation and the statistical sample size.
We anticipate that changes of temperature or energy by about $10\%$ should be large enough to obtain conclusive experimental results \cite{Rauer2018} ($> 5\sigma$) where one can be confident about, e.g., observing heat flow or cooling in a given system.

\begin{figure*}
      \includegraphics[trim = 3.5cm 7cm 2.7cm 7cm, clip, width = 0.99\linewidth]{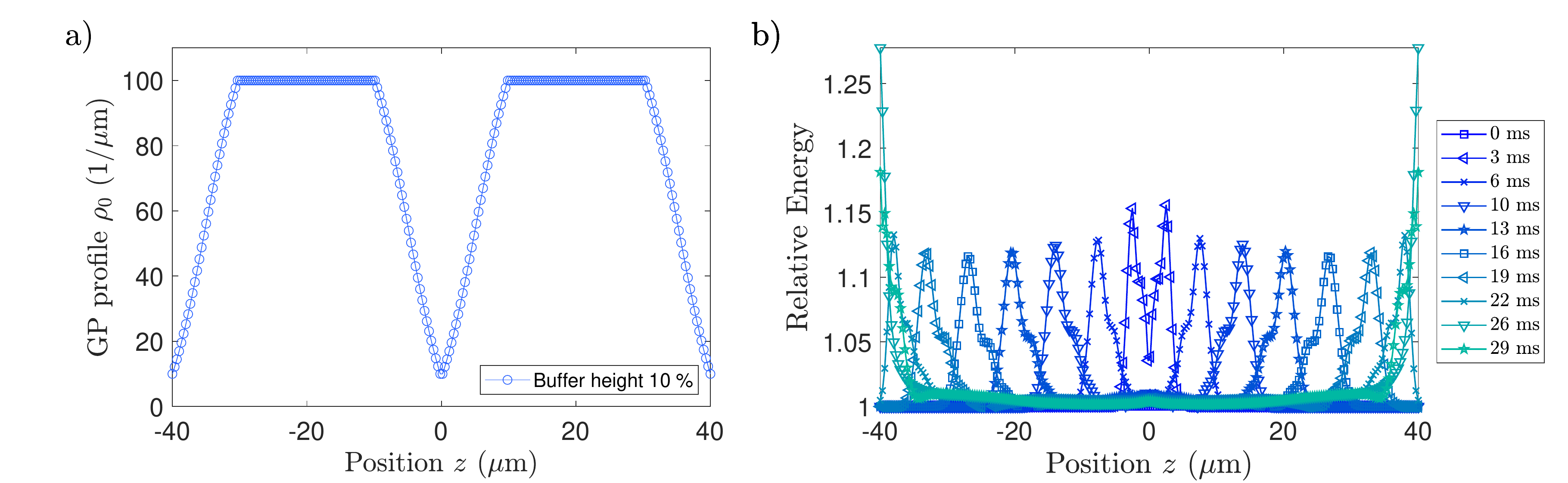}
\label{fig:sub1_corrs}
    \caption{{Operating a valve between two identical and independent thermal quasi-condensates.}
\emph{(a): Gross-Pitaevskii profiles.} 
        We consider two quasi-condensates which are homogeneous in the bulk but their density falls off towards zero at their edges.
        At position $z=0$ there is initially the boundary condition that in our effective model at a single point implements the separation between the two systems.
        As the systems become coupled the energy can tunnel between the two systems through this point. Throughout on line plots of real-space quantities bullets indicate the discretization lattice used in the simulation while the continuous lines are merely a guide to the eye.
    \emph{(b): Dynamics of energy density.} We plot $\di E(z) / \di z$ defined in Eq.~\eqref{eq:E_z} for different times during the coupling of two quasi-condensates.
    Initially, the energy density in each quasi-condensate is uniform, and we use that value to normalize the plotted values. During the coupling, localized energy is injected at the interface of the two systems and travels ballistically away in form of wave-packets, which increase the energy density by $\approx 15\%$.}
    \label{fig:merge_equal_density}
\end{figure*}

\section{Numerical studies of quantum thermodynamic primitives}
\label{Sec:three}

As sketched in Fig.~\ref{fig:ThermalMachines} above, \gv{the piston and the valve are building blocks that allow to construct} a refrigeration cycle.
In this section, we present results on the numerical modelling of the individual quantum thermodynamic primitives involved.

Each QTP that we propose is modelled by a Hamiltonian of the form \eqref{eq:HamwithJmain} described in the previous section, which allows 
us to simulate the dynamics of phonons and to calculate corresponding energy changes in the system.
As the model is quadratic, our simulations are done within the Gaussian framework and are computationally efficient. \nn{Moreover, this description allows us to efficiently evaluate information-theoretic entropies of such systems (e.g. relative entropy) which are relevant for thermodynamics of finite-sized quantum systems.}

Our model allows us to derive core predictions in the framework \nn{described in Section \ref{sec:1dQFT}.} In our simulations we use parameters that fit state-of-the-art experiments of 1D quasi-condensates performed on the Atom Chip platform.
More generally, our proposal \nn{is embedded in} the broader framework of thermodynamics with multi-mode Gaussian states, with Gaussian operations modelling the action of external system control.

\subsection{Coupling and decoupling two quasi-condensates: A valve}\label{subsec:valve}

Adjusting the external potential makes it possible to split the gas in two parts or merge at will \cite{MenottiAnglingCiracZollerSplitting}. We then study energy and correlation changes during the merging process.
A simple model is considered, where two quasi-condensates are coupled via a small buffer region.
Specifically, we consider a bipartite system, with each part $A$ and $B$ initially thermal and \nn{approximately} homogeneous, the two parts being separated by a buffer region of negligible size $\ell\sim \xi_\text{h}$ so that phonons cannot tunnel.
The Hamiltonian in Eq.~\eqref{eq:HamwithJmain} is specified by the Gross-Pitaevskii profile, which we choose with a shape according to Fig.~\ref{fig:merge_equal_density} (a).
Lastly, we specify Neumann boundary conditions (NBCs) at the edges. The density profiles that we choose have precisely the scope of smoothening further the boundary conditions in our implementation via a discretized lattice model. 

Denoting $\nGP^A$ and $\nGP^B$ as Gross-Pitaevskii profiles of parts $A$ and $B$ respectively, the initial Hamiltonian of the full system reads
\begin{equation}
\hat H_{A|B} =\hat H[\nGP^A]+\hat H[\nGP^B]\ ,
\end{equation}
where the tiny separation at the interface is modelled by the Hamiltonian $\hat H_{A|B}$ having in total 4 NBCs, two at the edges and two in the middle.
Next, we define the joint system to have a
 profile 
\begin{align}
 \nGP^{AB}(z)=
  \begin{cases}
  \nGP^A(z) & z\in A ,\\
    \nGP^B(z) & z\in B \ ,
  \end{cases}
\end{align}
implementing the ``gluing'' of the profiles.
Thus, in our minimal modelling  approach, we neglect the precise spatial details of experimental control necessary to switch from two independent systems to the coupled case as we assume that they are close and only a microscopic change is necessary for removing the small buffer region.
With that, we can take the final Hamiltonian of the merged systems to be
\begin{equation}
\hat H_{AB} = \hat H[\nGP^{AB}]= \hat H_{A|B}+\hat H_{\rm int}.
\end{equation}
This joint Hamiltonian has only 2 NBCs, and there is an interaction $\hat H_{\rm int}$ between $A$ and $B$. 
Due to this coupling, the thermal state of $\hat H_{AB}$, in contrast with that of $\hat H_{A|B}$, contains correlations between $A$ and $B$.

Note that \gv{during this evolution the boundary conditions at the interface change dynamically.}
We handle this boundary condition issue by interpolating linearly between the uncoupled Hamiltonian with 4 NBCs and the coupled Hamiltonian with 2 NBCs.
Thus, we model the time-resolved dynamics of the merging protocol by the time-dependent Hamiltonian
\def \Tm{t_\text{merge}}
\be
\begin{aligned}
  \hat H_{A-B}(t) &= \left(1-\frac t\Tm\right)  \hat H_{A|B} + \frac t \Tm \hat H_{AB} \ ,
\end{aligned}
\ee
within $t\in [0,t_{\rm merge}]$.
Here we model the situation that the change in the external potential makes the density profiles become smoothly interpolated.
Note also that we perform a lattice discretization to compute the physical quantities of interest (see Appendix \ref{app:QTPsimdetails}) and in this framework mixing boundary conditions is well-defined. 

\nn{Subsequently, we consider two independent thermal quasi-condensates, and adapt initial conditions that are natural for experiments, where evaporative cooling yields a thermal distribution of phonons with temperatures $T_A=T_B=\SI{50}{\nano\kelvin}$ at initial time $t=0$.}
Thermal states are defined with respect to a given Hamiltonian $\hat H$, and the density matrix reads
\begin{equation}
   \hat \gamma_T[\hat H] := \mathcal Z^{-1} e^{-\hat H/(k_B T)},
\end{equation}
where $\mathcal Z =\trace(e^{-\hat H/(k_B T)})$ is the partition function and $k_B$ is the Boltzmann constant.
We use Gross-Pitaevskii profiles with peak density $\nGP^A=\nGP^B= 100$ \SI{}{atoms / \micro\meter}, smoothly falling off towards smaller values at the edges, see Fig.~\ref{fig:merge_equal_density}~(a).
These choices reflect typical experiments realized in a box trap of size $L=\SI{50}{\micro\meter}$ with $N_\text{at}=5000$.
The fall-off at the edges according to the erf-function has been chosen phenomenologically 
- any trap that is not infinitely strong will lead to a smooth fall-off at the edges.

In Fig.~\ref{fig:merge_equal_density} (b) we show numerical results for a  linear ramp with merging time $\Tm= \SI{40}{\milli\second}$.
This is a relatively long time-scale,  chosen to demonstrate that excitations can be reflected at the edge and start returning towards the interface.
Initially, energy is distributed homogeneously in $A$ and $B$ so we present the energy distribution relative to that value.
This relative measure is employed throughout, since it allows us to disregard the cut-off dependent shift coming from zero-point fluctuations. In fact, our effective Hamiltonian is not normal-ordered but instead regularized by the healing length $\xi_\text{h}=\hbar /(mc)$ of the system (note that the cutoff $\Delta z$ in our numerical simulations is \nn{smaller} than the healing length).
As anticipated, merging two systems via tunnel coupling induces excitations in form of counter-propagating wave-packets\mg{, see Ref.~\cite{Wang_2015} for a detailed experimental and theoretical study of the dynamics of such excitations}. The wave-packets travel with the respective speed of sound, which in typical experiments on the Atom Chip platform is $c\approx\SI{2}{\micro\meter/ \milli\second}$ \cite{Rauer2018}.
The simulation predicts that the wave-packets increase the local energy by a sizeable amount of $\sim 15\%$.
This may cause system dynamics to deviate from the \gv{linearized approximation}. Nevertheless, the higher-order terms 
should have only the effect of dispersing the wave-packets.
According to our simulations, the amount of injected excitations is higher if systems are coupled at peak density, see Appendix \ref{app:Mergingdetails}.
This is because in the lattice approximation we are adding an off-diagonal coupling between the two edges of $A$ and $B$ that scales \nn{$\propto \nGP(z=0)$} with density. Therefore, merging is ``softer'' if it occurs at a lower density value.
Physically speaking, it is more stable to couple two sensitive systems harbouring gapless excitations through diluted regions compared to at peak density.
\begin{figure}
    \centering
    \includegraphics[trim = 2.7cm 2cm 3cm 2.4cm, width=\linewidth]{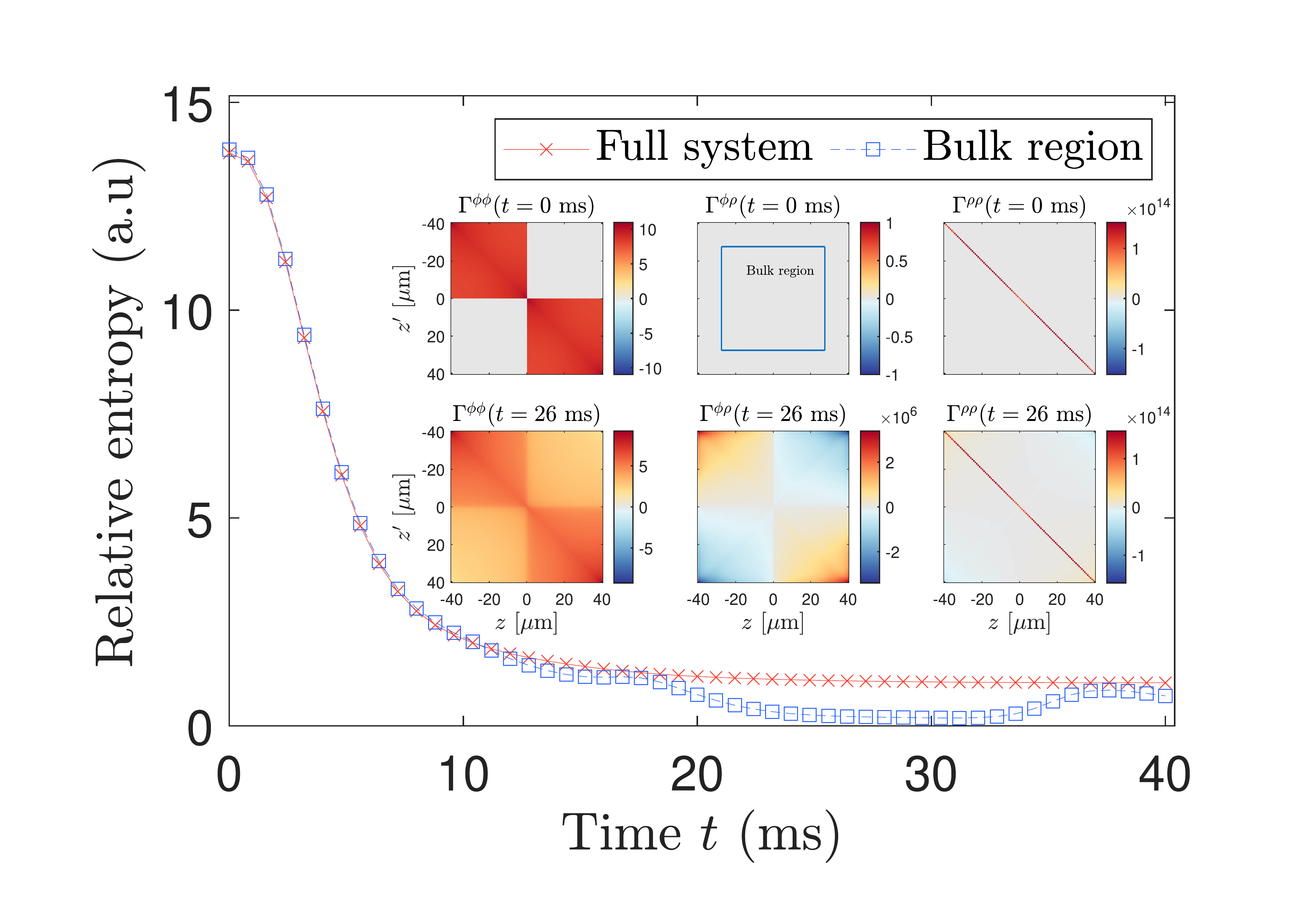}
    
    \caption{{Correlations before and after merging.}
    The initial covariance matrix 
    $\Gamma(t=\SI{0}{\milli\second})$ (inset top) is 
    characterized by phase fluctuations 
    $\Gamma^{\phi\phi}=2C^{\phi\phi}$ ranging only over the individual 
    systems, no cross correlations between phase and 
    density operators  $\Gamma^{\phi\rho}\equiv 0$, and 
    density fluctuations $\Gamma^{\rho\rho}=2C^{\rho\rho}$ being 
    essentially diagonal. When heat excitations reach 
    the edges, the covariance matrix 
    $\Gamma(t=\SI{26}{\milli\second})$ (inset bottom) 
    restricted to the bulk region of the system agrees 
    with the thermal covariance 
    matrix of the joint 
    Hamiltonian: phase fluctuations $\Gamma^{\phi\phi}$ 
    become uniform over the joint system, in the bulk 
    of the system cross correlations vanish 
    $\Gamma^{\phi\rho}\approx 0$, while density 
    fluctuations  $\Gamma^{\rho\rho}$ are diagonal. 
    Quantitatively, we plot the relative entropy of the time-dependent covariance matrix with respect to that of the thermal state of the coupled Hamiltonian at $T=\SI{50}{\nano\kelvin}$ and observe that it decreases rapidly over around \SI{10}{\milli\second}.
   Due to the presence of the heat wave-packets, the relative entropy for the full system (red crosses) does not converge to zero over time, while for the covariance matrix restricted to the bulk region (blue squares) essentially vanishes at around $t=\SI{26}{\milli\second}$ and then increases again. }
\label{fig:Relent}
\end{figure}

It is instructive to analyze the correlations of the coupled state during the merging.
As shown in Fig.~\ref{fig:Relent}, we find that initially there are no correlations between $A$ and $B$ and hence we see that two independent thermal quasi-condensates are not thermal with respect to the joint Hamiltonian.
During merging, the parts become coupled and the established correlations drive the state towards being close to the joint thermal state, see Appendix \ref{app:Mergingdetails} for more details.
Interestingly, after the first traversal time, i.e., when a local excitation at the merging interface has traveled to the edges, the joint system is already close to being thermal in the bulk (cf. inset of  Fig.~\ref{fig:Relent}).

The observation that the merged parts become jointly thermal can be further 
quantified by evaluating the \emph{relative entropy}, given for any two states by
$S(\hat \gamma \| \hat \sigma) = \trace(\hat \gamma(\log \hat \gamma - \log \hat \sigma))$. 
Evaluating this with respect to a thermal state yields 
\begin{equation}
    S(\hat \gamma \| \hat \gamma_T[\hat H]) := \left(F(\hat \gamma)-F(\hat \gamma_T[\hat H])\right)/(k_B T) \ge 0,
\end{equation}
where $F(\hat \sigma)= \trace(\hat H \hat \sigma) - k_B T S(\hat \sigma)$ is the {\it free energy} of the state relative to the ambient temperature $T$ and the Hamiltonian $\hat H$. Here $S(\hat \varrho)=-\trace(\hat \varrho \log \hat \varrho)$
is the von Neumann entropy.
Notably, the relative entropy is zero if and only if the two covariance matrices are the same (see Appendix~\ref{app:GaussianApp} for further details).
This makes it a \nn{strong} measure of deviation from thermal equilibrium. 
Finally, this measure can be computed also for \emph{reduced} density matrices which then captures how systems are similar \emph{locally}.

In order to check if the merging QTP is intensive we calculate the relative entropy of the state evolving during merging with respect to the thermal state of the coupled Hamiltonian at $T=\SI{50}{\nano\kelvin}$.
Initially, the relative entropy decreases rapidly, reflecting the ongoing thermalization around the interface of the two systems, where the correlations are being established.
For the whole system the relative entropy does not reach zero and levels off to a constant value within about \SI{10}{\milli\second}. 
This is due to the wave-packets being always present in the system, hence the impossibility for the entire system to be in thermal equilibrium.
If we consider the reduced covariance matrix describing only the bulk middle region, we see that around \SI{20}{\milli\second} the relative entropy drops essentially to zero.
This means that once the excitations leave the window of observation, the system left behind agrees in that region with the (joint) thermal state.
Finally, for longer times the wave-packets come back to the bulk and allow for detecting an out-of-equilibrium component of the state.

\gv{We expect that features observed in this numerical study should remain true even under perturbations to the model and thus that temperature for locally merged systems is intensive is a generic feature of this QTP.
This is because perturbations are not expected to change the character of low-energy excitations so the spectrum should remain approximately linear and a local change of the Hamiltonian should generically create a localized surplus of energy propagating through the system with the speed of sound.
}

In the case presented here, we have shown an example where there has been no net heat flow between two systems.
The next section shows how to enable heat flow between two systems, by performing work from outside, thereby creating an effective temperature difference.
\gv{As an outlook, in Sec.~\ref{sec:corrandanomalous} we also present the case of two initially different temperatures, 
observing that non-Markovianity effects in this case are even more pronounced and can potentially lead to further 
\nn{interesting} effects, such as \emph{anomalous heat-flow}. }

\subsection{Compressing and decompressing: A piston}

In this subsection, 
we see how external control \gv{which compresses or expands the gas, enables a \nn{condensate} to function as a piston.
The external control will effectively} perform work on the quasi-condensate, increasing or decreasing its energy depending on the change in volume. This is similar to thermodynamics of an ideal gas with the difference that we are considering a quantum many-body system. 
\gv{Experimentally,} operations for this QTP have already been implemented with use of shortcuts to adiabacity (see Ref.~\cite{rohringer2015non} where the extension of the  profile has been stably modified).
\begin{figure*}[t]
    \centering
    \includegraphics[trim = 7cm 7cm 5cm 6cm, clip,  width=0.99\linewidth]{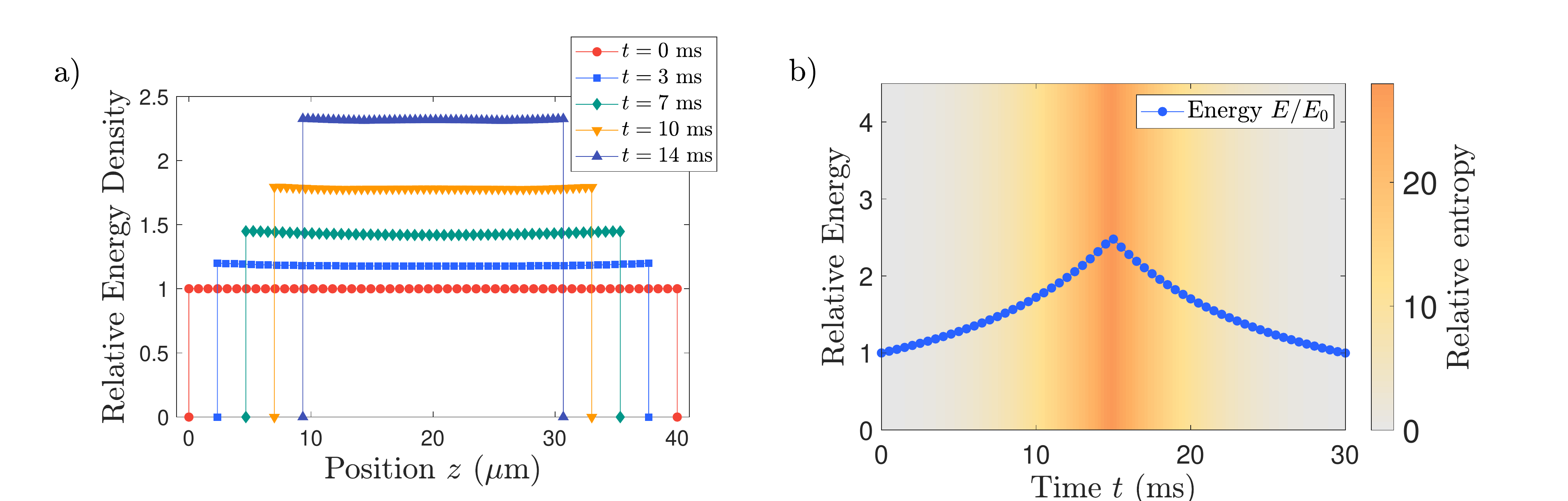}
    
    \caption{{Single stroke of a piston.} A condensate of length $\SI{40}{\micro\meter}$, initially thermal and homogeneous is compressed to half length within $\SI{15}{\milli\second}$ and then re-expanded to its initial length in the same time. \emph{(a): Energy density during compression}. The piston keeps a homogeneous energy density that increases when compressed due to increasing pressure of the gas.
    \emph{(b): (Non-)equilibrium properties of piston}. We plot over time the total energy relative to its initial value (blue dots) and the relative entropy to the closest thermal state as a colour gradient during the compression and decompression.
    The piston goes out of equilibrium, as the relative entropy to the closest thermal state increases during compression.
    The reverse happens when the piston decompresses -- it is again fully thermal at the initial energy and temperature at the end.}

    \label{fig:piston}
\end{figure*}

\begin{figure*}[t]
\centering
  \centering
  \includegraphics[trim = 2.5cm 7cm 3.5cm 6cm, clip, width=0.99\linewidth]{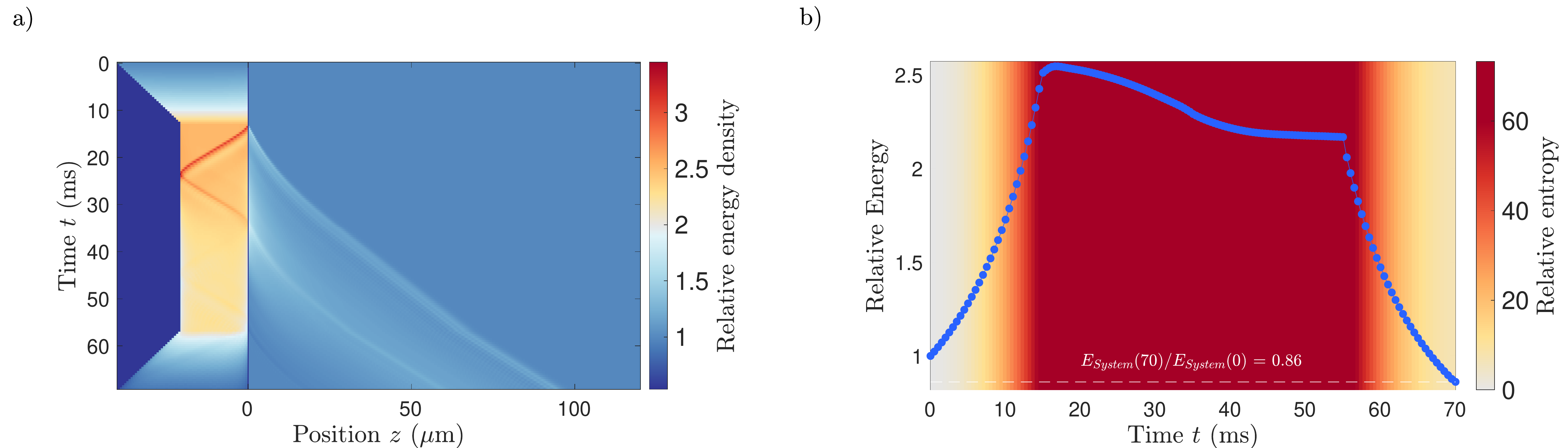}
  \caption{{Heat flow between the piston and bath.} We consider the piston and bath being two initially independent condensates, with the bath being three times larger in size. The piston on the left is compressed to half of its original size. We then couple the piston to the bath at $t = \SI{15}{\milli\second}$ and start decoupling them right after. \emph{(a): Energy density over time.} We plot the energy as a colour gradient in a space-time grid. The coupling between the two parts introduces the propagation of wave-packets at the speed of sound, which is higher in the piston, due to the higher density resulting from  compression. \emph{(b): Energy dynamics in non-equilibrium.} We plot the ratio of average energy versus initial energy in the piston over time. We observe that it first increases strongly, while decreasing to a value that is less than 1,
  just before the piston starts coupling with our system of interest. This is what will allow  us to cool the system with a full Otto cycle. We also plot the relative entropy to the best fit thermal state as a colour gradient in the background and observe that during coupling the system goes strongly out of equilibrium, while returning to be close to equilibrium at the end.}\label{fig:couple_split_piston}
\end{figure*}
Here, we propose a model to describe what happens to phonons when the confining trap (space occupied by the gas) changes.
Let the length of a uniform system change continuously over time 
in the sense that a homogeneous Gross-Pitaevskii profile $\nGP$ with support of length $L$ changes to  $\nGP(t)$ with corresponding length $L(t)$.
The operation is assumed to preserve the atom number $N_\text{atoms} = \nGP L$ so that
\begin{align}
    \nGP(t) =  {\nGP} \frac {L(0)} {L(t)}\ .
    \end{align}
This time-dependent Gross-Pitaevskii profile assumes that the change in volume is slow so that a homogeneous system remains homogeneous at all times.
Under this assumption, the Hamiltonian \eqref{eq:HamwithJmain} parametrized by a time-dependent Gross-Pitaevskii profile $\nGP(t)$
    \begin{align}\label{eq:Ht_compression}
    \hat H(t) = \hat H[ \nGP(t)] 
\end{align} 
describes the phonons during the size change. 
Using Eq.~\eqref{eq:Ht_compression}, the integration in \eqref{eq:HamwithJmain}  ranges over the time-dependent length of the system $L(t)$.
In the lattice approximation this is implemented by discretizing the Hamiltonian at each time considered, and identifying the respective cells at consecutive times as they change only infinitesimally.
It is also possible to consider formulating the procedure using a fixed representation of momentum mode and time-dependent eigenmode wave-functions~\cite{PhysRevA.99.053615}. 

In the homogeneous case by a change of the integration variable we can write the time-dependent Hamiltonian as 
\mg{
\begin{equation}
\hat H(t)
=\hspace{-0.1cm} \int_0^{L(0)} \hspace{-0.3cm} \di z \biggl[\frac{\hbar^2 \lambda^2(t) \nGP}{2m}\left(\partial_{z}\pp\right)^2+\frac g 2 \lambda(t)  \delta \hat \nu^2 + h J\nGP  \pp^2 \biggr]\ ,
\label{eq:H_quench2main}
\end{equation}
}
where we have also defined a rescaled density fluctuation field $\delta \hat  \nu = \dd /\lambda(t)$ in order to preserve the canonical commutation relations. I.e., this way we have $[\delta \hat \nu(z),\pp(z^\prime)]=\i \delta(z-z^\prime)$.
Here we made the integration limits explicit and changed the frame so that the length of the system is effectively constant but the Hamiltonian density becomes time-dependent due to the dimensionless ratio
\begin{align}
  \lambda(t)=\frac {L(0)} {L(t)}\ .
\end{align}
We observe that if the system stays homogeneous, then \gv{the time-dependent Hamiltonian \eqref{eq:H_quench2main} has the same momentum eigenmodes at all times $t$, but they} become squeezed.
We should hence expect that compressing introduces \emph{squeezing} of phase and density quadratures. \gv{See Appendix \ref{app:Compressiondetails} for an extended discussion, including the numerical implementation of the compression model and see also Ref.~\cite{PhysRevA.99.053615} for a related study.}

With this model we can simulate the functioning of a piston: In Fig.~\ref{fig:piston} we show the results of a simulation of a single stroke.
It is moreover possible to check whether the piston remains thermal during the process.
We first observe that the energy density stays homogeneously distributed at all times  (cf. Fig.~\ref{fig:piston} (a)). Moreover, it changes in relation to volume: as shown in Fig.~\ref{fig:piston} (b), the total energy increases and comes back to the initial value during the stroke of the piston.
Nevertheless, a more refined check involving the relative entropy shows that the system is not at thermal equilibrium at all times.
In particular, at a sequence of times during the evolution we evaluate the relative entropy between the time-dependent state and the thermal states corresponding to the system Hamiltonian.
The thermal state with the lowest relative entropy gives then the effective fit for the temperature.
It is clear that if the time-dependent state remains thermal at all times, then there will be a temperature for which the relative entropy vanishes.
However, we find that this value is strictly positive, which indicates that the piston is away from thermal equilibrium during the compression-decompression process, and returns to thermal equilibrium only when reaching its original length.
This effect can be naturally explained by the presence of squeezing in the system, but we focus here on the thermodynamic aspects of the model and refer to Ref.~\cite{PhysRevA.99.053615} for a discussion of the dynamical Casimir effect.

We can now use the compression QTP in order to enable heat flow between two systems.
In Fig.~\ref{fig:couple_split_piston}, 
we show the steps (1-2) of the Otto cycle that have been  sketched in Fig.~\ref{fig:ThermalMachines}, i.e., 
we compress the piston, couple it to the bath and after decoupling decompress it back to its initial state.
As before, piston and bath are initially both thermal. They also have the same overall shape of the Gross-Pitaevskii profile, with the only difference that the bath is larger than the piston.
As it has been shown above, coupling two systems with the same temperatures does not lead to heat flow.
However, after the piston is compressed its energy is higher and so is its effective temperature.
This creates an effective temperature difference between piston and bath which, \gv{using then the valve QTP, enables heat flow from the piston to the bath.}
After this heat flow is completed we close the valve and decompress the piston to its initial length, and note from Fig.~\ref{fig:couple_split_piston} (a) that it becomes colder than it has been initially.
 Fig.~\ref{fig:couple_split_piston} (a) shows the results of this protocol plotting the full spatio-temporal dynamics of energy density. In Fig.~\ref{fig:couple_split_piston} (b) we show that the compressed piston couples to the bath with effectively squeezed modes, so that the two systems are not at thermal equilibrium while the valve is open.
Nevertheless energy in the piston decreases, due to heat flowing into the bath which is seen in Fig.~\ref{fig:couple_split_piston} (a) in form of a light-colour stripe entering the bath.
Finally, we find that the total energy in the piston decreases to a lower value than initially, thus we conclude that the piston has been overall cooled down.
At the end of the protocol the decompression undoes the squeezing of the modes and the piston essentially comes back approximately to thermal equilibrium, signified by a low relative entropy to a thermal state.

Summarizing, we have performed work on the piston which therefore allowed us to enable heat flow between condensates. By composing the compression QTP with the open valve QTP, we demonstrate that is is possible to deposit some of the piston's energy into the bath.

\section{Composing quantum thermodynamic primitives to build  a quantum field refrigerator}
\label{sec:Otto}
\begin{figure*}
    \centering
\includegraphics[ trim = 1cm 1.5cm 0cm 2cm,clip, width=0.99\textwidth]{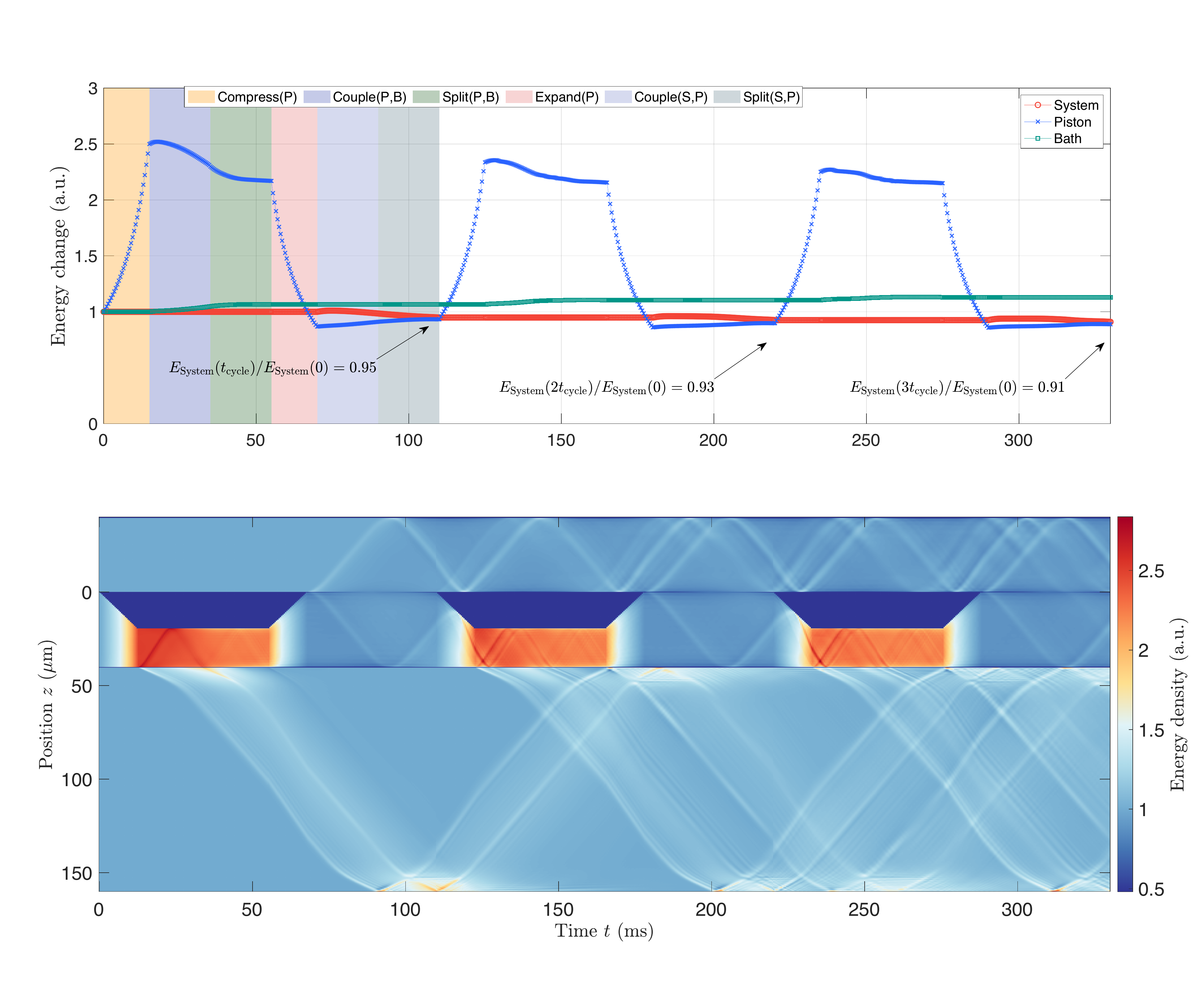}
\caption{
{\bf Top: Quantum field refrigerator.} 
The QFM is initialized in thermal equilibrium and equal density, i.e., the system, piston and bath only differ in length which is \SI{40},  \SI{40}, \SI{120}{\micro\meter} respectively.
We run the Otto cycle by compressing the piston (15 ms), depositing heat in the bath (40 ms) and then expanding the piston again (15 ms).
The cooling begins at around 70ms by coupling the initially thermal system to the cooled piston.
The systems exchange energy by the physical mechanism of the valve described in Section \ref{subsec:valve}.
After the final splitting of the system and piston, we find that the system cools down, while the quantum field refrigerator extracts $\sim$ 5\% of the system's initial energy.
This drop in energy is large enough to be detected by existing experimental read-out methods. In this plot, one observes that further cycles continue to contribute to cooling of the system, but only in very small amounts.
 The currently used parameters are probably non-optimal, and we anticipate improvements of the refrigeration efficiency via optimal control. This, however, will depend on the modelling of other details in the quantum simulation of this cycle.
 {\bf Bottom: Time and space resolved energy dynamics during the operation of the QFM.} From the top we show the system, piston (which changes in size) and bath.
 Whenever a valve QTP is operated, wavepackets are injected and multiple reflections in each system can occur.
The principal wavepacket in the bath is timed to arrive at the interface to the piston at around \SI{160}{\milli\second} when the valve is closing, so that the piston energy is not further increased. The overall amount of energy in the bath increases, which is due to the presence of multiple wave-packets.
 It is noteworthy that, depending on scheduling, the larger among two coupled systems can take up two wave-packets. Hence, considering the piston to be substantially larger than the system could allow to remove all excitations from the system.
 This is also why the bath takes up most of the wavepackets.
 }
\label{fig:Otto_energies}
\end{figure*}
\mg{The challenge one faces studying cold atomic gases experimentally is that all methods of cooling eventually always reach a limit once the temperature is small enough.
In the ultra-cold regime the last resort is to let some of the atoms escape the trap.
Ideally one would like that \nn{1) those particles leaving the system to be pre-selected such that they carry above-average energy, and 2) the gas left behind re-thermalizes~\cite{davis1995analytical}}.
These two ingredients make up evaporative cooling. However, in one-dimensional systems they cease to apply, due to the change in scattering properties~\cite{MazetsBreakdown,Breakdown2,andreev1980hydrodynamics,buchhold2015kinetic}.
Nevertheless, in Ref.~\cite{Rauer2016a} the effect of letting atoms escape by applying an additional rf-field has been explored in the 1D \nn{regime: at} extremely cold temperatures it has been demonstrated that cooling continues, and its limits have  also been  quantitatively mapped out.

The intensities and detuning of the applied rf-field ensures the energy-independent loss of atoms \cite{Rauer2016a}.
This observation has suggested a successful modelling approach to the process by the phononic Hamiltonian \eqref{eq:H_quench} whose density parameter $\nGP$ decreases over time according to the atom loss rate.
Within this model the energy gets decreased due to this change in the Hamiltonian.
By assuming that the uniform atom loss process is sufficiently slow, the dynamics of phononic modes 
has been used to theoretically explain why the system is left approximately in thermal equilibrium with a decreasing temperature \cite{Rauer2016a}, see Refs.~\cite{GrisinsUniformLoss,QuantumCarusotto,EvaporationQunatum} for additional discussions.

By an analytical treatment which assumed that 1) the atom loss process is adiabatically slow, 2) only phonons rather than particle-like excitations are involved and 3) shot-noise has a negligible contribution, this model leads to the relation 
\begin{align}
    T'/T\approx (\nGP'/\nGP)^{3/2} \ .
    \label{eq:scaling}
\end{align}
In harmonic confinement, the peak density is $\max (\nGP)\propto N_\text{atoms}^{2/3}$, which together with Eq.~\eqref{eq:scaling} yields qualitative agreement with experimental observations: the coldest temperature reached is found to depend linearly on the number of atoms, 
\begin{equation}\label{eq:fund_limit_cooling}
T\propto N_\text{atoms}.
\end{equation} 
Thus, the temperature of a condensate can be lowered by allowing for more atom losses; however, this dilutes the system and cannot be continued indefinitely, otherwise quasi-condensate properties will be lost \cite{Petrov00}.
Eq. \eqref{eq:scaling} is also valid for a box-like confinement, where the temperature dependence on atom number is expected to be non-linear.

Refs.~\cite{Rauer2016a,Schweigler_thesis} give representative values for cooling in a harmonic trap.
For state-of-the-art data reached with box-like confinement \cite{Rauer2018}, $N_\text{atoms}\approx 5000$ confined atoms can form an approximately homogeneous condensate of about $L=\SI{50}{\micro\meter}$, and the estimated temperature is $T\approx\SI{50}{\nano\kelvin}$~\cite{quantumreadout,Rauer2018}.
Summarizing, uniform atom losses do lead to cooling, but this does not follow the usual mechanism of re-thermalization via scattering as typically seen in evaporative cooling. Rather, this is a direct consequence of decreasing the density parameter in the phononic Hamiltonian.

In general, the effective barrier of Eq.~\eqref{eq:fund_limit_cooling} seems hard to overcome. 
The current achievable lowest temperature for a fixed prescribed density on the system is limited by the initial density and temperature of the gas accessible from previous stages of cooling (laser operated); and this initial density cannot be infinitely large, given the constraint of operating in the quasi-condensate regime.
Moreover, the evaporative cooling will eventually either exhaust the available atoms diluting the system (effectively leaving the quasi-condensate regime), or completely lose its efficiency in the sense that the evaporation has negligible cooling effect due to infinite thermalization time. \nn{One therefore requires novel cooling methods to overcome this impasse.}

\subsection{Cooling by escaping atoms within QTP framework}

In this section, we point out that cooling by uniform atom losses, which is the state-of-art cooling technique for one-dimensional gases in the lowest temperature regimes, can be conceptually captured in the QTP framework.
In particular, consider a sequential concatenation of a dilution of the system and possibly coupling to a cold bath. The continuous dilution that arises from atoms escaping the system can be conceptually modelled by the piston and valve QTPs. 
Indeed, if during cooling we have the gas of $N_\text{atoms}$ atoms uniformly occupying the interval of length $L$, then after one particle escapes the linear density will change according to 
\begin{equation}
    \nGP=\frac{N_\text{atoms}}{L} \quad \rightarrow \quad \nGP'=\frac{N_\text{atoms}-1}{L}.
\end{equation}
However, the same can be achieved by the system behaving like a piston of size $L$ expanding by $\Delta L$, such that the linear density changes according to 
\begin{equation}
   \nGP=\frac{N_\text{atoms}}{L} \quad \mapsto \quad\nGP''=\frac{N_\text{atoms}}{L+\Delta L} .
\end{equation}
To complete the description we can impose $\Delta L$ to be such that $\nGP''$ equals $\nGP'$.
Additionally, we imagine placing a valve to be positioned at $x=L$ at the edge of the piston, so that when it expands to $L+\Delta L$, atoms exit the valve into vacuum, and after we close it the remaining system has the same density $\nGP'$ and the length is $L$, exactly as in evaporative cooling.
In other words, in our modelling, the piston lowers the density by operating at constant particle number $N_\text{atoms}$ and varying length $L$; while in evaporative-like cooling, the change of density occurs at constant length $L$ but varying particle number $N_\text{atoms}$.
However, on the level of intensive thermodynamical quantities, both are the same and this observation is implemented by the fiducial valve shutting of the $\Delta L$ portion of the expanded piston.

As a remark, the QTP framework can be also used to capture evaporative cooling involving a re-thermalization process.
When evaporative cooling is most effective in its operation, only the atoms that individually carry above average energy leave the system.
This way of cooling is more efficient as each escaping atom carries on average more energy than an atom remaining in the system does. 
This can be modelled in the QTP framework by opening and closing a valve coupling the system to a cold bath.
The heat flux and timing jointly govern the exchange of energy between the system and bath; they should be chosen such that the lowering of the system's energy is the same as an evaporating atom would do.
}
\subsection{\mg{Cooling by atom number dilution with a subsequent recompression to restore atom density}}

\mg{
Cooling facilitated by atoms escaping the trap irreversibly dilutes the system.
As explained above lowering the temperature of  the atoms at a given density being fixed is the right way to compare different cooling approaches.
As anticipated in Fig.~\ref{fig:ThermalMachines} running refrigeration cycles like in a machine can be expected to lead to cooling without changing the atom density.
However it is also true that in a QFM as presented in Fig.~\ref{fig:ThermalMachines} there are two sub-systems, acting as the piston and bath, which are constituted by a sizeable amount of atoms.
The question then arises: Can any advantage be gained when aiming to cool at prescribed density in simply evaporating these systems?

While such a question is quite general, let us discuss it by formulating a representative protocol whose analysis will suggest an overall answer.
First of all, if the dilution has to have any effect on the system we must allow for contact with the sub-system that we 
\je{would like} to cool at constant density.
One way to achieve that is to consider the entire system (system, piston, 
and bath in Fig.~\ref{fig:ThermalMachines}) to be uniform, then cool it down by dilution implemented by the escaping atoms, and then use the piston QTP to compress the system back again to restore the density to the initial value.
The idea here is that the evaporation of the amount of atoms taken up by the piston and bath should lead to cooling and after the compression the system should have the prescribed density.

However, we can anticipate that this effect will not lead to overall cooling.
This is because the dilution cools down the system by reducing the density via the atom number but the compression heats the system up as should be in a gas and has been discussed in Fig.~\ref{fig:piston}.
Intuitively, on the phononic level this is seen by noticing that increasing the density,  implemented by reducing the volume of the system, changes the Hamiltonian which associates a larger energetic penalty to phase fluctuations.
In other words, we first  in a time-dependent fashion change the linear density to same value by reducing the numerator (atom number) in its definition and then increase the density back to the initial value by decreasing the denumerator (system length).
As long as we are in the phononic regime it does not matter which process changes the density -- the modelling will be the same and both processes, that is cooling by atom losses and recompression, admit the same modelling using the phononic Hamiltonian so one should expect that they are mutually complementary.

In Fig.~\ref{fig:piston} we show that a stroke by compression and recompression is effectively reversible in that the overall phononic energy returns to its initial level.
In the model it does not matter whether the density is changed by changing the atom number or the length of the system.
For this reason, we expect the reversibility of the phononic energy change to be also valid when combining the dilution process by uniform atom losses to cool down with the piston QTP to restore the density.
This can be verified in a future experiment to lay the ground for implementing a quantum field refrigerator based on the QFM involving the much more sophisticated approach using cycles and composing many QTPs together.

On the theoretical grounds supplemented by the empirical knowledge drawn from past experiments the case for refrigeration via the QFM seems to be clear: Reducing the entropy per particle in a sub-system of a cold atoms system should be achieved by moving this entropy to a bath as in a QFM.
Having said that, considering other interesting variants of combining processes such as atom losses and QTPs described here for problems of interest, cooling being one particular example, is available experimentally and can be further explored in the future.
As we will show next, if one \je{aims} to achieve cooling in a systematic way it is advisable to run QTP cycles in a QFM as illustrated in Fig.~\ref{fig:ThermalMachines}.

}
\subsection{\mg{Quantum field refrigerator: QTP cycles for sequential cooling and reduction of entropy of a subsystem}}
In this section, we demonstrate how to compose the discussed primitives to perform a useful protocol, namely cooling. 
By simulating the quantum field refrigeration machine depicted in Fig.~\ref{fig:ThermalMachines} at this density and temperature, we find a cooling cycle where the system  temperature decreases, highlighting the usefulness of such a new active cooling protocol. 
The cycle works as follows.

\emph{(1)} \mg{The machine is initialized by setting a system, a piston, and a bath to their respective thermal equilibria.}

\emph{(2)} The first non-trivial thermodynamic transformation is the compression of the piston with a subsequent interaction with the bath. 
The work inserted to compress the piston enables heat flow as shown above in Fig.~\ref{fig:couple_split_piston}.

\emph{(3)} After \mg{decoupling} the piston from the bath, the piston is  expanded back to its initial length.
This aims to cool it down and when it subsequently interacts with the system it should take up some heat from it.

\emph{(4)} 
Finally, the piston and system are \mg{decoupled} again and the cycle can be repeated.

In Fig.~\ref{fig:Otto_energies}, we depict the energy changes of these three pieces of the QFM over the duration of the Otto refrigeration protocol obtained from a numerical simulation
\footnote{Note also that experiments with nuclear magnetic
resonance have been performed realizing a
quantum Otto heat engine operating under a reservoir at effective negative temperatures \cite{PhysRevLett.122.240602}.}. 
It can be seen that the piston first increases its energy due to compression ($\Tc= \SI{15}{\milli\second}$) and then lowers it during interaction with the bath and successive expansion ($\Tm+\Tc= \SI{35}{\milli\second}$). Finally, the piston increases again its energy when interacting with the system and then resizing to its original length (again $\Tm+\Tc= \SI{35}{\milli\second}$). Overall, at the end of the first cycle ($\Tcy= \SI{110}{\milli\second}$), the piston has slightly decreased in energy, while system and bath have consistently decreased and increased their energy, respectively. 
By performing three Otto cycles, we obtained cooling of \gv{$9\%$ in a total time of $\SI{330}{\milli\second}$, which gives us an estimate of the {\it cooling power of our QFM}. However, we also observe that such cooling power actually decreases in subsequent cycles, thus raising the question of the ultimate limits of cooling for this machine. 
We discuss this interesting aspect further in 
Sec.~\ref{sec:Conclusions_efficiency}.}

\subsection{Discussion of the engine: Our estimates vs. other prospects}
\mg{We considered rather conservative estimations for the parameters. 
Several ways to weaken the requirements can be explored in the experiment in order to obtain a higher cooling ratio. }
\emph{(i)} As shown in the bottom panel of Fig.~\ref{fig:Otto_energies} the piston has been compressed to half its length which ultimately limits the capacity of the machine to cool down. Performing more work and compressing \mg{the piston} more would allow for further cooling. 
\emph{(ii)} Modifying the \mg{barrier} height and various other aspects of our QFM model higher cooling ratios are possible as shown in Appendix~\ref{app:Ottodetails}.
These among others could be \mg{to reset} baths, coupling at higher density etc. which \mg{have} features that depend on the particular implementation and hence cannot be completely anticipated theoretically ahead of performing the experiment.
\emph{(iii)} Let us remark that for the sake of simplicity and also for analogy with the usual thermodynamic Otto engine, the piston is the only component that changes size during the protocol. 
However, one can think of more general scenarios in which the bath is expanded while piston is compressed, and afterwards, the system is compressed while piston is expanded -- after-all in the experiment it will be our goal to cool down the quasi-condensate more than \mg{it} is possible with existing methods and an unconventional quantum thermal machine with various elements changing their size would be helpful for \mg{this purpose}.
Summarizing, there are a lot of \mg{important points} one can consider when devising a QFM.
It is clear that once QTPs are realized, their conceptual clarity will be advantageous in order to appropriately compose them to achieve maximal possible cooling in the experiment.

\section{Discussions and further scope}\vspace{-0.3cm}
\label{sec:discussion}

While further developing the framework of QFMs and during the upcoming efforts to realize a QFM experimentally, numerous 
questions relating to the fundamental
physics of the system and technological implementation 
beyond the scope of this initial manuscript will have to be further investigated. 
Our discussions below highlight several aspects which could invite expertise from fields such as engineering and quantum control of out-of-equilibrium quantum many-body systems to become particularly useful.
Thinking ahead, the program of devising a QFM presented in this work is also expected to stimulate a range of further 
{theoretical investigations} in the field
of quantum thermodynamics
\cite{Topical,PerspectiveKurizki,christian_review,KosloffReview,MillenReview,Janet,Niedenzu2019quantized}. These will range from (experimentally inspired) studies of the {role of information} in quantum thermodynamics to prospects for further development of the  theory of quantum thermodynamics from a quantum information perspective. 

\subsection{The role of information in the QFM}\label{sec:corrandanomalous}     \vspace{-0.3cm}

If we could -- fictitiously -- precisely
measure the many-body eigenstates of our complete machine, we could in principle achieve 
complete control about the system. 
Needless to say, in a quantum many-body system this is impractical and we have to restrict ourselves to physically relevant, local, few-body observables and a finite set of their 
correlations. Ref.~\cite{Schweigler2017} provides an overview on how far one can presently experimentally go in such endeavors.
These limitations will define what we can possibly know about the system and what we can hence make elaborate use of -- and what we are bound not to be able to know and therefore need to ignore. 
In this section, we highlight several important aspects of accessing information/correlations and observing their roles in such a many-body QFM. The manipulation of one-dimensional quasi-condensates via relatively simple yet highly controlled thermodynamic processes in the deep quantum regime seems to be an ideal test-bed for such considerations.

\subsubsection{Correlations 
\je{and}
\nn{anomalous heat flow}}\vspace{-0.3cm}

An interesting future direction is the exploration of the question \emph{how strongly are the elements of the QFM correlated}, how to quantify and control these correlations, and how to make use of them explicitly in the design of a QFM. 
The coupling and de-coupling of two interacting many-body systems, i.e., the operation of the valve QTP, is a direct way to induce correlations or even entangle the two. 
The canonical example thereby is the double well, that has a physics similar to a beam-splitter in quantum optics. 
When the de-coupling is slower than the time scale given by the interaction energy, the two systems will build up \emph{quantum correlations}, which persist even if they are separated~\cite{Jo2007c,Esteve2008,Berrada13}.  
An indication that this also works for the excitations in a many-body system described by an effective quantum field theory is the observation of number squeezing in the modes created by slow splitting
~\cite{Langen2015}. 

\begin{figure}
    \centering
    \includegraphics[width=\linewidth]{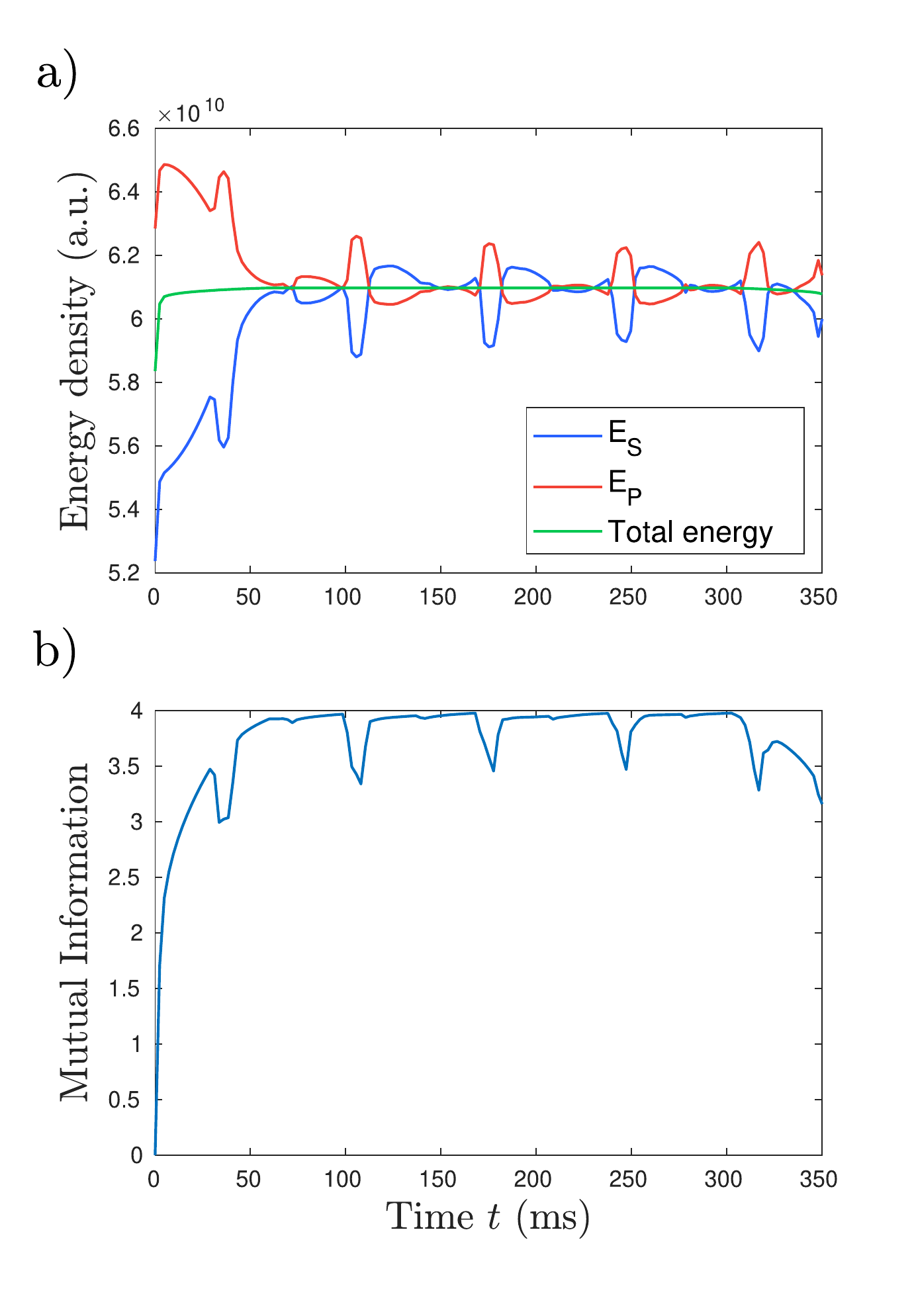}
    \caption{Two systems 30 $\mu$m and 40 $\mu$m long start uncoupled with temperatures of 50 nK and 60 nK, respectively. They are 
    merged in the first 60 ms of the evolution; after that the coupled systems evolve for 240 ms and then are decoupled during 60 ms.
    a) Energy flow between the two condensates. After merging there is heat flowing in both directions. b) Mutual information of the 
    two condensates. The mutual information increases once the two condensates start interacting; after that it decreases for a short
    time and increases again to its maximum value, this happens when the energy is maximum in the condensate with a hotter initial 
    temperature (and minimum for the other condensate), which shows that the mutual information is \nn{correlated} with the reversal of heat 
    flow.
    }
    \label{fig:aflow}
\end{figure}

\nn{The engineering of such correlations is an important question especially in the context of \emph{work extraction}
\cite{NegativeEntropy},
since they may produce interesting dynamics. In particular, with the exhaustion of correlations, instead of inputting extra 
energy/work into the system, one can induce a reverse in what is called the ``thermodynamic arrow of time'', referring to a reverse 
in the direction of heat flow between two systems. Such a phenomenon is commonly referred to as \emph{anomalous heat flow}
\cite{Jennings2010,partovi2008entanglement,  Jennings2010,Jevtic2012,DelRio2016, micadei2019reversing}. Proof-of-principle experiments between qubits have
been demonstrated, which involve the particular engineering of specific unitary processes to address a fixed, two-dimensional energy subspace \cite{micadei2019reversing}. \mg{There also exists experiments studying thermodynamic spin currents which are blocked by the initial state preparation \cite{Husmann8563}.
This blocking is anomalous but not in the sense that the current is reversed, for which correlations must be engineered appropriately. 
Thus an  anomalous reversal of  heat flow with a detailed experimental evaluation of the role of correlations in this process has yet to be worked out in  detail for complex many-body systems in the quantum regime. This is important an important question as}
it is not clear whether global, macroscopic operations are enough to generate i) the right correlations, and ii) dynamics that allow 
the emergence of such behaviour.

Our simulations, on the other hand, predict that the process of merging two condensates creates 
the desired effect of creating correlations that will \gv{potentially} lead to anomalous heat flows (see Fig.~\ref{fig:aflow}). We can quantify the amount of generated 
correlations by computing the mutual information between condensates $S$ and $P$, which is defined by
\begin{equation}
I(S:P) = S(\hat\rho_S) + S(\hat\rho_P) - S(\hat\rho_{SP}),
\end{equation}
where we recall that $S(\cdot)$ is the von Neumann entropy of the quantum system. 
\gv{Note that non-monotonous behaviour of the mutual information can be also used as a signature of non-Markovianity~\cite{RevModPhys.88.021002,SusanaNoMarkov}}.
For the Gaussian states in our study, these quantities are directly computable given the covariance matrices (see Appendix \ref{app:GaussianApp}). 
\gv{Furthermore, this can be also accessed} in the experiments via tomographic data. We see from our simulations that the idle evolution of the joint many-body condensate is sufficient to produce periodic oscillations in the mutual information, in which a similar oscillatory behaviour in the direction of heat flow 
(similar to an AC current) can be observed. It remains to verify how much of the change in mutual information is directly responsible for the reversal of heat flow. Not only this is a fundamentally interesting aspect to study by itself, but its natural presence in the working of the thermal machine also raises the question if one can use this heat flow to our advantage. \gv{For example, it is known that
with correlations there is also the possibility of providing a way of implementing the extraction of macroscopic work 
probabilistically from a heat bath~\cite{boes2019passing}. } 

\gv{In this light, it would be interesting, also} \mg{in relation to earlier experimental works on cooling cold atomic gases with sequential operations \cite{Brantut713}, \gv{to reveal such quantum aspects of protocols of this type in near-future Atom Chip experiments}.}}
Refs.~\cite{Gring2012,Langen2015} \je{have} 
uncovered signatures of  quantum noise and squeezing during longitudinal splitting of the 
quasi-condensates.
This is an exciting indication that it can be possible to  reveal (with statistical significance) the presence of entanglement under similar conditions, e.g.,
quantum correlations between eigen-modes reflecting the effect of various perturbations that can be applied. \nn{The detailed study and controlled usage of these phenomena is therefore one of the future directions of immediate interest, which our platform of interest has a natural advantage of studying.}

\nn{For the simulations \nn{of the full quantum fridge (Figure \ref{fig:Otto_energies})} we currently assume that \emph{de-phasing} occurs after we split systems, i.e., the correlations between the elements of the QFM are modelled to be lost every time splitting is completed. This should be understood as establishing a reference, first-case study where temperature fluctuations and dephasing due to long cycle times render the effect of correlations on the QFM operation to be small.
This should then be compared with experiments, in order to understand the extent of how correlations influence the machine performance. Moreover, further lowering the currently accessible temperatures  will allow to enter the few phonon regime in which quantum vacuum fluctuations will certainly become manifest.
These features are closely connected to entanglement in real space~\cite{anders2007entanglement,anderswinter2008} because the phononic vacuum is entangled in real space as it can be understood via arguments from conformal field theory~\cite{calabrese2004entanglement}.
In this regime, the thermal coherence length $\lambda_T$ will be comparable to the system size and phase correlations will decay polynomially instead of exponentially.}

\subsubsection{Non-Markovian effects}\vspace{-0.3cm}
\nn{Besides anomalous heat flow, there are more generic non-Markovian effects, which our system can be used as a observational test-bed} 
\cite{Markov,PhysRevLett.105.050403,RevModPhys.88.021002,SusanaNoMarkov} on thermodynamic operations.
Such dynamics originate
from the intermediate size of the 
bath so that a back-flow of information occurs.
The presumably 
principal source of non-Markovianity 
is hinted at in Fig.~\ref{fig:Otto_energies}, 
where we see that the wave-packets injected by operating the valve get reflected from the boundaries of the system and {come back} to the position of their origin in finite time, in 
fierce violation of any meaningful Markov approximation. 
Notably, this effect should be expected to hold also in presence of weak non-Gaussian perturbations as  various Atom Chip experiments have already experimentally demonstrated that these features remain intact in close to integrable situations, also in presence of non-trivial trap geometries.

In most works on quantum thermodynamics \cite{gour2015resource,qtSai},
an infinite bath is considered, but it is unclear under which conditions these modelling assumptions would be valid for the intermediate-sized baths in a QFM experiment.
The studies of local recurrences can be seen as entry points to interesting theoretical studies of the possible repercussions of wave-packets returning back to their origin in \emph{finite} time. 

Loss of information ultimately 
proceeds through de-phasing of collective excitations.
For quasi-condensates these are phonons ~\cite{Bistritzer2007,Kitagawa2011,Gring2012}, the de-phased state emerges in a light-cone fashion~\cite{Langen2013}, and is described by a generalized Gibbs ensemble~\cite{Langen2015}, i.e., different modes can have effectively different temperatures determined by the state preparation.
The long time behaviour depends on the spectrum of these collective  modes.  
If the atoms are confined to a box shaped trap, then the phonon frequencies become commensurate, i.e., $\omega_k = 
\pi c k/L$, with $k=1,2,\ldots$ being the mode index, and \emph{recurrences} are observable at short times~\cite{Geiger2014,Rauer2018}. 
This effect is a distinct source of non-Markovianity from the localized wave-packets returning to their origin in finite time and can occur even in a \emph{homogeneous} system.
As detailed in Ref.~\cite{quantumreadout} the recurrence is a recurrence of the squeezed  (momentum) modes, where each mode $k$ is represented by an ellipse in phase space rotating around the origin with frequency $\omega_k$ and all ellipses realign their axes as soon as the slowest $k=1$ mode rotates by a full angle. 
In that moment the $k=2$ mode will have made additionally one more full turn, and similarly higher modes too. I.e., due to the linear spectrum all modes realign.
This pertains to eigenmode populations and the state in real-space can be homogeneous during the dynamics.
This, however, does not occur in a harmonic longitudinal confinement with trap frequency $\omega_{||}$, where the eigenfrequencies are non-linear $\omega_k = \omega_{||} \sqrt{k(k+1)/2}$~\cite{Petrov00} and are incommensurate. Still, 
when the entire system is engineered to be captured by few collective commensurate modes, 
non-Markovian behaviour and significant memory effects can dominate the system dynamics.

Let us illustrate that with an example: 
The role of the reservoir in a thermal machine cycle will strongly depend on the design of the mode spectrum and on when the ``contacts'' take place. I.e., the timing of the valve QTPs will matter. 
If the re-coupling is in between recurrences, the reservoir will appear de-phased and with seemingly no memory of what happened during the previous cycle.
However, the system is coherent: By changing the timing the valve coupling can occur at the time of the recurrence and the reservoir can appear to have memory of what happened during a previous cycle, and hence be a  non-Markovian bath. 
Designing the longitudinal confinement in each part of the thermal machine will allow us to have in principle (nearly) full control of the memory of selected states in the thermal machine at later times.
This will allow us to design and probe a large variety of interesting Markovian and non-Markovian situations~\cite{Pezzutto2016,Hofer2017,Gonzalez2017,Uzdin2016, Observation}.

\subsubsection{Finite-size effects \nn{due to energy fluctuations}}\vspace{-0.3cm}
Individual realizations of the experiment are subjected to non-negligible thermal fluctuations. A particularly interesting question lies in observing the predictions related to \textit{finite-size effects} derived in various theoretical frameworks of quantum thermodynamics.
Our systems are small, and therefore can be heavily influenced by fluctuations in energy. \nn{Moreover, we are interested in a single-shot process of cooling, namely to run the machine for at most a few cycles for a single initial preparation; as opposed to preparing a large amount of identical condensates and seeking to cool them only on average.
The performance of machines in such a single-shot setting} has typically been captured by additional ``thermodynamic laws'' which are distinct from the standard laws that are valid in the thermodynamic limit.
Such ``laws'' are essentially constraints which have been phrased in terms of \emph{(i)} generalized free energies in the context of a resource-theoretic language of quantum thermodynamics~\cite{brandao2015second}, \emph{(ii)} fine-grained Jarzynski equalities~\cite{alhambra2016fluctuating} or  \emph{(iii)} other measures specifically tailored for Gaussian systems~\cite{serafini2019gaussian}. 
These are
intricate and important theoretical descriptions. 
But make-or-break questions for the significance
of such pictures presumably are the following ones:
Can we observe their predictions? 
Specifically, how 
relevant are they to characterize the potentials and limits of practical thermodynamic protocols such as the cooling scheme proposed in this work? 
Much remains to be explored in this direction for quantum many-body systems in contrast to other physical settings where specific ideas have been proposed~\cite{halpern2018fundamental}.

\subsection{Efficiency of quantum machines\nn{: Notions of work and performance versus theoretical limits}}\label{sec:Conclusions_efficiency}
\mhu{Turning our attention to the notion of efficiency of quantum machines, we \je{would like} to connect the expected performance of our proposal to \je{limits} set in the literature. A couple of comments are in order before we begin this discussion. First of all, there are different notions of efficiency that one could discuss: On the one hand, the \emph{quantum} efficiency would compare how much work is drawn from the quantum system in order to implement the machine operation. For a fridge that would be the coefficient of performance, simply given by the quotient of the heat removed from the target system and the work performed by the piston. On the other hand, the \emph{complete} efficiency would be the quotient of heat removed by total work invested in keeping the machine running, i.e., including the power drawn by the computers, DMD and other physical machinery that is needed to keep the system running as a whole. As the cost of control is generically orders of magnitude above the energy scale of the system, 
any complete evaluation of efficiency of a controlled quantum engine (such as we propose) would not \nn{be very meaningful}, since running this machine as an engine to generate work would be futile: Much more work would have to be put into the control as one could possibly expect to gain. The quantum efficiency on the other hand, doesn't have a great operational meaning unless supplemented by further context: \nn{from a pragmatic perspective, it is unclear }why should one care only about the work that is specifically done by the piston and ignore all the work that went into generating the field defining the piston in the first place. 

The main goal of the machine is to cool down a target system in ways that transcend the possibilities of purely classical refrigeration processes.
Here, an exciting direction is to instead explore experimentally the \emph{fundamental limits of cooling}, as they are usually captured in readings of the third law(s) of (quantum) thermodynamics. 
Adapting the terminology from recent works \cite{PhysRevLett.123.170605,clivaz2019unifying}, the QTP toolbox can be seen as \emph{coherent }building blocks for a thermal machine, whereas resource-theoretic operations are energy-incoherent. 
Nevertheless, coherent operations (i.e., time-dependent Hamiltonian control operations) and incoherent operations (i.e. resource-theoretic operations), if both given the same amount of complexity, can achieve similar performance (final energy) in terms of cooling of qubit systems~\cite{PhysRevLett.123.170605}.
Therefore, the fundamental bounds obtained from resource-theoretic frameworks in machine performance~\cite{brandao2015second,woods2019maximum,wilming2017third}, especially such as the third law derivations~\cite{wilming2017third,masanes2017general}, may  be a valuable benchmark. 

\nn{In all thermodynamical processes considered for our QFM in the future, an important issue} is the notion of work itself. Indeed, quantifying {work extraction} in the quantum regime can be treated by various theoretical frameworks of quantum thermodynamics and might yield different results depending on the definition used \cite{Niedenzu_2019}.
For the refrigeration cycle that we have proposed here, however, the useful output of the QFM is easy to assess, as shown in Fig.~\ref{fig:Otto_energies}.
This, however, is not as straightforward in general for other tasks that may be implemented with the QTPs that we have presented.
In that case, additional ideas for quantifying quantum work will have to be developed in accordance with our modelling involving exclusively unitary processes induced by time-dependent Hamiltonians. 
For example, since we always initialize the quasi-condensates in a thermal state, this process is similar to the standard setting of \emph{fluctuation relations}~\cite{PhysRevLett.78.2690}. 
However, the statements of work extraction provided by fluctuation relations involve initial and final projective energy measurements on the system which is \emph{not} directly measurable in experiments with quantum many-body systems.

QTPs can be used to perform work on systems and in the process we saw that this brings them out of equilibrium.
Therefore, in the \emph{resource-theoretic} framework of quantum thermodynamics~\cite{opphor,brandao2015second} they should be interpreted as being \emph{resourceful} and hence stand in contrast to \emph{free} operations and states which are usually studied in this formalism.
This highlights the \emph{gap} between this powerful, but abstract, framework with what is meaningfully achievable in experimental setups. 
So far, the energetic worth of non-thermal resource states 
has been studied in the context of distillation rates~\cite{brandao2013resource, chubb2018beyond}. Our platform will provide a testbed for  quantifications on the level of operations rather than states and might be potentially useful for practical settings.
}

\subsection{Experimental realization of a QFM}  \vspace{-0.3cm}
The first type of energy referring to known properties can in the widest sense be related to \emph{work}, the latter to \emph{heat} and entropy. Ultimately, it is the amount of information one has about the energy present in the system which decides whether it should be interpreted as heat or work \cite{Topical,WorkAndHeat,DelRio2015,perarnau2015,brunner2014,woods2019maximum,ng2017surpassing}. Aside from the conceptual issue of separating work and heat, much progress has also been made in theoretical quantum thermodynamics, in terms of predictions of how energy exchange in finite-sized quantum systems would occur.

In the present manuscript, we have deliberately focused on a simple and straightforward way 
of implementing a QFM in the form of a one-dimensional bosonic quantum gas which can be described with a Gaussian effective model.  
Having said that, there are many interesting directions that can be explored in order to extend our proposal.

\subsubsection{Non-Gaussian QFMs}  \vspace{-0.3cm}

We have
 left it open to what extent higher-order, non-Gaussian contributions will play a substantial role in the operation of the QFM in the experiment.
One example where these could potentially matter is when running the QFM with a long cycle time. 
This is because of de-phasing or damping effects which are not present in the TLL model can occur in real experiments already around \SI{50}{\milli\second} \cite{Rauer2018,quantumreadout}. We currently take into account effects of dephasing only whenever two systems are split, which is consistent with this timescale since merge-split protocol considered in our simulation is $\SI{40}{\milli\second}$. A key aspect of future investigations will 
be to comprehensively explore \emph{weak} non-Gaussian effects arising from such effects.
We expect many-body de-phasing to primarily have the effect of thermalizing the bath, but otherwise not obstructing the heat flow which occurs faster than the on-set of any de-phasing observed so far. Nevertheless, a detailed study will provide more substantial insights into this important aspect. 

The second interesting case is to notice that during the splitting and recombination, the mean density at the interface is low and there the \mg{ linearized phononic} description might break down.
This may lead higher-order interactions to become substantial and induce scattering of phonons around the interface.
In the experimental implementation, one can extract higher-order correlations and study non-Gaussian correlations \cite{Schweigler2017}.
If present, they can be studied by numerical field theoretic calculations \cite{PhysRevLett.121.110402} or compared with predictions based on fundamental relations in quantum thermodynamics~\cite{narasimhachar2019thermodynamic}.

Finally, the existing Atom Chip platform allows to controllably add \emph{sine-Gordon} interactions \cite{Schweigler2017,Gritsev07} and hence also non-Gaussian QFMs can be explored experimentally.
The sine-Gordon model is paradigmatic for our understanding of quantum field theory~\cite{Coleman75,Mandelstam,Thirring195891,Faddeev19781} thanks to its rich physics, e.g., excitations of finite mass and non-trivial topological properties.  
The experimental implementation ~\cite{Schweigler2017} following the quantum simulation proposal from Ref.~\cite{Gritsev07} has been realized using two longitudinally tunnel-coupled one-dimensional quasi-condensates.  
In this case the system should be described by \emph{relative} degrees of freedom, the relative phase $\pp_\text{rel}(z)$ and  density  $\dd_\text{rel}(z)$ fluctuation fields.
These fields are obtained by considering the difference of the respective fields of each of the condensates, see, e.g.,~Ref.~\cite{Rauer2018} for a detailed discussion in relation to a recent experiment.
Using interferometric measurements \cite{Schumm05,Rauer2018,van2018projective,Schweigler2017,Schweigler_thesis} correlation functions of the relative phase can be measured which allowed to substantiate that the physical system has been correctly described by the effective sine-Gordon Hamiltonian for two adjacent quasi-condensates
\begin{eqnarray} 
\begin{split}
\label{eq:SG}
\hat {H}_{\mathrm{sG}} =&\int \mathrm{d}z\biggl[\frac{\hbar^2 \nGP(z)}{4m}\left(\partial_z\pp_\text{rel}(z)\right)^2+g \dd_\text{rel}(z)^2  \biggr]  \\
&- \int{ \mathrm{d}z ~ 2 \hbar J \nGP \cos({\pp_\text{rel}(z)}) } \, ,
\end{split}
\label{eq:sG}
\end{eqnarray}
\gv{with effective field operators now capturing the} \emph{relative} phonon modes \gv{and interacting with a non-Gaussian cosine term}. 
Tuning of the tunnel coupling $J$ is possible experimentally which would allow to build  QFMs in various interaction regimes, ranging from a  system of non-interacting modes to a strongly correlated quantum system with topological excitations.  See~Ref.~\cite{Schmiedmayer2018} for further details in the context of thermal machines and Ref.~\cite{Schweigler2017} for a detailed experimental study of the many-body aspects of the model.

Summarizing, the two coupled one-dimensional quasi-condensates will allow us to build and study strongly correlated QFMs, where the degree of correlations (that is, the degree of higher-order correlation functions that are relevant) can be experimentally tuned. It is known that the 
time evolution of interacting local quantum systems is computationally
hard (technically speaking, it is BQP-complete in
worst-case complexity, referring to
bounded-error quantum polynomial time), and in practice computationally
demanding for classical computers for physically relevant problems. This applies
as well to the 
equilibrium processes involved in the operation of the QFM. 
While numerical studies may prove inefficient, the properties of these strongly correlated QFMs can be probed experimentally in detail by measurements of (higher-order) correlation functions~\cite{Schweigler2017,Zache2020}.

\subsubsection{Matter-wave interferometry of parallel QFMs} \vspace{-0.3cm}

An intriguing idea is to run machines in parallel.
This opens up the possibility to compare the operation of two identical machines by direct observation of \emph{matter-wave interferometry} \cite{cronin2009optics}. 
On the Atom Chip it is possible to conceive of two machines positioned side by side, parallel to each other (Fig.~\ref{fig:ThermalMachines} would then be the side-view of \emph{two} machines) and they would be identical in the sense of having the same initial state preparations and subsequent control operations implementing QTPs making up the Otto cycle. This can be done using well-established protocols of manipulating the gas using a longitudinal double-well and interference has been observed in this case in various situations \cite{Schumm05,Gring2012,Schweigler2017,Rauer2018,Zache2020,schweigler2020decay}.
Interferometry by its nature looks at \emph{relative} fluctuations and hence disregards classical disturbances in the operation which are identical for both systems and directly measures \emph{quantum} fluctuations.
Their appearance should be studied interferometrically for various initial states of the two systems including

\emph{(i)} two independent systems created by cooling two cold atomic clouds separately. 
This provides the base case to be compared to when studying more interesting initial states. 

\emph{(ii)} Two systems that are de-phased in a pre-thermalized state~\cite{Gring2012}.
In this case the temperature of relative degrees of freedom has been found not to be fully determined by the cooling process but rather to be related to the tunnel coupling $J$ in Eq.~(\ref{eq:sG}) present during state preparation. 
 
 \emph{(iii)} Two systems with (nearly) identical phonon modes with  strongly suppressed quantum noise in the relative degrees of freedom.  Such states have been achieved experimentally \cite{Langen2015,Berrada13} and can be further improved by optimal control of the splitting process \cite{Grond09}.
 
 Each of these approaches would prepare machines that would have distinct initial conditions and an experimental study would allow to gain insights on how these influence the operation of the QFMs.
 Observing features where cases \emph{(ii-iii)} would differ from the simple case of independent machines \emph{(i)} would then most likely require a non-classical explanation.

\subsubsection{Necessity of optimal control for operations of the QFM}           \vspace{-0.3cm}
In our present study  we 
have involved only very simple protocols to operate the different primitives building a QFM.  In a real world implementation, 
one would naturally like to speed up and optimize the different steps of a QFM. This should be in fact expected to be a crucial matter.
This can be done by implementing optimal control methods \cite{werschnik2007quantum,PhysRevLett.103.240501,PhysRevLett.106.190501,Koch_2016},
such as those reported for splitting a double well in Ref.~\cite{Grond09} or for the excitation process in Ref.~\cite{VanFrank2015a}.
Notably, some of the control theory is already established for the piston  QTP and has been successfully implemented \cite{rohringer2015non}.
There, a harmonic trapping potential has been considered and the extension of the Gross-Pitaevskii profile has stably been modified.
This involved the fact that modifying a harmonic trapping potential acts essentially as a \emph{lens} for the individual atoms making up the system, so not only one can compress them appropriately but also accurately decelerate them when needed.

\subsubsection{Diagnostic tools for QFMs}

Finally, a particularly important direction to study is the development of further {diagnostic techniques} for the system along the lines of recent developments \cite{Schumm05,van2018projective,PhysRevLett113.045303,Schweigler2017,Schweigler_thesis,quantumreadout,PhysRevA.98.043604,PhysRevLett.116.050402,PhysRevLett.105.230402,PhysRevLett.96.130403,PhysRevLett.106.230405,Imambekov09, Manz11,Aidelsburger2017Relax2d,schaff2014interferometry,Benchmarking}. We have discussed in Sec.~\ref{sec:QuntumFieldMachine_readout} and Appendix~\ref{sec:app_readout} the current experimental read-out capabilities 
and have proposed how to enhance them by novel variants of tomographic data analysis \cite{quantumreadout}.
Detailed \emph{monitoring of the QFM} will be crucial and novel \emph{hardware} solutions can aid that goal.

A particularly interesting possibility is to trap a three-dimensional condensate close to the one-dimensional QFM and use it as a sensing device.
Matter-wave interference between two systems of different dimensionality seems to be interesting in its own right offering to study a wealth of various physical phenomena \cite{van2018projective,schaff2014interferometry}.
In addition, it could be expected to provide additional read-out resources with the goal of circumventing the current imaging resolution limitations that are difficult to improve otherwise.
The implementation of this scheme would have the advantage of providing a direct measurement of the phase along of a single quasi-condensate in contrast of the indirect tomographic approach. See
Ref.~\cite{Aidelsburger2017Relax2d} for related work in this direction and additionally Refs.~\cite{Niedenzu2019quantized,yang2020cooling} for a discussion of possibilities for immersion cooling. 

Finally, let us remark about the possibility of performing \emph{non-destructive} measurements which are essential, e.g., for an analysis of a thermodynamical process using fluctuation relation theorems which involves a two-step measurement process on the same system.
Currently the measurements performed in experiments using the Atom Chip are destructive, see, e.g., Ref.~\cite{van2018projective} 
for a discussion of measurements following a time-of-flight expansion.
When experimenting with 1D systems destructive measurements are experimentally easier because of the small atom number.
In that case one can illuminate the complete system for read-out and every atom scatters many photons. Measuring in time of flight has the additional advantage that the atomic cloud which is initially only a few 100 nm in transverse size, can expand transversely to a size that is above the resolution limit of the imaging optics (as used in Refs.\ \cite{wildermuth2005microscopic,aigner2008long}). 
For long time of flight the atoms have moved away from the Atom Chip elements that have been used for the control in the experiment, thus reducing spurious light scattering that can contaminate the pictures.
Such a measurement is destructive in two ways: (1) the gas is released from the trap, (2) as the atoms are heated up so much by the light scattering that the low-energy quasi-condensate description for the atomic cloud is not valid anymore: the BEC evaporates unless the system is large \cite{andrews1996direct,saba2005light,freilich2010real,KuZwierlein_Vortex,VortexTrento,Seroka:s}.
Thus, non-destructive measurements in our system can not be performed by illuminating the atoms for read-out multiple times but rather in a less disruptive way, e.g., by \emph{out-coupling} of atoms \cite{saba2005light}.
Here, one would like to remove selectively atoms from a portion of the system and measure these projectively away from the system.
This has the advantage that the system will not be destroyed and the measurement can be repeated.
Additionally, imaging \emph{individual} out-coupled atoms allows to consider \emph{quantum limited} measurements~\cite{bucker2009single,bergschneider_Li}.
However, the mechanical effects of even the second order Zeeman effect in the strong magnetic field gradients of the chip traps, which are still on to keep the remaining system running, make these measurements more difficult.
Still, these engineering challenges could be overcome in near-term. 

For applications in quantum thermodynamics and fluctuation relation theorems, it should be noted that these measurements would be local in space (product measurements of commuting observables).
This is in contrast to many protocols assuming projective measurements in the entangled and non-local \emph{energy eigenbasis} --  in general it is not clear  how to achieve these experimentally  demanding requirements in quantum many-body systems.

As a final outlook, let us remark that measurements using outcoupled atoms could potentially allow for implementing error mitigation for the refrigeration QFM: When merging two systems, the number of excitations will be influenced by phase diffusion of the phase zero-mode and the measurement of the relative phase between the outcoupled atoms could allow to select the experimental runs that happen to have fewer excitations than the average realization. 
Assessing the back-action  on the system and the influence on the performance of the QFM in such a scheme is an interesting question for future study.

{\subsubsection{Relation of our proposal to other platforms}}

\je{At the heart of this work stands the design of a specific QFM based on continuous cold atomic programmable potentials, for good reason, as this
is a blueprint for a quantum thermal machine following the desiderata that we have laid out. In this sense, we see this specific choice rather as a strength of this proposal and not a weakness.} 

\je{Having said that, it should be clear that ideas of creating a toolbox of \emph{thermodynamic primitives} as macroscopic operations over quasi-condensates, reminiscent of operations acting on bulk systems in conventional thermodynamics, 
may well carry over to other cold atomic platforms,
stressing the generality of the approach taken. Specifically, for systems of 
cold atoms in optical lattices, digital mirror devices have been 
used to implement programmable potentials \cite{2dMBLScience,SingleSite}. In such settings, giving rise to programmable Bose-Hubbard dynamics, 
a coupling and decoupling giving rise to a valve and the compressing and decompressing realizing a piston could be realized, following
the general prescription of this work.

On a related but different note, we mention the relationship of the present proposal
to other proposals of quantum thermal machines or refrigerators that have been
put forward. Indeed, cold atomic quantum thermal machines have been considered~\cite{Niedenzu2019quantized}, investigating the refrigeration of an atomic cloud, but not following the mindset of operational primitives laid out here. There, the use of two atomic species has been suggested, in which one atomic species implements the working medium and the other implements two baths that are hot and cold, respectively. Our proposed system is simpler and therefore presumably more robust against experimental uncertainties and imperfections than that of Ref.\ \cite{Niedenzu2019quantized} which requires a very high level of control over the system and very precise fine tuning of experimental parameters. The recent work Ref.\ \cite{Widera} starts with a gas of Rubidium atoms cooled down to low temperatures, where the thermal machine, however, consists of  individual cesium atoms and does not operate in the quantum many-body regime.
Work already mentioned above demonstrates thermoelectricity in a fermionic ultra-cold 
atoms channel, connected to two reservoirs
\cite{Brantut713}.

Further away still from the setting
we consider here are proposed refrigeration schemes based on a phonon pumping mechanism in nano-mechanical systems \cite{Nano1,Nano2}. These schemes also aim at achieving cooling of a quantum system, albeit in a quite different way from the setting considered here. In that
work, non-interacting phonons are suggested to
provide the work fluid, in contrast to massive atoms that are in the focus of attention in the present work. The phonon number is not preserved and 
an nano-mechanical system is expected to be
open. This observation -- together with the 
fact that the piston is anticipated to be realized
as a traveling lattice perturbation acting as a semi-reflective barrier -- seems to come along
with substantial experimental challenges. There are also similarities, in that cycles involving 
three subsystems are being considered. In
the present work, the cycles are composed
of operational primitives involving massive
and potentially interacting atomic quantum
systems.
%
%
The most significant contribution of the present work is to carefully introduce and discuss these basic operations at hand of numerical results based on an accurate microscopic model, for which the
cyclic processes in quantum thermodynamics devised are more an example
than an aim in its own right.}\\

\section{Conclusions}

In this work, we have set out to devise a blueprint
for a genuine quantum thermal machine in  one-dimensional ultra-cold atomic gases, a platform that we propose to realize complex thermodynamic tasks.
We have proposed a quantum field machine (QFM) involving  phononic degrees of freedom described by an effective quantum field theory. When devising this blueprint,
resorting to a quantum mechanical description has been
crucial to reduce the physical description of the system to a point where the functioning of the machine can be easily grasped.
In order to 
provide guidance towards constructing thermal field
machines, we characterize a toolbox of thermodynamic primitives which are macroscopic operations over  quasi-condensates, reminiscent of operations acting on bulk systems in conventional thermodynamics. 
Our proposal puts forward a scheme for a refrigeration QFM that involves a system featuring quantum effects, a cold atomic gas, and the machine performs a useful task -- \emph{cooling of phononic quantum fields}.
In contrast to previous realizations of quantum engines this cannot be practically achieved by controlling every single degree of freedom of the system as there are just too many.
It goes without saying that this task is useful and we hence fully accommodated the three requirements that we have set in the outset of this work for a thermal machine to be a \emph{genuine} quantum machine.
We found that quantum effects present in our QFM  are currently detrimental to its cause: Operating the valve of the QFM  induces inevitable excitations adding thermal noise of reservoirs.  
This is rooted in the quantum effect of phase diffusion of  phase zero-modes and in the dynamical Casimir effect and features a detailed temporal structure thus far ignored. 
Remarkably, even after accounting for realistic ``imperfections''  expected in the experiment we predict notable cooling. 
As detailed, exploring further quantum features is possible, including \emph{(i)} at sufficiently low temperatures entanglement or zero-point fluctuations leading to sub-Poissonian noise when operating the valve \cite{Gring2012}, \emph{(ii)} non-Gaussian QFMs \cite{Schweigler2017}, \emph{(iii)} non-Markovian QFMs \cite{Rauer2018}, \emph{(iv)} parallel machines amenable to measurements using matter-wave interferometry \cite{Gring2012,Langen2015,Schweigler2017,Rauer2018,Pigneur2018}, \emph{(v)} quantum phase diffusion and phase-locking via Josephson oscillations \cite{Pigneur2018} \emph{(vi)} few phonon regime  similar to quantum optics in the few photon regime where the quantized nature of the energy spectrum becomes manifest and individual runs of the QFM will unavoidably fluctuate.

It is clear that this work constitutes only a commencing study of a research program of a larger scope. We perform classically
efficient numerical simulations, but 
calculations for a non-Gaussian QFM are expected to hit the computational complexity barrier: \je{It is key to our
work that the blueprint for a QFM devised here resorts to a 
quantum many-body regime, in contrast to work that aims at understanding single-atom heat machines \cite{PhysRevLett.120.170601}.}
Even though our operational principles and cycles are reminiscent of those of  classical heat engines, i.e., canonical thermodynamical transformations, we highlighted some interesting issues obstructing understanding the functioning of our QFM using resource theories.
We have encountered quantum features which in the future should, to the contrary of our current observations, be seen not as a burden but as a potential advantage: They should be used to  improve the performance of the thermal machine in the deep quantum regime.

Our theoretical model for a Gaussian QFM is expected to largely capture the qualitative operation of the QFM. The quantitative features may change and there is a rich number of entry points for non-Gaussian behaviour to set in. We expect their effect to be small and to not overhaul our predictions.  
Ultimately, whether this will play out to be true in reality can only be decided by performing an experiment.

We firmly believe that such a machine can and 
should be built which will deepen our understanding of thermodynamics in the quantum regime. 
Further progress in the {research field of 
quantum thermodynamics} necessitates the development of useful quantum machines to drive, motivate and guide the theoretical development of the corresponding laws, just as the advent of steam engines propelled the development of thermodynamics in the 19th century.
It is our hope that the roadmap laid out in this work will serve this cause well.

\section*{Acknowledgements}
We are grateful to Fred Jendrzejewski and Spyros Sotiriadis for useful discussions. 
 J.~E.~and J.~Schm. are supported by the 
DFG Research Unit FOR 2724 on ``Thermal  machines in the thermal world''.
J.~Sch., J.~E., and M.~H.~have also received  funds from the FQXi (FQXi-IAF19-03-S2)
 within the project
``Fueling quantum field machines with information'', for which the present effort is key.
M.~H.~and G.~V.~acknowledge funding from the Austrian Science Fund (FWF) through the START project Y879-N27 and the Lise-Meitner project M 2462-N 27, M.~H.~and J.~Sch.~acknowledge the ESQ  Discovery Grant ``\emph{Emergence of physical laws: From mathematical foundations to applications in many body physics}''. J.~Sab.~acknowledges funding from the Austrian Science Fund (FWF) through the DK CoQuS. J.~E.~acknowledges funding
from the DFG CRC 183 (Project A03), 
 the  European  Union's  Horizon2020  research  and innovation  programme  under  grant  agreement  No.~817482 (PASQuanS). 
 N.~N.\ acknowledges funding from the Alexander von Humboldt foundation, and the Nanyang Technological University, Singapore under its Nanyang Assistant Professorship Start Up Grant.
I.~M.~acknowledges the
support by
the Wiener Wissenschafts- und Technologiefonds (WWTF) via Grant No. MA16-066 (SEQUEX).
J.~Sab., M.~P., and Y.~O.\ thank \je{for} 
the support from Funda\c{c}\~{a}o para a Ci\^{e}ncia e a Tecnologia (Portugal), namely through project UIDB/EEA/50008/2020. 
J.~Sab.\ acknowledges the support from the 
DP-PMI and FCT (Portugal).

\newpage

\appendix
\begin{widetext}
\newpage

\section*{Appendix}

This appendix provides sections accompanying the discussion presented in the main text as follows.
We begin by giving in Appendix~\ref{sec:app_readout} more details about the precise quantities that can be measured in experiments on the Atom Chip and discuss how to connect these to thermodynamical quantities.
Next, in Appendix~\ref{app:GaussianApp} we summarize the essential ingredients of the bosonic Gaussian formalism which is the analytical base for the numerical code that produced our results.
Finally, Appendix~\ref{app:QTPsimdetails} provides an extended discussion on simulation details, including precise formulation of the lattice approximation employed in the code, extended description of the valve QTP (including how to compute the energy density or compare to the continuum limit in the scenario of sudden (quench) merging), of the piston QTP (including additional discussion of the model, compression dynamics and details of coupling inhomogenous QFTs after compression) and finally we discuss different relaxations of parameter constraints that have yielded almost $30\%$ cooling ratio.

\section{Experimentally monitoring thermodynamic transformations in phononic quantum simulators}
\label{sec:app_readout}

Let us begin by discussing which quantities, if measured experimentally, would reveal insights about the thermodynamic transformations in the system.
We then proceed by explaining what are the direct experimental observables and how to connect to the desirable thermodynamical observables.
In experiments, one should distinguish the cases of having a single quasi-condensates and two which are adjacent.
First let us discuss the former case where we have access to measurements of the atom numbers locally by transversal density absorption imaging (from the side).
These numbers will be ultimately binned together due to finite resolution.
The recovered atom number per bin $N_i(z_j,t)$ will fluctuate randomly between realizations $i$ and will give spatially resolved data where $z_j$ can be measured in steps of about $\Delta z_{\rm res}=\SI{2}{\micro\meter}$ on the Atom Chip.
The quantities obtained for this lattice can be compared to theory by convoluting the continuum quantities by a Gaussian function with $\sigma_{\rm res}=\SI{3}{\micro\meter}$ and evaluating \cite{quantumreadout,Schweigler_thesis}.
After taking this data at a given time $t$  one can obtain  the density fluctuations as follows.
The empirical mean of the observable random variable $N_i(z_j,t)$  gives access to the 
Gross-Pitaevskii profile
\begin{equation}\label{eq:rho_0z}
    \nGP(z_j,t) \approx \frac 1 M \sum_{i=1}^M N_i(z_j,t).
   \end{equation} 
After subtracting these values from the individual realizations and squaring the shifted random variable we obtain the estimator
\begin{align}
    \Gamma^{\rho\rho}(z_j,z_{j'},t_i) \approx \frac 1 M \sum_{i=1}^M \left( N_i(z_j,t_i)-\nGP(z)\right)  \left(N_i(z_j',t_i)-\nGP(z)\right)\ .
\end{align}
Indeed, what we obtain by this is nothing else than the estimate the second moments of density fluctuations  away from the mean density of the quasi-condensates.
The on-site correlation gives information about the energy of phonons.
\gv{As discussed in Sec.~\ref{sec:Implementation} in the main text,} the total energy in the system can be obtained by considering the formal expression for expectation value of the Hamiltonian
\begin{align}
    \langle \hat H[\nGP]\rangle =\int
\di z \biggl[&\frac{\hbar^2 \nGP(z)}{2m}\langle\left(\partial_z\pp(z)\right)^2\rangle+\frac g 2 \langle\dd(z)^2 \rangle \biggr] \ .
\end{align}
Of course in the experiment one can only measure at discrete positions but what we can do is try to obtain this quantity via a finite Riemann sum, specifically in the density sector we find
\begin{align}
    E_\rho(t) = \frac g 2 \int \di z  \langle \dd(z)^2\rangle \approx \frac g 2 \sum_j C^{\rho\rho}(z_j,z_{j},t_i) \Delta z_\text{res}  \ .
\end{align}
By considering the summand in this expression we get access to the energy density for the density fluctuations.
Studying how it changes in time between different pixel positions $z_j$ will then give information about the dynamics of the energy of density fluctuations in the system.

For the single quasi-condensate, as explained in the main text, it is not possible to measure directly the phase fluctuations.
This is important, however, in order to assess the energy contribution coming from the gradient of the phase operator. This information can be obtained via a tomographic approach  by studying the velocities of wave-packets going through the system, as demonstrated recently~\cite{quantumreadout}.
The basic idea is that the phononic Hamiltonian can be put to a normal form
\begin{align}
\hat H=  \sum_{k>0}  \frac{\hbar \omega_k} 2 (\p_k^2+\d_k^2) + \frac g 2 \d_0^2\ 
\end{align}
using the eigenmode operators $\p_k,\d_k$ that depend on the Gross-Pitaevskii profile arising from cosine eigenfunctions in the homogeneous case.
We then find that the dynamics of the density fluctuation operator reveals information about the phase operator by means of the relation
\begin{align}
  \d_k(t)=\cos(\omega_k t )\d_k +\sin(\omega_kt)\p_k\ .
  \label{eq:p_t}
\end{align}
Exploiting this expression
to relate observables at different times and using the analysis and reconstruction methods developed in Ref.~\cite{quantumreadout} should then give access to the second moments of the phase fluctuations.
Specifically, one would reconstruct the second moments  $C^{\phi\phi}_{k,k'}=\langle \p_k\p_{k'}\rangle$ of eigenmodes $k,k'$ and we can obtain the total energy contained in the phase sector by simply summing
\begin{align}
    E_\phi(t) = \frac 12\sum_{k>0} \hbar\omega_k C^{\phi\phi}_{k,k}(t)\ .
\end{align}
Additionally, one can translate the second moments of the eigenmodes to real space after performing the derivative on the eigenfunctions which should give the local information about the energy.
When considering two condensates one has access to interferometric measurements of the relative phase fluctuations $\pp_\text{rel}=\pp_1-\pp_2$ between two quasi-condensates and based on non-equilibrium variations of that observable, relative density fluctuations $\dd_\text{rel}=\dd_1-\dd_2$ have been
reconstructed in Ref.~\cite{quantumreadout}.
Density absorption is still available to measure desnsity fluctuations of the common degrees of freedom  $\dd_\text{com}=\dd_1+\dd_2$ but is usually less revealing.

\section{Gaussian models in the simulations of QTPs}\label{app:GaussianApp}

The continuous Hamiltonian given in Eq.~\eqref{eq:H_quench} can be appropriately discretized, which we explain in Section \ref{app:Discretization}. The system can then be described in terms of quadrature operators, in particular, one can describe the quantum states and dynamics with the Gaussian framework of covariance matrices and symplectic transformations. In this section, we present a short summary of
the formalism of Gaussian quantum information, see, e.g., Refs.\ \cite{GaussianQuantumInfo,Continuous} for more complete reviews on the subject. 

We consider bosonic systems of $N$ bosonic modes, associated with quadratures
\be
{\bf \hat X} := (\hat q_1,\hat q_2,\dots,\hat q_N,\hat p_1,
\hat p_2,\dots,\hat p_N)^T 
\ee
that can be seen as the $N$ position and momentum
operators, respectively.
The canonical commutation relations can be captured 
as $[\hat X_l,\hat X_m]=i
\Omega_{l,m}$
for $l,m=1,\dots, N$,
giving rise to the symplectic form
\begin{equation}\label{eq:symplectic_mtrx}
    \Omega=\left( 
    \begin{matrix}
    0 & \id \\ 
    -\id & 0 
    \end{matrix}\right). 
\end{equation}
Given a density matrix $\hat\gamma$, 
we define the {\it vector of mean values} $\bar{{\bf X}}:=\aver{\bf \hat X}_{\hat\gamma}=\trace(\hat \gamma{\bf \hat X})$:
these are the first moments of the set of quadrature operators $\hat X$ corresponding to the quantum state. The second moments can be 
collected in the {\it covariance matrix} with
entries 
\begin{equation}
\Gamma_{i,j}:=\aver{\hat X_i \hat X_j +\hat X_j \hat X_i}_{\hat\gamma} -2\aver{\hat X_i}_{\hat\gamma}\aver{\hat X_j}_{\hat\gamma}\ .  
\end{equation}
For a single mode, namely $N=1$,
the diagonal elements of $\Gamma$ are simply the two variances $\Gamma_{1,1}=2(\Delta \hat q_1)^2_{\hat\gamma}$ and $\Gamma_{2,2}=2(\Delta \hat p_1)^2_{\hat\gamma}$. The 
single constraint for the real-valued matrix to correspond to a physical state is given by the {\it Heisenberg uncertainty relation}, which can be 
concisely written as a semi-definite constraint as
\begin{equation}
\Gamma + i\Omega\geq 0 .
\end{equation}
Of key importance in this work are bosonic 
Gaussian states.
A general Gaussian state of $N$ modes is fully
described by the vector of mean values  and the covariance matrix corresponding to all modes. 
Gaussian states are ubiquitous in 
physical systems. For example, thermal states
${\hat\gamma}_\beta[\hat H]=\exp(-\beta \hat H)/\trace(\exp(-\beta \hat H))$ are Gaussian whenever the Hamiltonian $\hat H$ is 
quadratic in the field operators, which again is a
very common situation in many physical settings. 
In condensed matter physics and in quantum field
theory, such a situation would be referred to as being
\emph{non-interacting}.
Generally, 
every Gaussian state with full support \footnote{This means that the density matrix has no zero eigenvalue.} can be written in a form resembling thermal states of quadratic Hamiltonians, namely there exists a 
$H$ such that
\be\label{eq:faithGauss}
{\hat\gamma}[H]= \frac 1 Z \exp\left( -\tfrac 1 2 ({\bf \hat X}-{\bf \bar X})^T H ({\bf \hat X}-{\bf \bar X}) \right) ,\qquad H=\left(\begin{matrix}
H_{qq} & H_{qp} \\ 
H_{pq} & H_{pp} 
\end{matrix}\right) ,
\ee
where $H$
is a real positive semi-definite 
$2N\times 2N$ matrix written in block form for clarity and 
\be
Z=\trace\left[\exp\left( -\tfrac 1 2 ({\bf \hat X}-{\bf \bar X})^T H ({\bf \hat X}-{\bf \bar X}) \right) \right] =
\sqrt{\det((\Gamma +i\Omega)/2)}
\ee
is the normalization, which can be fully determined by the covariance matrix of the Gaussian state $\Gamma$. 
The relation between $\Gamma$ and the matrix $H$ appearing in the expression above is
\be
\begin{aligned}\label{eq:relHgammaandback}
H&=2i\Omega \ {\rm arcoth}(i\Gamma \Omega), \qquad
\Gamma&= i\Omega  \coth(i\Omega H/2).
\end{aligned}
\ee
In turn, any generic quadratic (Hermitian) Hamiltonian can be written similarly as above, i.e., with $H$ being a real positive-semi-definite $2N\times 2N$ matrix. Thus, as a difference with respect to the above matrix appearing in the expression for faithful Gaussian states, a generic quadratic Hamiltonian can also contain zero eigenvalues (and need not to be diagonalizable). 

The (Gaussian) unitary evolution corresponding to the time-independent quadratic Hamiltonian 
translates into the symplectic transformation acting on the covariance matrix, given by
\be\label{eq:symplmatunit}
G(t)=  \exp(\Omega H t) ,
\ee
such that the evolved covariance matrix is $\Gamma (t) = G(t) \Gamma(0) G(t)^T$. A similar relation holds for the evolution with time-dependent Hamiltonians, see for example the discussion on the QTP primitives in Sec.~\ref{app:QTPsimdetails}.
Thus, in the framework of Gaussian states and operations one can work directly with just the mean vector and the covariance matrix, since they jointly contain all the information that characterizes the Gaussian state. 
In particular, given a quadratic Hamiltonian $\hat H=\sum_{k,l} H_{k,l} \hat X_k \hat X_l $, the average energy of a state $\hat\gamma$ can be easily computed as
\be
E_\gamma = \trace(\hat H \hat\gamma)=\sum_{k,l} H_{k,l}
\trace(\hat\gamma \hat X_k \hat X_l)   = \sum_{k,l} H_{k,l} \left(\tfrac 1 2 \Gamma_{k,l}+\aver{X_l}_\gamma \aver{X_k}_\gamma\right).
\ee
The covariance matrix and the Hamiltonian matrix can be put into normal form by symplectic transformations, which read
\be
\begin{aligned}\label{eq:normal_modes}
\Gamma &= M \left( \bigoplus_{k}\gamma_k \id_2 \right) M^T , \qquad
H = M \left( \bigoplus_{k}\omega_k \id_2 \right) M^T ,
\end{aligned}
\ee
where $M$ is a symplectic matrix and the $\{\gamma_k \}$ (respectively $\{\omega_k \}$) are the {\it symplectic eigenvalues} and are the eigenvalues of $|i\Omega \Gamma|$ (respectively $|i\Omega H|$). 
Clearly, the symplectic eigenvalues of $\Gamma$ and $H$ are related to each other in the same relation as Eq.~\eqref{eq:relHgammaandback}, e.g., for a thermal covariance matrix at inverse temperature $\beta^{-1} = k_B T$, we have 
\be\label{eq:sympeigcovtherm}
d_k=\coth(\beta \omega_k/2),
\ee
which is the usual relation between the normal mode frequencies $\omega_k$ of a harmonic oscillator Hamiltonian and the normal covariances of its thermal state. Note that by identifying $\gamma_k = 2\aver{n_k}+1$, this agrees with the Bose-Einstein number distribution formula 
\begin{equation}
\aver{\hat n_k}=e^{-\beta \omega_k}/(1- e^{-\beta \omega_k}).
\end{equation}
The von Neumann entropy of a quantum state $\rho$ can be also directly computed from its covariance matrix $\Gamma$, and in particular just from its symplectic eigenvalues (as is true for every unitarily invariant quantity). In fact, recall the definition 
\be
    S(\hat\gamma):=-\trace(\hat\gamma \log \hat\gamma) ,
\ee
and that it is invariant under unitaries. By considering the density matrix expressed as in Eq.~(\ref{eq:faithGauss}), we notice that we can first apply local unitaries (namely displacement operators) so to put ${\bf \bar X}_\rho=0$. Then, by taking the matrix logarithm, we find 
the expression for the von Neumann entropy of a (faithful) Gaussian state to be
\be\label{eq:vnentrforfaith}
S(\hat\gamma[H])= \frac 1 2 \log \det\left(\frac{\Gamma +i\Omega}{2}\right) + \frac 1 2 \sum_{k,l}{\rm arcoth}(i\Gamma \Omega)_{k,l}(i\Omega \Gamma)_{l,k} ,
\ee
and in terms the symplectic eigenvalues of the covariance matrix it reads
\be\label{eq:entrfromsimpl}
S(\hat\gamma[H])=\sum_{k=1}^N 
    \left[ \left(\frac{d_{k} +1} 2 \right)\log \left(\frac{d_{k} +1} 2 \right)
    -\left(\frac{d_{k} -1} 2 \right)\log \left(\frac{d_{k} -1} 2 \right) \right] .
\ee
For a thermal covariance matrix with $\beta>0$, 
we can rewrite this expression in terms of normal mode frequencies:
\be\label{eq:1modeentropy}
S(\beta) = \displaystyle\sum_k \left[\frac{\beta \omega_k e^{-\beta \omega_k}}{1- e^{-\beta \omega_k}} - \log \big( 1- e^{-\beta \omega_k} \big) \right] .
\ee
Recall that $F(\cdot)= \trace(H \cdot) - S(\cdot)/\beta$ is the \emph{non-equilibrium free energy} of the state relative to its surrounding ambient temperature $\beta^{-1}$ and its corresponding Hamiltonian $H$.
In the case of thermal states, the free energy is given as
\be\label{eq:freenergymodes}
F(\beta) = \trace(\hat H \hat \gamma_\beta[ \hat H])-\beta^{-1} S(\beta)=\beta^{-1}\displaystyle\sum_k \log (1- e^{-\beta \omega_k}) . 
\ee
Given two faithful Gaussian states $\hat\gamma$ and $\hat \sigma$, each on $N$ bosonic modes, described by covariance matrices $\Gamma$ and
$\Upsilon$ respectively, it is also easy to compute their 
relative entropy according to
\begin{equation}\label{eq:relentapp}
    S(\hat \gamma\|\hat \sigma) = -S(\hat \gamma) 
    - \trace(\hat \gamma\log\hat \sigma) = \trace\left(\hat \gamma(\log \hat \gamma-\log\hat \sigma)\right) ,
\end{equation}
essentially because again it is easy to compute the logarithm of such states. Since the first term is nothing but the negative von-Neumann entropy that can be computed according to Eq.\ (\ref{eq:entrfromsimpl}), we can see this by just considering the second term. By considering the form (\ref{eq:faithGauss}) of faithful Gaussian states,
we obtain
\be
\trace(\hat \gamma\log\hat \sigma) = -\log Z_\sigma + \trace\left[\hat \gamma \left( -\tfrac 1 2 ({\bf \hat X}-{\bf \bar X_\sigma})^T H_\sigma ({\bf \hat X}-{\bf \bar X_\sigma}) \right)\right] ,
\ee
where we have simply used the fact that logarithm and exponential of a matrix are inverse functions.
We can also simplify further the above expression and write it just in terms of (combinations of) covariance matrices elements as
\be
-\trace(\hat \gamma\log\hat \sigma) = \frac 1 2 \log \det(({\Upsilon} +i\Omega)/2) + \frac 1 4 \sum_{k,l}{\Upsilon}_{k,l} (H_\sigma)_{k,l} + \frac 1 2 ({\bf \bar X_\sigma}-{\bf \bar X_\rho})^T H_\sigma ({\bf \bar X_\sigma}-{\bf \bar X_\rho}) ,
\ee
which leads to 
\be
S(\hat \gamma\|\hat \sigma) = \frac 1 2 \left[ \log \left(\frac{\det(({\Upsilon} +i\Omega)/2)}{\det((\Gamma +i\Omega)/2)} \right) + \frac 1 2 \sum_{k,l}\Gamma_{k,l} (H_\sigma-H_\rho)_{k,l} + ({\bf \bar X_\sigma}-{\bf \bar X_\rho})^T H_\sigma ({\bf \bar X_\sigma}-{\bf \bar X_\rho}) \right],
\ee
where we have also used the expression (\ref{eq:vnentrforfaith}) for the von Neumann entropy. Note once more that the matrices $H_\rho$ and $H_\sigma$ can be also directly obtained from $\Gamma$ and $\Upsilon$ respectively through Eq.~(\ref{eq:relHgammaandback}).
Another useful expression can be written down, containing explicitly the symplectic eigenvalues of the two covariance matrices. For that we notice that the logarithm of the partition function $Z_\sigma$ can be also expressed as
\be
\log Z_\sigma= \frac 1 2 \sum_k \log\left((\upsilon^2_k -1)/2 \right) ,
\ee
where $\{\upsilon_k\}$ are the symplectic eigenvalues of $\Upsilon$.
Thus, we can write
\be
S(\hat \gamma\|\hat \sigma) = - S(\hat \gamma) + \sum_k \log\left((\upsilon^2_k -1)/2 \right)+ \frac 1 4 \sum_{k,l}\Gamma_{k,l} (H_\sigma)_{k,l} + \frac 1 2 ({\bf \bar X_\sigma}-{\bf \bar X_\rho})^T H_\sigma ({\bf \bar X_\sigma}-{\bf \bar X_\rho}) ,
\ee
where we can also use the expression (\ref{eq:entrfromsimpl}) for $S(\rho)$. Finally, note that if $\sigma$ is a true thermal state of a Hamiltonian $H$ at inverse temperature $\beta>0$, then for any state $\rho$, we have
\be
S(\hat \gamma\|\hat \sigma) = \beta (F(\hat \gamma)-F(\hat \sigma) ).
\ee

\section{Details of the QTP simulations}\label{app:QTPsimdetails}

\subsection{Lattice discretization scheme}
\label{app:Discretization}
Here we define a lattice version of the phononic Hamiltonian, obtained by discretising the interval $[-\RD,\RD]$ into $\Npxl$ 
pixels, each of size $\Delta z=2\RD/\Npxl$ \cite{Mora03,JavanainenPhononApproach}. This is particularly important to make numerical calculations, especially for the case of non-homogeneous external potentials.
Fixing $\Npxl$, for $i=1,\ldots, \Npxl+1$ the coordinates of the discretization lattice read $z_i = -\RD + 2 \RD \frac{i-1} {\Npxl}$, and we define discretization pixels which are the closed intervals $p_i=[z_i,z_{i+1}]$ for $i=1,\dots,\Npxl$.
We then introduce the discretized version of density and phase operators as the integration of the field operators via 
\begin{align}
 \discpp_i =  {\frac{ 1 }{\Delta z} }\int_{p_i} \di z\ \pp( z),
\qquad \discdd_{i} = {\frac{ 1 }{\Delta z} }\int_{p_i} \di z\  \dd( z),
\label{eq:x_discrete2}
\end{align}
with $|p_i|= \Delta z= {2 \RD}/{\Npxl}$. 
These discretized operators yields a vector of canonical coordinates 
\begin{align}
{\bf \hat X} =\left(\discdd_1\ldots\discdd_{\Npxl},\discpp_1,\ldots \discpp_{\Npxl}\right)^T,
\end{align}
satisfying (re-scaled) bosonic canonical commutation relations  
\begin{align}
[\hat X_j, \hat X_k]=\i  \Omega_{j,k}/\Delta z\ ,
\end{align}
where $\Omega$ is defined in Eq.~\eqref{eq:symplectic_mtrx}.
As explained in Ref.~\cite{quantumreadout}, in the continuum limit $\Npxl \rightarrow\infty$, the right-hand side will yield a Dirac delta because $1/\Delta z$.

To discretise the model, we follow Ref.~\cite{Mora03} and consider the geometric mean 
\begin{align}
\eta_i = \sqrt{{ \nGP(z_i)  \nGP(z_{i+1})}}
\end{align}
for $i=1,\dots,\Npxl$. The discretization of the effective model will be a quadratic operator in the discretised modes $\discpp_i$ and $\discdd_i$. At the lowest order approximation, one obtains 
\begin{align}\label{eq:discretHamilt}
\hat H&\approx \Delta z \sum_{i=1}^{\Npxl-1} \frac{\hbar^2\eta_i }{2m}\left[\frac{ \discpp_i-\discpp_{i+1}}{\Delta z}\right]^2 +\Delta z \sum_{i=1}^{\Npxl} \frac{g(z_i)} 2 \left(\discdd_i{}\right)^2  
  =:\hat H_{\Npxl} .
\end{align}
Note that so far in the main text, we have suppressed for simplicity the possible spatial dependence of the coupling constant $g$, which is true for a homogeneous quasi-condensate and in general has little influence.
In general $g$ depends on the Gross-Pitaevskii profile,
\begin{align}g(z) =\hbar \omega_\perp a_s {(2 + 3 a_s \nGP(z))}/{ (1+2a_s\nGP(z))^{3/2}}
\end{align}
where $\omega_\perp$ is the radial trapping frequency and  $a_s$ is the scattering length \cite{Rauer2018,Salasnich2002}.
This dependency on the spatial coordinate $z$ has been included in our numerical simulations.
From this, we obtain the matrix representation of the above Hamiltonian
\begin{align}
   H &= \frac12 \cdot {\bf \hat X}^\t \left( H_{\rho\rho}[g,\Delta z] \oplus H_{\phi\phi}[\nGP,\Delta z] \right) {\bf \hat X} \ ,\label{eq:HamLapp}\\
    H_{\rho\rho}[\Delta z] &=  \Delta z \cdot {\rm diag} \big(g(z_1),g(z_2),\cdots,g(z_{N_c})\big)\ ,\\
    H_{\phi\phi}[\nGP,\Delta z] &= \frac{\hbar^2 }{m \Delta z } 
  \begin{pmatrix} 
  \eta_1 &-\eta_1\\
  -\eta_1 &\eta_1+\eta_2 &-\eta_2\\
  &&\ddots\\
 &  & -\eta_{\Npxl-2} &\eta_{\Npxl-2}+\eta_{\Npxl-1} &-\eta_{\Npxl-1}\\
& && -\eta_{\Npxl-1} &\eta_{\Npxl-1}
  \end{pmatrix}+2 \hbar \  \diag\left(J(z_1)\eta(z_1),\ldots, J(z_\Npxl)\eta(z_\Npxl)\right),
\end{align}
where we have used the functional notation $H_{\phi\phi}[\nGP, \Delta z]$ to emphasize that these couplings depend on the mean-field density profile and the size of the pixels.
We additionally added a small term $\propto J$ which is meant to regularize the zero-mode. This way, all computations are made with fully-supported Gaussian states so that numerical instabilities do not occur.
Physically, it can be interpreted as adding a small mass term of the type $\hat H_J = h J \int {\rm d} z \nGP(z) \pp(z)^2$ and we have checked that, as long as the coupling is chosen to be around $J\approx 0.01$ the dynamics is not affected in the times scales of \SI{300}{\milli\second} that we have in mind. See also Sec.~\ref{app:zero_mode} below for a more extended discussion.

Starting from a set of canonical coordinates $\discq$, then for a symplectic $M\in\R^{2\Npxl\times 2\Npxl}$, 
i.e., fulfilling
\begin{equation}
 M\, \Omega\, M^T = \Omega, \end{equation}
we have that ${\bf \hat{r}} = M\discq$ will again denote a vector of canonically commuting operators if 
which can be seen by explicitly checking that $\hat{r}$ again fulfills  $[\hat{r}_j,\hat{r}_k ] = \i \Omega_{j,k} / \Delta z$.

We can then diagonalize our Hamiltonian as follows: First, we use the symplectic matrix
\begin{equation}
 M_1 = \begin{pmatrix}
 \sqrt{H_{\rho\rho}^{-1}}&0\\ 0&\sqrt{H_{\rho\rho}} \end{pmatrix} = M_1^T,
 \end{equation}
since $H_{\rho\rho}$ is diagonal. Then, we have
\begin{equation}
 M_1^{T} H M_1 = \id_{\Npxl} \oplus \left(\sqrt{H_{\rho\rho}^{-1}} H_{\phi\phi} \sqrt{H_{\rho\rho}}\right) \ =: \id_{\Npxl} \oplus \tilde H_{\phi\phi},
\end{equation}
where $\tilde H_{\phi\phi}$ is the matrix of the phase couplings in the new coordinates, which is real and symmetric, and therefore can be diagonalized by an orthogonal transformation $O$ with $\tilde H_\phi= O\Sigma O^T$. 
 Here, $\Sigma$ is diagonal and we assume that all zero eigenvalues are sorted to the first $\Npxl^0 \geq0$ positions, i.e., $\Sigma = 0_{\Npxl^0}\oplus \tilde{\Sigma}$ with $\tilde{\Sigma}> 0 $ diagonal and we define the eigenfrequencies $\omega$ via $\tilde{\Sigma}^{1/2} = \mathrm{diag}(\omega_{\Npxl^0+1},\dots,\omega_{\Npxl})$.
 With the diagonal matrix  $\Sigma_\phi = \id_{\Npxl^0} \oplus \tilde{\Sigma}$ and the transformation
 \begin{equation}
  M_2 = \begin{pmatrix}O\Sigma_\phi^{1/4}&0\\0& O\Sigma_\phi^{-1/4}\end{pmatrix}
 \end{equation}
 we obtain
 \begin{equation}
  M_2^TM_1^T H M_1 M_2 = (\id_{\Npxl^0} \oplus \tilde \Sigma^{1/2}) \oplus (0_{\Npxl^0} \oplus \tilde \Sigma^{1/2}) \ .
 \end{equation}
That is, in the canonical coordinates ${\bf \hat{r}} = (\hat Q_1,\dots,\hat Q_{\Npxl},\hat P_1,\dots,\hat P_{\Npxl})= \sqrt{\Delta z}(M_1 M_2)^{-1}\discq$ we have that the Hamiltonian in Eq.~\eqref{eq:HamLapp} takes the form
\begin{equation}\label{eq:discrdecoupled}
 \hat H = \frac12 \sum_{j = 1}^{\Npxl^0} \hat Q_j^{2} + \frac12\sum_{j=\Npxl^0+1}^{\Npxl} \omega_j (\hat P_j^{2}+\hat Q_j^{2}) .
\end{equation}
Finally, we can define creation/annihilation operators  $(\hat c^\dagger_j,\hat c_j)$ for each normal mode from the relation
\begin{equation}
    \hat P_j^{2}+ \hat Q_j^{2} = 2\hat c_j^\dagger \hat c_j + \hat \id  .
\end{equation}
 Note that the new coordinates satisfy true canonical commutation relations $[\hat Q_k,\hat P_l] = i \delta_{k,l}$ and consequently we also have
\begin{equation}
    [\hat c_k, \hat c_l^\dagger] = \delta_{k,l} ,
\end{equation}
for all $k,l$.
However, our original discretized field operators satisfy re-scaled commutation relations.
This means that the symplectic matrix corresponding to the evolution with $\hat H_{\Npxl}$ in the original coordinates is given by Eq.~(\ref{eq:symplmatunit}), where the symplectic form is re-scaled, namely $\Omega\mapsto (\Delta z)^{-1} \Omega$.

Thermal states of the above Hamiltonian have covariance matrices of the form $\Gamma = \Gamma_{\rho\rho} \oplus \Gamma_{\phi\phi}$, and can be also explicitly computed from the normal modes and the corresponding symplectic transformation, namely Eqns.~\eqref{eq:normal_modes} and \eqref{eq:sympeigcovtherm}. The expression is somewhat complicated for the general case, but for the special case of homogeneous systems (which we will be interested in) $H_{\rho \rho} = \kappa \id$ with $\kappa:= \Delta z g$, 
we get
\be
\Gamma_\beta =  \frac 1 {\sqrt{\kappa}} H_{\phi\phi}^{1/2} \oplus \sqrt{\kappa} H_{\phi\phi}^{-1/2} +\frac 1 {\sqrt{\kappa}}(H_{\phi\phi}^{1/2} T) \oplus \sqrt{\kappa}(H_{\phi\phi}^{-1/2}T) ,
\ee
where 
\be
T:=2\left( \exp(2\beta \sqrt{\kappa}H_{\phi\phi}^{1/2}) -\id \right)^{-1}. 
\ee
As discussed in the previous Sec.~\ref{app:GaussianApp}, diagonalizing the Hamiltonian in terms of normal modes, the covariance matrix becomes also diagonal with symplectic eigenvalues given by Eq.~(\ref{eq:sympeigcovtherm}). From these symplectic eigenvalues one can also write down the the (von Neumann) entropy and the free energy as in Eqs.~(\ref{eq:1modeentropy}, \ref{eq:freenergymodes}).

\subsection{Details of merging and splitting}\label{app:Mergingdetails}

In this section, we provide an extended discussion of the merge/split primitive. This is a three-step process, involving two condensates $A$ and $B$ with lengths $L_A$ and $L_B$ and densities $\nGP^A$ and $\nGP^B$, and consisting in
\begin{itemize}  \setlength\itemsep{0em}
    \item[(a)] merging the two initially independent condensates during a time $t_{\rm merge}$,
    \item[(b)] letting them evolve with the fully merged Hamiltonian for a time $t_{\rm evolve}$, and
    \item[(c)]  splitting the joint condensate back into two parts $A$ and $B$, with the same lengths as the initials, during a time $t_{\rm split}$.
\end{itemize}

\subsubsection{Merging}
For the merging process, we encounter 
a time dependent Hamiltonian $\hat H_{A-B}(t)$ such that 
\begin{equation}
\hat H_{A-B}(0)= \hat H_{\Npxl^A}[\nGP^A]+\hat H_{\Npxl^B}[\nGP^B] , 
\label{eq:H_uncoupled}
\end{equation}
where our Hamiltonians are given by the lattice model in Eq.~(\ref{eq:discretHamilt}) and (keeping constant the small distance cutoff $\Delta z$) are functionals of the initial mean-field density profiles of the two condensates. 
Note that since we \je{would like to} to couple the two systems, we \je{require} them to have a consistent momentum cutoff $(\Delta z)^A=(\Delta z)^B=\Delta z$ (so that waves traveling across quasi-condensates with same atom density in the simulation should not change in speed due to the different discretization), and consequently their number of pixels will be in the same proportion as their lengths, i.e., 
\begin{equation}
    \Npxl^A=L_A \Delta z = \Npxl^B  L_A /L_B .
\end{equation}
The coupling matrix of the uncoupled Hamiltonian  in Eq.~\eqref{eq:H_uncoupled} is by
\begin{align}
H_{\rho\rho,A|B} &= H_{\rho\rho, A} \oplus H_{\rho\rho,B}, \qquad
H_{\phi\phi,A|B} = H_{\phi\phi, A} \oplus H_{\phi\phi,B} \ .
\end{align}

To merge the condensates, an interaction Hamiltonian is switched on, so that the joint Hamiltonian as in Eq.~\eqref{eq:HamLapp} has a matrix representation given by
\begin{align}
H_{\rho\rho,AB} &= H_{\rho\rho, A} \oplus H_{\rho\rho,B}, \qquad
H_{\phi\phi,AB} = H_{\phi\phi, A} \oplus H_{\phi\phi,B} + \frac{t}{t_{\rm merge}} H_{\rm int}, 
\end{align}
where the interaction matrix is given by 
\begin{equation}
    (H_{\rm int})_{i,j} = \frac{\hbar^2 }{2m \Delta z } \eta_{\Npxl^A} \left(
    \delta_{\Npxl^A,i} \delta_{\Npxl^A,j}
    +\delta_{\Npxl^A+1, i} \delta_{\Npxl^A+1,j}
    -\delta_{\Npxl^A, i} \delta_{\Npxl^A+1,j} - \delta_{\Npxl^A+1,i} \delta_{\Npxl^A,j}
    \right)\ . 
\end{equation}
with 
$\eta_{\Npxl^A}:=\sqrt{\rho^A(\Npxl^A)\cdot\rho^B(1)}$.
Note that this interaction contains also the local terms in the boundary region $[\Npxl^A,\Npxl^A+1]$.
We hence see that the couplings during the merging are given by 
\begin{equation}
H_{A-B}(t) =(1- \tfrac{t}{t_\text{merge}})H_{A|B} +\tfrac{t}{t_\text{merge}}H_{AB}.
\end{equation}

\begin{figure}
    \includegraphics[trim=7cm 19cm 4.5cm 20cm,width=0.85\textwidth]{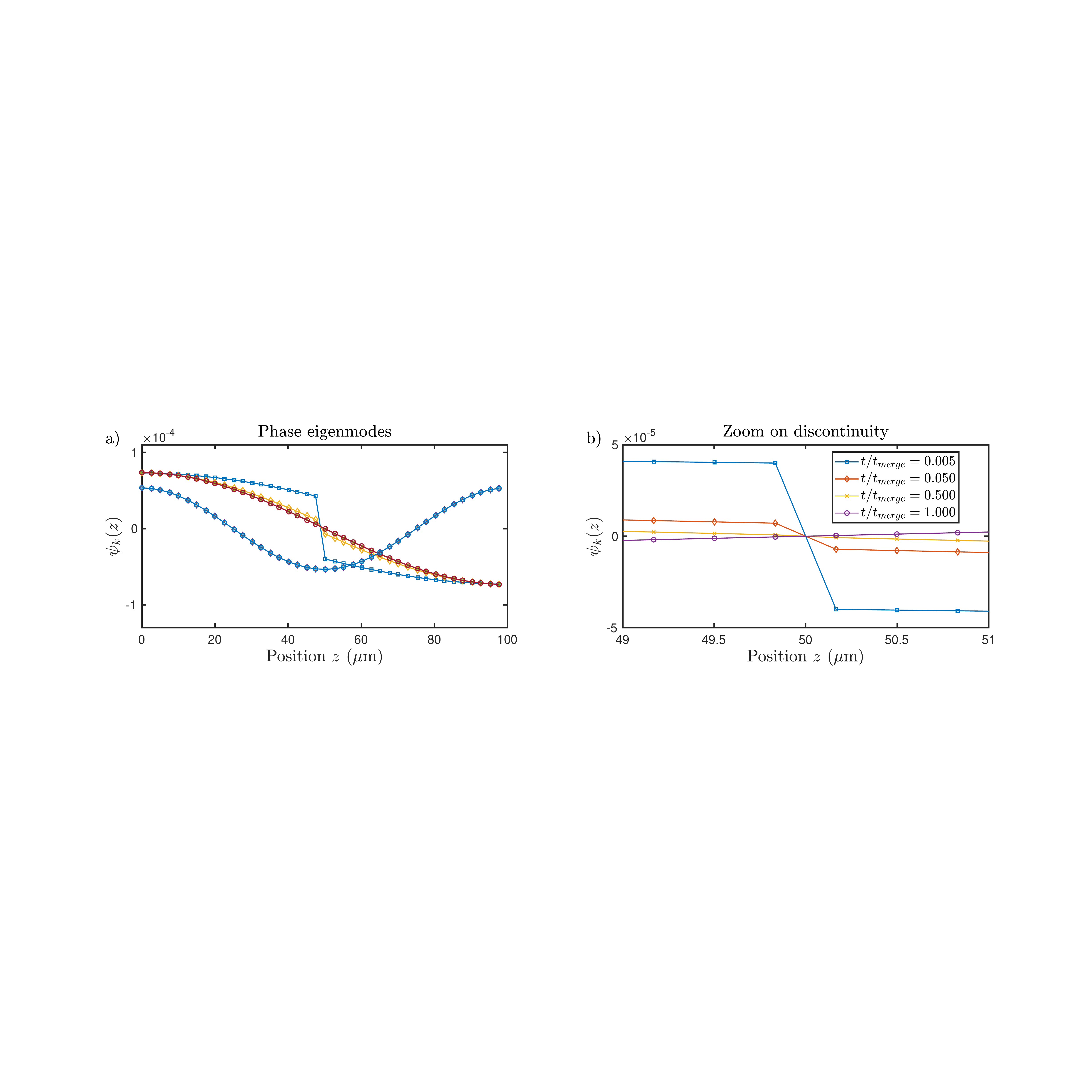}
    \caption{Snapshots of the phase and density eigenmode functions for the first and second lowest modes, taken at different Trotter steps $t/\Tm$ during merging for a fully homogeneous profile.}
    \label{fig:eigenmodes_homogeneous_trotterized}
\end{figure}
\begin{figure*}
 \includegraphics[trim=0cm 0cm 1cm 2cm,clip,width = \textwidth]{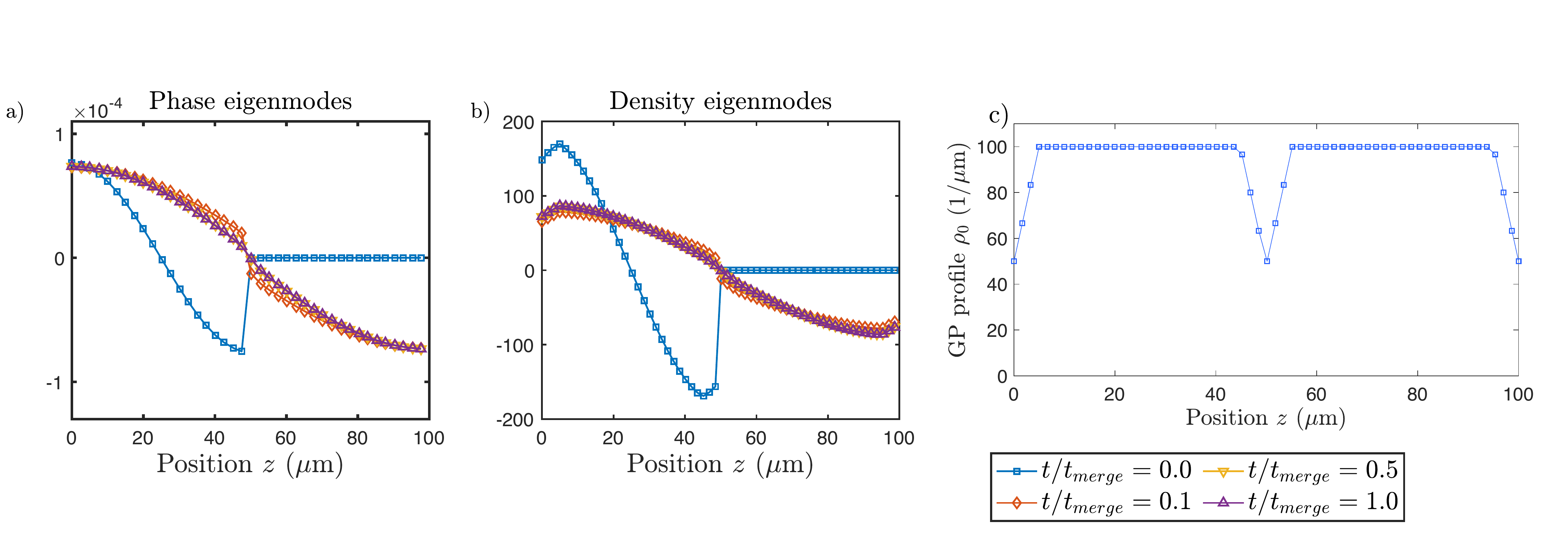}
    \caption{{Snapshots of the phase and density eigenmode functions for the two lowest lying modes taken at different Trotter steps $t/\Tm$ during merging for a homogeneous profile with a trapeze-like buffer region.}
    \emph{(a)} The odd modes during the coupling have a discontinuity of varying strength which diminishes as the merging proceeds.
    In contrast, there is little influence of the merging on the even modes as they can be obtained by connecting the odd modes of the individual uncoupled systems.
    \emph{(b)} The discontinuity is sharp, changing suddenly from one pixel to another.}
    \label{fig:eigenmodes_trapeze_trotterized}
\end{figure*}

For the numerical implementation, we also discretize the time evolution so that we divide the $[0,t_{\rm merge}]$ time interval into $\Ntro$ steps of duration $\Delta t = t_{\rm merge}/\Ntro$. Then, the symplectic evolution matrix reads
\begin{equation}
    G_{\rm merge}(t_{\rm merge}) = \displaystyle\prod_{j=1}^{\Ntro} \exp\left( \Omega H_{A-B}(t_j)/\Delta z \right) ,
\end{equation}
where $H_{A-B}(t) = H_{\Npxl^A}+ H_{\Npxl^B} + \frac j {\Ntro} H_{\rm int}$. 
Examples of eigenmodes for the time-dependent Hamiltonian are plotted in Figs.~\ref{fig:eigenmodes_homogeneous_trotterized}, \ref{fig:eigenmodes_trapeze_trotterized} for a homogeneous Gross-Pitaevskii profile with or without a trapeze-like buffer region.
 We find that mode functions that are odd in $z$ hybridize via a jump which gets smoothened during the merging while,  mode functions which are even in $z$ get glued automatically.
    We see also that all modes have a local extremum at the boundary which means we have Neumann boundary conditions.

In the main text we have shown the results for a model of the quasi-condensates where the Gross-Pitaevskii profile falls off smoothly from its peak value in the bulk to a lower value on the edges.
In principle, it is possible to consider the effective model to be constant everywhere, whereas the edge of the condensate (where excitations get reflected) can be modelled by the boundary conditions.
However, this abstraction turns out to be too simplistic. Fig.~\ref{fig:high_modes_merging} provides a demonstration of what occurs in such a scenario. Since the process is simulated via the merging of the boundary conditions of the two condensates, in particular occurring at a \emph{single} pixel, it is hence \emph{independent of the momentum cut-off}. As a result, momenta at all scales are populated, however, this does not faithfully capture the physics of the continuum model, since the dispersion relation is not linear. In order to avoid this, it is therefore necessary for the model to resolve details of the coupling zone. 

\begin{figure}
    \centering

    \includegraphics[width = 0.99\linewidth, trim=5cm 6cm 2.5cm 7.5cm,clip]{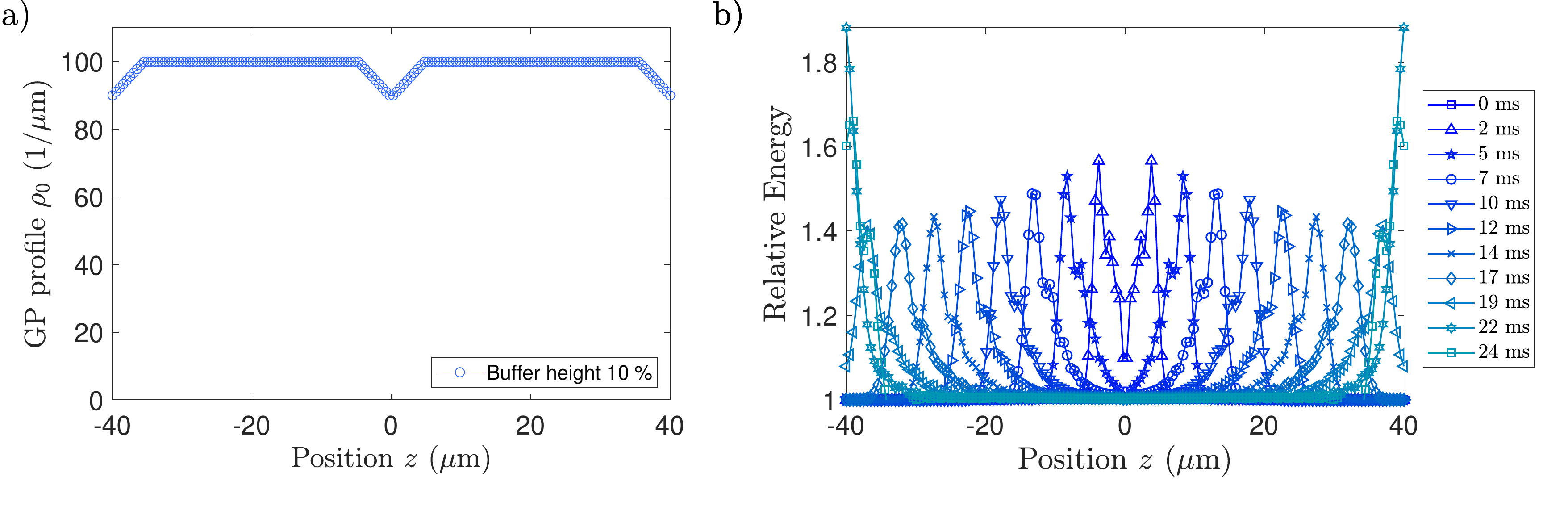}
    \caption{{The presence of high momentum modes in a merging model with non-extensive buffer region between condensates}. 
    \emph{ a)} We show the merging scenario similar to the figure in the main text with the difference that that the coupling zone is much smaller and ends more abruptly at $90\%$ of the peak value.
     \emph{ b)} We now find that the transport of the excitations is \emph{dispersive} which can be seen by the peaks of wave-packets falling down as their propagation.
     As the dynamics is modelled to be unitary and the Hamiltonian doesn't change in the bulk this means that the energy the wave-packets carry stays constant while being broadened.
         In lattice theories, dispersion can be proven analytically if short wavelengths are present in the state \cite{gge}.
    }
    \label{fig:high_modes_merging}
\end{figure}

\subsubsection{Idle evolution}
In-between merging and splitting, one can allow some idle evolution time $t_{\rm evolve}$ in which the joint system evolves with the fully coupled Hamiltonian. This can be applied with a single symplectic matrix, since the Hamiltonian is time-independent. Wavepackets injected during the previous merging process will travel ballistically through the entire joint system (as long as we taken care to remain in the regime where high-momentum modes are negligible and the linear dispersive relation holds).
\subsubsection{Splitting}
Finally, we implement the splitting procedure by a time-dependent Hamiltonian reversing the linear interpolation that has been discussed for merging.
In the numerical simulation the  covariance matrix of the $A$ and $B$ quasi-condensates after merging, idle evolution and splitting would have the form
\be
\Gamma_{A-B}(t_{\rm tot}) = G_{\rm split} G_{\rm evolve}G_{\rm merge} \Gamma_{A-B}(0)G_{\rm merge}^T G_{\rm evolve}^T G_{\rm split}^T\ .
\ee
For reasons discussed in the main text, whenever we simulate the Otto cycle, we neglect the correlations between the two parts of the split condensate at the end of the process. 
In other words, we project the final covariance matrix into the direct sum of the two local covariance matrices for systems $A$ and $B$, i.e., at the end of the full protocol, $t = t_{\rm tot} = t_{\rm merge}+t_{\rm evolve}+t_{\rm split}$, by setting
\be
\Gamma_{A-B}^{\rm fin}(t_{\rm tot}) = \Gamma_A(t_{\rm tot}) \oplus \Gamma_B(t_{\rm tot}),
\ee
where $\Gamma_A(t_{\rm tot})$ is the submatrix of $\Gamma_{A-B}(t_{\rm tot}) $ corresponding to the subsystem $A$ and $\Gamma_B(t_{\rm tot})$ corresponds to $B$.

\subsubsection{Energy density injected during merging QTP}
Given the quadratic Hamiltonian over time $\hat H_{A-B}(t)$ we would like to study also the spatial distribution of the energy. In the discretized models, it is natural to study the energy per pixel $z$, namely 
\begin{align}
 E(z,t)= E_z(t) \Delta z = \tfrac 1 2 \left(  H_{A-B}(t) \Gamma_{A-B}(t) \right)_{z,z} \Delta z ,
\end{align}
where the notation $(\cdot)_{z,z}$ refers to the diagonal matrix element at pixel $z$.
Note that if $\Delta z$ is constant, this amounts to just computing the quantities $E_z(t)$.
Plotting $E_z$ over pixel positions is then a way of visualizing which regions in space have more energy than others. Doing this over varying times can show us how energy flows over time from one part of the system to the other.
For example, in Fig.~\ref{fig:high_modes_merging}, we observe that merging the two systems amounts to inserting energy at their boundary continuously over the merging time (or in discrete bits at each Trotter time step).
This energy then flows through the system at speed of sound velocity (which is $c=\sqrt{g \nGP/m}\propto \sqrt{\nGP}$), reaching the external boundaries and then bouncing back toward the center. Thus, in particular, if the ratio between the coupling time $t_{\rm merge}$ and the length of a system (say $A$) is chosen such that 
\be
c=L_A/t_{\rm merge} ,
\ee
then the energy perturbation precisely reaches the external boundary of system $A$. Similarly for system $B$. Clearly, then, when the two lengths $L_A$ and $L_B$ are not equal the energy flow cannot be synchronized so that the perturbation wave bounces back to the interface from both external walls at the same time.
During the idle evolution time $t_{\rm evolve}$ no additional energy is injected, but the energy flow continues.
Finally, during the splitting process some energy is taken away from the system, again continuously over the splitting time and at the interface between the two parts. However, the total amount of energy taken back during the splitting is in general lower than the one inserted during merging. Hence, the total energy inserted during the entire protocol is always non-negative, and the amount is smaller given a protocol with longer time. 

As a last comment, we note that the fact that our simulations use discretized space and time also implies that, besides the fact that energy is injected and ejected at the interface in discretized bits over $t_{\rm merge}$ and $t_{\rm split}$, the energy flow also takes places in pixels over time steps. In particular, all of this imposes us once more for consistency to make sure that the two coupled systems have the same small distance cut-off $\Delta z$, which also ensures that the lengths in the two systems are in the same ratios between their number of pixels.
This issue becomes particularly important when a compression/expansion QTP takes place before merging since in that case, as we are going to discuss in detail in the next subsection, the cut-off $\Delta z$ changes in time.

\subsection{Regularization of the zero-mode: Phase locking via excitation tunneling}
\label{app:zero_mode}

Here we discuss more in detail the additional complications arising from the zero-modes of the phononic model and how to regularize them, in order to avoid instabilities during coupling. 
The mode expansion of Eq.~\eqref{eq:H_quench} reads 
\begin{align}
    \hat H_\text{P}[\nGP]=  \sum_{k>0}&\hbar \frac{\omega_k} 2 (\p_k^2+\d_k^2) + \frac g 2 \d^2_\text{ZM} ,
\end{align}
where $\omega_k$ are the eigen-frequencies of the phase and momentum eigenmodes $\pp_k,\dd_k$ and there is a special mode, called the zero-mode, $\dd_\text{ZM} \propto \int \di z \d(z)$ which is different from $k>0$ eigenmodes as the canonically conjugate quantity $\pp_\text{ZM} \propto \int \di z \p(z)$ does not appear in the Hamiltonian, i.e., it does not cost energy.
The zero mode has the interpretation of total momentum frame of the excitations~\cite{lewenstein1996quantum,JavanainenWilkens,LeggettComment,JavanainenReply}.

This mode expansion can be found in the continuum limit by solving the set of partial differential equation associated to the Heisenberg equations of motion, namely
\begin{equation}\label{eq:sturmliouvillecont}
\begin{gathered}
\left\{
\begin{array}{c}
     \partial_t \delta \varrho(z,t) =  \frac{\hbar}{m} \partial_z \left( \nGP(z) \partial_z \varphi(z,t) \right) \\
    \partial_t \varphi(z,t) =-\frac g \hbar \delta \varrho (z,t) 
\end{array}
\right. \imply \partial_t^2 \varphi_k(z,t)=- \frac{g}{m} \partial_z \left( \nGP(z) \partial_z \varphi_k(z) \right) ,
    \end{gathered}
\end{equation} 
and, as usual, for the $k>0$ modes we can look for solutions of the type $\varphi_k(z,t)=\varphi_k(z)e^{i\omega_k t}$, so that Eq.~\eqref{eq:sturmliouvillecont} becomes a Sturm-Liouville problem 
\begin{equation}
    \omega_k^2 \varphi_k(z) = -\frac{g}{m} \partial_z \left( \nGP(z) \partial_z \varphi_k(z) \right)  ,
\end{equation}
and similarly for $\delta \varrho(z,t)$. We can then find solutions which form an orthonormal basis with respect to the scalar product 
\begin{equation}
\langle f,g\rangle:= \int \de z f(z) g(z) ,
\end{equation}
i.e., we have $\langle \varphi_k(z), \varphi_l(z) \rangle =\langle \delta \varrho_k(z), \delta \varrho_l(z) \rangle = \delta_{k,l}$, where $\delta_{k,l}$ is the Kronecker delta. 

However, besides those one can also find a solution with $\omega_k=0$, which gives rise to the zero mode with quadrature operators denoted by$(\dd_\text{ZM}$ and $\pp_\text{ZM})$. 
These are necessary for the set of eigenmode functions to be complete and we can expand the field operators as 
\begin{equation}
    \begin{aligned}
    \dd(z,t) &= \dd_\text{ZM} + \sum_{k>0} \sqrt{\frac{\hbar \omega_k}{g}} \delta \varrho_k (z)(e^{i\omega_k t} \hat a_k^\dagger + e^{-i\omega_k t} \hat a_k), \\
    \pp(z,t) &= \pp_\text{ZM} -\frac g \hbar t\dd_\text{ZM} -i\sum_{k>0} \sqrt{\frac{g}{\hbar \omega_k}} \varphi_k (z)(e^{i\omega_k t} \hat a_k^\dagger - e^{-i\omega_k t} \hat a_k) ,
    \end{aligned} 
\end{equation}
and we define eigenmode operators at $t=0$ (with $k>0$) from the relations 
\begin{eqnarray}
\dd_k &=& \sqrt{\frac{\hbar \omega_k}{g}} (\hat a_k^\dagger + \hat a_k),\qquad
\pp_k = -i \sqrt{\frac{g}{\hbar \omega_k}} 
(\hat a_k^\dagger - \hat a_k), 
\end{eqnarray}
such that they obey canonical commutation relations $[\dd_k,\pp_l]=i \delta_{k,l}$ for all $k,l$.

Let us now consider the time evolution when coupling two systems governed by the Hamiltonian 
\begin{equation}
\hat H_{A-B}(t) = \left(1-\tfrac t\Tm\right)  \hat H_{A|B} + \tfrac t \Tm \hat H_{AB}
\end{equation}
for $t\in[0,\Tm]$. Note that now at each instant $t$ this Hamiltonian has implicitly different boundary conditions at the interface $z=0$. See also Figs.~\ref{fig:eigenmodes_homogeneous_trotterized},\ref{fig:eigenmodes_trapeze_trotterized} where the eigenmode functions of this time-dependent Hamiltonian are shown at different times $t/\Tm$ for the discretized model.

Thus, we see that, \gv{while coupling,} the zero-modes \gv{of the two systems} will hybridize to form the joint zero-mode and one mode that costs energy.
However, this energy cost will cause the coupled system to have enormous energy if the original phase zero-modes are non-trivially populated, which leads to an unstable time-evolution.
In this situation the \gv{lowest order phononic} model is not anymore a good approximation to the Lieb-Liniger model \eqref{eq:HfullBEC} as the density fluctuations may no longer be small.

Nevertheless, one can refine the model considered here to reflect more accurately the corresponding physical process: energy will change continuously, since when we couple the systems by ramping down the separation barrier, there will be an additional term in the Hamiltonian, representing tunneling between the condensates.
The density phase expansion of this term will additionally give rise to a term of the type $\hbar J\cos(\Delta\pp)$ penalizing phase fluctuations $\Delta\pp=\pp_L-\pp_R$ ranging over the interface.
The action of this term is to induce \emph{phase-locking} between the two condensates being merged together, see Refs.~\cite{Rauer2018,Schweigler_thesis,Pigneur2018,Pigneur2018PRA} for experimental discussions and references therein for the theoretical overview.
The large coupling expansion of this term motivates the effective model we used in the numerical simulations
\begin{align}
    \hat H[\nGP]&=\hat H_\text{P}[\nGP]+   h \int\di z \   J(z) \nGP(z) \pp(z)^2 \ .
    \label{eq:H_JJJ}
\end{align}

\subsubsection{Analytical derivation of gapping-out
the zero-mode in the homogeneous phase-locking model}

\begin{figure*}[t]\centering
 \includegraphics[trim= 2cm 10cm 2cm 8cm, width=\textwidth]{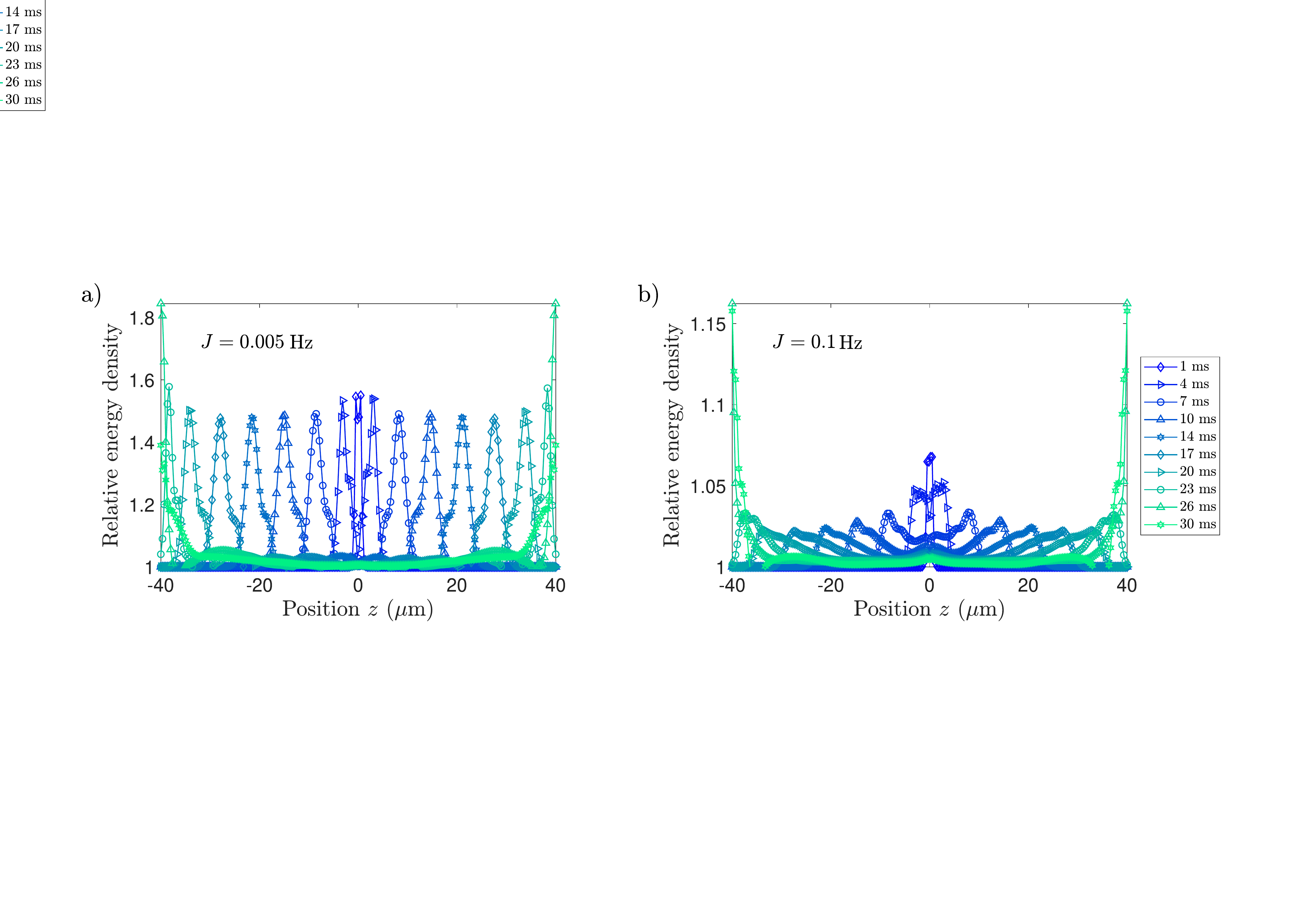}
    \caption{{Coupling of two quasi-condensates for different initial phase-locking.} 
    Similarly to the main text we consider the phase locking to act with constant strength along each of the condensates and show the influence of other values of $J$ on the outcome of merging. 
    \emph{(a):} For a low value of $J = \SI{0.005}{Hz}$ there is substantially more excitations compared to the value $J = \SI{0.01}{Hz}$ used for all plots presented in the main text.
    \emph{(b):} On the other hand for larger values of phase-locking such as $J = \SI{0.1}{Hz}$ the excitations become suppressed as the phase zero-mode acquires a larger energetic penalty and its initial thermal second moments are smaller.
    Note, that when increasing the tunnel coupling $J$ further, one expects a non-Gaussian regime due to non-negligible interactions stemming from the full cosine potential \cite{Schweigler2017}.
    }
    \label{fig:merge_equal_density_different_J}
\end{figure*}

\begin{figure*}[h!]\centering
 \includegraphics[ width=\textwidth]{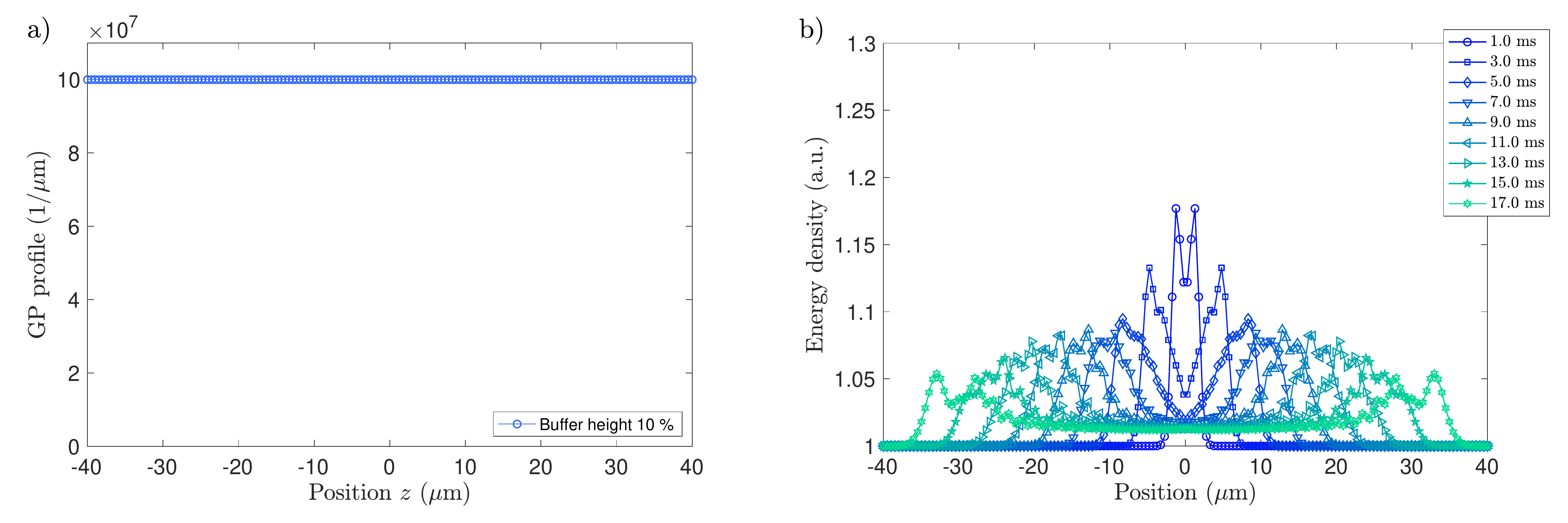}
    \caption{{Merging of  homogeneous systems where the zero mode is artificially removed from the evolution.}
    Using the eigenmode decomposition for $J=0$ we set $\dd_0=\pp_0=0$ in the  Hamiltonian that governs the merging and compute the initial state using a pseudo-inverse disregarding the zero-mode. We see a behaviour, both qualitative and quantitative, similar to merging with a regularization coupling chosen as $J=\SI{0.01}{\hertz}$.}
    \label{fig:merge_equal_density_no_zeromode}
\end{figure*}
In the experiment the phase-locking term will be acting around the interface.
For the case of large extension of this coupling (or two side-ways coupled systems \cite{Rauer2018,Whitlock03}) it is instructive to  consider $J = \text{const}$ throughout the condensates.
In this case additionally taking $\nGP= \text{const}$ we can analytically see that this term effectively gaps out the phase zero-mode.
One way to see this it by noticing that this term amounts to add a (small) ``mass'' term to Eq.~\eqref{eq:sturmliouvillecont}, leading to the modified Sturm-Liouville problem 
\begin{equation}
    \omega_k^2 \varphi_k(z) = - \frac{g}{m} \partial_z \left( \nGP(z) \partial_z \varphi_k(z) \right) + 2J \rho(z) \varphi_k(z) ,
\end{equation}
which effectively removes the zero-mode. 

Let us show this specifically in the case when all coupling constants do not vary over the condensate of length $L$, 
i.e., 
$g(z)=g, \rho_0(z)=\rho_0$ and $J(z) = J$.
The Hamiltonian then reads
\begin{align}
    \hat H &=\int_{0}^{L} dz \left[\frac{\hbar^2 \nGP}{2m}(\partial_z \pp(z))^2+\frac g 2 \dd(z)^2 + \hbar J\nGP \pp(z)^2\right],
\end{align}
and has no zero-modes unless $J=0$.
In this case the eigenfrequencies read 
\begin{align}
    \omega_k =  \frac{\pi c k}{L}
\end{align}
with the speed of sound given by $\sqrt{g\nGP / m}$.
To bring the Hamiltonian to the normal form, we define the squeezing constants
\begin{align}
    \alpha_k = \sqrt{ \frac{\hbar \omega_k}{g}} +\delta_{k,0}
\end{align}
from which we define for $k>0$
\begin{align}
\dd_k = \alpha_k\sqrt{\frac 2L} \int_0^L \di z \cos(\pi k z/L ) \d(z)\quad\text{and} \quad \pp_k = \alpha_k^{-1}\sqrt{\frac 2L }\int_0^L \di z \cos(\pi k z/L ) \p(z)
\end{align}
and
\begin{align}
\dd_0 := \sqrt{\frac 1 {2L}} \int_0^L \di z  \d(z)\quad\text{and} \quad \pp_0 := \sqrt{\frac 1 {2L}}\int_0^L \di z  \p(z)
\end{align}
which stand out by having different normalization constants and would be the zero-mode operators for $J=0$. 
Using standard trigonometric integrals we find
\begin{align}
  \int_{0}^Ldz \dd(z)^2&
=\sum_{k=0}^\infty \alpha_k^2 \d_k^2
\quad\text{and}\quad
\int_{0}^Ldz (\partial_z \pp(z))^2 = \sum_{k>0} \frac {\pi^2 k^2}{L^2\alpha^2_k} \p_k^2  
\quad\text{and}\quad
\int_{0}^Ldz \pp(z)^2=\sum_{k=0}^\infty \alpha_k^{-2} \p_k^2 \ .
\end{align}
Therefore,
\begin{align}
\hat H
&=\sum_{k=1}^\infty\frac{\hbar \omega_k}{2} \left[ \d_k^2 + \p_k^2\right] +\frac g 2 \d_0^2+\hbar J\nGP\sum_{k=1}^\infty \alpha_k^{-2}\p_k^2+\hbar J \nGP \pp_0^2 \ .
\end{align}
Further defining
\begin{align}
\zeta_k := \frac{4 gJ\nGP}{\hbar^2\omega_k^2}
\end{align}
for $k>0$ we obtain the form
\begin{align}
\hat H=\sum_{k=1}^\infty\frac{\hbar \omega_k}{2} \left[ \d_k^2 + (1+\zeta_k )\p_k^2\right] + \frac g 2\d_0^2 +\hbar J\nGP \p_0^2 \ .
\end{align}
Thus, the $k=0$ eigenmode of $\hat H$ has the eigenfrequency $\omega_{k=0} = \hbar \sqrt{ g J\nGP }$ and is not a zero-mode when $J\neq 0$.
We also see that there is additionally a squeezing interaction which decays for $k\rightarrow\infty$.
Fig.~\ref{fig:merge_equal_density_different_J} shows plots of 
merging for different values of $J$ when assuming that the phase-locking term acts homogeneously in space.
Fig.~\ref{fig:merge_equal_density_no_zeromode} show merging obtained by artificially removing the zero-mode, in order to highlight its contribution to the excitations present during the merging.

\subsubsection{Justification of the phase-locking model}
Finally we provide a justification for the phenomenological model above.
The argument will be based on the theoretical observation from Ref.~\cite{KaganeEtAl} that a potential barrier is effectively transparent for low-frequency excitations.
This hints that we can phase-lock systems in order to reduce the impact of excitations coming from zero-mode coupling and once this is done one can reduce the barrier further to increase heat transmission.

We consider a quasi-condensate of mean-density $\nGP$ in a box of length $2L\gg 2a$ (below we set $L \rightarrow \infty$ for simplicity) with a barrier extending from $z=-a$ to $z=a$ and having a finite height which exceeds the chemical potential by $U_B$.
Moreover, we assume that tunnel coupling between zero modes of the 
left and right quasi-condensates is negligible. 
This means that the background density and low-energy excitations feel a hard wall at $z=-a$ for the left quasi-condensate and at $z=a$ for the right quasi-condensate. 
For concreteness, let us focus on the left quasi-condensate, and it is clear that similar considerations apply also for the right one.
The background solution in the bulk (far from the leftmost end) is $\Psi _{0,L} (z)=\sqrt{\nGP}\tanh [-( z+a)/\xi_h]$, $z<-a$, where  
$\xi_h =\hbar /(mc)$ is the healing length. 
Considering the first order in matter field fluctuation  $\hat \Psi_L =  \Psi_{0,L}\hat \id+\delta\hat\Psi_L $  
around the full stationary solution the Hamiltonian term corresponding to atom scattering becomes (neglecting a costant term)
\begin{align}
  \hat V_L = 2 g \int _{-\infty}^{-a} \di z \, 
|\Psi_{0,L}|^2 \delta \hat \Psi_L ^\dag (z) \delta \hat \Psi_L (z) \ .
\end{align}
This term, because of the large gradient of $\Psi_{0,L}$ in the two bulks, 
couples low-energy excitations to high-energy ones. 
The former can be represented as 
$\delta \hat \Psi_L (z,t)\tanh [-( z+a)/\xi_h]$, where $\delta \hat \Psi_L(z,t)$ 
is subject to Neumann 
boundary conditions at $z=-a$.
The factor $\tanh [-( z+a)/\xi_h]$ follows from considering the adiabatic solution of the time-dependent GPE for excitations with a frequency 
much lower than $g\nGP/\hbar $ and makes the fluctuation vanishing at the wall. 
Let us now consider the propagation of high-energy excitations.
The high-energy, particle-like excitations propagating from the left ($>$) or from the 
right ($<$) are parametrized with the following set of orthogonal functions:
\begin{align}
\psi _k^>(z)\sim \left \{ \begin{array}{ll} 
e^{ikz}+i \sin \beta _k e^{i\alpha _k}e^{-ikz},& z<-a \\ 
\cos \beta _k e^{i\alpha _k }e^{ikz}, & z>a\end{array}\right. ~~~, \qquad \psi _k^<(z)\sim \left \{ \begin{array}{ll} 
\cos \beta _k e^{i\alpha _k }e^{-ikz},& z<-a \\ 
e^{-ikz}+i \sin \beta _k e^{i\alpha _k}e^{ikz},& z>a\end{array}\right. ~. 
    \end{align}
Here, $\alpha _k, \, \beta _k$ parametrize the transmission and reflection amplitudes ($k>0$)
and we also have $\langle \psi _k^\varsigma |\psi _{k^\prime }^{\varsigma ^\prime }\rangle = 
\delta _{\varsigma ^\prime ,\varsigma }\delta (k^\prime -k)$,~~~${\varsigma ^\prime , \, \varsigma}=~>,\, <$. 

We can expect that $\cos \beta _k$ rapidly increases from almost 0 to almost 1, when $k$ approaches 
$q_B =\sqrt{2mU_B}/\hbar $. 
We apply a perturbative approach, whit the Hamiltonian with hard walls at $z=\pm a$ being the unperturbed Hamiltonian 
and the Hamiltonian with the barrier of a finite height being the perturbed one. The second-order approximation 
yields the following term coupling low-energy excitation fields in the left and right quasicondensates: 
\begin{align}
\hat H_{LR}=-\int _{-\infty }^{-a}dz \int _a^\infty dz^\prime \, {\cal J}(z,z^\prime )[\delta \hat \Psi ^\dag _L(z)
\delta \hat \Psi _R(z^\prime)  + \mathrm{H.c.}],
\end{align}

where the effective coupling coefficient is
\begin{align}
{\cal J}(z,z^\prime )=\frac {4(g\nGP)^2}{U_B}
\tanh ^3\left( -\frac {z+a}{\xi _h}\right) \tanh ^3\left( \frac {z^\prime -a}{\xi _h}\right) 
\int _{q_B}^\infty \frac {dk}{\pi }\, \frac{q_B^2}{k^2}\{ \cos [k(z+z^\prime )]+\cos [k(z-z^\prime )-\alpha _k-\beta _k]\} .  
\end{align}




In the harmonic approximation, we replace 
$\delta \hat \Psi_R^\dag \delta \hat \Psi_L \approx \frac 12 \rho_0 (\hat \varphi_L -\hat \varphi_R)^2$ (after neglecting density fluctuations~\cite{popov2001functional}) obtaining
\begin{align}
\hat H_{LR} 
&\approx {\rm const} + \frac 1 2 \int _{-\infty }^{-a} dz\int ^{\infty }_a dz^\prime  \, 
{\cal J}(z,z^\prime )\nGP(\pp(z)-\pp(z') )^2
\\
& \approx \frac 1 2 \int _{-\infty }^{-a} dz \, 
\tilde{\cal J}(z )\nGP \pp(z)^2+\frac 1 2 \int ^{\infty }_a dz \tilde{\cal J}(z ) \nGP\pp(z)^2
\ -\int _{-\infty }^{-a} dz\int ^{\infty }_a dz^\prime  \, 
{\cal J}(z,z^\prime )\nGP \pp(z)\pp(z') ,
\end{align}
that motivates the phenomenological model \eqref{eq:H_JJJ}. 
Note that in \eqref{eq:H_JJJ} we further neglect the last interaction term $-\int _{-\infty }^{-a} dz\int ^{\infty }_a dz^\prime  \, 
{\cal J}(z,z^\prime ) \nGP \pp(z)\pp(z')$.

\subsubsection{Phase diffusion after removing the phase-locking interaction}

We consider a thermal state with $J\neq0$ with full support and finite energy penalty on the $k=0$ eigen-mode (for $J=0$ it is the phase zero-mode).
We thus have $\langle \p_0^2\rangle\propto k_BT $ and $\langle \d_0^2\rangle\propto k_BT$ similar to the ordinary $k>0$ modes.
We then perform a quench to $J=0$, which means $\p_0\rightarrow \p_\text{ZM}$ and $\d_0\rightarrow \d_\text{ZM}$, and observe how the phase zero-mode grows given by the equation
\begin{align}\label{eq:zm_diffusion}
    \langle \pp_\text{ZM}^2(t)\rangle=\langle \pp_0^2\rangle+\frac{g^2t^2}{\hbar^2}\langle \dd_0^2\rangle\ .
\end{align}
Fig.~\ref{fig:zeromode_diffusion} demonstrates the effect of taking into account zero-mode phase diffusion during the merging process. 
\begin{figure}
    \centering
    \includegraphics[trim=2.5cm 2cm 0cm 2cm,clip,width=\textwidth]{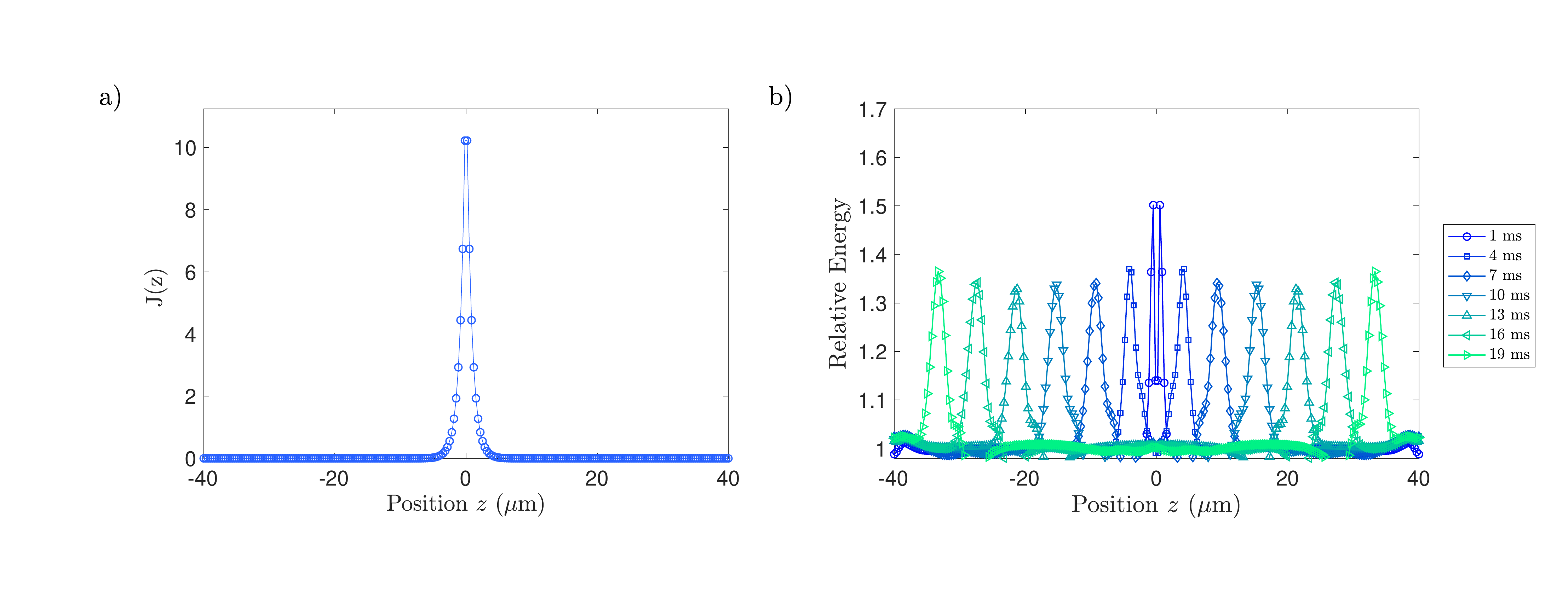}
    \caption{\textbf{The contribution of zero-mode phase diffusion to energy excitations during merging.} Two systems are prepared in thermal, phase-locked states according to the localized coupling $J$ shown in panel \emph{(a)} with overall strength similar to the value used in the main text $\overline J = \frac{1}{L}\int \di z J(z) \approx \SI{0.25}{\hertz}$. 
    The magnitude of $J$ at the interface can be tuned in experiments by the barrier parameters. After the preparation phase, $J$ is then quenched to 0, signifying a decoupling of the two systems into independent, gapless Luttinger liquids, where the zero phase mode has no contribution to energy and diffuses according to Eq.~\eqref{eq:zm_diffusion} for a total time of $\SI{25}{\milli\second}$. When the systems are again merged, as shown in panel \emph{(b)}, we see that large excitations can potentially be induced due to the diffusion of the zero mode. In order to minimize the energy of excitations  one should choose a large $J$ in the beginning (meaning a stronger phase-locking during preparation), and design the cycle times to be shorter.
    The amount of excitations here is an over-estimate as we did not include in the modelling the possibility of  phase-locking the condensates before merging, this process could involve strong correlations via the Josephson junction and could counter-act phase diffusion.
    }
    \label{fig:zeromode_diffusion}
\end{figure}
\subsection{Sudden merging in the continuous QFT limit and additional checks of the numerical simulation}

Let us also briefly discuss here how do our simulations compare with the continuum limit $\Delta z \rightarrow 0$. 
Essentially, besides the fact that the field operators themselves have the appropriate continuum limit, in the static case we are also interested in recovering the spectrum and the eigenfunctions of the Hamiltonian \eqref{eq:H_quench} to some extent. 
In particular, let us consider two types of density profiles $\nGP(z)$ that are piece-wise constant functions: 
(1) two disconnected parts of lengths $L_A$ and $L_B$ on intervals $[-L_A,0)$ and $(0,L_B]$, where the interface is at $z=0$, which corresponds to the Hamiltonian $\hat H_{A|B}$ and (2) a single system with length $L_{AB}=L_A+L_B$ where the high wall at the interface has been removed, which corresponds to the Hamiltonian $\hat H_{AB}$.
We further impose Neumann (open) boundary conditions at all boundary  points, i.e., $\partial_z \varphi_k|_{-L_A}=\partial_z \delta \varrho_k|_{-L_A}=\partial_z \varphi_k|_{L_B}=\partial_z \delta \varrho_k|_{L_B}=0$, and similarly for the point $z=0$ in case (1).
 In such cases, solutions to Eq.~\eqref{eq:sturmliouvillecont} can be easily found on each interval and are given by usual oscillatory functions with a linear dispersion relation
\begin{equation}
    \omega_k(L) = \pi c k /L ,
\end{equation}
where $c=\sqrt{\nGP g/m}$ is the speed of sound, $k$ is an integer number and it also depends on the length of the corresponding interval $L\in \{L_A,L_B,L_{AB}\}$.
In case (1) we have the two solutions for $k>0$
\begin{align}
\delta \varrho_{2k-1}^{A|B} (z) &= \varphi_{2k-1}^{A|B}(z) =\left\{
\begin{array}{cl}
\sqrt{\frac 2 {L_A}} \cos\left(\pi k (z+L_A)/L_A \right)  &\mbox{for} \ z\in [-L_A,0] , \\
0 & \mbox{for} \ z\in (0,L_B],
\end{array}
\right. \\
\delta \varrho_{2k}^{A|B} (z) &= \varphi_{2k}^{A|B}(z) =\left\{
\begin{array}{cl}
0 & \mbox{for} \ z\in [-L_A,0], \\
\sqrt{\frac 2 {L_B}} \cos\left(\pi k (z-L_B)/L_B \right)  &\mbox{for} \ z\in (0,L_B] 
\end{array}
\right. ,
\end{align} 
with corresponding dispersion relations respectively $\omega_{2k-1} = \pi c k /L_A$ and $\omega_{2k} = \pi c k /L_B$. 
Note that in the case $L_A=L_B$ there is a degeneracy between even and odd modes.

For case (2), instead, we have the solutions
\begin{equation}\label{eq:eigenmodesHAB}
\delta \varrho_k^{AB} (z) = \varphi_k^{AB}(z) = \sqrt{\frac 2 {L_{AB}}} \cos\left(\pi k (z+L_A)/L_{AB} \right) ,
\end{equation}
with dispersion relation $\omega_k^{AB} = \pi c k /L_{AB}$.
The agreement of the dispersion relation and the profile of the eigenmode functions in the static case can be observed in Figs.~\ref{fig:eigenmodes_disjoint},\ref{fig:homogeneous_DR} where a comparison with the discretized homogeneous model with or without a trapeze-like buffer region is shown.

\begin{figure*}[h!]
    \centering
    \includegraphics[trim=2cm 9cm 3cm 9cm,clip,width=\textwidth]{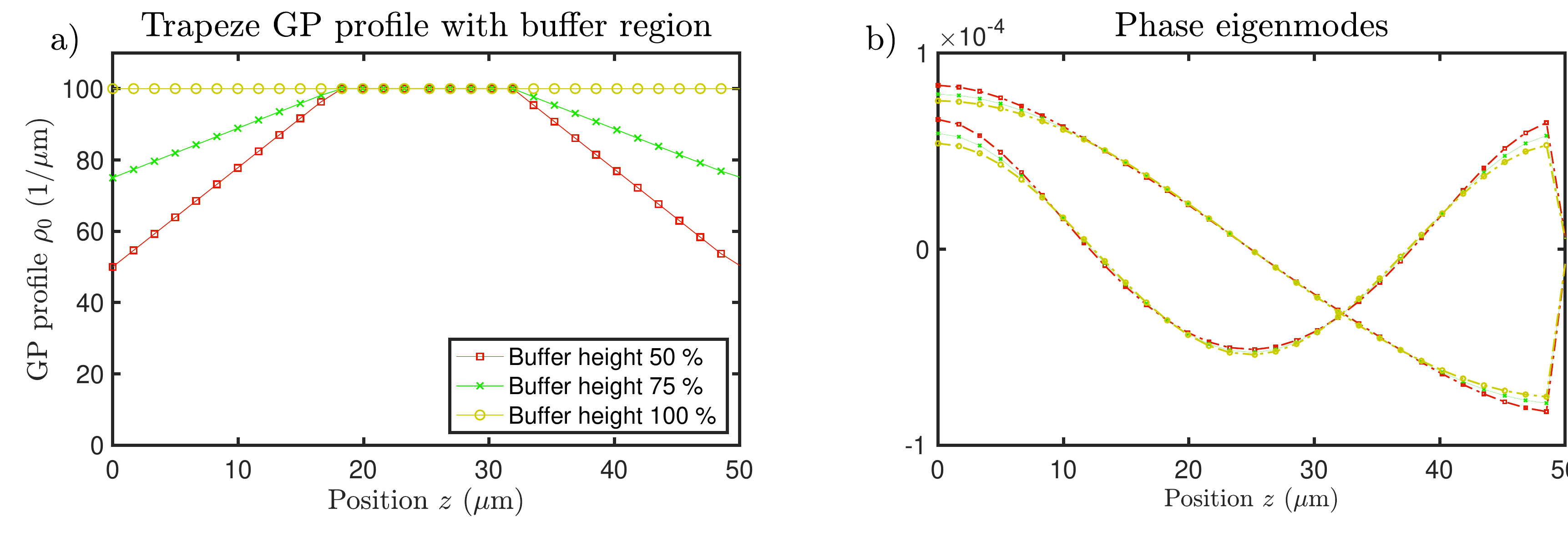} 
    \caption{{Influence of the buffer region on eigenmodes.} \emph{(a):} Using the lattice discretization various inhomogenous Gross-Pitaevskii profiles can be considered.
    \emph{(b):}
    Phase eigenmodes for $k=1$ and $k=2$ of disjoint Hamiltonian $H_{A| B}$. Only the half system with nonzero eigenmode functions is shown as they vanish outside of the support of the  profile.
    Qualitatively, all modes retain their oscillatory nature, though at the edges, where the inhomogeneity is the largest, there is a systematic change in the wave-functions.
    }
    \label{fig:eigenmodes_disjoint}
\end{figure*}

\begin{figure}[h!]
    \begin{minipage}{0.475\textwidth}
\includegraphics[trim=3.5cm 8cm 4.5cm 8.5cm,clip,width=\linewidth]{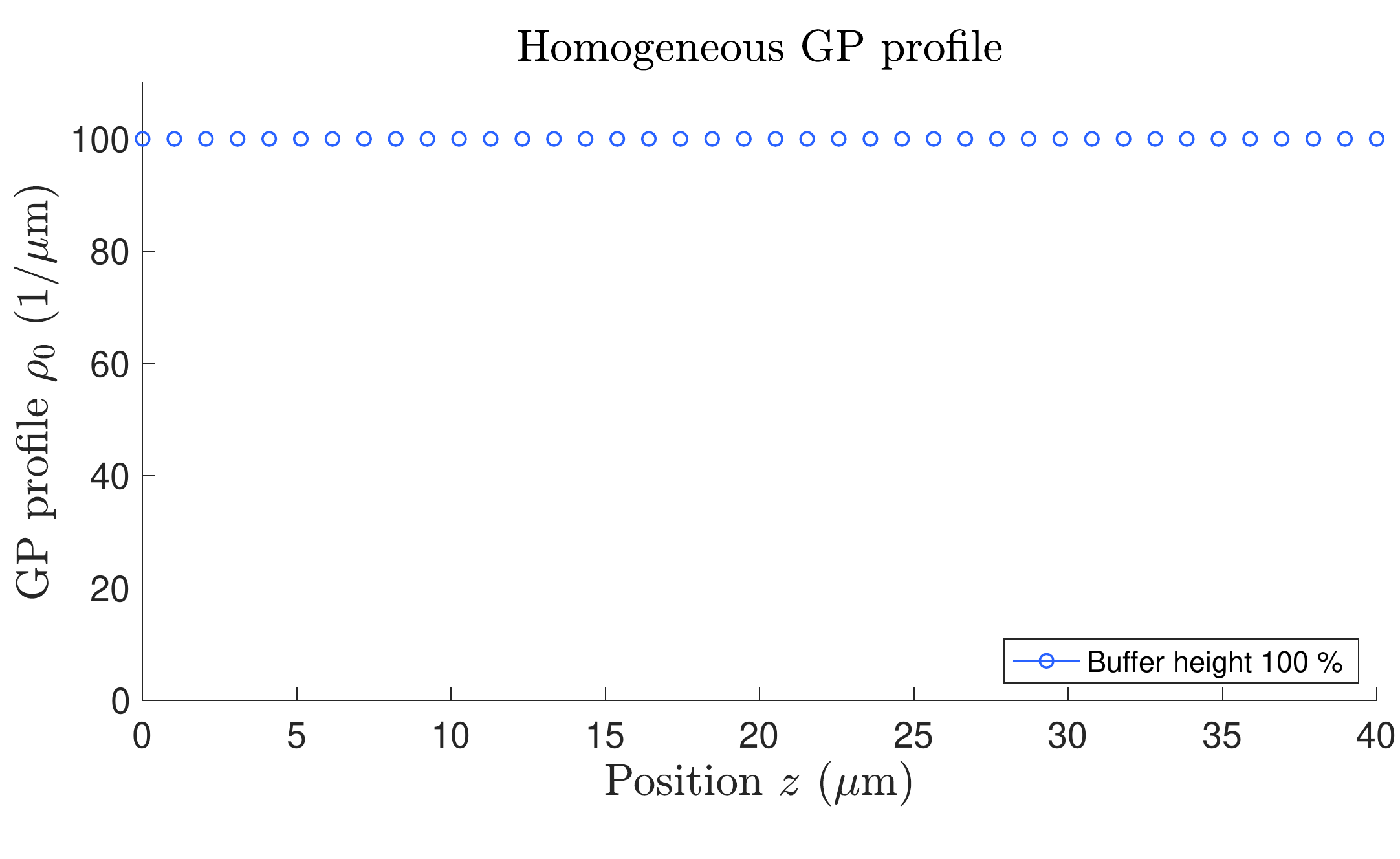}
\includegraphics[trim=1.8cm 4cm 2.4cm 1.5cm,clip,width=\linewidth]{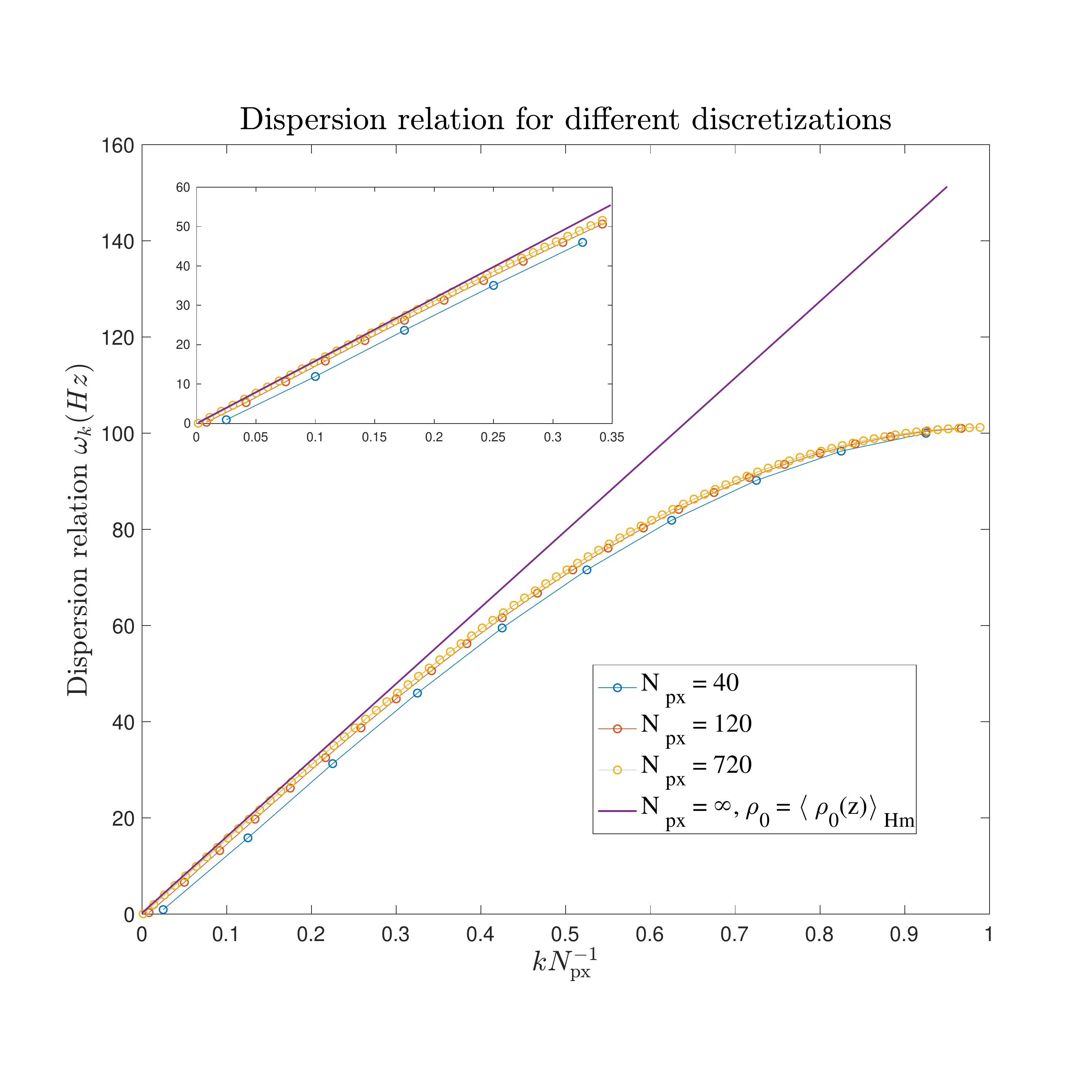}
\caption{{Dispersion relations for the homogeneous Gross-Pitaevskii profile.} 
For a homogeneous profile \emph{(top)} we find that for high momentum modes the dispersion relation is no longer approximately linear due to the lattice discretization.
This leads to the dispersion of the wave packets during for example merging. The inset shows that approximately 30$\%$ of the low-energy modes already gives rise to a good approximation to the continuum limit, especially for discretizations above the order of $\sim 100$ pixels, which is the resolution at where our simulations have been performed
(see also Fig.\ \ref{TPGP}).}
\label{fig:homogeneous_DR}
    \end{minipage}
    \hfill
        \begin{minipage}{0.48\textwidth}
    \includegraphics[trim=3.7cm 8cm 4.5cm 8.8cm,clip,width=\linewidth]{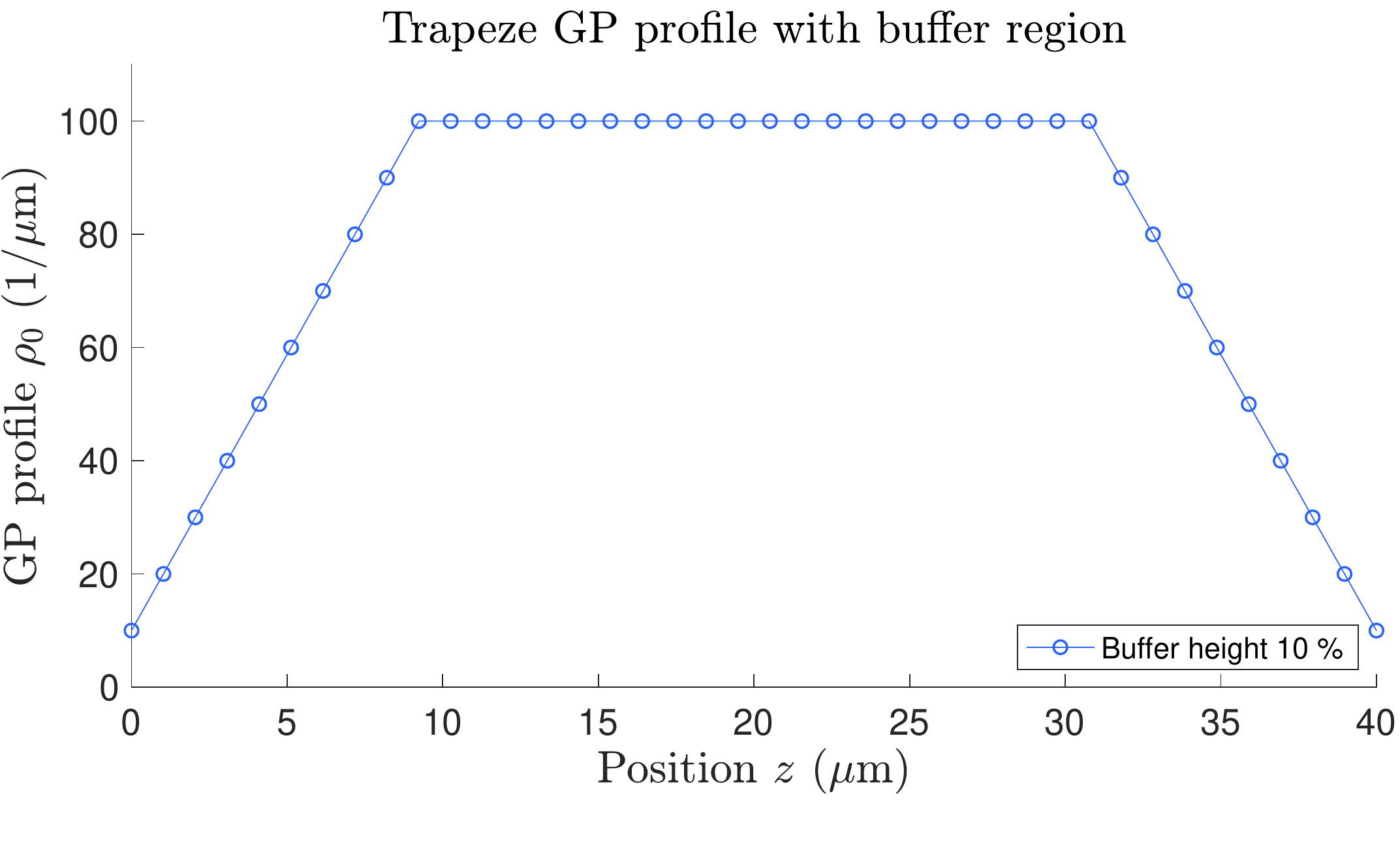}
    \includegraphics[trim=1.8cm 4cm 2.4cm 1.5cm,clip,width=\linewidth]{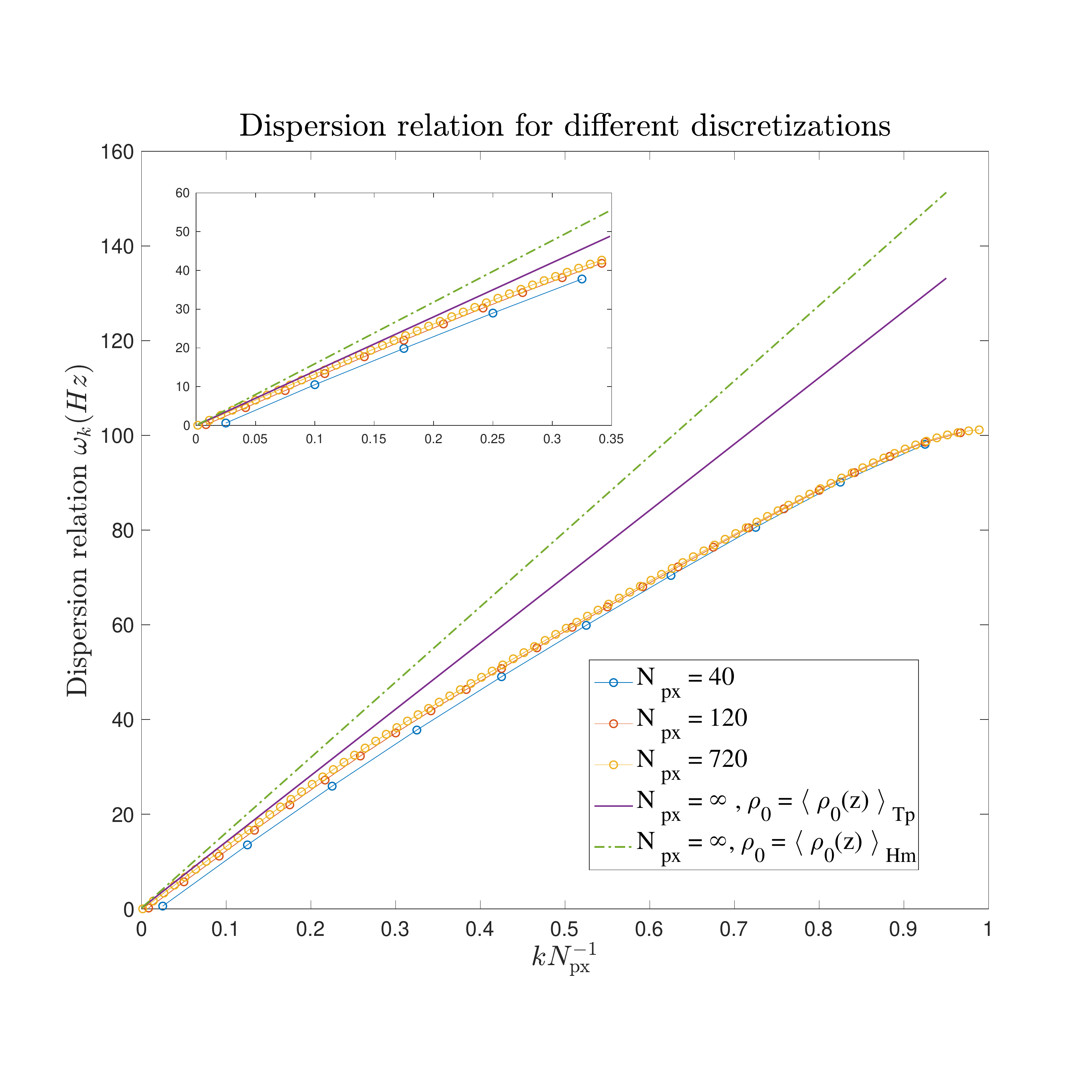}
\caption{{Dispersion relations for the trapeze Gross-Pitaevskii profile.}
When compared to Fig.~\ref{fig:homogeneous_DR}, the dispersion relation for higher modes is closer to a linear curve, especially for high momentum modes. This is why when using such a trapeze GP model when putting condensates in heat contact, we observe wave packets undergoing significantly less dispersion. Since less atoms are considered in this condensate due to the trapeze-shaped profile
(Tp), the low-lying energy modes are better approximated by the continuum limit assuming a \textit{homogeneous} condensate with  profile at $\rho_0 = \langle \rho_0(z)\rangle_{\rm Tp} = {\rm const} $.}
\label{TPGP}
\end{minipage}
\end{figure}

Afterwards, let us try to compare the dynamics of the merging QTP with its continuous
quantum field limit. 
First of all, we observe that the initial state in the continuous QFT, i.e., the thermal state of the QFT limit of the split Hamiltonian, would have the spectrum of the covariance matrix given by $\aver{(\pp_k)^2+(\dd_k)^2} = 2\aver{\hat n_k}+1$, where $\aver{\hat n_k}=1/(\exp(\beta \omega_k)-1)$ are the normal mode occupation numbers, given by the usual Bose-Einstein distribution. Then, from the fact that for our initial state we have
$\aver{\hat a_k^\dagger \hat a_l^\dagger + \hat a_k \hat a_l}=0$ and $\aver{\hat a_k^\dagger \hat a_l + \hat a_k \hat a_l^\dagger}=(2\aver{\hat n_k}+1)\delta_{k,l}$, we obtain for the initial real space correlation matrix
\begin{equation}\label{eq:initcovQFT}
C_{A-B}(z,z^\prime,t=0) =C^{\rho\rho}(z,z^\prime) \oplus C^{\phi\phi}(z,z^\prime)  =\sum_{k>0} \frac{\hbar \omega_k}{g} \delta \varrho_k(z)\delta \varrho_k(z^\prime)(2\aver{\hat n_k}+1) \oplus \sum_{k>0} \frac{g}{\hbar \omega_k} \varphi_k(z)\varphi_k(z^\prime)(2\aver{\hat n_k}+1),
\end{equation}
where here and in the following discussion we discard the zero mode, since in the simulations we have regularized it as discussed in Sec.~\ref{app:zero_mode}.

Clearly, the result in Eq.~\eqref{eq:initcovQFT} is very similar to the discretized case, but contains small differences in the normal mode frequencies and in the functional form of 
the normal modes with respect to the real space modes. 
For the energy density we thus obtain 
\begin{align}
\frac {\mathrm{d}E(z,0)}{\mathrm{d}z}  &=
\frac{\hbar^2 \nGP(z)}{2m}\partial_{z_1}\partial_{z_2}C_{A-B}^{\phi\phi}(z_1,z_2,t=0)\big|_{z_1=z_2=z}+\frac g 2 C_{A-B}^{\rho\rho}(z,z,t=0) \\
&=\hbar \sum_{k>0}  \left(  \frac{\nGP}{2m}  \frac{g}{\omega_k} (\partial_z \varphi_k (z))^2  + \frac g 2  \frac{\omega_k}{g} \delta \varrho_k^2(z) \right) (2\aver{\hat n_k}+1) \\
&=\left\{
\begin{array}{c}
\frac{\hbar}{L_A}\sum_{k \ odd} \omega_k  (\aver{\hat n_k}+1/2) \quad \mbox{for} \ z \in [-L_A,0) \\
\frac{\hbar}{L_B}\sum_{k \ even} \omega_k  (\aver{\hat n_k}+1/2)
\quad \mbox{for} \ z \in (0,L_B] ,
\end{array}
\right.
\end{align}
where in the first equality we have used 
$\aver{\hat a_k^\dagger \hat a_l + \hat a_k \hat a_l^\dagger}=(2\aver{\hat n_k}+1)\delta_{k,l}$
and in the second equality we used that
\begin{equation}
\begin{aligned}
\frac{\nGP}{2m} \frac{g}{\omega_k} (\partial_z \varphi_k (z))^2  + \frac g 2  \frac{\omega_k}{g} \delta \varrho_k^2(z)&=\frac 1 {L_A} \frac{\omega_k} 2 \qquad \mbox{for $k$ odd}, \\
\frac{\nGP}{2m} \frac{g}{\omega_k} (\partial_z \varphi_k (z))^2  + \frac g 2  \frac{\omega_k}{g} \delta \varrho_k^2(z)&=\frac 1 {L_B} \frac{\omega_k} 2 \qquad \mbox{for k even}
\end{aligned}
\end{equation}
for all $z$ respectively in $[-L_A,0)$ and $(0,L_B]$, and we have that the functions are zero otherwise.
Let us now consider the time-dependent interaction. The energy density at time $t>0$ during this evolution is calculated as
\begin{equation}
\begin{gathered}\label{eq:deltaEoft}
\frac {\mathrm{d}E(z,t)}{\mathrm{d}z}  =
\frac{\hbar^2 \nGP(z)}{2m}\partial_{z_1}\partial_{z_2}C_{A-B}^{\phi\phi}(z_1,z_2,t)\big|_{z_1=z_2=z}+\frac g 2 C_{A-B}^{\rho\rho}(z,z,t)  ,
  \end{gathered}
\end{equation}
where now we need the diagonal blocks of the correlation matrix at time $t$, namely $C_{A-B}^{\phi\phi}(z,z^\prime,t) = \langle \pp(z,t)\pp(z^\prime,t)\rangle$ and $C_{A-B}^{\rho\rho}(z,z^\prime,t) = \langle \dd(z,t)\dd(z^\prime,t)\rangle$, which, in turn, can be calculated from the 
instantaneous eigenmode functions at time $t$, that are given essentially by solving Eq.~(\ref{eq:sturmliouvillecont}), but now with different boundary conditions at the interface point $z=0$.
See Figs.~\ref{fig:eigenmodes_homogeneous_trotterized},\ref{fig:eigenmodes_trapeze_trotterized} for a plot of the lowest lying eigenmode functions in the discretized model.

Specifically, given the eigenmode functions $\varphi^{(t)}_k(z)$ and $\delta \varrho^{(t)}_k(z)$ of the Hamiltonian at time $t$, together with the corresponding eigenmode frequencies $\omega^{(t)}_k$,
we can find the time-evolved field operators at time $t$ as
\begin{equation}
\pp(z,t) = -i \sum_{k} \sqrt{\frac{g}{\hbar \omega^{(t)}_k}}  \varphi^{(t)}_k(z) \left(e^{i \omega^{(t)}_k t} \hat t_k^\dagger - 
  e^{-i \omega^{(t)}_k t}  \hat t_k \right) ,
\end{equation}
where $\hat t_k$ and $\hat t_k^\dagger$ are the instantaneous creation/annihilation, obtained with a (real) Bogoliubov
transformation
\begin{equation}
\hat t_k=\sum_l  u_{k,l} \hat a_l + v_{k,l} \hat a^\dagger_l  ,
\end{equation}
from those at $t=0$. The Bogoliubov coefficients are obtained by imposing that the operators $\pp(z,t=0)$ and $\dd(z,t=0)$ coincide with the initial ones, i.e.,
\begin{equation}
  -i \sum_{k} \sqrt{\frac{g}{\hbar \omega^{(t)}_k}}  \varphi^{(t)}_k(z) \sum_{l}(u_{k,l}-v_{k,l})\left(\hat a_l^\dagger - 
   \hat a_l \right) = -i \sum_{k} \sqrt{\frac{g}{\hbar \omega^{(0)}_k}}  \varphi^{(0)}_k(z) \left(\hat a_k^\dagger - 
   \hat a_k \right) ,
\end{equation}
and can be extracted from the scalar products between the intial and the instantaneous eigenmode functions:
\begin{equation}
\begin{aligned}
u_{k,l}-v_{k,l}&=\sqrt{\frac{\omega^{(t)}_k}{ \omega^{(0)}_l}} \langle \varphi^{(0)}_l(z), \varphi^{(t)}_k(z)\rangle \qquad 
u_{k,l}+v_{k,l}=\sqrt{\frac{ \omega^{(0)}_l}{\omega^{(t)}_k}} \langle \delta \varrho^{(0)}_l(z),\delta \varrho^{(t)}_k(z)\rangle ,
\end{aligned}
\end{equation}
where the relation on the right comes from a similar condition on the $\delta \varrho_k$ eigenfunctions.
Thus, substituting all of the above relations, the evolved correlation matrices can be obtained through the formulas
\begin{equation}\label{eq:bogoCVt}
\begin{aligned}
C_{A-B}^{\phi\phi}(z,z^\prime,t) &= \sum_{k,l,r} (2\aver{\hat n_k}+1) \frac{g}{\hbar \sqrt{\omega^{(t)}_l \omega^{(t)}_r}}
\varphi^{(t)}_l(z) \varphi^{(t)}_r(z^\prime) 2\left( (u_{l,k}u_{r,k}+v_{l,k}v_{r,k})\cos\left( (\omega^{(t)}_r - \omega^{(t)}_l) t\right) \right. \\ 
&-\left. (u_{l,k}v_{r,k}+v_{l,k}u_{r,k}) \cos\left((\omega^{(t)}_r + \omega^{(t)}_l) t\right) \right) 
, 
\end{aligned}
\end{equation}
and
\begin{equation}
    \begin{aligned}
C_{A-B}^{\rho\rho}(z,z^\prime,t) &= \sum_{k,l,r} (2\aver{\hat n_k}+1)\frac{\hbar \sqrt{\omega_l^{(t)} \omega^{(t)}_r}}{g}\delta \varrho^{(t)}_l(z) \delta \varrho^{(t)}_r(z^\prime) 2\left( (u_{l,k}u_{r,k}+v_{l,k}v_{r,k})\cos\left( (\omega^{(t)}_r - \omega^{(t)}_l) t\right) \right. \\ 
&+\left. (u_{l,k}v_{r,k}+v_{l,k}u_{r,k}) \cos\left((\omega^{(t)}_r + \omega^{(t)}_l) t\right) \right) , 
\end{aligned}
\end{equation}
and finally we obtain the expression for the energy density by plugging all of this into Eq.~\eqref{eq:deltaEoft}
\begin{equation}\label{eq:deltaEoftbogo}
\begin{aligned}
\frac {\mathrm{d}E(z,t)}{\mathrm{d}z}  = \hbar
\sum_{k,l,r} (\aver{\hat n_k}+\tfrac 1 2) &\left[ S^{(t)}_{l,r}(z) \left(u_{l,k}u_{r,k}+v_{l,k}v_{r,k})\cos\left( (\omega^{(t)}_r - \omega^{(t)}_l) t\right) \right) \right. \\
&\left. + D^{(t)}_{l,r}(z)
\left(u_{l,k}v_{r,k}+v_{l,k}u_{r,k})\cos\left( (\omega^{(t)}_r + \omega^{(t)}_l) t\right) \right)
\right] ,
\end{aligned}
\end{equation}
where, to shorten the notation, we have defined the quantities
\begin{equation}\label{eq:SDquant}
    \begin{aligned}
    S^{(t)}_{l,r}(z)&:=\left(\frac{\nGP(z) g}{m \sqrt{\omega^{(t)}_l \omega^{(t)}_r}}(\partial_{z}\varphi^{(t)}_l(z)) (\partial_{z}\varphi^{(t)}_r(z)) +  \sqrt{\omega_l^{(t)} \omega^{(t)}_r} \delta \varrho^{(t)}_l(z) \delta \varrho^{(t)}_r(z) \right) ,\\
    D^{(t)}_{l,r}(z)&:=\left(\sqrt{\omega_l^{(t)} \omega^{(t)}_r} \delta \varrho^{(t)}_l(z) \delta \varrho^{(t)}_r(z)-\frac{ \nGP(z) g}{m \sqrt{\omega^{(t)}_l \omega^{(t)}_r}}(\partial_{z}\varphi^{(t)}_l(z)) (\partial_{z}\varphi^{(t)}_r(z)) \right) 
    \end{aligned}
\end{equation}
that depend only on the instantaneous eigenfunctions and eigenfrequencies.
Hence, to calculate the energy density at time $t$ we just need the additional calculation of the Bogoliubov coefficients.

As an illustrative example, let us now consider the case $L_A=L_B=L$ and in which we quench directly to the full $\hat H_{AB}$ at 
$t=0$. 
In such case we have that the eigenmode functions at $t=0^+$ are given by (\ref{eq:eigenmodesHAB}) with $L_{AB}=2L$ and 
the corresponding eigenfrequencies are $\omega^{(+)}_k = \pi c k /2L$ with the same sound velocity $c$, which are just half of the corresponding odd frequencies at $t=0^-$. Note, however, that at $t=0^-$ there is a degeneracy, such that the $\omega_k$ with odd $k$ have the same values as the even $k$.
Thus, $\omega^{(+)}_k$ coincide with the even eigenfrequencies at $t=0^-$. The quantities (\ref{eq:SDquant}) at time $t=0^+$ read 
\begin{equation}
\begin{aligned}
    S^{(+)}_{l,r}(z) &= \frac{\pi c}{2L} \sqrt{l r} \sin\left(\pi (l+r)(z+L)/2L \right) ,\\
    D^{(+)}_{l,r}(z) &= \frac{\pi c}{2L} \sqrt{l r} \cos\left(\pi (l+r) (z+L)/2L \right) .
    \end{aligned}
\end{equation}
Furthermore, the Bogoliubov coefficients satisfy
\begin{equation}
\begin{aligned}
u_{2k-1,l}&=\frac 1 2 \frac{\omega^{(+)}_l+\omega^{(0)}_k}{\sqrt{\omega^{(0)}_k\omega^{(+)}_l}} \ ,
\quad O_{k,l} = \frac{2k+l}{\sqrt{2kl}} \ , \quad O_{k,l} = u_{2k,l}  ,\\
v_{2k-1,l}&=\frac 1 2 \frac{\omega^{(+)}_l-\omega^{(0)}_k}{\sqrt{\omega^{(0)}_k\omega^{(+)}_l}} \ ,
\quad O_{k,l} = \frac{2k-l}{\sqrt{2kl}} \ ,\quad  O_{k,l} = v_{2k,l}, 
\end{aligned}
\end{equation}
where 
\begin{equation}
O_{k,l}=\frac{\sqrt 2} L \int_0^L \de z \cos(\pi k z/L)\cos(\pi l z/2L)=\frac{\sqrt 2}{\pi (k+l/2)} \left(\frac{k\sin\left((k-l/2)\pi \right)}{k-l/2}+(-1)^k\sin\left(l\pi/2\right) \right)
\end{equation}
are the scalar products between the eigenfunctions at $t=0^-$ and $t=0^+$. Plugging all of this into Eq.~\eqref{eq:deltaEoftbogo} we finally obtain the energy density at time $t>0$ as 
\begin{equation}
\begin{aligned}
\frac {\mathrm{d}E(z,t)}{\mathrm{d}z}  = \hbar
\sum_{k \text{ even}}\sum_{r,l} (\aver{\hat n_k}+\tfrac 1 2) \frac{\pi c}{2kL} &\Big[ (4k^2+lr) \sin\left(\pi (l+r) (z+L)/2L \right) \cos\left( \pi c t (r-l)/2L\right) \\ 
+&(4k^2-lr) \cos\left(\pi (l+r) (z+L)/2L \right) \cos\left( \pi c t (r+l)/2L\right)
\Big] ,
\end{aligned}
\end{equation}
and we can see that this expression reflects a superposition of waves traveling at speed of sound $c$, and, in particular, there is no dispersion.

\subsection{Details of compression and expansion}\label{app:Compressiondetails}

Here we give a more detailed discussion of the approximations that have been  involved in formulating the piston model in the main text.
We consider the Lieb-Liniger model for the gas trapped in a box of changing size from $L(0)$ to $L(t)$.
After the standard phononic expansion $\hat \Psi = \sqrt{\hat \rho}e^{i\hat\theta}$ in the long wave-length limit the Hamiltonian can be approximated as
\begin{align}\label{eq:LLhamcomprapp}
\hat H_\text{LL}&\approx\int_0^L \di z \biggl[\frac{\hbar^2}{2m}(\partial_{z}\hat\theta) \hat \rho (\partial_{z}\hat\theta) +\frac g2 {\hat \rho}^2  \biggr].
\end{align}
We next split the operators around the classical hydrodynamical solutions, specifically we introduce the density fluctuations $\hat \rho = \nGP + \dd$ and phase fluctuations $\hat \theta  = \varphi +\pp$.


The classical phase can be interpreted as the velocity-potential by means of the equation $v = \hbar\partial_z\varphi /m $.
In a simple case where only one wall is moving we have 
that the classical hydrodynamic equations
\begin{align}
    \frac{\partial \rho}{\partial t } +\frac{\partial}{\partial z}(\rho v) &= 0, \\
    \frac{\partial v}{\partial t } + v \frac{\partial v}{\partial z} &= -\frac{g}{m}\frac{\partial \rho}{\partial z} ,
\end{align}
have solution given by 
\begin{align}
    \nGP(t)= \frac{N}{L(t)},\quad v(z,t) = z \frac{\dot{L}(t)}{L(t)}.
\end{align}
In this case we find that the velocity depends on the position and matches the velocity of the moving wall at the boundary, namely that $v(z=0) = 0$ and $v(z=L(t)) = \dot{L}(t)$.
This solution is obtained 
in the long-wavelength limit and neglecting the acceleration of the walls \cite{RMP_Stringari}. A similar solution can be obtained also in the case of both walls moving, with Neumann boundary conditions at each wall.

By integrating the velocity we obtain the classical phase field $\varphi$ which we next use to linearize the Hamiltonian \eqref{eq:LLhamcomprapp}. We thus obtaining the model
\begin{align}\label{eq:LLhamcomprappreal}
\hat H(t) &=\int_0^{L(t)} \di z \biggl[\frac{\hbar^2 \nGP(z,t) }{2m}\left(\partial_{z}\pp\right)^2+\frac g 2 \dd^2 
+
\frac{\hbar \dot{L}(t)z}{2L(t)}[\dd(\partial_z\pp)+(\partial_z\pp) \dd]
\biggr] \ .
\end{align}
If the evolution is slow (adiabatic), a lattice model with the fixed number of sites can serve as a good approximation to the discrete-value representation of the continuous system. 
Thus, in the following we neglect the second cross-coupling term between phases and densities, so to model a quasi-static case where the Gross-Pitaevskii profile gets compressed very slowly. Then, it is also illustrative to observe explicitly how this process works in an infinitesimal step-wise fashion. 
The infinitesimal length change is
\begin{align}
  L\rightarrow L_\epsilon &= (1+\epsilon)L ,
\end{align}
and, correspondingly, a homogeneous Gross-Pitaevskii profile $\nGP$ changes to  $\nGP(\epsilon) = (1+\epsilon)^{-1}\nGP$.
Then, the Hamiltonian after the size change reads
\begin{align}\label{eq:LLhamcomprappepsilon}
\hat H_\epsilon&=\int_0^{L_\epsilon} \di z \biggl[\frac{\hbar^2 \nGP}{2m(1+\epsilon)}\left(\partial_{z}\pp \right)^2+\frac g 2 \dd^2  \biggr] ,
\end{align}
which is Eq.~\eqref{eq:LLhamcomprappreal} without the last term.
Thus we observe that if a Gross-Pitaevskii profiles changes slowly in length then the phonons are described by a similar Hamiltonian, only with modified couplings. \gv{Note that here we did not consider explicitly the phase-locking term $\hat H_J$, however, since it has a linear dependence on the density, it does not change while changing the total length. In the main text instead, we wrote down the full Hamiltonian with the additional (unmodified) phase-locking term, which is also what we considered in our simulations.}

In the lattice model, we perform a similar procedure, but work fully in real space, this time with the Hamiltonian as a functional of both mean-field density and the small-distance cutoff. 
Starting from the discretized Hamiltonian $\hat H_{\Npxl}[\nGP,\Delta z]$ of a single condensate with $\Npxl$ pixels, length $L= \Npxl \Delta z$, and density $\nGP$, we perform at each step a small length change $L\mapsto L_\epsilon= \Npxl \Delta \zeta$, corresponding to a renormalization
\be\label{eq:trotterizedcomp}
H_{\Npxl}[\nGP,\Delta z] \mapsto  H_{\Npxl}[\nGP(\epsilon),\Delta \zeta] =   H_{\rho \rho}(1+\epsilon) \oplus  H_{\phi \phi}/(1+\epsilon)^2 \ ,
\ee
where we have used that $\nGP(\epsilon)=\nGP/(1+\epsilon)$ and $\Delta \zeta = (1+\epsilon) \Delta z$.
Thus, we see that we are implementing a discretized version of the Hamiltonian
\eqref{eq:LLhamcomprappepsilon}.

Then, in order to complete the full length change $\Delta L$ in a time $t_{\rm comp}=\Ntro \Delta t$, where $\Delta t$ is a small time interval and $\Ntro$ is the total number of Trotter steps, at each discrete time step we perform an inifinitesimal length change, such that 
\be
\epsilon = \Delta L /\Ntro .
\ee
What we get is the state of the phonons after compressing by a finite amount. 
This assumes that the phonons always see a quasi-static background metric, that is their dynamical time-scales are much faster than how we compress the condensate.
We observe that a sufficiently slow compression will not mix much between the modes and there will be thermal squeezing of the phonons.
We also see that the energy will in fact change. This is expected, since we are performing work on the system by compressing it which means it should increase in energy. The compression protocol is therefore our main way to realize a piston, where one may actively perform/extract work on a condensate by changing its length, and therefore its energy density and effective temperature.

\begin{figure}
    \centering
    \includegraphics[clip, trim = 2.2cm 7.8cm 2cm 7.3cm, width =0.88 \textwidth]{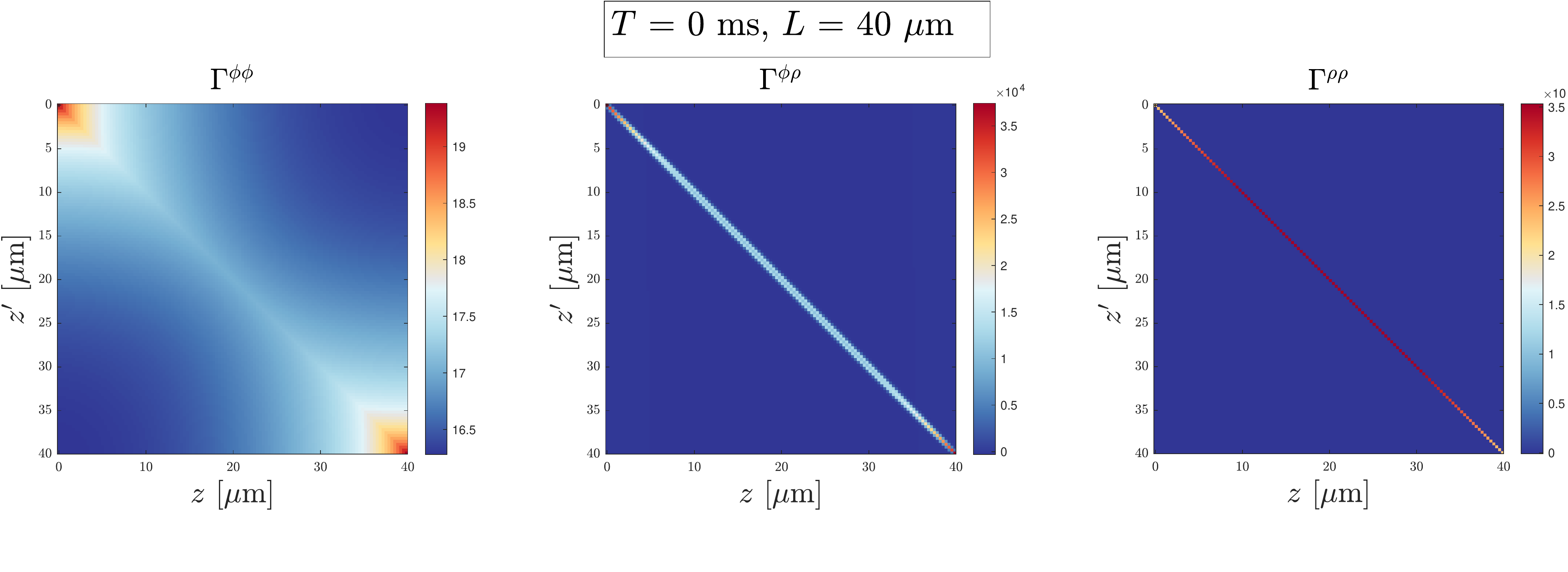}
    \includegraphics[clip, trim = 2.2cm 7.8cm 2cm 7.3cm, width =0.88 \textwidth]{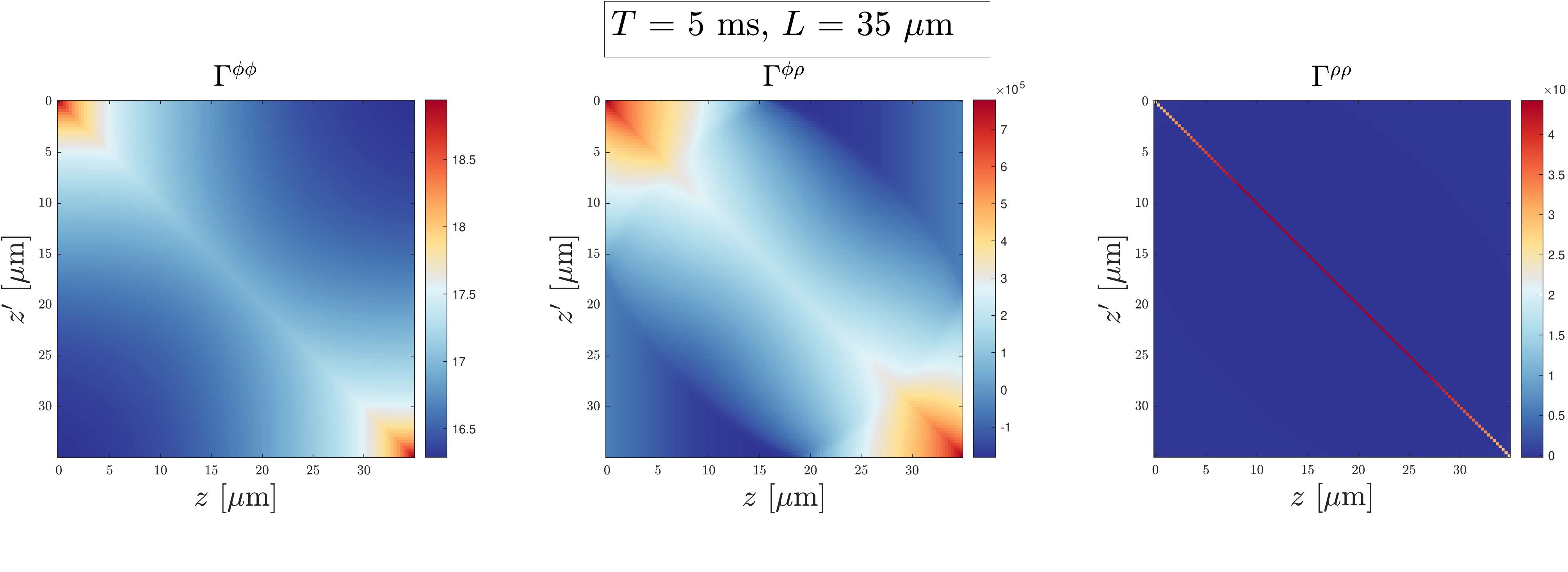}
    \includegraphics[clip, trim = 2.2cm 7.8cm 2cm 7.3cm, width =0.88 \textwidth]{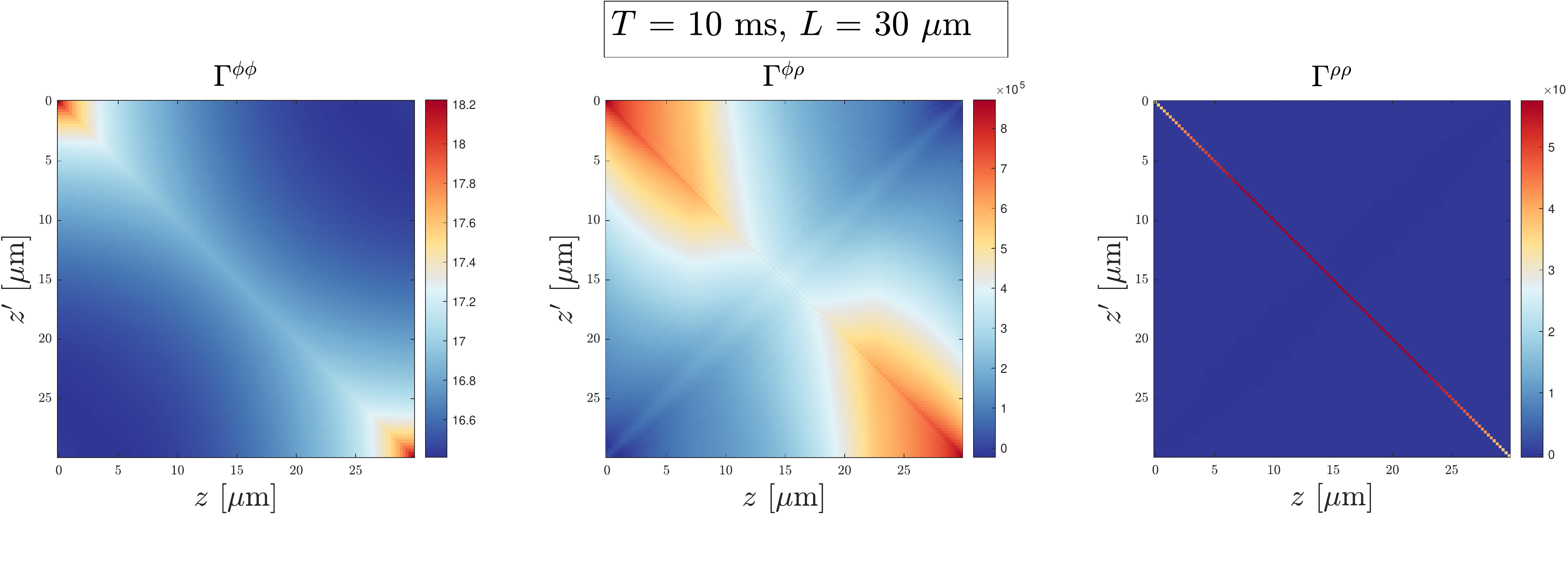}
    \includegraphics[clip, trim = 2.2cm 8cm 2cm 7.3cm, width =0.88 \textwidth]{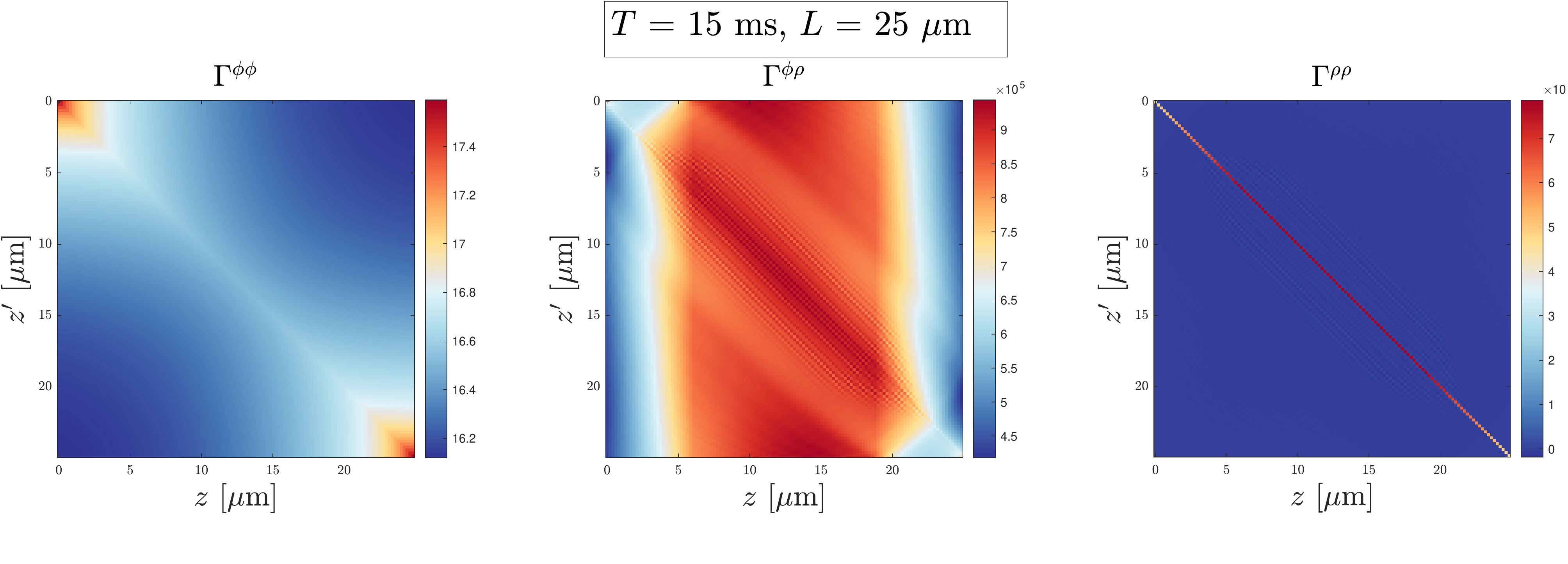}
    \caption{Covariance matrix of the piston while being compressed over a period of \SI{20}{\milli\second} from $L(0)=\SI{40}{\micro\meter}$ to $L(t_{\rm comp})=\SI{20}{\micro\meter}$.
    The squeezing of eigenmodes can be also seen in real space as shown here: 
    One finds that the overall magnitude of phase fluctuations decreases while for the density fluctuations it increases.
    The cross correlations do not contribute to the energy of the piston but their presence signifies that the system is not thermal during compression.
    }
    \label{fig:CVM_piston_compression}
\end{figure}

\subsubsection{Renormalizing the cutoff during compression QTP} 

Let us now discuss a technical detail arising in the compression and 
expansion QTP (see Fig.\ \ref{fig:CVM_piston_compression}). An implicit difference between the initial and final Hamiltonians of a compression/expansion step is that the continuous field theory should be defined
in the time-dependent line $[0,L(t)]$.
In principle, we can also make a change of the integration variable $z\mapsto \zeta = z L(0)/L(t)$, such that the theory is defined with a constant length.
However, a subtle issue arises: the field commutation relations $[\dd(z),\pp(z^\prime)]=\i\delta(z-z^\prime)$ depend on the coordinate $z$; thus a rescaling of the coordinate must be compensated by a corresponding rescaling of the density fluctuation field, in order to maintain the correct commutation relations.
Then, calling $\lambda(t)=L(0)/L(t)$ we define the transformation
\begin{equation}
    \begin{aligned}
    \zeta &= \lambda(t) z  , \\
    \delta \hat \nu &= \dd/ \lambda(t)  ,
    \end{aligned}
\end{equation}
such that the Hamiltonian \eqref{eq:LLhamcomprappreal} becomes
\begin{equation}\label{eq:compHamBogtransf}
    \hat H =\int_0^{L(0)} \di \zeta \biggl[\frac{\hbar^2 \nGP(z,0) \lambda^2(t) }{2m}\left(\partial_{\zeta}\pp\right)^2+\frac g 2 \lambda(t) \delta \hat \nu^2 \biggr] ,
\end{equation}
which effectively amounts to a renormalization of the line differential as
\be
\di z \mapsto \di \zeta = \lambda(t) \di z  ,
\ee
at the same time ensuring that the fields satisfy the correct commutation relations:
\begin{equation}
   [\dd(\zeta),\pp(\zeta^\prime)]= \i \delta(\zeta-\zeta^\prime)/\lambda(t) \ \Rightarrow \ [\delta \hat \nu(\zeta), \pp(\zeta^\prime)]=\i \delta(\zeta-\zeta^\prime) .
\end{equation}
Note that by making this field transformation, the full Hamiltonian \eqref{eq:LLhamcomprappreal} 
is transformed in such a way that the time derivatives of its parameters disappear. Therefore, the Hamiltonian \eqref{eq:compHamBogtransf} can be approximated by a lattice model without restrictions on the rate of change of parameters, i.e., no assumption about adiabaticity is required anymore. However, one has still to be careful with defining correctly the new rescaled density-fluctuation field in the discretized model.

A similar issue arises also working directly in the discretized version of \eqref{eq:LLhamcomprappepsilon}: By fixing the number of pixels and just rescale the cut-off $\Delta z$ at each Trotter step we are changing its effective momentum cut-off. Concretely, if we keep the number of pixels we see that the discretization length $\Delta \zeta$ has changed according to 
\begin{align}
    \Delta \zeta = \frac{L_\epsilon} L \Delta z .
\end{align}
It is important to stress once more that the covariance matrices satisfy the Heisenberg constraint that depends on $\Delta z$. Thus, we begin with a covariance matrix $\Gamma$ that satisfies
\begin{align}
\Gamma +\frac 1 {\Delta z}\i \Omega \geq 0 ,
\end{align}
but after size change it should satisfy
\begin{align}
    \Gamma +\frac 1 {\Delta \zeta}\i \Omega \geq 0 .
\end{align}
However, the natural way to implement the compression is, as we discussed above, to apply a symplectic transformation $G(\epsilon)=\exp(\Omega H_\epsilon/\Delta \zeta)$ that preserves the symplectic form and hence does not allow to switch between the Heisenberg cones with $\Delta z \mapsto \Delta \zeta$.
The way to implement the latter switch is to multiply the condition of the second cone and find that
\begin{align}
    \Gamma +\frac 1 {\Delta z}\i \Omega \geq 0 \Leftrightarrow \frac{\Delta z}{\Delta \zeta}\Gamma +\frac 1 {\Delta \zeta}\i \Omega \geq 0.
\end{align}
Hence we can now do the compression by setting
\begin{align}
    \Gamma(t_{\rm comp}) = \frac{\Delta z}{\Delta \zeta} G(t_{\rm comp}) \Gamma(0) G^T(t_{\rm comp})
\end{align}
where $G(t_{\rm comp})$ implements the Trotterized evolution from Eq.~\eqref{eq:trotterizedcomp}.
This covariance matrix will satisfy the Heisenberg relation at the target discretization length.
This is not anymore just a sympletic transformation, but an affinely symplectic transformation which preserves the symplectic form up to an overall pre-factor.

\subsection{Achieving larger cooling in the Otto cycle}\label{app:Ottodetails}
In the main text, we have shown how to concatenate the QTPs introduced, in order to operate a refrigerator to cool down part of the system. The remaining question is then how can we optimize the transfer of energy from the system to the bath (via piston), by tuning the various parameters that we have, such as $t_{\rm merge}$, $t_{\rm split}$, $t_{\rm comp}$, $L_{P(B,S)}$ etc. We discuss the effects and therefore the strategy of choice for some of the parameters below:

\begin{enumerate}\setlength{\itemsep}{0em}
    \item \textbf{Initial lengths (and length ratios) of system, piston and bath.} The lengths of each machine compartment determines their heat capacity. For example, a larger piston would be able to absorb (or lose) more heat when interacting with the system (bath). The size of the bath would largely determine how strong the non-Markovian effects are, especially since wave-packets are traveling ballistically in the condensate. For example, in our simulations the bath is only 3 times larger, which is a realistic figure when considering implementations. According to Fig.~\ref{fig:Otto_energies}, the wave-packets induced in the bath at the piston-bath interface has already travelled to the other bath edge and returned to the interface during the second cycle of piston-bath interaction, effectively making the process non-Markovian. In Fig.~\ref{fig:otto_reset}, we see a simulation where one effectively simulates a Markovian bath (and piston) by reinitializating them before every new cycle.
    \item \textbf{Compression ratio of piston.} It is clear that the more compression the piston undergoes, the more work is injected into the refrigerator. This causes a larger effective temperature difference between piston and bath, thereby inducing a larger amount of heat flow between them, which in turn increases the capability of the piston to later absorb heat while interacting with the system. While in classical scenarios the piston stays in equilibrium while gradually increasing in temperature, our model of compression as discussed in Section \ref{sec:piston} is akin to squeezing, and therefore the higher the amount of compression, the further we expect the system goes out of equilibrium, which is seen in our simulations.
    \item \textbf{Compression ratio of bath and system.} In the protocol we presented, for simplicity, the bath and system never undergo any change in length. However, if we imagine the three condensates on a chip, whenever we compress the piston, this leaves additional room for the bath to expand. Such an additional step, if undertaken, will further increase the temperature gradient and therefore facilitate heat flow.
    \item \textbf{Total duration when merging and splitting two systems.} Suppose two systems are connected and heat flow occurs due to an effective temperature gradient. How would one design the protocol to allow a maximum amount of net heat flow? Naturally, one expects that in the long time limit, energy will be equally distributed throughout the joint system, i.e., 
    they thermalize. However, we are interested mostly in finite time scales. Therefore, in practice, the most relevant parameters to set are the timings of merging, with respect to the lengths of the interacting systems. Moreover, the energy input during merging is non-negligible due to the relatively small sizes of each system. To overcome this, for example, one could time the protocol so that when we split the condensates again at the end, the wave-packets come back to the interface and then are taken out of the system due to the change in Hamiltonian. This can be done because we know the speed of sound in the condensate, concretely, it becomes natural to set $t_{\rm p} = L_p/c$, where $t_{\rm p}$ is a relevant time scale of the piston process. This illustrates the role of information in such a process: although a lot of energy may be injected during merging, the information about this energy is preserved, and therefore it can be suitably retrieved (instead of being irreversibly lost into other degrees of freedom). 
    \item \textbf{Further refinements when considering the Gross-Pitaevskii profile of condensates.} We have seen this in the case of putting two systems into heat contact. When a single condensate sits in the trap, the bulk region has a roughly uniform density, which is why one usually considers the fairly good approximation of a homogeneous $\rho_0$. The situation becomes more complicated when two such systems are merged: ideally, we \je{would like} the contact interface to have large atom density as well, so that heat transport is maximized. However, we saw from the simulations that this induced extremely high momentum modes which may cause us to observe more dispersion, and furthermore the Luttinger liquid analysis may no longer be useful in such regimes. On the other hand, having a small contact interface such as shown in the trapeze profile would imply that heat flow occurs more slowly in finite time scales. 
\end{enumerate}

\begin{figure}[h]
    \centering
    \includegraphics[trim = 1cm 3.3cm 0cm 5cm, clip, width=0.99\textwidth]{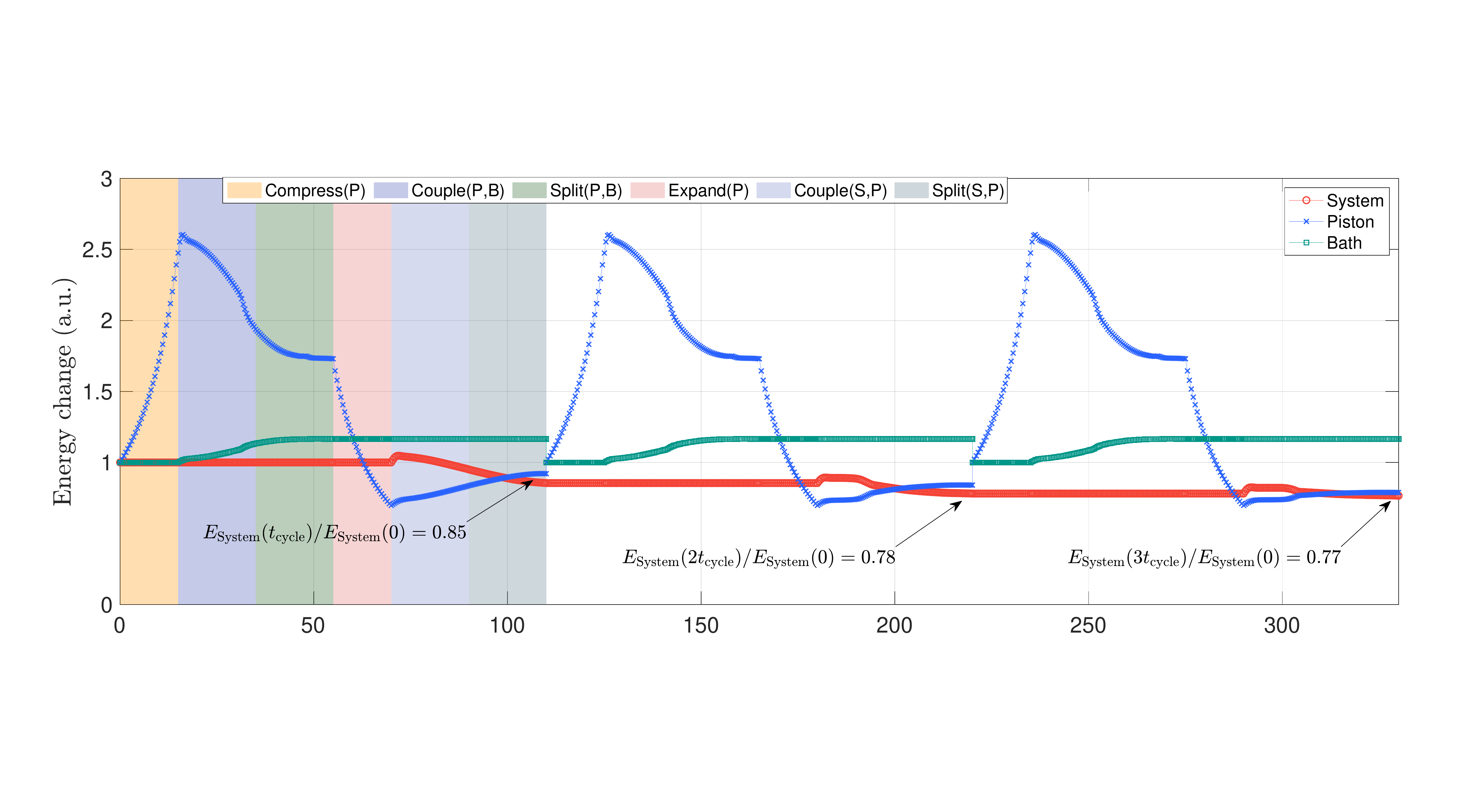}
    \includegraphics[trim = 1cm 6cm 1.3cm 7.5cm, clip, width=0.99\textwidth]{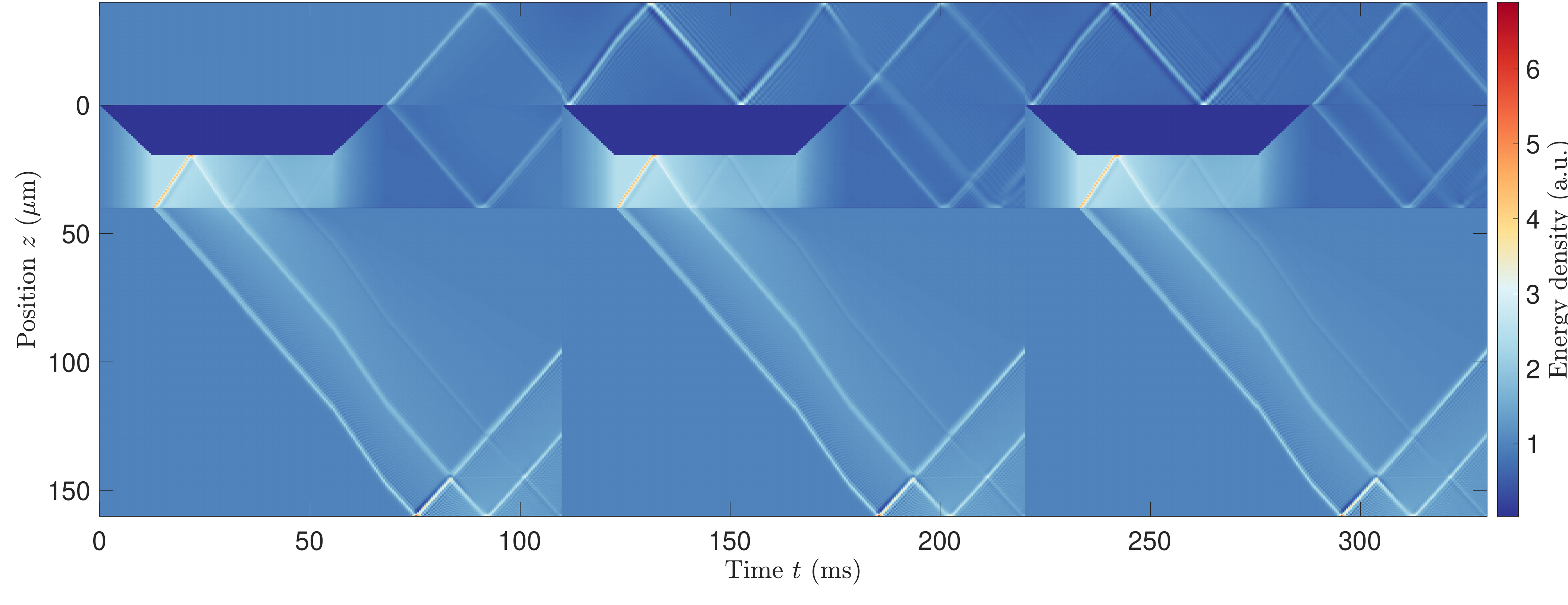} 
\caption{In this figure, we show the Otto cycle energy changes of system, piston and bath with a different setting: $t_{\rm couple} = \SI{20}{\milli\second}$, $t_{\rm split} = t_{\rm comp} = \SI{20}{\milli\second}$, $L_{\rm piston} = L_{\rm system} = \SI{40}{\micro\meter}$, and $L_{\rm bath}= \SI{120}{\micro\meter}$. On one hand, the Gross-Pitaevskii profile of the condensates at the edge drop off only slightly to about 0.8 of the peak value, which allows for more heat flow to occur between condensates during finite time, with the cost of injecting higher momentum modes into the simulation. On the other hand, after each cycle we reset the piston and the bath to its original state. This reinitialization, while challenging to perform in experiments, allow fresh thermal resources to be brought into the QFM and therefore allows us to achieve more cooling in subsequent cycles. Despite having a similar qualitative behaviour as in Fig. \ref{fig:Otto_energies}, it is evident that the various specific parameters governing each of the primitives will affect the final cooling efficiency of the engine. }
    \label{fig:otto_reset}
\end{figure}

\end{widetext}


\begin{thebibliography}{169}%
\makeatletter
\providecommand \@ifxundefined [1]{%
 \@ifx{#1\undefined}
}%
\providecommand \@ifnum [1]{%
 \ifnum #1\expandafter \@firstoftwo
 \else \expandafter \@secondoftwo
 \fi
}%
\providecommand \@ifx [1]{%
 \ifx #1\expandafter \@firstoftwo
 \else \expandafter \@secondoftwo
 \fi
}%
\providecommand \natexlab [1]{#1}%
\providecommand \enquote  [1]{``#1''}%
\providecommand \bibnamefont  [1]{#1}%
\providecommand \bibfnamefont [1]{#1}%
\providecommand \citenamefont [1]{#1}%
\providecommand \href@noop [0]{\@secondoftwo}%
\providecommand \href [0]{\begingroup \@sanitize@url \@href}%
\providecommand \@href[1]{\@@startlink{#1}\@@href}%
\providecommand \@@href[1]{\endgroup#1\@@endlink}%
\providecommand \@sanitize@url [0]{\catcode `\\12\catcode `\$12\catcode
  `\&12\catcode `\#12\catcode `\^12\catcode `\_12\catcode `\%12\relax}%
\providecommand \@@startlink[1]{}%
\providecommand \@@endlink[0]{}%
\providecommand \url  [0]{\begingroup\@sanitize@url \@url }%
\providecommand \@url [1]{\endgroup\@href {#1}{\urlprefix }}%
\providecommand \urlprefix  [0]{URL }%
\providecommand \Eprint [0]{\href }%
\providecommand \doibase [0]{http://dx.doi.org/}%
\providecommand \selectlanguage [0]{\@gobble}%
\providecommand \bibinfo  [0]{\@secondoftwo}%
\providecommand \bibfield  [0]{\@secondoftwo}%
\providecommand \translation [1]{[#1]}%
\providecommand \BibitemOpen [0]{}%
\providecommand \bibitemStop [0]{}%
\providecommand \bibitemNoStop [0]{.\EOS\space}%
\providecommand \EOS [0]{\spacefactor3000\relax}%
\providecommand \BibitemShut  [1]{\csname bibitem#1\endcsname}%
\let\auto@bib@innerbib\@empty
\bibitem [{\citenamefont {Goold}\ \emph {et~al.}(2016)\citenamefont {Goold},
  \citenamefont {Huber}, \citenamefont {Riera}, \citenamefont {del Rio},\ and\
  \citenamefont {Skrzypczyk}}]{Topical}%
  \BibitemOpen
  \bibfield  {author} {\bibinfo {author} {\bibfnamefont {J.}~\bibnamefont
  {Goold}}, \bibinfo {author} {\bibfnamefont {M.}~\bibnamefont {Huber}},
  \bibinfo {author} {\bibfnamefont {A.}~\bibnamefont {Riera}}, \bibinfo
  {author} {\bibfnamefont {L.}~\bibnamefont {del Rio}}, \ and\ \bibinfo
  {author} {\bibfnamefont {P.}~\bibnamefont {Skrzypczyk}},\ }\bibfield  {title}
  {\enquote {\bibinfo {title} {The role of quantum information in
  thermodynamics - a topical review},}\ }\href {\doibase
  10.1088/1751-8113/49/14/143001} {\bibfield  {journal} {\bibinfo  {journal}
  {J. Phys. A}\ }\textbf {\bibinfo {volume} {49}},\ \bibinfo {pages} {143001}
  (\bibinfo {year} {2016})}\BibitemShut {NoStop}%
\bibitem [{\citenamefont {Kurizki}\ \emph {et~al.}(2015)\citenamefont
  {Kurizki}, \citenamefont {Shahmoon},\ and\ \citenamefont
  {Zwick}}]{PerspectiveKurizki}%
  \BibitemOpen
  \bibfield  {author} {\bibinfo {author} {\bibfnamefont {G.}~\bibnamefont
  {Kurizki}}, \bibinfo {author} {\bibfnamefont {E.}~\bibnamefont {Shahmoon}}, \
  and\ \bibinfo {author} {\bibfnamefont {A.}~\bibnamefont {Zwick}},\ }\bibfield
   {title} {\enquote {\bibinfo {title} {Thermal baths as quantum resources:
  More friends than foes?}}\ }\href {\doibase 10.1088/0031-8949/90/12/128002}
  {\bibfield  {journal} {\bibinfo  {journal} {Phys. Scr.}\ }\textbf {\bibinfo
  {volume} {90}},\ \bibinfo {pages} {128002} (\bibinfo {year}
  {2015})}\BibitemShut {NoStop}%
\bibitem [{\citenamefont {Gogolin}\ and\ \citenamefont
  {Eisert}(2016)}]{christian_review}%
  \BibitemOpen
  \bibfield  {author} {\bibinfo {author} {\bibfnamefont {C.}~\bibnamefont
  {Gogolin}}\ and\ \bibinfo {author} {\bibfnamefont {J.}~\bibnamefont
  {Eisert}},\ }\bibfield  {title} {\enquote {\bibinfo {title} {Equilibration,
  thermalisation, and the emergence of statistical mechanics in closed quantum
  systems},}\ }\href {\doibase 10.1088/0034-4885/79/5/056001} {\bibfield
  {journal} {\bibinfo  {journal} {Rep. Prog. Phys.}\ }\textbf {\bibinfo
  {volume} {79}},\ \bibinfo {pages} {56001} (\bibinfo {year}
  {2016})}\BibitemShut {NoStop}%
\bibitem [{\citenamefont {Kosloff}(2013)}]{KosloffReview}%
  \BibitemOpen
  \bibfield  {author} {\bibinfo {author} {\bibfnamefont {R.}~\bibnamefont
  {Kosloff}},\ }\bibfield  {title} {\enquote {\bibinfo {title} {Quantum
  thermodynamics},}\ }\href {\doibase 10.3390/e15062100} {\bibfield  {journal}
  {\bibinfo  {journal} {Entropy}\ }\textbf {\bibinfo {volume} {15}},\ \bibinfo
  {pages} {2100} (\bibinfo {year} {2013})}\BibitemShut {NoStop}%
\bibitem [{\citenamefont {Millen}\ and\ \citenamefont
  {Xuereb}(2016)}]{MillenReview}%
  \BibitemOpen
  \bibfield  {author} {\bibinfo {author} {\bibfnamefont {J.}~\bibnamefont
  {Millen}}\ and\ \bibinfo {author} {\bibfnamefont {A.}~\bibnamefont
  {Xuereb}},\ }\bibfield  {title} {\enquote {\bibinfo {title} {Perspective on
  quantum thermodynamics},}\ }\href {\doibase 10.1088/1367-2630/18/1/011002}
  {\bibfield  {journal} {\bibinfo  {journal} {New J. Phys.}\ }\textbf {\bibinfo
  {volume} {18}},\ \bibinfo {pages} {011002} (\bibinfo {year}
  {2016})}\BibitemShut {NoStop}%
\bibitem [{\citenamefont {Vinjanampathy}\ and\ \citenamefont
  {Anders}(2016{\natexlab{a}})}]{Janet}%
  \BibitemOpen
  \bibfield  {author} {\bibinfo {author} {\bibfnamefont {S.}~\bibnamefont
  {Vinjanampathy}}\ and\ \bibinfo {author} {\bibfnamefont {J.}~\bibnamefont
  {Anders}},\ }\bibfield  {title} {\enquote {\bibinfo {title} {Quantum
  thermodynamics},}\ }\href {\doibase 10.1080/00107514.2016.1201896} {\bibfield
   {journal} {\bibinfo  {journal} {Contemp. Phys.}\ }\textbf {\bibinfo {volume}
  {57}},\ \bibinfo {pages} {545} (\bibinfo {year}
  {2016}{\natexlab{a}})}\BibitemShut {NoStop}%
\bibitem [{\citenamefont {Niedenzu}\ \emph
  {et~al.}(2019{\natexlab{a}})\citenamefont {Niedenzu}, \citenamefont {Mazets},
  \citenamefont {Kurizki},\ and\ \citenamefont
  {Jendrzejewski}}]{Niedenzu2019quantized}%
  \BibitemOpen
  \bibfield  {author} {\bibinfo {author} {\bibfnamefont {W.}~\bibnamefont
  {Niedenzu}}, \bibinfo {author} {\bibfnamefont {I.}~\bibnamefont {Mazets}},
  \bibinfo {author} {\bibfnamefont {G.}~\bibnamefont {Kurizki}}, \ and\
  \bibinfo {author} {\bibfnamefont {F.}~\bibnamefont {Jendrzejewski}},\
  }\bibfield  {title} {\enquote {\bibinfo {title} {Quantized refrigerator for
  an atomic cloud},}\ }\href {\doibase 10.22331/q-2019-06-28-155} {\bibfield
  {journal} {\bibinfo  {journal} {{Quantum}}\ }\textbf {\bibinfo {volume}
  {3}},\ \bibinfo {pages} {155} (\bibinfo {year}
  {2019}{\natexlab{a}})}\BibitemShut {NoStop}%
\bibitem [{\citenamefont {Ro{\ss}nagel}\ \emph {et~al.}(2016)\citenamefont
  {Ro{\ss}nagel}, \citenamefont {Dawkins}, \citenamefont {Tolazzi},
  \citenamefont {Abah}, \citenamefont {Lutz}, \citenamefont {Schmidt-Kaler},\
  and\ \citenamefont {Singer}}]{Rossnagel14}%
  \BibitemOpen
  \bibfield  {author} {\bibinfo {author} {\bibfnamefont {J.}~\bibnamefont
  {Ro{\ss}nagel}}, \bibinfo {author} {\bibfnamefont {S.}~\bibnamefont
  {Dawkins}}, \bibinfo {author} {\bibfnamefont {N.}~\bibnamefont {Tolazzi}},
  \bibinfo {author} {\bibfnamefont {O.}~\bibnamefont {Abah}}, \bibinfo {author}
  {\bibfnamefont {E.}~\bibnamefont {Lutz}}, \bibinfo {author} {\bibfnamefont
  {F.}~\bibnamefont {Schmidt-Kaler}}, \ and\ \bibinfo {author} {\bibfnamefont
  {K.}~\bibnamefont {Singer}},\ }\bibfield  {title} {\enquote {\bibinfo {title}
  {A single-atom heat engine},}\ }\href@noop {} {\bibfield  {journal} {\bibinfo
   {journal} {Science}\ }\textbf {\bibinfo {volume} {352}},\ \bibinfo {pages}
  {325} (\bibinfo {year} {2016})}\BibitemShut {NoStop}%
\bibitem [{\citenamefont {von Lindenfels}\ \emph {et~al.}(2019)\citenamefont
  {von Lindenfels}, \citenamefont {Gr\"ab}, \citenamefont {Schmiegelow},
  \citenamefont {Kaushal}, \citenamefont {Schulz}, \citenamefont {Mitchison},
  \citenamefont {Goold}, \citenamefont {Schmidt-Kaler},\ and\ \citenamefont
  {Poschinger}}]{PhysRevLett.123.080602}%
  \BibitemOpen
  \bibfield  {author} {\bibinfo {author} {\bibfnamefont {D.}~\bibnamefont {von
  Lindenfels}}, \bibinfo {author} {\bibfnamefont {O.}~\bibnamefont {Gr\"ab}},
  \bibinfo {author} {\bibfnamefont {C.~T.}\ \bibnamefont {Schmiegelow}},
  \bibinfo {author} {\bibfnamefont {V.}~\bibnamefont {Kaushal}}, \bibinfo
  {author} {\bibfnamefont {J.}~\bibnamefont {Schulz}}, \bibinfo {author}
  {\bibfnamefont {M.~T.}\ \bibnamefont {Mitchison}}, \bibinfo {author}
  {\bibfnamefont {J.}~\bibnamefont {Goold}}, \bibinfo {author} {\bibfnamefont
  {F.}~\bibnamefont {Schmidt-Kaler}}, \ and\ \bibinfo {author} {\bibfnamefont
  {U.~G.}\ \bibnamefont {Poschinger}},\ }\bibfield  {title} {\enquote {\bibinfo
  {title} {Spin heat engine coupled to a harmonic-oscillator flywheel},}\
  }\href {\doibase 10.1103/PhysRevLett.123.080602} {\bibfield  {journal}
  {\bibinfo  {journal} {Phys. Rev. Lett.}\ }\textbf {\bibinfo {volume} {123}},\
  \bibinfo {pages} {080602} (\bibinfo {year} {2019})}\BibitemShut {NoStop}%
\bibitem [{\citenamefont {Horne}\ \emph {et~al.}(2020)\citenamefont {Horne},
  \citenamefont {Yum}, \citenamefont {Dutta}, \citenamefont {H{\"a}nggi},
  \citenamefont {Gong}, \citenamefont {Poletti},\ and\ \citenamefont
  {Mukherjee}}]{Haenggi}%
  \BibitemOpen
  \bibfield  {author} {\bibinfo {author} {\bibfnamefont {N.~V.}\ \bibnamefont
  {Horne}}, \bibinfo {author} {\bibfnamefont {D.}~\bibnamefont {Yum}}, \bibinfo
  {author} {\bibfnamefont {T.}~\bibnamefont {Dutta}}, \bibinfo {author}
  {\bibfnamefont {P.}~\bibnamefont {H{\"a}nggi}}, \bibinfo {author}
  {\bibfnamefont {J.}~\bibnamefont {Gong}}, \bibinfo {author} {\bibfnamefont
  {D.}~\bibnamefont {Poletti}}, \ and\ \bibinfo {author} {\bibfnamefont
  {M.}~\bibnamefont {Mukherjee}},\ }\bibfield  {title} {\enquote {\bibinfo
  {title} {Single-atom energy-conversion device with a quantum load},}\ }\href
  {\doibase 10.1038/s41534-020-0264-6} {\bibfield  {journal} {\bibinfo
  {journal} {npj Quant. Inf.}\ }\textbf {\bibinfo {volume} {6}},\ \bibinfo
  {pages} {37} (\bibinfo {year} {2020})}\BibitemShut {NoStop}%
\bibitem [{\citenamefont {Ono}\ \emph {et~al.}(2020)\citenamefont {Ono},
  \citenamefont {Shevchenko}, \citenamefont {Mori}, \citenamefont {Moriyama},\
  and\ \citenamefont {Nori}}]{PhysRevLett.125.166802}%
  \BibitemOpen
  \bibfield  {author} {\bibinfo {author} {\bibfnamefont {K.}~\bibnamefont
  {Ono}}, \bibinfo {author} {\bibfnamefont {S.~N.}\ \bibnamefont {Shevchenko}},
  \bibinfo {author} {\bibfnamefont {T.}~\bibnamefont {Mori}}, \bibinfo {author}
  {\bibfnamefont {S.}~\bibnamefont {Moriyama}}, \ and\ \bibinfo {author}
  {\bibfnamefont {F.}~\bibnamefont {Nori}},\ }\bibfield  {title} {\enquote
  {\bibinfo {title} {Analog of a quantum heat engine using a single-spin
  qubit},}\ }\href {\doibase 10.1103/PhysRevLett.125.166802} {\bibfield
  {journal} {\bibinfo  {journal} {Phys. Rev. Lett.}\ }\textbf {\bibinfo
  {volume} {125}},\ \bibinfo {pages} {166802} (\bibinfo {year}
  {2020})}\BibitemShut {NoStop}%
\bibitem [{\citenamefont {Pekola}(2015)}]{NewReferenceKSPekola}%
  \BibitemOpen
  \bibfield  {author} {\bibinfo {author} {\bibfnamefont {J.~P.}\ \bibnamefont
  {Pekola}},\ }\bibfield  {title} {\enquote {\bibinfo {title} {Towards quantum
  thermodynamics in electronic circuits},}\ }\href {\doibase 10.1038/nphys3169}
  {\bibfield  {journal} {\bibinfo  {journal} {Nature Phys.}\ }\textbf {\bibinfo
  {volume} {11}},\ \bibinfo {pages} {118--123} (\bibinfo {year}
  {2015})}\BibitemShut {NoStop}%
\bibitem [{\citenamefont {Klatzow}\ \emph {et~al.}(2019)\citenamefont
  {Klatzow}, \citenamefont {Becker}, \citenamefont {Ledingham}, \citenamefont
  {Weinzetl}, \citenamefont {Kaczmarek}, \citenamefont {Saunders},
  \citenamefont {Nunn}, \citenamefont {Walmsley}, \citenamefont {Uzdin},\ and\
  \citenamefont {Poem}}]{PhysRevLett.122.110601}%
  \BibitemOpen
  \bibfield  {author} {\bibinfo {author} {\bibfnamefont {J.}~\bibnamefont
  {Klatzow}}, \bibinfo {author} {\bibfnamefont {J.~N.}\ \bibnamefont {Becker}},
  \bibinfo {author} {\bibfnamefont {P.~M.}\ \bibnamefont {Ledingham}}, \bibinfo
  {author} {\bibfnamefont {C.}~\bibnamefont {Weinzetl}}, \bibinfo {author}
  {\bibfnamefont {K.~T.}\ \bibnamefont {Kaczmarek}}, \bibinfo {author}
  {\bibfnamefont {D.~J.}\ \bibnamefont {Saunders}}, \bibinfo {author}
  {\bibfnamefont {J.}~\bibnamefont {Nunn}}, \bibinfo {author} {\bibfnamefont
  {I.~A.}\ \bibnamefont {Walmsley}}, \bibinfo {author} {\bibfnamefont
  {R.}~\bibnamefont {Uzdin}}, \ and\ \bibinfo {author} {\bibfnamefont
  {E.}~\bibnamefont {Poem}},\ }\bibfield  {title} {\enquote {\bibinfo {title}
  {Experimental demonstration of quantum effects in the operation of
  microscopic heat engines},}\ }\href {\doibase 10.1103/PhysRevLett.122.110601}
  {\bibfield  {journal} {\bibinfo  {journal} {Phys. Rev. Lett.}\ }\textbf
  {\bibinfo {volume} {122}},\ \bibinfo {pages} {110601} (\bibinfo {year}
  {2019})}\BibitemShut {NoStop}%
\bibitem [{\citenamefont {Rauer}(2019)}]{Rauer19}%
  \BibitemOpen
  \bibfield  {author} {\bibinfo {author} {\bibfnamefont {B.}~\bibnamefont
  {Rauer}},\ }\emph {\bibinfo {title} {Non-equilibrium dynamics beyond
  dephasing recurrences and loss induced cooling in one-dimensional Bose
  gases}},\ \href@noop {} {Ph.D. thesis} (\bibinfo {year} {2019})\BibitemShut
  {NoStop}%
\bibitem [{\citenamefont {Schweigler}\ \emph {et~al.}(2021)\citenamefont
  {Schweigler}, \citenamefont {Gluza}, \citenamefont {Tajik}, \citenamefont
  {Sotiriadis}, \citenamefont {Cataldini}, \citenamefont {Ji}, \citenamefont
  {M{\o}ller}, \citenamefont {Sabino}, \citenamefont {Rauer}, \citenamefont
  {Eisert},\ and\ \citenamefont {J.}}]{schweigler2020decay}%
  \BibitemOpen
  \bibfield  {author} {\bibinfo {author} {\bibfnamefont {T.}~\bibnamefont
  {Schweigler}}, \bibinfo {author} {\bibfnamefont {M.}~\bibnamefont {Gluza}},
  \bibinfo {author} {\bibfnamefont {M.}~\bibnamefont {Tajik}}, \bibinfo
  {author} {\bibfnamefont {S.}~\bibnamefont {Sotiriadis}}, \bibinfo {author}
  {\bibfnamefont {F.}~\bibnamefont {Cataldini}}, \bibinfo {author}
  {\bibfnamefont {S.-C.}\ \bibnamefont {Ji}}, \bibinfo {author} {\bibfnamefont
  {F.~S.}\ \bibnamefont {M{\o}ller}}, \bibinfo {author} {\bibfnamefont
  {J.}~\bibnamefont {Sabino}}, \bibinfo {author} {\bibfnamefont
  {B.}~\bibnamefont {Rauer}}, \bibinfo {author} {\bibfnamefont
  {J.}~\bibnamefont {Eisert}}, \ and\ \bibinfo {author} {\bibfnamefont
  {S.}~\bibnamefont {J.}},\ }\bibfield  {title} {\enquote {\bibinfo {title}
  {{Decay and recurrence of non-Gaussian correlations in a quantum many-body
  system}},}\ }\href {\doibase 10.1038/s41567-020-01139-2} {\bibfield
  {journal} {\bibinfo  {journal} {Nature Phys.}\ }\textbf {\bibinfo {volume}
  {17}},\ \bibinfo {pages} {559} (\bibinfo {year} {2021})}\BibitemShut
  {NoStop}%
\bibitem [{\citenamefont {Schmiedmayer}(2018)}]{Schmiedmayer2018}%
  \BibitemOpen
  \bibfield  {author} {\bibinfo {author} {\bibfnamefont {J.}~\bibnamefont
  {Schmiedmayer}},\ }\bibfield  {title} {\enquote {\bibinfo {title}
  {{One-dimensional atomic superfluids as a model system for quantum
  thermodynamics}},}\ }in\ \href {\doibase 10.1007/978-3-319-99046-0_34} {\emph
  {\bibinfo {booktitle} {Thermodynamics in the Quantum Regime}}},\ \bibinfo
  {editor} {edited by\ \bibinfo {editor} {\bibfnamefont {F.}~\bibnamefont
  {Binder}}, \bibinfo {editor} {\bibfnamefont {L.~A.}\ \bibnamefont {Correa}},
  \bibinfo {editor} {\bibfnamefont {C.}~\bibnamefont {Gogolin}}, \bibinfo
  {editor} {\bibfnamefont {J.}~\bibnamefont {Anders}}, \ and\ \bibinfo {editor}
  {\bibfnamefont {G.}~\bibnamefont {Adesso}}}\ (\bibinfo  {publisher} {Springer
  Nature},\ \bibinfo {year} {2018})\ Chap.~\bibinfo {chapter} {34}, pp.\
  \bibinfo {pages} {823--851},\ \Eprint {http://arxiv.org/abs/1805.11539}
  {arXiv:1805.11539} \BibitemShut {NoStop}%
\bibitem [{\citenamefont {Gluza}\ \emph {et~al.}(2020)\citenamefont {Gluza},
  \citenamefont {Schweigler}, \citenamefont {Rauer}, \citenamefont {Krumnow},
  \citenamefont {Schmiedmayer},\ and\ \citenamefont {Eisert}}]{quantumreadout}%
  \BibitemOpen
  \bibfield  {author} {\bibinfo {author} {\bibfnamefont {M.}~\bibnamefont
  {Gluza}}, \bibinfo {author} {\bibfnamefont {T.}~\bibnamefont {Schweigler}},
  \bibinfo {author} {\bibfnamefont {B.}~\bibnamefont {Rauer}}, \bibinfo
  {author} {\bibfnamefont {C.}~\bibnamefont {Krumnow}}, \bibinfo {author}
  {\bibfnamefont {J.}~\bibnamefont {Schmiedmayer}}, \ and\ \bibinfo {author}
  {\bibfnamefont {J.}~\bibnamefont {Eisert}},\ }\bibfield  {title} {\enquote
  {\bibinfo {title} {Quantum read-out for cold atomic quantum simulators},}\
  }\href@noop {} {\bibfield  {journal} {\bibinfo  {journal} {Comm. Phys,}\
  }\textbf {\bibinfo {volume} {3}},\ \bibinfo {pages} {12} (\bibinfo {year}
  {2020})}\BibitemShut {NoStop}%
\bibitem [{\citenamefont {Cazalilla}(2004)}]{Cazalilla}%
  \BibitemOpen
  \bibfield  {author} {\bibinfo {author} {\bibfnamefont {M.~A.}\ \bibnamefont
  {Cazalilla}},\ }\bibfield  {title} {\enquote {\bibinfo {title} {Bosonizing
  one-dimensional cold atomic gases},}\ }\href@noop {} {\bibfield  {journal}
  {\bibinfo  {journal} {J. Phys. B}\ }\textbf {\bibinfo {volume} {37}},\
  \bibinfo {pages} {S1} (\bibinfo {year} {2004})}\BibitemShut {NoStop}%
\bibitem [{\citenamefont {Giamarchi}(2004)}]{Giamarchi2004}%
  \BibitemOpen
  \bibfield  {author} {\bibinfo {author} {\bibfnamefont {T.}~\bibnamefont
  {Giamarchi}},\ }\href@noop {} {\emph {\bibinfo {title} {Quantum physics in
  one dimension}}}\ (\bibinfo  {publisher} {Clarendon Press},\ \bibinfo
  {address} {Oxford},\ \bibinfo {year} {2004})\BibitemShut {NoStop}%
\bibitem [{\citenamefont {Schweigler}\ \emph {et~al.}(2017)\citenamefont
  {Schweigler}, \citenamefont {Kasper}, \citenamefont {Erne}, \citenamefont
  {Mazets}, \citenamefont {Rauer}, \citenamefont {Cataldini}, \citenamefont
  {Langen}, \citenamefont {Gasenzer}, \citenamefont {Berges},\ and\
  \citenamefont {Schmiedmayer}}]{Schweigler2017}%
  \BibitemOpen
  \bibfield  {author} {\bibinfo {author} {\bibfnamefont {T.}~\bibnamefont
  {Schweigler}}, \bibinfo {author} {\bibfnamefont {V.}~\bibnamefont {Kasper}},
  \bibinfo {author} {\bibfnamefont {S.}~\bibnamefont {Erne}}, \bibinfo {author}
  {\bibfnamefont {I.~E.}\ \bibnamefont {Mazets}}, \bibinfo {author}
  {\bibfnamefont {B.}~\bibnamefont {Rauer}}, \bibinfo {author} {\bibfnamefont
  {F.}~\bibnamefont {Cataldini}}, \bibinfo {author} {\bibfnamefont
  {T.}~\bibnamefont {Langen}}, \bibinfo {author} {\bibfnamefont
  {T.}~\bibnamefont {Gasenzer}}, \bibinfo {author} {\bibfnamefont
  {J.}~\bibnamefont {Berges}}, \ and\ \bibinfo {author} {\bibfnamefont
  {J.}~\bibnamefont {Schmiedmayer}},\ }\bibfield  {title} {\enquote {\bibinfo
  {title} {{Experimental characterization of a quantum many-body system via
  higher-order correlations}},}\ }\href {\doibase 10.1038/nature22310}
  {\bibfield  {journal} {\bibinfo  {journal} {Nature}\ }\textbf {\bibinfo
  {volume} {545}},\ \bibinfo {pages} {323--326} (\bibinfo {year}
  {2017})}\BibitemShut {NoStop}%
\bibitem [{\citenamefont {Mora}\ and\ \citenamefont {Castin}(2003)}]{Mora03}%
  \BibitemOpen
  \bibfield  {author} {\bibinfo {author} {\bibfnamefont {C.}~\bibnamefont
  {Mora}}\ and\ \bibinfo {author} {\bibfnamefont {Y.}~\bibnamefont {Castin}},\
  }\bibfield  {title} {\enquote {\bibinfo {title} {{Extension of Bogoliubov
  theory to quasicondensates}},}\ }\href {\doibase 10.1103/PhysRevA.67.053615}
  {\bibfield  {journal} {\bibinfo  {journal} {Phys. Rev. A}\ }\textbf {\bibinfo
  {volume} {67}},\ \bibinfo {pages} {053615} (\bibinfo {year}
  {2003})}\BibitemShut {NoStop}%
\bibitem [{\citenamefont {Popov}(2001)}]{popov2001functional}%
  \BibitemOpen
  \bibfield  {author} {\bibinfo {author} {\bibfnamefont {V.~N.}\ \bibnamefont
  {Popov}},\ }\href@noop {} {\emph {\bibinfo {title} {Functional integrals in
  quantum field theory and statistical physics}}},\ Vol.~\bibinfo {volume} {8}\
  (\bibinfo  {publisher} {Springer Science \& Business Media},\ \bibinfo {year}
  {2001})\BibitemShut {NoStop}%
\bibitem [{\citenamefont {Gritsev}\ \emph {et~al.}(2007)\citenamefont
  {Gritsev}, \citenamefont {Polkovnikov},\ and\ \citenamefont
  {Demler}}]{Gritsev07}%
  \BibitemOpen
  \bibfield  {author} {\bibinfo {author} {\bibfnamefont {V.}~\bibnamefont
  {Gritsev}}, \bibinfo {author} {\bibfnamefont {A.}~\bibnamefont
  {Polkovnikov}}, \ and\ \bibinfo {author} {\bibfnamefont {E.}~\bibnamefont
  {Demler}},\ }\bibfield  {title} {\enquote {\bibinfo {title} {Linear response
  theory for a pair of coupled one-dimensional condensates of interacting
  atoms},}\ }\href {\doibase 10.1103/PhysRevB.75.174511} {\bibfield  {journal}
  {\bibinfo  {journal} {Phys. Rev. B}\ }\textbf {\bibinfo {volume} {75}},\
  \bibinfo {pages} {174511} (\bibinfo {year} {2007})}\BibitemShut {NoStop}%
\bibitem [{\citenamefont {Folman}\ \emph {et~al.}(2000)\citenamefont {Folman},
  \citenamefont {Kr\"uger}, \citenamefont {Cassettari}, \citenamefont {Hessmo},
  \citenamefont {Maier},\ and\ \citenamefont {Schmiedmayer}}]{Folman2000}%
  \BibitemOpen
  \bibfield  {author} {\bibinfo {author} {\bibfnamefont {R.}~\bibnamefont
  {Folman}}, \bibinfo {author} {\bibfnamefont {P.}~\bibnamefont {Kr\"uger}},
  \bibinfo {author} {\bibfnamefont {D.}~\bibnamefont {Cassettari}}, \bibinfo
  {author} {\bibfnamefont {B.}~\bibnamefont {Hessmo}}, \bibinfo {author}
  {\bibfnamefont {T.}~\bibnamefont {Maier}}, \ and\ \bibinfo {author}
  {\bibfnamefont {J.}~\bibnamefont {Schmiedmayer}},\ }\bibfield  {title}
  {\enquote {\bibinfo {title} {Controlling cold atoms using nanofabricated
  surfaces: Atom chips},}\ }\href {\doibase 10.1103/PhysRevLett.84.4749}
  {\bibfield  {journal} {\bibinfo  {journal} {Phys. Rev. Lett.}\ }\textbf
  {\bibinfo {volume} {84}},\ \bibinfo {pages} {4749--4752} (\bibinfo {year}
  {2000})}\BibitemShut {NoStop}%
\bibitem [{\citenamefont {Folman}\ \emph {et~al.}(2002)\citenamefont {Folman},
  \citenamefont {Krüger}, \citenamefont {Schmiedmayer}, \citenamefont
  {Denschlag},\ and\ \citenamefont {Henkel}}]{FOLMAN2002263}%
  \BibitemOpen
  \bibfield  {author} {\bibinfo {author} {\bibfnamefont {R.}~\bibnamefont
  {Folman}}, \bibinfo {author} {\bibfnamefont {P.}~\bibnamefont {Krüger}},
  \bibinfo {author} {\bibfnamefont {J.}~\bibnamefont {Schmiedmayer}}, \bibinfo
  {author} {\bibfnamefont {J.}~\bibnamefont {Denschlag}}, \ and\ \bibinfo
  {author} {\bibfnamefont {C.}~\bibnamefont {Henkel}},\ }\bibfield  {title}
  {\enquote {\bibinfo {title} {Microscopic atom optics: From wires to an atom
  chip},}\ \ }(\bibinfo  {publisher} {Academic Press},\ \bibinfo {year}
  {2002})\ pp.\ \bibinfo {pages} {263 -- 356}\BibitemShut {NoStop}%
\bibitem [{\citenamefont {Reichel}\ and\ \citenamefont
  {Vuletic}(2011)}]{reichel2011atom}%
  \BibitemOpen
  \bibfield  {author} {\bibinfo {author} {\bibfnamefont {J.}~\bibnamefont
  {Reichel}}\ and\ \bibinfo {author} {\bibfnamefont {V.}~\bibnamefont
  {Vuletic}},\ }\href@noop {} {\emph {\bibinfo {title} {Atom chips}}}\
  (\bibinfo  {publisher} {John Wiley \& Sons},\ \bibinfo {year}
  {2011})\BibitemShut {NoStop}%
\bibitem [{\citenamefont {Tajik}\ \emph {et~al.}(2019)\citenamefont {Tajik},
  \citenamefont {Rauer}, \citenamefont {Schweigler}, \citenamefont {Cataldini},
  \citenamefont {{a}o Sabino}, \citenamefont {M{\o}ller}, \citenamefont {Ji},
  \citenamefont {Mazets},\ and\ \citenamefont {Schmiedmayer}}]{Tajik2019}%
  \BibitemOpen
  \bibfield  {author} {\bibinfo {author} {\bibfnamefont {M.}~\bibnamefont
  {Tajik}}, \bibinfo {author} {\bibfnamefont {B.}~\bibnamefont {Rauer}},
  \bibinfo {author} {\bibfnamefont {T.}~\bibnamefont {Schweigler}}, \bibinfo
  {author} {\bibfnamefont {F.}~\bibnamefont {Cataldini}}, \bibinfo {author}
  {\bibfnamefont {J.}~\bibnamefont {{a}o Sabino}}, \bibinfo {author}
  {\bibfnamefont {F.~S.}\ \bibnamefont {M{\o}ller}}, \bibinfo {author}
  {\bibfnamefont {S.-C.}\ \bibnamefont {Ji}}, \bibinfo {author} {\bibfnamefont
  {I.~E.}\ \bibnamefont {Mazets}}, \ and\ \bibinfo {author} {\bibfnamefont
  {J.}~\bibnamefont {Schmiedmayer}},\ }\bibfield  {title} {\enquote {\bibinfo
  {title} {Designing arbitrary one-dimensional potentials on an atom chip},}\
  }\href {\doibase 10.1364/OE.27.033474} {\bibfield  {journal} {\bibinfo
  {journal} {Opt. Express}\ }\textbf {\bibinfo {volume} {27}},\ \bibinfo
  {pages} {33474--33487} (\bibinfo {year} {2019})}\BibitemShut {NoStop}%
\bibitem [{\citenamefont {Petrov}\ \emph {et~al.}(2000)\citenamefont {Petrov},
  \citenamefont {Shlyapnikov},\ and\ \citenamefont {Walraven}}]{Petrov00}%
  \BibitemOpen
  \bibfield  {author} {\bibinfo {author} {\bibfnamefont {D.~S.}\ \bibnamefont
  {Petrov}}, \bibinfo {author} {\bibfnamefont {G.~V.}\ \bibnamefont
  {Shlyapnikov}}, \ and\ \bibinfo {author} {\bibfnamefont {J.~T.~M.}\
  \bibnamefont {Walraven}},\ }\bibfield  {title} {\enquote {\bibinfo {title}
  {{Regimes of quantum degeneracy in trapped 1D gases}},}\ }\href@noop {}
  {\bibfield  {journal} {\bibinfo  {journal} {Phys. Rev. Lett.}\ }\textbf
  {\bibinfo {volume} {85}},\ \bibinfo {pages} {3745--3749} (\bibinfo {year}
  {2000})}\BibitemShut {NoStop}%
\bibitem [{\citenamefont {Lewenstein}\ and\ \citenamefont
  {You}(1996)}]{lewenstein1996quantum}%
  \BibitemOpen
  \bibfield  {author} {\bibinfo {author} {\bibfnamefont {M.}~\bibnamefont
  {Lewenstein}}\ and\ \bibinfo {author} {\bibfnamefont {L.}~\bibnamefont
  {You}},\ }\bibfield  {title} {\enquote {\bibinfo {title} {Quantum phase
  diffusion of a bose-einstein condensate},}\ }\href@noop {} {\bibfield
  {journal} {\bibinfo  {journal} {Phys. Rev. Lett.}\ }\textbf {\bibinfo
  {volume} {77}},\ \bibinfo {pages} {3489} (\bibinfo {year}
  {1996})}\BibitemShut {NoStop}%
\bibitem [{\citenamefont {Javanainen}\ and\ \citenamefont
  {Wilkens}(1997)}]{JavanainenWilkens}%
  \BibitemOpen
  \bibfield  {author} {\bibinfo {author} {\bibfnamefont {J.}~\bibnamefont
  {Javanainen}}\ and\ \bibinfo {author} {\bibfnamefont {M.}~\bibnamefont
  {Wilkens}},\ }\bibfield  {title} {\enquote {\bibinfo {title} {{Phase and
  phase diffusion of a split Bose-Einstein condensate}},}\ }\href {\doibase
  10.1103/PhysRevLett.78.4675} {\bibfield  {journal} {\bibinfo  {journal}
  {Phys. Rev. Lett.}\ }\textbf {\bibinfo {volume} {78}},\ \bibinfo {pages}
  {4675--4678} (\bibinfo {year} {1997})}\BibitemShut {NoStop}%
\bibitem [{\citenamefont {Leggett}\ and\ \citenamefont
  {Sols}(1998)}]{LeggettComment}%
  \BibitemOpen
  \bibfield  {author} {\bibinfo {author} {\bibfnamefont {A.~J.}\ \bibnamefont
  {Leggett}}\ and\ \bibinfo {author} {\bibfnamefont {F.}~\bibnamefont {Sols}},\
  }\bibfield  {title} {\enquote {\bibinfo {title} {Comment on ``phase and phase
  diffusion of a split bose-einstein condensate''},}\ }\href {\doibase
  10.1103/PhysRevLett.81.1344} {\bibfield  {journal} {\bibinfo  {journal}
  {Phys. Rev. Lett.}\ }\textbf {\bibinfo {volume} {81}},\ \bibinfo {pages}
  {1344--1344} (\bibinfo {year} {1998})}\BibitemShut {NoStop}%
\bibitem [{\citenamefont {Javanainen}\ and\ \citenamefont
  {Wilkens}(1998)}]{JavanainenReply}%
  \BibitemOpen
  \bibfield  {author} {\bibinfo {author} {\bibfnamefont {J.}~\bibnamefont
  {Javanainen}}\ and\ \bibinfo {author} {\bibfnamefont {M.}~\bibnamefont
  {Wilkens}},\ }\bibfield  {title} {\enquote {\bibinfo {title} {Javanainen and
  wilkens reply:},}\ }\href {\doibase 10.1103/PhysRevLett.81.1345} {\bibfield
  {journal} {\bibinfo  {journal} {Phys. Rev. Lett.}\ }\textbf {\bibinfo
  {volume} {81}},\ \bibinfo {pages} {1345--1345} (\bibinfo {year}
  {1998})}\BibitemShut {NoStop}%
\bibitem [{\citenamefont {Rauer}\ \emph {et~al.}(2018)\citenamefont {Rauer},
  \citenamefont {Erne}, \citenamefont {Schweigler}, \citenamefont {Cataldini},
  \citenamefont {Tajik},\ and\ \citenamefont {Schmiedmayer}}]{Rauer2018}%
  \BibitemOpen
  \bibfield  {author} {\bibinfo {author} {\bibfnamefont {B.}~\bibnamefont
  {Rauer}}, \bibinfo {author} {\bibfnamefont {S.}~\bibnamefont {Erne}},
  \bibinfo {author} {\bibfnamefont {T.}~\bibnamefont {Schweigler}}, \bibinfo
  {author} {\bibfnamefont {F.}~\bibnamefont {Cataldini}}, \bibinfo {author}
  {\bibfnamefont {M.}~\bibnamefont {Tajik}}, \ and\ \bibinfo {author}
  {\bibfnamefont {J.}~\bibnamefont {Schmiedmayer}},\ }\bibfield  {title}
  {\enquote {\bibinfo {title} {Recurrences in an isolated quantum many-body
  system},}\ }\href {\doibase 10.1126/science.aan7938} {\bibfield  {journal}
  {\bibinfo  {journal} {Science}\ }\textbf {\bibinfo {volume} {359}},\ \bibinfo
  {pages} {307--310} (\bibinfo {year} {2018})}\BibitemShut {NoStop}%
\bibitem [{\citenamefont {Schweigler}(2019{\natexlab{a}})}]{Schweigler_thesis}%
  \BibitemOpen
  \bibfield  {author} {\bibinfo {author} {\bibfnamefont {T.}~\bibnamefont
  {Schweigler}},\ }\emph {\bibinfo {title} {Correlations and dynamics of
  tunnel-coupled one-dimensional Bose gases}},\ \href@noop {} {Ph.D. thesis}
  (\bibinfo {year} {2019}{\natexlab{a}})\BibitemShut {NoStop}%
\bibitem [{\citenamefont {Kagan}\ \emph {et~al.}(2003)\citenamefont {Kagan},
  \citenamefont {Kovrizhin},\ and\ \citenamefont {Maksimov}}]{KaganeEtAl}%
  \BibitemOpen
  \bibfield  {author} {\bibinfo {author} {\bibfnamefont {Y.}~\bibnamefont
  {Kagan}}, \bibinfo {author} {\bibfnamefont {D.~L.}\ \bibnamefont
  {Kovrizhin}}, \ and\ \bibinfo {author} {\bibfnamefont {L.~A.}\ \bibnamefont
  {Maksimov}},\ }\bibfield  {title} {\enquote {\bibinfo {title} {{Anomalous
  tunneling of phonon excitations between two Bose-Einstein condensates}},}\
  }\href {\doibase 10.1103/PhysRevLett.90.130402} {\bibfield  {journal}
  {\bibinfo  {journal} {Phys. Rev. Lett.}\ }\textbf {\bibinfo {volume} {90}},\
  \bibinfo {pages} {130402} (\bibinfo {year} {2003})}\BibitemShut {NoStop}%
\bibitem [{\citenamefont {Menotti}\ \emph {et~al.}(2001)\citenamefont
  {Menotti}, \citenamefont {Anglin}, \citenamefont {Cirac},\ and\ \citenamefont
  {Zoller}}]{MenottiAnglingCiracZollerSplitting}%
  \BibitemOpen
  \bibfield  {author} {\bibinfo {author} {\bibfnamefont {C.}~\bibnamefont
  {Menotti}}, \bibinfo {author} {\bibfnamefont {J.~R.}\ \bibnamefont {Anglin}},
  \bibinfo {author} {\bibfnamefont {J.~I.}\ \bibnamefont {Cirac}}, \ and\
  \bibinfo {author} {\bibfnamefont {P.}~\bibnamefont {Zoller}},\ }\bibfield
  {title} {\enquote {\bibinfo {title} {{Dynamic splitting of a Bose-Einstein
  condensate}},}\ }\href {\doibase 10.1103/PhysRevA.63.023601} {\bibfield
  {journal} {\bibinfo  {journal} {Phys. Rev. A}\ }\textbf {\bibinfo {volume}
  {63}},\ \bibinfo {pages} {023601} (\bibinfo {year} {2001})}\BibitemShut
  {NoStop}%
\bibitem [{\citenamefont {Gring}\ \emph {et~al.}(2012)\citenamefont {Gring},
  \citenamefont {Kuhnert}, \citenamefont {Langen}, \citenamefont {Kitagawa},
  \citenamefont {Rauer}, \citenamefont {Schreitl}, \citenamefont {Mazets},
  \citenamefont {Smith}, \citenamefont {Demler},\ and\ \citenamefont
  {Schmiedmayer}}]{Gring2012}%
  \BibitemOpen
  \bibfield  {author} {\bibinfo {author} {\bibfnamefont {M.}~\bibnamefont
  {Gring}}, \bibinfo {author} {\bibfnamefont {M.}~\bibnamefont {Kuhnert}},
  \bibinfo {author} {\bibfnamefont {T.}~\bibnamefont {Langen}}, \bibinfo
  {author} {\bibfnamefont {T.}~\bibnamefont {Kitagawa}}, \bibinfo {author}
  {\bibfnamefont {B.}~\bibnamefont {Rauer}}, \bibinfo {author} {\bibfnamefont
  {M.}~\bibnamefont {Schreitl}}, \bibinfo {author} {\bibfnamefont {I.~E.}\
  \bibnamefont {Mazets}}, \bibinfo {author} {\bibfnamefont {D.~A.}\
  \bibnamefont {Smith}}, \bibinfo {author} {\bibfnamefont {E.}~\bibnamefont
  {Demler}}, \ and\ \bibinfo {author} {\bibfnamefont {J.}~\bibnamefont
  {Schmiedmayer}},\ }\bibfield  {title} {\enquote {\bibinfo {title}
  {{Relaxation and prethermalization in an isolated quantum system.}}}\ }\href
  {\doibase 10.1126/science.1224953} {\bibfield  {journal} {\bibinfo  {journal}
  {Science}\ }\textbf {\bibinfo {volume} {337}},\ \bibinfo {pages} {1318--22}
  (\bibinfo {year} {2012})}\BibitemShut {NoStop}%
\bibitem [{\citenamefont {Carusotto}\ \emph {et~al.}(2010)\citenamefont
  {Carusotto}, \citenamefont {Balbinot}, \citenamefont {Fabbri},\ and\
  \citenamefont {Recati}}]{carusotto2010density}%
  \BibitemOpen
  \bibfield  {author} {\bibinfo {author} {\bibfnamefont {I.}~\bibnamefont
  {Carusotto}}, \bibinfo {author} {\bibfnamefont {R.}~\bibnamefont {Balbinot}},
  \bibinfo {author} {\bibfnamefont {A.}~\bibnamefont {Fabbri}}, \ and\ \bibinfo
  {author} {\bibfnamefont {A.}~\bibnamefont {Recati}},\ }\bibfield  {title}
  {\enquote {\bibinfo {title} {Density correlations and analog dynamical
  casimir emission of bogoliubov phonons in modulated atomic bose-einstein
  condensates},}\ }\href@noop {} {\bibfield  {journal} {\bibinfo  {journal}
  {Europ. Phys. J. D}\ }\textbf {\bibinfo {volume} {56}},\ \bibinfo {pages}
  {391--404} (\bibinfo {year} {2010})}\BibitemShut {NoStop}%
\bibitem [{\citenamefont {Michael}\ \emph {et~al.}(2019)\citenamefont
  {Michael}, \citenamefont {Schmiedmayer},\ and\ \citenamefont
  {Demler}}]{PhysRevA.99.053615}%
  \BibitemOpen
  \bibfield  {author} {\bibinfo {author} {\bibfnamefont {M.~H.}\ \bibnamefont
  {Michael}}, \bibinfo {author} {\bibfnamefont {J.}~\bibnamefont
  {Schmiedmayer}}, \ and\ \bibinfo {author} {\bibfnamefont {E.}~\bibnamefont
  {Demler}},\ }\bibfield  {title} {\enquote {\bibinfo {title} {From the moving
  piston to the dynamical casimir effect: Explorations with shaken
  condensates},}\ }\href {\doibase 10.1103/PhysRevA.99.053615} {\bibfield
  {journal} {\bibinfo  {journal} {Phys. Rev. A}\ }\textbf {\bibinfo {volume}
  {99}},\ \bibinfo {pages} {053615} (\bibinfo {year} {2019})}\BibitemShut
  {NoStop}%
\bibitem [{\citenamefont {Chen}\ \emph {et~al.}(2019)\citenamefont {Chen},
  \citenamefont {Watanabe}, \citenamefont {Yu}, \citenamefont {Guan},\ and\
  \citenamefont {del Campo}}]{chen2019interaction}%
  \BibitemOpen
  \bibfield  {author} {\bibinfo {author} {\bibfnamefont {Y.-Y.}\ \bibnamefont
  {Chen}}, \bibinfo {author} {\bibfnamefont {G.}~\bibnamefont {Watanabe}},
  \bibinfo {author} {\bibfnamefont {Y.-C.}\ \bibnamefont {Yu}}, \bibinfo
  {author} {\bibfnamefont {X.-W.}\ \bibnamefont {Guan}}, \ and\ \bibinfo
  {author} {\bibfnamefont {A.}~\bibnamefont {del Campo}},\ }\bibfield  {title}
  {\enquote {\bibinfo {title} {An interaction-driven many-particle quantum heat
  engine and its universal behavior},}\ }\href@noop {} {\bibfield  {journal}
  {\bibinfo  {journal} {npj Quant. Inf.}\ }\textbf {\bibinfo {volume} {5}},\
  \bibinfo {pages} {1--6} (\bibinfo {year} {2019})}\BibitemShut {NoStop}%
\bibitem [{\citenamefont {Jaramillo}\ \emph {et~al.}(2016)\citenamefont
  {Jaramillo}, \citenamefont {Beau},\ and\ \citenamefont {del
  Campo}}]{Jaramillo_2016}%
  \BibitemOpen
  \bibfield  {author} {\bibinfo {author} {\bibfnamefont {J.}~\bibnamefont
  {Jaramillo}}, \bibinfo {author} {\bibfnamefont {M.}~\bibnamefont {Beau}}, \
  and\ \bibinfo {author} {\bibfnamefont {A.}~\bibnamefont {del Campo}},\
  }\bibfield  {title} {\enquote {\bibinfo {title} {Quantum supremacy of
  many-particle thermal machines},}\ }\href {\doibase
  10.1088/1367-2630/18/7/075019} {\bibfield  {journal} {\bibinfo  {journal}
  {New J. Phys.}\ }\textbf {\bibinfo {volume} {18}},\ \bibinfo {pages} {075019}
  (\bibinfo {year} {2016})}\BibitemShut {NoStop}%
\bibitem [{\citenamefont {Cazalilla}\ \emph {et~al.}(2011)\citenamefont
  {Cazalilla}, \citenamefont {Citro}, \citenamefont {Giamarchi}, \citenamefont
  {Orignac},\ and\ \citenamefont {Rigol}}]{Cazalilla11}%
  \BibitemOpen
  \bibfield  {author} {\bibinfo {author} {\bibfnamefont {M.~A.}\ \bibnamefont
  {Cazalilla}}, \bibinfo {author} {\bibfnamefont {R.}~\bibnamefont {Citro}},
  \bibinfo {author} {\bibfnamefont {T.}~\bibnamefont {Giamarchi}}, \bibinfo
  {author} {\bibfnamefont {E.}~\bibnamefont {Orignac}}, \ and\ \bibinfo
  {author} {\bibfnamefont {M.}~\bibnamefont {Rigol}},\ }\bibfield  {title}
  {\enquote {\bibinfo {title} {One dimensional bosons: From condensed matter
  systems to ultracold gases},}\ }\href {\doibase 10.1103/RevModPhys.83.1405}
  {\bibfield  {journal} {\bibinfo  {journal} {Rev. Mod. Phys.}\ }\textbf
  {\bibinfo {volume} {83}},\ \bibinfo {pages} {1405--1466} (\bibinfo {year}
  {2011})}\BibitemShut {NoStop}%
\bibitem [{\citenamefont {Langen}\ \emph {et~al.}(2013)\citenamefont {Langen},
  \citenamefont {Geiger}, \citenamefont {Kuhnert}, \citenamefont {Rauer},\ and\
  \citenamefont {Schmiedmayer}}]{Langen2013}%
  \BibitemOpen
  \bibfield  {author} {\bibinfo {author} {\bibfnamefont {T.}~\bibnamefont
  {Langen}}, \bibinfo {author} {\bibfnamefont {R.}~\bibnamefont {Geiger}},
  \bibinfo {author} {\bibfnamefont {M.}~\bibnamefont {Kuhnert}}, \bibinfo
  {author} {\bibfnamefont {B.}~\bibnamefont {Rauer}}, \ and\ \bibinfo {author}
  {\bibfnamefont {J.}~\bibnamefont {Schmiedmayer}},\ }\bibfield  {title}
  {\enquote {\bibinfo {title} {{Local emergence of thermal correlations in an
  isolated quantum many-body system}},}\ }\href {\doibase 10.1038/nphys2739}
  {\bibfield  {journal} {\bibinfo  {journal} {Nature Phys.}\ }\textbf {\bibinfo
  {volume} {9}},\ \bibinfo {pages} {640--643} (\bibinfo {year}
  {2013})}\BibitemShut {NoStop}%
\bibitem [{\citenamefont {Langen}\ \emph {et~al.}(2015)\citenamefont {Langen},
  \citenamefont {Erne}, \citenamefont {Geiger}, \citenamefont {Rauer},
  \citenamefont {Schweigler}, \citenamefont {Kuhnert}, \citenamefont
  {Rohringer}, \citenamefont {Mazets}, \citenamefont {Gasenzer},\ and\
  \citenamefont {Schmiedmayer}}]{Langen2015}%
  \BibitemOpen
  \bibfield  {author} {\bibinfo {author} {\bibfnamefont {T.}~\bibnamefont
  {Langen}}, \bibinfo {author} {\bibfnamefont {S.}~\bibnamefont {Erne}},
  \bibinfo {author} {\bibfnamefont {R.}~\bibnamefont {Geiger}}, \bibinfo
  {author} {\bibfnamefont {B.}~\bibnamefont {Rauer}}, \bibinfo {author}
  {\bibfnamefont {T.}~\bibnamefont {Schweigler}}, \bibinfo {author}
  {\bibfnamefont {M.}~\bibnamefont {Kuhnert}}, \bibinfo {author} {\bibfnamefont
  {W.}~\bibnamefont {Rohringer}}, \bibinfo {author} {\bibfnamefont {I.~E.}\
  \bibnamefont {Mazets}}, \bibinfo {author} {\bibfnamefont {T.}~\bibnamefont
  {Gasenzer}}, \ and\ \bibinfo {author} {\bibfnamefont {J.}~\bibnamefont
  {Schmiedmayer}},\ }\bibfield  {title} {\enquote {\bibinfo {title}
  {{Experimental observation of a generalized Gibbs ensemble}},}\ }\href
  {\doibase 10.1126/science.1257026} {\bibfield  {journal} {\bibinfo  {journal}
  {Science}\ }\textbf {\bibinfo {volume} {348}},\ \bibinfo {pages} {207--211}
  (\bibinfo {year} {2015})}\BibitemShut {NoStop}%
\bibitem [{\citenamefont {Yang}\ \emph {et~al.}(2017)\citenamefont {Yang},
  \citenamefont {Chen}, \citenamefont {Zheng}, \citenamefont {Sun},
  \citenamefont {Dai}, \citenamefont {Guan}, \citenamefont {Yuan},\ and\
  \citenamefont {Pan}}]{yangTLL}%
  \BibitemOpen
  \bibfield  {author} {\bibinfo {author} {\bibfnamefont {B.}~\bibnamefont
  {Yang}}, \bibinfo {author} {\bibfnamefont {Y.-Y.}\ \bibnamefont {Chen}},
  \bibinfo {author} {\bibfnamefont {Y.-G.}\ \bibnamefont {Zheng}}, \bibinfo
  {author} {\bibfnamefont {H.}~\bibnamefont {Sun}}, \bibinfo {author}
  {\bibfnamefont {H.-N.}\ \bibnamefont {Dai}}, \bibinfo {author} {\bibfnamefont
  {X.-W.}\ \bibnamefont {Guan}}, \bibinfo {author} {\bibfnamefont {Z.-S.}\
  \bibnamefont {Yuan}}, \ and\ \bibinfo {author} {\bibfnamefont {J.-W.}\
  \bibnamefont {Pan}},\ }\bibfield  {title} {\enquote {\bibinfo {title}
  {{Quantum criticality and the Tomonaga-Luttinger liquid in one-dimensional
  Bose gases}},}\ }\href {\doibase 10.1103/PhysRevLett.119.165701} {\bibfield
  {journal} {\bibinfo  {journal} {Phys. Rev. Lett.}\ }\textbf {\bibinfo
  {volume} {119}},\ \bibinfo {pages} {165701} (\bibinfo {year}
  {2017})}\BibitemShut {NoStop}%
\bibitem [{\citenamefont {Grimm}\ and\ \citenamefont
  {Ovchinnikov}(1987)}]{dipole_trap}%
  \BibitemOpen
  \bibfield  {author} {\bibinfo {author} {\bibfnamefont {R.}~\bibnamefont
  {Grimm}}\ and\ \bibinfo {author} {\bibfnamefont {Y.~B.}\ \bibnamefont
  {Ovchinnikov}},\ }\bibfield  {title} {\enquote {\bibinfo {title} {{Optical
  dipole traps for neutral atoms}},}\ }\href@noop {} {\  (\bibinfo {year}
  {1987})},\ \Eprint {http://arxiv.org/abs/9902072v1} {arXiv:9902072v1}
  \BibitemShut {NoStop}%
\bibitem [{\citenamefont {Aidelsburger}\ \emph {et~al.}(2017)\citenamefont
  {Aidelsburger}, \citenamefont {Ville}, \citenamefont {Saint-Jalm},
  \citenamefont {Nascimb\`ene}, \citenamefont {Dalibard},\ and\ \citenamefont
  {Beugnon}}]{Aidelsburger2017Relax2d}%
  \BibitemOpen
  \bibfield  {author} {\bibinfo {author} {\bibfnamefont {M.}~\bibnamefont
  {Aidelsburger}}, \bibinfo {author} {\bibfnamefont {J.~L.}\ \bibnamefont
  {Ville}}, \bibinfo {author} {\bibfnamefont {R.}~\bibnamefont {Saint-Jalm}},
  \bibinfo {author} {\bibfnamefont {S.}~\bibnamefont {Nascimb\`ene}}, \bibinfo
  {author} {\bibfnamefont {J.}~\bibnamefont {Dalibard}}, \ and\ \bibinfo
  {author} {\bibfnamefont {J.}~\bibnamefont {Beugnon}},\ }\bibfield  {title}
  {\enquote {\bibinfo {title} {Relaxation dynamics in the merging of $n$
  independent condensates},}\ }\href {\doibase 10.1103/PhysRevLett.119.190403}
  {\bibfield  {journal} {\bibinfo  {journal} {Phys. Rev. Lett.}\ }\textbf
  {\bibinfo {volume} {119}},\ \bibinfo {pages} {190403} (\bibinfo {year}
  {2017})}\BibitemShut {NoStop}%
\bibitem [{\citenamefont {Ha}\ \emph {et~al.}(2015)\citenamefont {Ha},
  \citenamefont {Clark}, \citenamefont {Parker}, \citenamefont {Anderson},\
  and\ \citenamefont {Chin}}]{dmd_ha_roton}%
  \BibitemOpen
  \bibfield  {author} {\bibinfo {author} {\bibfnamefont {L.-C.}\ \bibnamefont
  {Ha}}, \bibinfo {author} {\bibfnamefont {L.~W.}\ \bibnamefont {Clark}},
  \bibinfo {author} {\bibfnamefont {C.~V.}\ \bibnamefont {Parker}}, \bibinfo
  {author} {\bibfnamefont {B.~M.}\ \bibnamefont {Anderson}}, \ and\ \bibinfo
  {author} {\bibfnamefont {C.}~\bibnamefont {Chin}},\ }\bibfield  {title}
  {\enquote {\bibinfo {title} {Roton-maxon excitation spectrum of bose
  condensates in a shaken optical lattice},}\ }\href {\doibase
  10.1103/PhysRevLett.114.055301} {\bibfield  {journal} {\bibinfo  {journal}
  {Phys. Rev. Lett.}\ }\textbf {\bibinfo {volume} {114}},\ \bibinfo {pages}
  {055301} (\bibinfo {year} {2015})}\BibitemShut {NoStop}%
\bibitem [{\citenamefont {Zupancic}\ \emph {et~al.}(2016)\citenamefont
  {Zupancic}, \citenamefont {Preiss}, \citenamefont {Ma}, \citenamefont
  {Lukin}, \citenamefont {Tai}, \citenamefont {Rispoli}, \citenamefont
  {Islam},\ and\ \citenamefont {Greiner}}]{dmd_Zupancic}%
  \BibitemOpen
  \bibfield  {author} {\bibinfo {author} {\bibfnamefont {P.}~\bibnamefont
  {Zupancic}}, \bibinfo {author} {\bibfnamefont {P.~M.}\ \bibnamefont
  {Preiss}}, \bibinfo {author} {\bibfnamefont {R.}~\bibnamefont {Ma}}, \bibinfo
  {author} {\bibfnamefont {A.}~\bibnamefont {Lukin}}, \bibinfo {author}
  {\bibfnamefont {M.~E.}\ \bibnamefont {Tai}}, \bibinfo {author} {\bibfnamefont
  {M.}~\bibnamefont {Rispoli}}, \bibinfo {author} {\bibfnamefont
  {R.}~\bibnamefont {Islam}}, \ and\ \bibinfo {author} {\bibfnamefont
  {M.}~\bibnamefont {Greiner}},\ }\bibfield  {title} {\enquote {\bibinfo
  {title} {Ultra-precise holographic beam shaping for microscopic quantum
  control},}\ }\href {\doibase 10.1364/OE.24.013881} {\bibfield  {journal}
  {\bibinfo  {journal} {Opt. Express}\ }\textbf {\bibinfo {volume} {24}},\
  \bibinfo {pages} {13881--13893} (\bibinfo {year} {2016})}\BibitemShut
  {NoStop}%
\bibitem [{\citenamefont {Eckel}\ \emph {et~al.}(2018)\citenamefont {Eckel},
  \citenamefont {Kumar}, \citenamefont {Jacobson}, \citenamefont {Spielman},\
  and\ \citenamefont {Campbell}}]{EckelCampbellPhysRevX.8.021021}%
  \BibitemOpen
  \bibfield  {author} {\bibinfo {author} {\bibfnamefont {S.}~\bibnamefont
  {Eckel}}, \bibinfo {author} {\bibfnamefont {A.}~\bibnamefont {Kumar}},
  \bibinfo {author} {\bibfnamefont {T.}~\bibnamefont {Jacobson}}, \bibinfo
  {author} {\bibfnamefont {I.~B.}\ \bibnamefont {Spielman}}, \ and\ \bibinfo
  {author} {\bibfnamefont {G.~K.}\ \bibnamefont {Campbell}},\ }\bibfield
  {title} {\enquote {\bibinfo {title} {{A rapidly expanding Bose-Einstein
  condensate: An expanding universe in the lab}},}\ }\href {\doibase
  10.1103/PhysRevX.8.021021} {\bibfield  {journal} {\bibinfo  {journal} {Phys.
  Rev. X}\ }\textbf {\bibinfo {volume} {8}},\ \bibinfo {pages} {021021}
  (\bibinfo {year} {2018})}\BibitemShut {NoStop}%
\bibitem [{\citenamefont {Henderson}\ \emph {et~al.}(2009)\citenamefont
  {Henderson}, \citenamefont {Ryu}, \citenamefont {MacCormick},\ and\
  \citenamefont {Boshier}}]{dmd_boshier}%
  \BibitemOpen
  \bibfield  {author} {\bibinfo {author} {\bibfnamefont {K.}~\bibnamefont
  {Henderson}}, \bibinfo {author} {\bibfnamefont {C.}~\bibnamefont {Ryu}},
  \bibinfo {author} {\bibfnamefont {C.}~\bibnamefont {MacCormick}}, \ and\
  \bibinfo {author} {\bibfnamefont {M.~G.}\ \bibnamefont {Boshier}},\
  }\bibfield  {title} {\enquote {\bibinfo {title} {{Experimental demonstration
  of painting arbitrary and dynamic potentials for Bose{\textendash}Einstein
  condensates}},}\ }\href {\doibase 10.1088/1367-2630/11/4/043030} {\bibfield
  {journal} {\bibinfo  {journal} {New J. Phys.}\ }\textbf {\bibinfo {volume}
  {11}},\ \bibinfo {pages} {043030} (\bibinfo {year} {2009})}\BibitemShut
  {NoStop}%
\bibitem [{\citenamefont {Amico}\ \emph {et~al.}(2020)\citenamefont {Amico}
  \emph {et~al.}}]{atomtronics_review}%
  \BibitemOpen
  \bibfield  {author} {\bibinfo {author} {\bibfnamefont {L.}~\bibnamefont
  {Amico}} \emph {et~al.},\ }\bibfield  {title} {\enquote {\bibinfo {title}
  {{Roadmap on atomtronics}},}\ }\href@noop {} {\  (\bibinfo {year} {2020})},\
  \Eprint {http://arxiv.org/abs/2008.04439} {arXiv:2008.04439
  [cond-mat.quant-gas]} \BibitemShut {NoStop}%
\bibitem [{\citenamefont {Rohringer}\ \emph {et~al.}(2015)\citenamefont
  {Rohringer}, \citenamefont {Fischer}, \citenamefont {Steiner}, \citenamefont
  {Mazets}, \citenamefont {Schmiedmayer},\ and\ \citenamefont
  {Trupke}}]{rohringer2015non}%
  \BibitemOpen
  \bibfield  {author} {\bibinfo {author} {\bibfnamefont {W.}~\bibnamefont
  {Rohringer}}, \bibinfo {author} {\bibfnamefont {D.}~\bibnamefont {Fischer}},
  \bibinfo {author} {\bibfnamefont {F.}~\bibnamefont {Steiner}}, \bibinfo
  {author} {\bibfnamefont {I.~E.}\ \bibnamefont {Mazets}}, \bibinfo {author}
  {\bibfnamefont {J.}~\bibnamefont {Schmiedmayer}}, \ and\ \bibinfo {author}
  {\bibfnamefont {M.}~\bibnamefont {Trupke}},\ }\bibfield  {title} {\enquote
  {\bibinfo {title} {Non-equilibrium scale invariance and shortcuts to
  adiabaticity in a one-dimensional bose gas},}\ }\href@noop {} {\bibfield
  {journal} {\bibinfo  {journal} {Scientific Rep.}\ }\textbf {\bibinfo {volume}
  {5}},\ \bibinfo {pages} {9820} (\bibinfo {year} {2015})}\BibitemShut
  {NoStop}%
\bibitem [{\citenamefont {Gritsev}\ \emph {et~al.}(2010)\citenamefont
  {Gritsev}, \citenamefont {Barmettler},\ and\ \citenamefont
  {Demler}}]{Gritsev_2010}%
  \BibitemOpen
  \bibfield  {author} {\bibinfo {author} {\bibfnamefont {V.}~\bibnamefont
  {Gritsev}}, \bibinfo {author} {\bibfnamefont {P.}~\bibnamefont {Barmettler}},
  \ and\ \bibinfo {author} {\bibfnamefont {E.}~\bibnamefont {Demler}},\
  }\bibfield  {title} {\enquote {\bibinfo {title} {Scaling approach to quantum
  non-equilibrium dynamics of many-body systems},}\ }\href {\doibase
  10.1088/1367-2630/12/11/113005} {\bibfield  {journal} {\bibinfo  {journal}
  {New J. Phys.}\ }\textbf {\bibinfo {volume} {12}},\ \bibinfo {pages} {113005}
  (\bibinfo {year} {2010})}\BibitemShut {NoStop}%
\bibitem [{\citenamefont {Wang}\ \emph {et~al.}(2015)\citenamefont {Wang},
  \citenamefont {Kumar}, \citenamefont {Jendrzejewski}, \citenamefont {Wilson},
  \citenamefont {Edwards}, \citenamefont {Eckel}, \citenamefont {Campbell},\
  and\ \citenamefont {Clark}}]{Wang_2015}%
  \BibitemOpen
  \bibfield  {author} {\bibinfo {author} {\bibfnamefont {Y.-H.}\ \bibnamefont
  {Wang}}, \bibinfo {author} {\bibfnamefont {A.}~\bibnamefont {Kumar}},
  \bibinfo {author} {\bibfnamefont {F.}~\bibnamefont {Jendrzejewski}}, \bibinfo
  {author} {\bibfnamefont {R.~M.}\ \bibnamefont {Wilson}}, \bibinfo {author}
  {\bibfnamefont {M.}~\bibnamefont {Edwards}}, \bibinfo {author} {\bibfnamefont
  {S.}~\bibnamefont {Eckel}}, \bibinfo {author} {\bibfnamefont {G.~K.}\
  \bibnamefont {Campbell}}, \ and\ \bibinfo {author} {\bibfnamefont {C.~W.}\
  \bibnamefont {Clark}},\ }\bibfield  {title} {\enquote {\bibinfo {title}
  {Resonant wavepackets and shock waves in an atomtronic {SQUID}},}\ }\href
  {\doibase 10.1088/1367-2630/17/12/125012} {\bibfield  {journal} {\bibinfo
  {journal} {New J. Phys.}\ }\textbf {\bibinfo {volume} {17}},\ \bibinfo
  {pages} {125012} (\bibinfo {year} {2015})}\BibitemShut {NoStop}%
\bibitem [{\citenamefont {Booker}\ \emph {et~al.}(2020)\citenamefont {Booker},
  \citenamefont {Bu{\v{c}}a},\ and\ \citenamefont {Jaksch}}]{dmd_kumar_2020}%
  \BibitemOpen
  \bibfield  {author} {\bibinfo {author} {\bibfnamefont {C.}~\bibnamefont
  {Booker}}, \bibinfo {author} {\bibfnamefont {B.}~\bibnamefont {Bu{\v{c}}a}},
  \ and\ \bibinfo {author} {\bibfnamefont {D.}~\bibnamefont {Jaksch}},\
  }\bibfield  {title} {\enquote {\bibinfo {title} {Non-stationarity and
  dissipative time crystals: spectral properties and finite-size effects},}\
  }\href {\doibase 10.1088/1367-2630/ababc4} {\bibfield  {journal} {\bibinfo
  {journal} {New J. Phys.}\ }\textbf {\bibinfo {volume} {22}},\ \bibinfo
  {pages} {085007} (\bibinfo {year} {2020})}\BibitemShut {NoStop}%
\bibitem [{\citenamefont {Eckel}\ \emph {et~al.}(2014)\citenamefont {Eckel},
  \citenamefont {Lee}, \citenamefont {Jendrzejewski}, \citenamefont {Murray},
  \citenamefont {Clark}, \citenamefont {Lobb}, \citenamefont {Phillips},
  \citenamefont {Edwards},\ and\ \citenamefont
  {Campbell}}]{eckel2014hysteresis}%
  \BibitemOpen
  \bibfield  {author} {\bibinfo {author} {\bibfnamefont {S.}~\bibnamefont
  {Eckel}}, \bibinfo {author} {\bibfnamefont {J.~G.}\ \bibnamefont {Lee}},
  \bibinfo {author} {\bibfnamefont {F.}~\bibnamefont {Jendrzejewski}}, \bibinfo
  {author} {\bibfnamefont {N.}~\bibnamefont {Murray}}, \bibinfo {author}
  {\bibfnamefont {C.~W.}\ \bibnamefont {Clark}}, \bibinfo {author}
  {\bibfnamefont {C.~J.}\ \bibnamefont {Lobb}}, \bibinfo {author}
  {\bibfnamefont {W.~D.}\ \bibnamefont {Phillips}}, \bibinfo {author}
  {\bibfnamefont {M.}~\bibnamefont {Edwards}}, \ and\ \bibinfo {author}
  {\bibfnamefont {G.~K.}\ \bibnamefont {Campbell}},\ }\bibfield  {title}
  {\enquote {\bibinfo {title} {Hysteresis in a quantized superfluid
  ‘atomtronic’circuit},}\ }\href@noop {} {\bibfield  {journal} {\bibinfo
  {journal} {Nature}\ }\textbf {\bibinfo {volume} {506}},\ \bibinfo {pages}
  {200--203} (\bibinfo {year} {2014})}\BibitemShut {NoStop}%
\bibitem [{\citenamefont {Schemmer}\ \emph {et~al.}(2018)\citenamefont
  {Schemmer}, \citenamefont {Johnson},\ and\ \citenamefont
  {Bouchoule}}]{PhysRevA.98.043604}%
  \BibitemOpen
  \bibfield  {author} {\bibinfo {author} {\bibfnamefont {M.}~\bibnamefont
  {Schemmer}}, \bibinfo {author} {\bibfnamefont {A.}~\bibnamefont {Johnson}}, \
  and\ \bibinfo {author} {\bibfnamefont {I.}~\bibnamefont {Bouchoule}},\
  }\bibfield  {title} {\enquote {\bibinfo {title} {Monitoring squeezed
  collective modes of a one-dimensional bose gas after an interaction quench
  using density-ripple analysis},}\ }\href {\doibase
  10.1103/PhysRevA.98.043604} {\bibfield  {journal} {\bibinfo  {journal} {Phys.
  Rev. A}\ }\textbf {\bibinfo {volume} {98}},\ \bibinfo {pages} {043604}
  (\bibinfo {year} {2018})}\BibitemShut {NoStop}%
\bibitem [{\citenamefont {Fang}\ \emph {et~al.}(2016)\citenamefont {Fang},
  \citenamefont {Johnson}, \citenamefont {Roscilde},\ and\ \citenamefont
  {Bouchoule}}]{PhysRevLett.116.050402}%
  \BibitemOpen
  \bibfield  {author} {\bibinfo {author} {\bibfnamefont {B.}~\bibnamefont
  {Fang}}, \bibinfo {author} {\bibfnamefont {A.}~\bibnamefont {Johnson}},
  \bibinfo {author} {\bibfnamefont {T.}~\bibnamefont {Roscilde}}, \ and\
  \bibinfo {author} {\bibfnamefont {I.}~\bibnamefont {Bouchoule}},\ }\bibfield
  {title} {\enquote {\bibinfo {title} {{Momentum-space correlations of a
  one-dimensional Bose gas}},}\ }\href {\doibase
  10.1103/PhysRevLett.116.050402} {\bibfield  {journal} {\bibinfo  {journal}
  {Phys. Rev. Lett.}\ }\textbf {\bibinfo {volume} {116}},\ \bibinfo {pages}
  {050402} (\bibinfo {year} {2016})}\BibitemShut {NoStop}%
\bibitem [{\citenamefont {Armijo}\ \emph {et~al.}(2010)\citenamefont {Armijo},
  \citenamefont {Jacqmin}, \citenamefont {Kheruntsyan},\ and\ \citenamefont
  {Bouchoule}}]{PhysRevLett.105.230402}%
  \BibitemOpen
  \bibfield  {author} {\bibinfo {author} {\bibfnamefont {J.}~\bibnamefont
  {Armijo}}, \bibinfo {author} {\bibfnamefont {T.}~\bibnamefont {Jacqmin}},
  \bibinfo {author} {\bibfnamefont {K.~V.}\ \bibnamefont {Kheruntsyan}}, \ and\
  \bibinfo {author} {\bibfnamefont {I.}~\bibnamefont {Bouchoule}},\ }\bibfield
  {title} {\enquote {\bibinfo {title} {Probing three-body correlations in a
  quantum gas using the measurement of the third moment of density
  fluctuations},}\ }\href {\doibase 10.1103/PhysRevLett.105.230402} {\bibfield
  {journal} {\bibinfo  {journal} {Phys. Rev. Lett.}\ }\textbf {\bibinfo
  {volume} {105}},\ \bibinfo {pages} {230402} (\bibinfo {year}
  {2010})}\BibitemShut {NoStop}%
\bibitem [{\citenamefont {Esteve}\ \emph {et~al.}(2006)\citenamefont {Esteve},
  \citenamefont {Trebbia}, \citenamefont {Schumm}, \citenamefont {Aspect},
  \citenamefont {Westbrook},\ and\ \citenamefont
  {Bouchoule}}]{PhysRevLett.96.130403}%
  \BibitemOpen
  \bibfield  {author} {\bibinfo {author} {\bibfnamefont {J.}~\bibnamefont
  {Esteve}}, \bibinfo {author} {\bibfnamefont {J.-B.}\ \bibnamefont {Trebbia}},
  \bibinfo {author} {\bibfnamefont {T.}~\bibnamefont {Schumm}}, \bibinfo
  {author} {\bibfnamefont {A.}~\bibnamefont {Aspect}}, \bibinfo {author}
  {\bibfnamefont {C.~I.}\ \bibnamefont {Westbrook}}, \ and\ \bibinfo {author}
  {\bibfnamefont {I.}~\bibnamefont {Bouchoule}},\ }\bibfield  {title} {\enquote
  {\bibinfo {title} {Observations of density fluctuations in an elongated bose
  gas: Ideal gas and quasicondensate regimes},}\ }\href {\doibase
  10.1103/PhysRevLett.96.130403} {\bibfield  {journal} {\bibinfo  {journal}
  {Phys. Rev. Lett.}\ }\textbf {\bibinfo {volume} {96}},\ \bibinfo {pages}
  {130403} (\bibinfo {year} {2006})}\BibitemShut {NoStop}%
\bibitem [{\citenamefont {Jacqmin}\ \emph {et~al.}(2011)\citenamefont
  {Jacqmin}, \citenamefont {Armijo}, \citenamefont {Berrada}, \citenamefont
  {Kheruntsyan},\ and\ \citenamefont {Bouchoule}}]{PhysRevLett.106.230405}%
  \BibitemOpen
  \bibfield  {author} {\bibinfo {author} {\bibfnamefont {T.}~\bibnamefont
  {Jacqmin}}, \bibinfo {author} {\bibfnamefont {J.}~\bibnamefont {Armijo}},
  \bibinfo {author} {\bibfnamefont {T.}~\bibnamefont {Berrada}}, \bibinfo
  {author} {\bibfnamefont {K.~V.}\ \bibnamefont {Kheruntsyan}}, \ and\ \bibinfo
  {author} {\bibfnamefont {I.}~\bibnamefont {Bouchoule}},\ }\bibfield  {title}
  {\enquote {\bibinfo {title} {Sub-poissonian fluctuations in a 1d bose gas:
  From the quantum quasicondensate to the strongly interacting regime},}\
  }\href {\doibase 10.1103/PhysRevLett.106.230405} {\bibfield  {journal}
  {\bibinfo  {journal} {Phys. Rev. Lett.}\ }\textbf {\bibinfo {volume} {106}},\
  \bibinfo {pages} {230405} (\bibinfo {year} {2011})}\BibitemShut {NoStop}%
\bibitem [{\citenamefont {Imambekov}\ \emph {et~al.}(2009)\citenamefont
  {Imambekov}, \citenamefont {Mazets}, \citenamefont {Petrov}, \citenamefont
  {Gritsev}, \citenamefont {Manz}, \citenamefont {Hofferberth}, \citenamefont
  {Schumm}, \citenamefont {Demler},\ and\ \citenamefont
  {Schmiedmayer}}]{Imambekov09}%
  \BibitemOpen
  \bibfield  {author} {\bibinfo {author} {\bibfnamefont {A.}~\bibnamefont
  {Imambekov}}, \bibinfo {author} {\bibfnamefont {I.~E.}\ \bibnamefont
  {Mazets}}, \bibinfo {author} {\bibfnamefont {D.~S.}\ \bibnamefont {Petrov}},
  \bibinfo {author} {\bibfnamefont {V.}~\bibnamefont {Gritsev}}, \bibinfo
  {author} {\bibfnamefont {S.}~\bibnamefont {Manz}}, \bibinfo {author}
  {\bibfnamefont {S.}~\bibnamefont {Hofferberth}}, \bibinfo {author}
  {\bibfnamefont {T.}~\bibnamefont {Schumm}}, \bibinfo {author} {\bibfnamefont
  {E.}~\bibnamefont {Demler}}, \ and\ \bibinfo {author} {\bibfnamefont
  {J.}~\bibnamefont {Schmiedmayer}},\ }\bibfield  {title} {\enquote {\bibinfo
  {title} {Density ripples in expanding low-dimensional gases as a probe of
  correlations},}\ }\href {\doibase 10.1103/PhysRevA.80.033604} {\bibfield
  {journal} {\bibinfo  {journal} {Phys. Rev. A}\ }\textbf {\bibinfo {volume}
  {80}},\ \bibinfo {pages} {033604} (\bibinfo {year} {2009})}\BibitemShut
  {NoStop}%
\bibitem [{\citenamefont {Manz}(2011)}]{Manz11}%
  \BibitemOpen
  \bibfield  {author} {\bibinfo {author} {\bibfnamefont {S.}~\bibnamefont
  {Manz}},\ }\emph {\bibinfo {title} {{Density correlations of expanding
  one-dimensional Bose gases}}},\ \href@noop {} {Ph.D. thesis},\ \bibinfo
  {school} {Vienna University of Technology} (\bibinfo {year}
  {2011})\BibitemShut {NoStop}%
\bibitem [{\citenamefont {Schumm}\ \emph {et~al.}(2005)\citenamefont {Schumm},
  \citenamefont {Hofferberth}, \citenamefont {Andersson}, \citenamefont
  {Wildermuth}, \citenamefont {Groth}, \citenamefont {Bar-Joseph},
  \citenamefont {Schmiedmayer},\ and\ \citenamefont {Kruger}}]{Schumm05}%
  \BibitemOpen
  \bibfield  {author} {\bibinfo {author} {\bibfnamefont {T.}~\bibnamefont
  {Schumm}}, \bibinfo {author} {\bibfnamefont {S.}~\bibnamefont {Hofferberth}},
  \bibinfo {author} {\bibfnamefont {L.~M.}\ \bibnamefont {Andersson}}, \bibinfo
  {author} {\bibfnamefont {S.}~\bibnamefont {Wildermuth}}, \bibinfo {author}
  {\bibfnamefont {S.}~\bibnamefont {Groth}}, \bibinfo {author} {\bibfnamefont
  {I.}~\bibnamefont {Bar-Joseph}}, \bibinfo {author} {\bibfnamefont
  {J.}~\bibnamefont {Schmiedmayer}}, \ and\ \bibinfo {author} {\bibfnamefont
  {P.}~\bibnamefont {Kruger}},\ }\bibfield  {title} {\enquote {\bibinfo {title}
  {{Matter-wave interferometry in a double well on an atom chip}},}\ }\href
  {\doibase 10.1038/nphys125} {\bibfield  {journal} {\bibinfo  {journal}
  {Nature Phys.}\ }\textbf {\bibinfo {volume} {1}},\ \bibinfo {pages} {57--62}
  (\bibinfo {year} {2005})}\BibitemShut {NoStop}%
\bibitem [{\citenamefont {van Nieuwkerk}\ \emph {et~al.}(2018)\citenamefont
  {van Nieuwkerk}, \citenamefont {Schmiedmayer},\ and\ \citenamefont
  {Essler}}]{van2018projective}%
  \BibitemOpen
  \bibfield  {author} {\bibinfo {author} {\bibfnamefont {Y.~D.}\ \bibnamefont
  {van Nieuwkerk}}, \bibinfo {author} {\bibfnamefont {J.}~\bibnamefont
  {Schmiedmayer}}, \ and\ \bibinfo {author} {\bibfnamefont {F.~H.~L.}\
  \bibnamefont {Essler}},\ }\bibfield  {title} {\enquote {\bibinfo {title}
  {Projective phase measurements in one-dimensional bose gases},}\ }\href@noop
  {} {\bibfield  {journal} {\bibinfo  {journal} {Scipost Phys.}\ }\textbf
  {\bibinfo {volume} {5}},\ \bibinfo {pages} {046} (\bibinfo {year}
  {2018})}\BibitemShut {NoStop}%
\bibitem [{\citenamefont {Hofferberth}\ \emph {et~al.}(2008)\citenamefont
  {Hofferberth}, \citenamefont {Lesanovsky}, \citenamefont {Schumm},
  \citenamefont {Imambekov}, \citenamefont {Gritsev}, \citenamefont {Demler},\
  and\ \citenamefont {Schmiedmayer}}]{Hofferberth08}%
  \BibitemOpen
  \bibfield  {author} {\bibinfo {author} {\bibfnamefont {S.}~\bibnamefont
  {Hofferberth}}, \bibinfo {author} {\bibfnamefont {I.}~\bibnamefont
  {Lesanovsky}}, \bibinfo {author} {\bibfnamefont {T.}~\bibnamefont {Schumm}},
  \bibinfo {author} {\bibfnamefont {A.}~\bibnamefont {Imambekov}}, \bibinfo
  {author} {\bibfnamefont {V.}~\bibnamefont {Gritsev}}, \bibinfo {author}
  {\bibfnamefont {E.}~\bibnamefont {Demler}}, \ and\ \bibinfo {author}
  {\bibfnamefont {J.}~\bibnamefont {Schmiedmayer}},\ }\bibfield  {title}
  {\enquote {\bibinfo {title} {Probing quantum and thermal noise in an
  interacting many-body system},}\ }\href {\doibase 10.1038/nphys941}
  {\bibfield  {journal} {\bibinfo  {journal} {Nature Phys.}\ }\textbf {\bibinfo
  {volume} {4}},\ \bibinfo {pages} {489--495} (\bibinfo {year}
  {2008})}\BibitemShut {NoStop}%
\bibitem [{\citenamefont {Schweigler}(2019{\natexlab{b}})}]{Schweigler19}%
  \BibitemOpen
  \bibfield  {author} {\bibinfo {author} {\bibfnamefont {T.}~\bibnamefont
  {Schweigler}},\ }\emph {\bibinfo {title} {{Correlations and dynamics of
  tunnel-coupled one-dimensional Bose gases}}},\ \href@noop {} {Ph.D. thesis}
  (\bibinfo {year} {2019}{\natexlab{b}})\BibitemShut {NoStop}%
\bibitem [{\citenamefont {Davis}\ \emph {et~al.}(1995)\citenamefont {Davis},
  \citenamefont {Mewes},\ and\ \citenamefont {Ketterle}}]{davis1995analytical}%
  \BibitemOpen
  \bibfield  {author} {\bibinfo {author} {\bibfnamefont {K.~B.}\ \bibnamefont
  {Davis}}, \bibinfo {author} {\bibfnamefont {M.-O.}\ \bibnamefont {Mewes}}, \
  and\ \bibinfo {author} {\bibfnamefont {W.}~\bibnamefont {Ketterle}},\
  }\bibfield  {title} {\enquote {\bibinfo {title} {An analytical model for
  evaporative cooling of atoms},}\ }\href@noop {} {\bibfield  {journal}
  {\bibinfo  {journal} {Applied Physics B}\ }\textbf {\bibinfo {volume} {60}},\
  \bibinfo {pages} {155--159} (\bibinfo {year} {1995})}\BibitemShut {NoStop}%
\bibitem [{\citenamefont {Mazets}\ \emph {et~al.}(2008)\citenamefont {Mazets},
  \citenamefont {Schumm},\ and\ \citenamefont
  {Schmiedmayer}}]{MazetsBreakdown}%
  \BibitemOpen
  \bibfield  {author} {\bibinfo {author} {\bibfnamefont {I.~E.}\ \bibnamefont
  {Mazets}}, \bibinfo {author} {\bibfnamefont {T.}~\bibnamefont {Schumm}}, \
  and\ \bibinfo {author} {\bibfnamefont {J.}~\bibnamefont {Schmiedmayer}},\
  }\bibfield  {title} {\enquote {\bibinfo {title} {Breakdown of integrability
  in a quasi-1d ultracold bosonic gas},}\ }\href {\doibase
  10.1103/PhysRevLett.100.210403} {\bibfield  {journal} {\bibinfo  {journal}
  {Phys. Rev. Lett.}\ }\textbf {\bibinfo {volume} {100}},\ \bibinfo {pages}
  {210403} (\bibinfo {year} {2008})}\BibitemShut {NoStop}%
\bibitem [{\citenamefont {Tan}\ \emph {et~al.}(2010)\citenamefont {Tan},
  \citenamefont {Pustilnik},\ and\ \citenamefont {Glazman}}]{Breakdown2}%
  \BibitemOpen
  \bibfield  {author} {\bibinfo {author} {\bibfnamefont {S.}~\bibnamefont
  {Tan}}, \bibinfo {author} {\bibfnamefont {M.}~\bibnamefont {Pustilnik}}, \
  and\ \bibinfo {author} {\bibfnamefont {L.~I.}\ \bibnamefont {Glazman}},\
  }\bibfield  {title} {\enquote {\bibinfo {title} {Relaxation of a high-energy
  quasiparticle in a one-dimensional bose gas},}\ }\href {\doibase
  10.1103/PhysRevLett.105.090404} {\bibfield  {journal} {\bibinfo  {journal}
  {Phys. Rev. Lett.}\ }\textbf {\bibinfo {volume} {105}},\ \bibinfo {pages}
  {090404} (\bibinfo {year} {2010})}\BibitemShut {NoStop}%
\bibitem [{\citenamefont {Andreev}(1980)}]{andreev1980hydrodynamics}%
  \BibitemOpen
  \bibfield  {author} {\bibinfo {author} {\bibfnamefont {A.}~\bibnamefont
  {Andreev}},\ }\bibfield  {title} {\enquote {\bibinfo {title} {The
  hydrodynamics of two-and one-dimensional liquids},}\ }\href@noop {}
  {\bibfield  {journal} {\bibinfo  {journal} {JETP}\ }\textbf {\bibinfo
  {volume} {51}},\ \bibinfo {pages} {1038} (\bibinfo {year}
  {1980})}\BibitemShut {NoStop}%
\bibitem [{\citenamefont {Buchhold}\ and\ \citenamefont
  {Diehl}(2015)}]{buchhold2015kinetic}%
  \BibitemOpen
  \bibfield  {author} {\bibinfo {author} {\bibfnamefont {M.}~\bibnamefont
  {Buchhold}}\ and\ \bibinfo {author} {\bibfnamefont {S.}~\bibnamefont
  {Diehl}},\ }\bibfield  {title} {\enquote {\bibinfo {title} {Kinetic theory
  for interacting luttinger liquids},}\ }\href@noop {} {\bibfield  {journal}
  {\bibinfo  {journal} {Europ. Phys. J. D}\ }\textbf {\bibinfo {volume} {69}},\
  \bibinfo {pages} {1--20} (\bibinfo {year} {2015})}\BibitemShut {NoStop}%
\bibitem [{\citenamefont {Rauer}\ \emph {et~al.}(2016)\citenamefont {Rauer},
  \citenamefont {Gri\ifmmode~\check{s}\else \v{s}\fi{}ins}, \citenamefont
  {Mazets}, \citenamefont {Schweigler}, \citenamefont {Rohringer},
  \citenamefont {Geiger}, \citenamefont {Langen},\ and\ \citenamefont
  {Schmiedmayer}}]{Rauer2016a}%
  \BibitemOpen
  \bibfield  {author} {\bibinfo {author} {\bibfnamefont {B.}~\bibnamefont
  {Rauer}}, \bibinfo {author} {\bibfnamefont {P.}~\bibnamefont
  {Gri\ifmmode~\check{s}\else \v{s}\fi{}ins}}, \bibinfo {author} {\bibfnamefont
  {I.~E.}\ \bibnamefont {Mazets}}, \bibinfo {author} {\bibfnamefont
  {T.}~\bibnamefont {Schweigler}}, \bibinfo {author} {\bibfnamefont
  {W.}~\bibnamefont {Rohringer}}, \bibinfo {author} {\bibfnamefont
  {R.}~\bibnamefont {Geiger}}, \bibinfo {author} {\bibfnamefont
  {T.}~\bibnamefont {Langen}}, \ and\ \bibinfo {author} {\bibfnamefont
  {J.}~\bibnamefont {Schmiedmayer}},\ }\bibfield  {title} {\enquote {\bibinfo
  {title} {Cooling of a one-dimensional bose gas},}\ }\href {\doibase
  10.1103/PhysRevLett.116.030402} {\bibfield  {journal} {\bibinfo  {journal}
  {Phys. Rev. Lett.}\ }\textbf {\bibinfo {volume} {116}},\ \bibinfo {pages}
  {030402} (\bibinfo {year} {2016})}\BibitemShut {NoStop}%
\bibitem [{\citenamefont {Gri\ifmmode~\check{s}\else \v{s}\fi{}ins}\ \emph
  {et~al.}(2016)\citenamefont {Gri\ifmmode~\check{s}\else \v{s}\fi{}ins},
  \citenamefont {Rauer}, \citenamefont {Langen}, \citenamefont {Schmiedmayer},\
  and\ \citenamefont {Mazets}}]{GrisinsUniformLoss}%
  \BibitemOpen
  \bibfield  {author} {\bibinfo {author} {\bibfnamefont {P.}~\bibnamefont
  {Gri\ifmmode~\check{s}\else \v{s}\fi{}ins}}, \bibinfo {author} {\bibfnamefont
  {B.}~\bibnamefont {Rauer}}, \bibinfo {author} {\bibfnamefont
  {T.}~\bibnamefont {Langen}}, \bibinfo {author} {\bibfnamefont
  {J.}~\bibnamefont {Schmiedmayer}}, \ and\ \bibinfo {author} {\bibfnamefont
  {I.~E.}\ \bibnamefont {Mazets}},\ }\bibfield  {title} {\enquote {\bibinfo
  {title} {Degenerate bose gases with uniform loss},}\ }\href {\doibase
  10.1103/PhysRevA.93.033634} {\bibfield  {journal} {\bibinfo  {journal} {Phys.
  Rev. A}\ }\textbf {\bibinfo {volume} {93}},\ \bibinfo {pages} {033634}
  (\bibinfo {year} {2016})}\BibitemShut {NoStop}%
\bibitem [{\citenamefont {Busch}\ \emph {et~al.}(2014)\citenamefont {Busch},
  \citenamefont {Carusotto},\ and\ \citenamefont
  {Parentani}}]{QuantumCarusotto}%
  \BibitemOpen
  \bibfield  {author} {\bibinfo {author} {\bibfnamefont {X.}~\bibnamefont
  {Busch}}, \bibinfo {author} {\bibfnamefont {I.}~\bibnamefont {Carusotto}}, \
  and\ \bibinfo {author} {\bibfnamefont {R.}~\bibnamefont {Parentani}},\
  }\bibfield  {title} {\enquote {\bibinfo {title} {Spectrum and entanglement of
  phonons in quantum fluids of light},}\ }\href {\doibase
  10.1103/PhysRevA.89.043819} {\bibfield  {journal} {\bibinfo  {journal} {Phys.
  Rev. A}\ }\textbf {\bibinfo {volume} {89}},\ \bibinfo {pages} {043819}
  (\bibinfo {year} {2014})}\BibitemShut {NoStop}%
\bibitem [{\citenamefont {Japha}\ \emph {et~al.}(1999)\citenamefont {Japha},
  \citenamefont {Choi}, \citenamefont {Burnett},\ and\ \citenamefont
  {Band}}]{EvaporationQunatum}%
  \BibitemOpen
  \bibfield  {author} {\bibinfo {author} {\bibfnamefont {Y.}~\bibnamefont
  {Japha}}, \bibinfo {author} {\bibfnamefont {S.}~\bibnamefont {Choi}},
  \bibinfo {author} {\bibfnamefont {K.}~\bibnamefont {Burnett}}, \ and\
  \bibinfo {author} {\bibfnamefont {Y.~B.}\ \bibnamefont {Band}},\ }\bibfield
  {title} {\enquote {\bibinfo {title} {Coherent output, stimulated quantum
  evaporation, and pair breaking in a trapped atomic bose gas},}\ }\href
  {\doibase 10.1103/PhysRevLett.82.1079} {\bibfield  {journal} {\bibinfo
  {journal} {Phys. Rev. Lett.}\ }\textbf {\bibinfo {volume} {82}},\ \bibinfo
  {pages} {1079--1083} (\bibinfo {year} {1999})}\BibitemShut {NoStop}%
\bibitem [{\citenamefont {Jo}\ \emph {et~al.}(2007)\citenamefont {Jo},
  \citenamefont {Shin}, \citenamefont {Will}, \citenamefont {Pasquini},
  \citenamefont {Saba}, \citenamefont {Ketterle}, \citenamefont {Pritchard},
  \citenamefont {Vengalattore},\ and\ \citenamefont {Prentiss}}]{Jo2007c}%
  \BibitemOpen
  \bibfield  {author} {\bibinfo {author} {\bibfnamefont {G.-B.}\ \bibnamefont
  {Jo}}, \bibinfo {author} {\bibfnamefont {Y.}~\bibnamefont {Shin}}, \bibinfo
  {author} {\bibfnamefont {S.}~\bibnamefont {Will}}, \bibinfo {author}
  {\bibfnamefont {T.~A.}\ \bibnamefont {Pasquini}}, \bibinfo {author}
  {\bibfnamefont {M.}~\bibnamefont {Saba}}, \bibinfo {author} {\bibfnamefont
  {W.}~\bibnamefont {Ketterle}}, \bibinfo {author} {\bibfnamefont {D.~E.}\
  \bibnamefont {Pritchard}}, \bibinfo {author} {\bibfnamefont {M.}~\bibnamefont
  {Vengalattore}}, \ and\ \bibinfo {author} {\bibfnamefont {M.}~\bibnamefont
  {Prentiss}},\ }\bibfield  {title} {\enquote {\bibinfo {title} {{Long phase
  coherence time and number squeezing of two Bose-Einstein condensates on an
  atom chip}},}\ }\href {\doibase 10.1103/PhysRevLett.98.030407} {\bibfield
  {journal} {\bibinfo  {journal} {Phys. Rev. Lett.}\ }\textbf {\bibinfo
  {volume} {98}},\ \bibinfo {pages} {030407} (\bibinfo {year}
  {2007})}\BibitemShut {NoStop}%
\bibitem [{\citenamefont {Est{\`{e}}ve}\ \emph {et~al.}(2008)\citenamefont
  {Est{\`{e}}ve}, \citenamefont {Gross}, \citenamefont {Weller}, \citenamefont
  {Giovanazzi},\ and\ \citenamefont {Oberthaler}}]{Esteve2008}%
  \BibitemOpen
  \bibfield  {author} {\bibinfo {author} {\bibfnamefont {J.}~\bibnamefont
  {Est{\`{e}}ve}}, \bibinfo {author} {\bibfnamefont {C.}~\bibnamefont {Gross}},
  \bibinfo {author} {\bibfnamefont {a.}~\bibnamefont {Weller}}, \bibinfo
  {author} {\bibfnamefont {S.}~\bibnamefont {Giovanazzi}}, \ and\ \bibinfo
  {author} {\bibfnamefont {M.~K.}\ \bibnamefont {Oberthaler}},\ }\bibfield
  {title} {\enquote {\bibinfo {title} {{Squeezing and entanglement in a
  Bose-Einstein condensate.}}}\ }\href {\doibase 10.1038/nature07332}
  {\bibfield  {journal} {\bibinfo  {journal} {Nature}\ }\textbf {\bibinfo
  {volume} {455}},\ \bibinfo {pages} {1216--1219} (\bibinfo {year}
  {2008})}\BibitemShut {NoStop}%
\bibitem [{\citenamefont {Berrada}\ \emph {et~al.}(2013)\citenamefont
  {Berrada}, \citenamefont {Frank}, \citenamefont {B\"ucker}, \citenamefont
  {Schumm}, \citenamefont {Schaff},\ and\ \citenamefont
  {Schmiedmayer}}]{Berrada13}%
  \BibitemOpen
  \bibfield  {author} {\bibinfo {author} {\bibfnamefont {T.}~\bibnamefont
  {Berrada}}, \bibinfo {author} {\bibfnamefont {S.~V.}\ \bibnamefont {Frank}},
  \bibinfo {author} {\bibfnamefont {R.}~\bibnamefont {B\"ucker}}, \bibinfo
  {author} {\bibfnamefont {T.}~\bibnamefont {Schumm}}, \bibinfo {author}
  {\bibfnamefont {J.-F.}\ \bibnamefont {Schaff}}, \ and\ \bibinfo {author}
  {\bibfnamefont {J.}~\bibnamefont {Schmiedmayer}},\ }\bibfield  {title}
  {\enquote {\bibinfo {title} {{Integrated Mach-Zehnder interferometer for
  Bose-Einstein condensates}},}\ }\href {\doibase 10.1038/ncomms3077}
  {\bibfield  {journal} {\bibinfo  {journal} {Nature Comm.}\ }\textbf {\bibinfo
  {volume} {4}},\ \bibinfo {pages} {2077} (\bibinfo {year} {2013})}\BibitemShut
  {NoStop}%
\bibitem [{\citenamefont {del Rio}\ \emph {et~al.}(2011)\citenamefont {del
  Rio}, \citenamefont {Aberg}, \citenamefont {Renner}, \citenamefont
  {Dahlsten},\ and\ \citenamefont {Vedral}}]{NegativeEntropy}%
  \BibitemOpen
  \bibfield  {author} {\bibinfo {author} {\bibfnamefont {L.}~\bibnamefont {del
  Rio}}, \bibinfo {author} {\bibfnamefont {J.}~\bibnamefont {Aberg}}, \bibinfo
  {author} {\bibfnamefont {R.}~\bibnamefont {Renner}}, \bibinfo {author}
  {\bibfnamefont {O.}~\bibnamefont {Dahlsten}}, \ and\ \bibinfo {author}
  {\bibfnamefont {V.}~\bibnamefont {Vedral}},\ }\bibfield  {title} {\enquote
  {\bibinfo {title} {The thermodynamic meaning of negative entropy},}\
  }\href@noop {} {\bibfield  {journal} {\bibinfo  {journal} {Nature}\ }\textbf
  {\bibinfo {volume} {474}},\ \bibinfo {pages} {61--63} (\bibinfo {year}
  {2011})}\BibitemShut {NoStop}%
\bibitem [{\citenamefont {Jennings}\ and\ \citenamefont
  {Rudolph}(2010)}]{Jennings2010}%
  \BibitemOpen
  \bibfield  {author} {\bibinfo {author} {\bibfnamefont {D.}~\bibnamefont
  {Jennings}}\ and\ \bibinfo {author} {\bibfnamefont {T.}~\bibnamefont
  {Rudolph}},\ }\bibfield  {title} {\enquote {\bibinfo {title} {{Entanglement
  and the thermodynamic arrow of time}},}\ }\href {\doibase
  10.1103/PhysRevE.81.061130} {\bibfield  {journal} {\bibinfo  {journal} {Phys.
  Rev. E}\ }\textbf {\bibinfo {volume} {81}},\ \bibinfo {pages} {061130}
  (\bibinfo {year} {2010})}\BibitemShut {NoStop}%
\bibitem [{\citenamefont {Partovi}(2008)}]{partovi2008entanglement}%
  \BibitemOpen
  \bibfield  {author} {\bibinfo {author} {\bibfnamefont {M.~H.}\ \bibnamefont
  {Partovi}},\ }\bibfield  {title} {\enquote {\bibinfo {title} {Entanglement
  versus stosszahlansatz: Disappearance of the thermodynamic arrow in a
  high-correlation environment},}\ }\href@noop {} {\bibfield  {journal}
  {\bibinfo  {journal} {Phys. Rev. E}\ }\textbf {\bibinfo {volume} {77}},\
  \bibinfo {pages} {021110} (\bibinfo {year} {2008})}\BibitemShut {NoStop}%
\bibitem [{\citenamefont {Jevtic}\ \emph {et~al.}(2012)\citenamefont {Jevtic},
  \citenamefont {Jennings},\ and\ \citenamefont {Rudolph}}]{Jevtic2012}%
  \BibitemOpen
  \bibfield  {author} {\bibinfo {author} {\bibfnamefont {S.}~\bibnamefont
  {Jevtic}}, \bibinfo {author} {\bibfnamefont {D.}~\bibnamefont {Jennings}}, \
  and\ \bibinfo {author} {\bibfnamefont {T.}~\bibnamefont {Rudolph}},\
  }\bibfield  {title} {\enquote {\bibinfo {title} {{Maximally and minimally
  correlated states attainable within a closed evolving system}},}\ }\href
  {\doibase 10.1103/PhysRevLett.108.110403} {\bibfield  {journal} {\bibinfo
  {journal} {Phys. Rev. Lett.}\ }\textbf {\bibinfo {volume} {108}},\ \bibinfo
  {pages} {110403} (\bibinfo {year} {2012})}\BibitemShut {NoStop}%
\bibitem [{\citenamefont {del Rio}\ \emph {et~al.}(2016)\citenamefont {del
  Rio}, \citenamefont {Hutter}, \citenamefont {Renner},\ and\ \citenamefont
  {Wehner}}]{DelRio2016}%
  \BibitemOpen
  \bibfield  {author} {\bibinfo {author} {\bibfnamefont {L.}~\bibnamefont {del
  Rio}}, \bibinfo {author} {\bibfnamefont {A.}~\bibnamefont {Hutter}}, \bibinfo
  {author} {\bibfnamefont {R.}~\bibnamefont {Renner}}, \ and\ \bibinfo {author}
  {\bibfnamefont {S.}~\bibnamefont {Wehner}},\ }\bibfield  {title} {\enquote
  {\bibinfo {title} {{Relative thermalization}},}\ }\href {\doibase
  10.1103/PhysRevE.94.022104} {\bibfield  {journal} {\bibinfo  {journal} {Phys.
  Rev. E}\ }\textbf {\bibinfo {volume} {94}},\ \bibinfo {pages} {022104}
  (\bibinfo {year} {2016})}\BibitemShut {NoStop}%
\bibitem [{\citenamefont {Micadei}\ \emph {et~al.}(2019)\citenamefont
  {Micadei}, \citenamefont {Peterson}, \citenamefont {Souza}, \citenamefont
  {Sarthour}, \citenamefont {Oliveira}, \citenamefont {Landi}, \citenamefont
  {Batalh{\~a}o}, \citenamefont {Serra},\ and\ \citenamefont
  {Lutz}}]{micadei2019reversing}%
  \BibitemOpen
  \bibfield  {author} {\bibinfo {author} {\bibfnamefont {K.}~\bibnamefont
  {Micadei}}, \bibinfo {author} {\bibfnamefont {J.~P.}\ \bibnamefont
  {Peterson}}, \bibinfo {author} {\bibfnamefont {A.~M.}\ \bibnamefont {Souza}},
  \bibinfo {author} {\bibfnamefont {R.~S.}\ \bibnamefont {Sarthour}}, \bibinfo
  {author} {\bibfnamefont {I.~S.}\ \bibnamefont {Oliveira}}, \bibinfo {author}
  {\bibfnamefont {G.~T.}\ \bibnamefont {Landi}}, \bibinfo {author}
  {\bibfnamefont {T.~B.}\ \bibnamefont {Batalh{\~a}o}}, \bibinfo {author}
  {\bibfnamefont {R.~M.}\ \bibnamefont {Serra}}, \ and\ \bibinfo {author}
  {\bibfnamefont {E.}~\bibnamefont {Lutz}},\ }\bibfield  {title} {\enquote
  {\bibinfo {title} {Reversing the direction of heat flow using quantum
  correlations},}\ }\href@noop {} {\bibfield  {journal} {\bibinfo  {journal}
  {Nature Comm.}\ }\textbf {\bibinfo {volume} {10}},\ \bibinfo {pages} {2456}
  (\bibinfo {year} {2019})}\BibitemShut {NoStop}%
\bibitem [{\citenamefont {Husmann}\ \emph {et~al.}(2018)\citenamefont
  {Husmann}, \citenamefont {Lebrat}, \citenamefont {H{\"a}usler}, \citenamefont
  {Brantut}, \citenamefont {Corman},\ and\ \citenamefont
  {Esslinger}}]{Husmann8563}%
  \BibitemOpen
  \bibfield  {author} {\bibinfo {author} {\bibfnamefont {D.}~\bibnamefont
  {Husmann}}, \bibinfo {author} {\bibfnamefont {M.}~\bibnamefont {Lebrat}},
  \bibinfo {author} {\bibfnamefont {S.}~\bibnamefont {H{\"a}usler}}, \bibinfo
  {author} {\bibfnamefont {J.-P.}\ \bibnamefont {Brantut}}, \bibinfo {author}
  {\bibfnamefont {L.}~\bibnamefont {Corman}}, \ and\ \bibinfo {author}
  {\bibfnamefont {T.}~\bibnamefont {Esslinger}},\ }\bibfield  {title} {\enquote
  {\bibinfo {title} {{Breakdown of the Wiedemann-Franz law in a unitary Fermi
  gas}},}\ }\href {\doibase 10.1073/pnas.1803336115} {\bibfield  {journal}
  {\bibinfo  {journal} {Proc. Natl. Ac. Sc.}\ }\textbf {\bibinfo {volume}
  {115}},\ \bibinfo {pages} {8563--8568} (\bibinfo {year} {2018})}\BibitemShut
  {NoStop}%
\bibitem [{\citenamefont {Breuer}\ \emph {et~al.}(2016)\citenamefont {Breuer},
  \citenamefont {Laine}, \citenamefont {Piilo},\ and\ \citenamefont
  {Vacchini}}]{RevModPhys.88.021002}%
  \BibitemOpen
  \bibfield  {author} {\bibinfo {author} {\bibfnamefont {H.-P.}\ \bibnamefont
  {Breuer}}, \bibinfo {author} {\bibfnamefont {E.-M.}\ \bibnamefont {Laine}},
  \bibinfo {author} {\bibfnamefont {J.}~\bibnamefont {Piilo}}, \ and\ \bibinfo
  {author} {\bibfnamefont {B.}~\bibnamefont {Vacchini}},\ }\bibfield  {title}
  {\enquote {\bibinfo {title} {{Colloquium: Non-Markovian dynamics in open
  quantum systems}},}\ }\href {\doibase 10.1103/RevModPhys.88.021002}
  {\bibfield  {journal} {\bibinfo  {journal} {Rev. Mod. Phys.}\ }\textbf
  {\bibinfo {volume} {88}},\ \bibinfo {pages} {021002} (\bibinfo {year}
  {2016})}\BibitemShut {NoStop}%
\bibitem [{\citenamefont {Rivas}\ \emph {et~al.}(2014)\citenamefont {Rivas},
  \citenamefont {Huelga},\ and\ \citenamefont {Plenio}}]{SusanaNoMarkov}%
  \BibitemOpen
  \bibfield  {author} {\bibinfo {author} {\bibfnamefont {A.}~\bibnamefont
  {Rivas}}, \bibinfo {author} {\bibfnamefont {S.~F.}\ \bibnamefont {Huelga}}, \
  and\ \bibinfo {author} {\bibfnamefont {M.~B.}\ \bibnamefont {Plenio}},\
  }\bibfield  {title} {\enquote {\bibinfo {title} {{Quantum non-Markovianity:
  characterization, quantification and detection}},}\ }\href {\doibase
  10.1088/0034-4885/77/9/094001} {\bibfield  {journal} {\bibinfo  {journal}
  {Rep. Prog. Phys.}\ }\textbf {\bibinfo {volume} {77}},\ \bibinfo {pages}
  {094001} (\bibinfo {year} {2014})}\BibitemShut {NoStop}%
\bibitem [{\citenamefont {Boes}\ \emph {et~al.}(2020)\citenamefont {Boes},
  \citenamefont {Gallego}, \citenamefont {Ng}, \citenamefont {Eisert},\ and\
  \citenamefont {Wilming}}]{boes2019passing}%
  \BibitemOpen
  \bibfield  {author} {\bibinfo {author} {\bibfnamefont {P.}~\bibnamefont
  {Boes}}, \bibinfo {author} {\bibfnamefont {R.}~\bibnamefont {Gallego}},
  \bibinfo {author} {\bibfnamefont {N.~H.}\ \bibnamefont {Ng}}, \bibinfo
  {author} {\bibfnamefont {J.}~\bibnamefont {Eisert}}, \ and\ \bibinfo {author}
  {\bibfnamefont {H.}~\bibnamefont {Wilming}},\ }\bibfield  {title} {\enquote
  {\bibinfo {title} {By-passing fluctuation theorems},}\ }\href@noop {}
  {\bibfield  {journal} {\bibinfo  {journal} {Quantum}\ }\textbf {\bibinfo
  {volume} {4}},\ \bibinfo {pages} {231} (\bibinfo {year} {2020})}\BibitemShut
  {NoStop}%
\bibitem [{\citenamefont {Brantut}\ \emph {et~al.}(2013)\citenamefont
  {Brantut}, \citenamefont {Grenier}, \citenamefont {Meineke}, \citenamefont
  {Stadler}, \citenamefont {Krinner}, \citenamefont {Kollath}, \citenamefont
  {Esslinger},\ and\ \citenamefont {Georges}}]{Brantut713}%
  \BibitemOpen
  \bibfield  {author} {\bibinfo {author} {\bibfnamefont {J.-P.}\ \bibnamefont
  {Brantut}}, \bibinfo {author} {\bibfnamefont {C.}~\bibnamefont {Grenier}},
  \bibinfo {author} {\bibfnamefont {J.}~\bibnamefont {Meineke}}, \bibinfo
  {author} {\bibfnamefont {D.}~\bibnamefont {Stadler}}, \bibinfo {author}
  {\bibfnamefont {S.}~\bibnamefont {Krinner}}, \bibinfo {author} {\bibfnamefont
  {C.}~\bibnamefont {Kollath}}, \bibinfo {author} {\bibfnamefont
  {T.}~\bibnamefont {Esslinger}}, \ and\ \bibinfo {author} {\bibfnamefont
  {A.}~\bibnamefont {Georges}},\ }\bibfield  {title} {\enquote {\bibinfo
  {title} {A thermoelectric heat engine with ultracold atoms},}\ }\href
  {\doibase 10.1126/science.1242308} {\bibfield  {journal} {\bibinfo  {journal}
  {Science}\ }\textbf {\bibinfo {volume} {342}},\ \bibinfo {pages} {713--715}
  (\bibinfo {year} {2013})}\BibitemShut {NoStop}%
\bibitem [{\citenamefont {Anders}\ and\ \citenamefont
  {Winter}(2007)}]{anders2007entanglement}%
  \BibitemOpen
  \bibfield  {author} {\bibinfo {author} {\bibfnamefont {J.}~\bibnamefont
  {Anders}}\ and\ \bibinfo {author} {\bibfnamefont {A.}~\bibnamefont
  {Winter}},\ }\bibfield  {title} {\enquote {\bibinfo {title} {Entanglement and
  separability of quantum harmonic oscillator systems at finite temperature},}\
  }\href@noop {} {\bibfield  {journal} {\bibinfo  {journal} {arXiv:0705.3026}\
  } (\bibinfo {year} {2007})}\BibitemShut {NoStop}%
\bibitem [{\citenamefont {Anders}(2008)}]{anderswinter2008}%
  \BibitemOpen
  \bibfield  {author} {\bibinfo {author} {\bibfnamefont {J.}~\bibnamefont
  {Anders}},\ }\bibfield  {title} {\enquote {\bibinfo {title} {Thermal state
  entanglement in harmonic lattices},}\ }\href {\doibase
  10.1103/PhysRevA.77.062102} {\bibfield  {journal} {\bibinfo  {journal} {Phys.
  Rev. A}\ }\textbf {\bibinfo {volume} {77}},\ \bibinfo {pages} {062102}
  (\bibinfo {year} {2008})}\BibitemShut {NoStop}%
\bibitem [{\citenamefont {Calabrese}\ and\ \citenamefont
  {Cardy}(2004)}]{calabrese2004entanglement}%
  \BibitemOpen
  \bibfield  {author} {\bibinfo {author} {\bibfnamefont {P.}~\bibnamefont
  {Calabrese}}\ and\ \bibinfo {author} {\bibfnamefont {J.}~\bibnamefont
  {Cardy}},\ }\bibfield  {title} {\enquote {\bibinfo {title} {Entanglement
  entropy and quantum field theory},}\ }\href@noop {} {\bibfield  {journal}
  {\bibinfo  {journal} {J. Stat. Mech.}\ }\textbf {\bibinfo {volume} {2004}},\
  \bibinfo {pages} {P06002} (\bibinfo {year} {2004})}\BibitemShut {NoStop}%
\bibitem [{\citenamefont {Wolf}\ \emph {et~al.}(2008)\citenamefont {Wolf},
  \citenamefont {Eisert}, \citenamefont {Cubitt},\ and\ \citenamefont
  {Cirac}}]{Markov}%
  \BibitemOpen
  \bibfield  {author} {\bibinfo {author} {\bibfnamefont {M.~M.}\ \bibnamefont
  {Wolf}}, \bibinfo {author} {\bibfnamefont {J.}~\bibnamefont {Eisert}},
  \bibinfo {author} {\bibfnamefont {T.~S.}\ \bibnamefont {Cubitt}}, \ and\
  \bibinfo {author} {\bibfnamefont {J.~I.}\ \bibnamefont {Cirac}},\ }\bibfield
  {title} {\enquote {\bibinfo {title} {{Assessing non-Markovian quantum
  dynamics}},}\ }\href@noop {} {\bibfield  {journal} {\bibinfo  {journal}
  {Phys. Rev. Lett.}\ }\textbf {\bibinfo {volume} {101}},\ \bibinfo {pages}
  {150402} (\bibinfo {year} {2008})}\BibitemShut {NoStop}%
\bibitem [{\citenamefont {Rivas}\ \emph {et~al.}(2010)\citenamefont {Rivas},
  \citenamefont {Huelga},\ and\ \citenamefont
  {Plenio}}]{PhysRevLett.105.050403}%
  \BibitemOpen
  \bibfield  {author} {\bibinfo {author} {\bibfnamefont {A.}~\bibnamefont
  {Rivas}}, \bibinfo {author} {\bibfnamefont {S.~F.}\ \bibnamefont {Huelga}}, \
  and\ \bibinfo {author} {\bibfnamefont {M.~B.}\ \bibnamefont {Plenio}},\
  }\bibfield  {title} {\enquote {\bibinfo {title} {{Entanglement and
  Non-Markovianity of quantum evolutions}},}\ }\href {\doibase
  10.1103/PhysRevLett.105.050403} {\bibfield  {journal} {\bibinfo  {journal}
  {Phys. Rev. Lett.}\ }\textbf {\bibinfo {volume} {105}},\ \bibinfo {pages}
  {050403} (\bibinfo {year} {2010})}\BibitemShut {NoStop}%
\bibitem [{\citenamefont {Gour}\ \emph {et~al.}(2015)\citenamefont {Gour},
  \citenamefont {M{\"u}ller}, \citenamefont {Narasimhachar}, \citenamefont
  {Spekkens},\ and\ \citenamefont {Halpern}}]{gour2015resource}%
  \BibitemOpen
  \bibfield  {author} {\bibinfo {author} {\bibfnamefont {G.}~\bibnamefont
  {Gour}}, \bibinfo {author} {\bibfnamefont {M.~P.}\ \bibnamefont
  {M{\"u}ller}}, \bibinfo {author} {\bibfnamefont {V.}~\bibnamefont
  {Narasimhachar}}, \bibinfo {author} {\bibfnamefont {R.~W.}\ \bibnamefont
  {Spekkens}}, \ and\ \bibinfo {author} {\bibfnamefont {N.~Y.}\ \bibnamefont
  {Halpern}},\ }\bibfield  {title} {\enquote {\bibinfo {title} {The resource
  theory of informational nonequilibrium in thermodynamics},}\ }\href@noop {}
  {\bibfield  {journal} {\bibinfo  {journal} {Physics Reports}\ }\textbf
  {\bibinfo {volume} {583}},\ \bibinfo {pages} {1--58} (\bibinfo {year}
  {2015})}\BibitemShut {NoStop}%
\bibitem [{\citenamefont {Vinjanampathy}\ and\ \citenamefont
  {Anders}(2016{\natexlab{b}})}]{qtSai}%
  \BibitemOpen
  \bibfield  {author} {\bibinfo {author} {\bibfnamefont {S.}~\bibnamefont
  {Vinjanampathy}}\ and\ \bibinfo {author} {\bibfnamefont {J.}~\bibnamefont
  {Anders}},\ }\bibfield  {title} {\enquote {\bibinfo {title} {Quantum
  thermodynamics},}\ }\href@noop {} {\bibfield  {journal} {\bibinfo  {journal}
  {Contemp. Phys.}\ }\textbf {\bibinfo {volume} {57}},\ \bibinfo {pages}
  {545--579} (\bibinfo {year} {2016}{\natexlab{b}})}\BibitemShut {NoStop}%
\bibitem [{\citenamefont {Bistritzer}\ and\ \citenamefont
  {Altman}(2007)}]{Bistritzer2007}%
  \BibitemOpen
  \bibfield  {author} {\bibinfo {author} {\bibfnamefont {R.}~\bibnamefont
  {Bistritzer}}\ and\ \bibinfo {author} {\bibfnamefont {E.}~\bibnamefont
  {Altman}},\ }\bibfield  {title} {\enquote {\bibinfo {title} {{Intrinsic
  dephasing in one-dimensional ultracold atom interferometers.}}}\ }\href
  {\doibase 10.1073/pnas.0608910104} {\bibfield  {journal} {\bibinfo  {journal}
  {Proc. Natl. Ac. Sc.}\ }\textbf {\bibinfo {volume} {104}},\ \bibinfo {pages}
  {9955--9} (\bibinfo {year} {2007})}\BibitemShut {NoStop}%
\bibitem [{\citenamefont {Kitagawa}\ \emph {et~al.}(2011)\citenamefont
  {Kitagawa}, \citenamefont {Imambekov}, \citenamefont {Schmiedmayer},\ and\
  \citenamefont {Demler}}]{Kitagawa2011}%
  \BibitemOpen
  \bibfield  {author} {\bibinfo {author} {\bibfnamefont {T.}~\bibnamefont
  {Kitagawa}}, \bibinfo {author} {\bibfnamefont {A.}~\bibnamefont {Imambekov}},
  \bibinfo {author} {\bibfnamefont {J.}~\bibnamefont {Schmiedmayer}}, \ and\
  \bibinfo {author} {\bibfnamefont {E.}~\bibnamefont {Demler}},\ }\bibfield
  {title} {\enquote {\bibinfo {title} {{The dynamics and prethermalization of
  one-dimensional quantum systems probed through the full distributions of
  quantum noise}},}\ }\href {\doibase 10.1088/1367-2630/13/7/073018} {\bibfield
   {journal} {\bibinfo  {journal} {New J. Phys.}\ }\textbf {\bibinfo {volume}
  {13}},\ \bibinfo {pages} {073018} (\bibinfo {year} {2011})}\BibitemShut
  {NoStop}%
\bibitem [{\citenamefont {Geiger}\ \emph {et~al.}(2014)\citenamefont {Geiger},
  \citenamefont {Langen}, \citenamefont {Mazets},\ and\ \citenamefont
  {Schmiedmayer}}]{Geiger2014}%
  \BibitemOpen
  \bibfield  {author} {\bibinfo {author} {\bibfnamefont {R.}~\bibnamefont
  {Geiger}}, \bibinfo {author} {\bibfnamefont {T.}~\bibnamefont {Langen}},
  \bibinfo {author} {\bibfnamefont {I.~E.}\ \bibnamefont {Mazets}}, \ and\
  \bibinfo {author} {\bibfnamefont {J.}~\bibnamefont {Schmiedmayer}},\
  }\bibfield  {title} {\enquote {\bibinfo {title} {{Local relaxation and
  light-cone-like propagation of correlations in a trapped one-dimensional Bose
  gas}},}\ }\href {\doibase 10.1088/1367-2630/16/5/053034} {\bibfield
  {journal} {\bibinfo  {journal} {New J. Phys.}\ }\textbf {\bibinfo {volume}
  {16}},\ \bibinfo {pages} {053034} (\bibinfo {year} {2014})}\BibitemShut
  {NoStop}%
\bibitem [{\citenamefont {Pezzutto}\ \emph {et~al.}(2016)\citenamefont
  {Pezzutto}, \citenamefont {Paternostro},\ and\ \citenamefont
  {Omar}}]{Pezzutto2016}%
  \BibitemOpen
  \bibfield  {author} {\bibinfo {author} {\bibfnamefont {M.}~\bibnamefont
  {Pezzutto}}, \bibinfo {author} {\bibfnamefont {M.}~\bibnamefont
  {Paternostro}}, \ and\ \bibinfo {author} {\bibfnamefont {Y.}~\bibnamefont
  {Omar}},\ }\bibfield  {title} {\enquote {\bibinfo {title} {{Implications of
  non-Markovian quantum dynamics for the Landauer bound}},}\ }\href {\doibase
  10.1088/1367-2630/18/12/123018} {\bibfield  {journal} {\bibinfo  {journal}
  {New J. Phys.}\ }\textbf {\bibinfo {volume} {18}},\ \bibinfo {pages} {123018}
  (\bibinfo {year} {2016})}\BibitemShut {NoStop}%
\bibitem [{\citenamefont {Hofer}\ \emph {et~al.}(2017)\citenamefont {Hofer},
  \citenamefont {Perarnau-Llobet}, \citenamefont {Miranda}, \citenamefont
  {Haack}, \citenamefont {Silva}, \citenamefont {Brask},\ and\ \citenamefont
  {Brunner}}]{Hofer2017}%
  \BibitemOpen
  \bibfield  {author} {\bibinfo {author} {\bibfnamefont {P.~P.}\ \bibnamefont
  {Hofer}}, \bibinfo {author} {\bibfnamefont {M.}~\bibnamefont
  {Perarnau-Llobet}}, \bibinfo {author} {\bibfnamefont {L.~D.~M.}\ \bibnamefont
  {Miranda}}, \bibinfo {author} {\bibfnamefont {G.}~\bibnamefont {Haack}},
  \bibinfo {author} {\bibfnamefont {R.}~\bibnamefont {Silva}}, \bibinfo
  {author} {\bibfnamefont {J.~B.}\ \bibnamefont {Brask}}, \ and\ \bibinfo
  {author} {\bibfnamefont {N.}~\bibnamefont {Brunner}},\ }\bibfield  {title}
  {\enquote {\bibinfo {title} {{Markovian master equations for quantum thermal
  machines: local versus global approach}},}\ }\href {\doibase
  10.1088/1367-2630/aa964f} {\bibfield  {journal} {\bibinfo  {journal} {New J.
  Phys.}\ }\textbf {\bibinfo {volume} {19}},\ \bibinfo {pages} {123037}
  (\bibinfo {year} {2017})}\BibitemShut {NoStop}%
\bibitem [{\citenamefont {Gonz{\'{a}}lez}\ \emph {et~al.}(2017)\citenamefont
  {Gonz{\'{a}}lez}, \citenamefont {Correa}, \citenamefont {Nocerino},
  \citenamefont {Palao}, \citenamefont {Alonso},\ and\ \citenamefont
  {Adesso}}]{Gonzalez2017}%
  \BibitemOpen
  \bibfield  {author} {\bibinfo {author} {\bibfnamefont {J.~O.}\ \bibnamefont
  {Gonz{\'{a}}lez}}, \bibinfo {author} {\bibfnamefont {L.~A.}\ \bibnamefont
  {Correa}}, \bibinfo {author} {\bibfnamefont {G.}~\bibnamefont {Nocerino}},
  \bibinfo {author} {\bibfnamefont {J.~P.}\ \bibnamefont {Palao}}, \bibinfo
  {author} {\bibfnamefont {D.}~\bibnamefont {Alonso}}, \ and\ \bibinfo {author}
  {\bibfnamefont {G.}~\bibnamefont {Adesso}},\ }\bibfield  {title} {\enquote
  {\bibinfo {title} {{Testing the validity of the local and global GKLS master
  equations on an exactly solvable model}},}\ }\href {\doibase
  10.1142/S1230161217400108} {\bibfield  {journal} {\bibinfo  {journal} {Open
  Sys. Inf. Dyn.}\ }\textbf {\bibinfo {volume} {24}},\ \bibinfo {pages}
  {1740010} (\bibinfo {year} {2017})}\BibitemShut {NoStop}%
\bibitem [{\citenamefont {Uzdin}\ \emph {et~al.}(2016)\citenamefont {Uzdin},
  \citenamefont {Levy},\ and\ \citenamefont {Kosloff}}]{Uzdin2016}%
  \BibitemOpen
  \bibfield  {author} {\bibinfo {author} {\bibfnamefont {R.}~\bibnamefont
  {Uzdin}}, \bibinfo {author} {\bibfnamefont {A.}~\bibnamefont {Levy}}, \ and\
  \bibinfo {author} {\bibfnamefont {R.}~\bibnamefont {Kosloff}},\ }\bibfield
  {title} {\enquote {\bibinfo {title} {{Quantum heat machines equivalence, work
  extraction beyond Markovianity, and strong coupling via heat exchangers}},}\
  }\href {\doibase 10.3390/e18040124} {\bibfield  {journal} {\bibinfo
  {journal} {Entropy}\ }\textbf {\bibinfo {volume} {18}},\ \bibinfo {pages}
  {124} (\bibinfo {year} {2016})}\BibitemShut {NoStop}%
\bibitem [{\citenamefont {Groeblacher}\ \emph {et~al.}(2015)\citenamefont
  {Groeblacher}, \citenamefont {Trubarov}, \citenamefont {Prigge},
  \citenamefont {Aspelmeyer},\ and\ \citenamefont {Eisert}}]{Observation}%
  \BibitemOpen
  \bibfield  {author} {\bibinfo {author} {\bibfnamefont {S.}~\bibnamefont
  {Groeblacher}}, \bibinfo {author} {\bibfnamefont {A.}~\bibnamefont
  {Trubarov}}, \bibinfo {author} {\bibfnamefont {N.}~\bibnamefont {Prigge}},
  \bibinfo {author} {\bibfnamefont {M.}~\bibnamefont {Aspelmeyer}}, \ and\
  \bibinfo {author} {\bibfnamefont {J.}~\bibnamefont {Eisert}},\ }\bibfield
  {title} {\enquote {\bibinfo {title} {{Observation of non-Markovian
  micro-mechanical Brownian motion}},}\ }\href {\doibase 10.1038/ncomms8606}
  {\bibfield  {journal} {\bibinfo  {journal} {Nature Comm.}\ }\textbf {\bibinfo
  {volume} {6}},\ \bibinfo {pages} {7606} (\bibinfo {year} {2015})}\BibitemShut
  {NoStop}%
\bibitem [{\citenamefont {Brandao}\ \emph {et~al.}(2015)\citenamefont
  {Brandao}, \citenamefont {Horodecki}, \citenamefont {Ng}, \citenamefont
  {Oppenheim},\ and\ \citenamefont {Wehner}}]{brandao2015second}%
  \BibitemOpen
  \bibfield  {author} {\bibinfo {author} {\bibfnamefont {F.}~\bibnamefont
  {Brandao}}, \bibinfo {author} {\bibfnamefont {M.}~\bibnamefont {Horodecki}},
  \bibinfo {author} {\bibfnamefont {N.}~\bibnamefont {Ng}}, \bibinfo {author}
  {\bibfnamefont {J.}~\bibnamefont {Oppenheim}}, \ and\ \bibinfo {author}
  {\bibfnamefont {S.}~\bibnamefont {Wehner}},\ }\bibfield  {title} {\enquote
  {\bibinfo {title} {The second laws of quantum thermodynamics},}\ }\href@noop
  {} {\bibfield  {journal} {\bibinfo  {journal} {Proc. Natl. Ac. Sc.}\ }\textbf
  {\bibinfo {volume} {112}},\ \bibinfo {pages} {3275--3279} (\bibinfo {year}
  {2015})}\BibitemShut {NoStop}%
\bibitem [{\citenamefont {Alhambra}\ \emph {et~al.}(2016)\citenamefont
  {Alhambra}, \citenamefont {Masanes}, \citenamefont {Oppenheim},\ and\
  \citenamefont {Perry}}]{alhambra2016fluctuating}%
  \BibitemOpen
  \bibfield  {author} {\bibinfo {author} {\bibfnamefont {{\'A}.~M.}\
  \bibnamefont {Alhambra}}, \bibinfo {author} {\bibfnamefont {L.}~\bibnamefont
  {Masanes}}, \bibinfo {author} {\bibfnamefont {J.}~\bibnamefont {Oppenheim}},
  \ and\ \bibinfo {author} {\bibfnamefont {C.}~\bibnamefont {Perry}},\
  }\bibfield  {title} {\enquote {\bibinfo {title} {Fluctuating work: From
  quantum thermodynamical identities to a second law equality},}\ }\href@noop
  {} {\bibfield  {journal} {\bibinfo  {journal} {Phys. Rev. X}\ }\textbf
  {\bibinfo {volume} {6}},\ \bibinfo {pages} {041017} (\bibinfo {year}
  {2016})}\BibitemShut {NoStop}%
\bibitem [{\citenamefont {Serafini}\ \emph {et~al.}(2020)\citenamefont
  {Serafini}, \citenamefont {Lostaglio}, \citenamefont {Longden}, \citenamefont
  {Shackerley-Bennett}, \citenamefont {Hsieh},\ and\ \citenamefont
  {Adesso}}]{serafini2019gaussian}%
  \BibitemOpen
  \bibfield  {author} {\bibinfo {author} {\bibfnamefont {A.}~\bibnamefont
  {Serafini}}, \bibinfo {author} {\bibfnamefont {M.}~\bibnamefont {Lostaglio}},
  \bibinfo {author} {\bibfnamefont {S.}~\bibnamefont {Longden}}, \bibinfo
  {author} {\bibfnamefont {U.}~\bibnamefont {Shackerley-Bennett}}, \bibinfo
  {author} {\bibfnamefont {C.-Y.}\ \bibnamefont {Hsieh}}, \ and\ \bibinfo
  {author} {\bibfnamefont {G.}~\bibnamefont {Adesso}},\ }\bibfield  {title}
  {\enquote {\bibinfo {title} {Gaussian thermal operations and the limits of
  algorithmic cooling},}\ }\href {\doibase 10.1103/PhysRevLett.124.010602}
  {\bibfield  {journal} {\bibinfo  {journal} {Phys. Rev. Lett.}\ }\textbf
  {\bibinfo {volume} {124}},\ \bibinfo {pages} {010602} (\bibinfo {year}
  {2020})}\BibitemShut {NoStop}%
\bibitem [{\citenamefont {Halpern}\ and\ \citenamefont
  {Limmer}(2020)}]{halpern2018fundamental}%
  \BibitemOpen
  \bibfield  {author} {\bibinfo {author} {\bibfnamefont {N.~Y.}\ \bibnamefont
  {Halpern}}\ and\ \bibinfo {author} {\bibfnamefont {D.~T.}\ \bibnamefont
  {Limmer}},\ }\bibfield  {title} {\enquote {\bibinfo {title} {Fundamental
  limitations on photoisomerization from thermodynamic resource theories},}\
  }\href@noop {} {\bibfield  {journal} {\bibinfo  {journal} {Phys. Rev. A}\
  }\textbf {\bibinfo {volume} {101}},\ \bibinfo {pages} {042116} (\bibinfo
  {year} {2020})}\BibitemShut {NoStop}%
\bibitem [{\citenamefont {Clivaz}\ \emph
  {et~al.}(2019{\natexlab{a}})\citenamefont {Clivaz}, \citenamefont {Silva},
  \citenamefont {Haack}, \citenamefont {Brask}, \citenamefont {Brunner},\ and\
  \citenamefont {Huber}}]{PhysRevLett.123.170605}%
  \BibitemOpen
  \bibfield  {author} {\bibinfo {author} {\bibfnamefont {F.}~\bibnamefont
  {Clivaz}}, \bibinfo {author} {\bibfnamefont {R.}~\bibnamefont {Silva}},
  \bibinfo {author} {\bibfnamefont {G.}~\bibnamefont {Haack}}, \bibinfo
  {author} {\bibfnamefont {J.~B.}\ \bibnamefont {Brask}}, \bibinfo {author}
  {\bibfnamefont {N.}~\bibnamefont {Brunner}}, \ and\ \bibinfo {author}
  {\bibfnamefont {M.}~\bibnamefont {Huber}},\ }\bibfield  {title} {\enquote
  {\bibinfo {title} {Unifying paradigms of quantum refrigeration: A universal
  and attainable bound on cooling},}\ }\href {\doibase
  10.1103/PhysRevLett.123.170605} {\bibfield  {journal} {\bibinfo  {journal}
  {Phys. Rev. Lett.}\ }\textbf {\bibinfo {volume} {123}},\ \bibinfo {pages}
  {170605} (\bibinfo {year} {2019}{\natexlab{a}})}\BibitemShut {NoStop}%
\bibitem [{\citenamefont {Clivaz}\ \emph
  {et~al.}(2019{\natexlab{b}})\citenamefont {Clivaz}, \citenamefont {Silva},
  \citenamefont {Haack}, \citenamefont {Brask}, \citenamefont {Brunner},\ and\
  \citenamefont {Huber}}]{clivaz2019unifying}%
  \BibitemOpen
  \bibfield  {author} {\bibinfo {author} {\bibfnamefont {F.}~\bibnamefont
  {Clivaz}}, \bibinfo {author} {\bibfnamefont {R.}~\bibnamefont {Silva}},
  \bibinfo {author} {\bibfnamefont {G.}~\bibnamefont {Haack}}, \bibinfo
  {author} {\bibfnamefont {J.~B.}\ \bibnamefont {Brask}}, \bibinfo {author}
  {\bibfnamefont {N.}~\bibnamefont {Brunner}}, \ and\ \bibinfo {author}
  {\bibfnamefont {M.}~\bibnamefont {Huber}},\ }\bibfield  {title} {\enquote
  {\bibinfo {title} {Unifying paradigms of quantum refrigeration: Fundamental
  limits of cooling and associated work costs},}\ }\href@noop {} {\bibfield
  {journal} {\bibinfo  {journal} {Phys. Rev. E}\ }\textbf {\bibinfo {volume}
  {100}},\ \bibinfo {pages} {042130} (\bibinfo {year}
  {2019}{\natexlab{b}})}\BibitemShut {NoStop}%
\bibitem [{\citenamefont {Woods}\ \emph {et~al.}(2019)\citenamefont {Woods},
  \citenamefont {Ng},\ and\ \citenamefont {Wehner}}]{woods2019maximum}%
  \BibitemOpen
  \bibfield  {author} {\bibinfo {author} {\bibfnamefont {M.~P.}\ \bibnamefont
  {Woods}}, \bibinfo {author} {\bibfnamefont {N.~H.~Y.}\ \bibnamefont {Ng}}, \
  and\ \bibinfo {author} {\bibfnamefont {S.}~\bibnamefont {Wehner}},\
  }\bibfield  {title} {\enquote {\bibinfo {title} {The maximum efficiency of
  nano heat engines depends on more than temperature},}\ }\href@noop {}
  {\bibfield  {journal} {\bibinfo  {journal} {Quantum}\ }\textbf {\bibinfo
  {volume} {3}},\ \bibinfo {pages} {177} (\bibinfo {year} {2019})}\BibitemShut
  {NoStop}%
\bibitem [{\citenamefont {Wilming}\ and\ \citenamefont
  {Gallego}(2017)}]{wilming2017third}%
  \BibitemOpen
  \bibfield  {author} {\bibinfo {author} {\bibfnamefont {H.}~\bibnamefont
  {Wilming}}\ and\ \bibinfo {author} {\bibfnamefont {R.}~\bibnamefont
  {Gallego}},\ }\bibfield  {title} {\enquote {\bibinfo {title} {Third law of
  thermodynamics as a single inequality},}\ }\href@noop {} {\bibfield
  {journal} {\bibinfo  {journal} {Phys. Rev. X}\ }\textbf {\bibinfo {volume}
  {7}},\ \bibinfo {pages} {041033} (\bibinfo {year} {2017})}\BibitemShut
  {NoStop}%
\bibitem [{\citenamefont {Masanes}\ and\ \citenamefont
  {Oppenheim}(2017)}]{masanes2017general}%
  \BibitemOpen
  \bibfield  {author} {\bibinfo {author} {\bibfnamefont {L.}~\bibnamefont
  {Masanes}}\ and\ \bibinfo {author} {\bibfnamefont {J.}~\bibnamefont
  {Oppenheim}},\ }\bibfield  {title} {\enquote {\bibinfo {title} {A general
  derivation and quantification of the third law of thermodynamics},}\
  }\href@noop {} {\bibfield  {journal} {\bibinfo  {journal} {Nature Comm.}\
  }\textbf {\bibinfo {volume} {8}},\ \bibinfo {pages} {14538} (\bibinfo {year}
  {2017})}\BibitemShut {NoStop}%
\bibitem [{\citenamefont {Niedenzu}\ \emph
  {et~al.}(2019{\natexlab{b}})\citenamefont {Niedenzu}, \citenamefont {Huber},\
  and\ \citenamefont {Boukobza}}]{Niedenzu_2019}%
  \BibitemOpen
  \bibfield  {author} {\bibinfo {author} {\bibfnamefont {W.}~\bibnamefont
  {Niedenzu}}, \bibinfo {author} {\bibfnamefont {M.}~\bibnamefont {Huber}}, \
  and\ \bibinfo {author} {\bibfnamefont {E.}~\bibnamefont {Boukobza}},\
  }\bibfield  {title} {\enquote {\bibinfo {title} {Concepts of work in
  autonomous quantum heat engines},}\ }\href {\doibase
  10.22331/q-2019-10-14-195} {\bibfield  {journal} {\bibinfo  {journal}
  {Quantum}\ }\textbf {\bibinfo {volume} {3}},\ \bibinfo {pages} {195}
  (\bibinfo {year} {2019}{\natexlab{b}})}\BibitemShut {NoStop}%
\bibitem [{\citenamefont {Jarzynski}(1997)}]{PhysRevLett.78.2690}%
  \BibitemOpen
  \bibfield  {author} {\bibinfo {author} {\bibfnamefont {C.}~\bibnamefont
  {Jarzynski}},\ }\bibfield  {title} {\enquote {\bibinfo {title}
  {Nonequilibrium equality for free energy differences},}\ }\href {\doibase
  10.1103/PhysRevLett.78.2690} {\bibfield  {journal} {\bibinfo  {journal}
  {Phys. Rev. Lett.}\ }\textbf {\bibinfo {volume} {78}},\ \bibinfo {pages}
  {2690--2693} (\bibinfo {year} {1997})}\BibitemShut {NoStop}%
\bibitem [{\citenamefont {Horodecki}\ and\ \citenamefont
  {Oppenheim}(2013)}]{opphor}%
  \BibitemOpen
  \bibfield  {author} {\bibinfo {author} {\bibfnamefont {M.}~\bibnamefont
  {Horodecki}}\ and\ \bibinfo {author} {\bibfnamefont {J.}~\bibnamefont
  {Oppenheim}},\ }\bibfield  {title} {\enquote {\bibinfo {title} {Fundamental
  limitations for quantum and nano thermodynamics},}\ }\href@noop {} {\bibfield
   {journal} {\bibinfo  {journal} {Nature Comm.}\ }\textbf {\bibinfo {volume}
  {4}},\ \bibinfo {pages} {2059} (\bibinfo {year} {2013})}\BibitemShut
  {NoStop}%
\bibitem [{\citenamefont {Brandao}\ \emph {et~al.}(2013)\citenamefont
  {Brandao}, \citenamefont {Horodecki}, \citenamefont {Oppenheim},
  \citenamefont {Renes},\ and\ \citenamefont {Spekkens}}]{brandao2013resource}%
  \BibitemOpen
  \bibfield  {author} {\bibinfo {author} {\bibfnamefont {F.~G. S.~L.}\
  \bibnamefont {Brandao}}, \bibinfo {author} {\bibfnamefont {M.}~\bibnamefont
  {Horodecki}}, \bibinfo {author} {\bibfnamefont {J.}~\bibnamefont
  {Oppenheim}}, \bibinfo {author} {\bibfnamefont {J.~M.}\ \bibnamefont
  {Renes}}, \ and\ \bibinfo {author} {\bibfnamefont {R.~W.}\ \bibnamefont
  {Spekkens}},\ }\bibfield  {title} {\enquote {\bibinfo {title} {Resource
  theory of quantum states out of thermal equilibrium},}\ }\href@noop {}
  {\bibfield  {journal} {\bibinfo  {journal} {Phys. Rev. Lett.}\ }\textbf
  {\bibinfo {volume} {111}},\ \bibinfo {pages} {250404} (\bibinfo {year}
  {2013})}\BibitemShut {NoStop}%
\bibitem [{\citenamefont {Chubb}\ \emph {et~al.}(2018)\citenamefont {Chubb},
  \citenamefont {Tomamichel},\ and\ \citenamefont
  {Korzekwa}}]{chubb2018beyond}%
  \BibitemOpen
  \bibfield  {author} {\bibinfo {author} {\bibfnamefont {C.~T.}\ \bibnamefont
  {Chubb}}, \bibinfo {author} {\bibfnamefont {M.}~\bibnamefont {Tomamichel}}, \
  and\ \bibinfo {author} {\bibfnamefont {K.}~\bibnamefont {Korzekwa}},\
  }\bibfield  {title} {\enquote {\bibinfo {title} {Beyond the thermodynamic
  limit: finite-size corrections to state interconversion rates},}\ }\href@noop
  {} {\bibfield  {journal} {\bibinfo  {journal} {Quantum}\ }\textbf {\bibinfo
  {volume} {2}},\ \bibinfo {pages} {108} (\bibinfo {year} {2018})}\BibitemShut
  {NoStop}%
\bibitem [{\citenamefont {Gallego}\ \emph {et~al.}(2016)\citenamefont
  {Gallego}, \citenamefont {Eisert},\ and\ \citenamefont
  {Wilming}}]{WorkAndHeat}%
  \BibitemOpen
  \bibfield  {author} {\bibinfo {author} {\bibfnamefont {R.}~\bibnamefont
  {Gallego}}, \bibinfo {author} {\bibfnamefont {J.}~\bibnamefont {Eisert}}, \
  and\ \bibinfo {author} {\bibfnamefont {H.}~\bibnamefont {Wilming}},\
  }\bibfield  {title} {\enquote {\bibinfo {title} {Thermodynamic work from
  operational principles},}\ }\href@noop {} {\bibfield  {journal} {\bibinfo
  {journal} {New J. Phys.}\ }\textbf {\bibinfo {volume} {18}},\ \bibinfo
  {pages} {103017} (\bibinfo {year} {2016})}\BibitemShut {NoStop}%
\bibitem [{\citenamefont {del Rio}\ \emph {et~al.}(2015)\citenamefont {del
  Rio}, \citenamefont {Kraemer},\ and\ \citenamefont {Renner}}]{DelRio2015}%
  \BibitemOpen
  \bibfield  {author} {\bibinfo {author} {\bibfnamefont {L.}~\bibnamefont {del
  Rio}}, \bibinfo {author} {\bibfnamefont {L.}~\bibnamefont {Kraemer}}, \ and\
  \bibinfo {author} {\bibfnamefont {R.}~\bibnamefont {Renner}},\ }\bibfield
  {title} {\enquote {\bibinfo {title} {{Resource theories of knowledge}},}\
  }\href {http://arxiv.org/abs/1511.08818} {\  (\bibinfo {year} {2015})},\
  \Eprint {http://arxiv.org/abs/1511.08818} {arXiv:1511.08818} \BibitemShut
  {NoStop}%
\bibitem [{\citenamefont {Perarnau-Llobet}\ \emph {et~al.}(2015)\citenamefont
  {Perarnau-Llobet}, \citenamefont {Hovhannisyan}, \citenamefont {Huber},
  \citenamefont {Skrzypczyk}, \citenamefont {Brunner},\ and\ \citenamefont
  {Acin}}]{perarnau2015}%
  \BibitemOpen
  \bibfield  {author} {\bibinfo {author} {\bibfnamefont {M.}~\bibnamefont
  {Perarnau-Llobet}}, \bibinfo {author} {\bibfnamefont {K.~V.}\ \bibnamefont
  {Hovhannisyan}}, \bibinfo {author} {\bibfnamefont {M.}~\bibnamefont {Huber}},
  \bibinfo {author} {\bibfnamefont {P.}~\bibnamefont {Skrzypczyk}}, \bibinfo
  {author} {\bibfnamefont {N.}~\bibnamefont {Brunner}}, \ and\ \bibinfo
  {author} {\bibfnamefont {A.}~\bibnamefont {Acin}},\ }\bibfield  {title}
  {\enquote {\bibinfo {title} {Extractable work from correlations},}\ }\href
  {\doibase 10.1103/PhysRevX.5.041011} {\bibfield  {journal} {\bibinfo
  {journal} {Phys. Rev. X}\ }\textbf {\bibinfo {volume} {5}},\ \bibinfo {pages}
  {041011} (\bibinfo {year} {2015})}\BibitemShut {NoStop}%
\bibitem [{\citenamefont {Brunner}\ \emph {et~al.}(2014)\citenamefont
  {Brunner}, \citenamefont {Huber}, \citenamefont {Linden}, \citenamefont
  {Popescu}, \citenamefont {Silva},\ and\ \citenamefont
  {Skrzypczyk}}]{brunner2014}%
  \BibitemOpen
  \bibfield  {author} {\bibinfo {author} {\bibfnamefont {N.}~\bibnamefont
  {Brunner}}, \bibinfo {author} {\bibfnamefont {M.}~\bibnamefont {Huber}},
  \bibinfo {author} {\bibfnamefont {N.}~\bibnamefont {Linden}}, \bibinfo
  {author} {\bibfnamefont {S.}~\bibnamefont {Popescu}}, \bibinfo {author}
  {\bibfnamefont {R.}~\bibnamefont {Silva}}, \ and\ \bibinfo {author}
  {\bibfnamefont {P.}~\bibnamefont {Skrzypczyk}},\ }\bibfield  {title}
  {\enquote {\bibinfo {title} {Entanglement enhances cooling in microscopic
  quantum refrigerators},}\ }\href@noop {} {\bibfield  {journal} {\bibinfo
  {journal} {Phys. Rev. E}\ }\textbf {\bibinfo {volume} {89}},\ \bibinfo
  {pages} {032115} (\bibinfo {year} {2014})}\BibitemShut {NoStop}%
\bibitem [{\citenamefont {Ng}\ \emph {et~al.}(2017)\citenamefont {Ng},
  \citenamefont {Woods},\ and\ \citenamefont {Wehner}}]{ng2017surpassing}%
  \BibitemOpen
  \bibfield  {author} {\bibinfo {author} {\bibfnamefont {N.~H.~Y.}\
  \bibnamefont {Ng}}, \bibinfo {author} {\bibfnamefont {M.~P.}\ \bibnamefont
  {Woods}}, \ and\ \bibinfo {author} {\bibfnamefont {S.}~\bibnamefont
  {Wehner}},\ }\bibfield  {title} {\enquote {\bibinfo {title} {Surpassing the
  carnot efficiency by extracting imperfect work},}\ }\href@noop {} {\bibfield
  {journal} {\bibinfo  {journal} {New J. Phys.}\ }\textbf {\bibinfo {volume}
  {19}},\ \bibinfo {pages} {113005} (\bibinfo {year} {2017})}\BibitemShut
  {NoStop}%
\bibitem [{\citenamefont {Kukuljan}\ \emph {et~al.}(2018)\citenamefont
  {Kukuljan}, \citenamefont {Sotiriadis},\ and\ \citenamefont
  {Takacs}}]{PhysRevLett.121.110402}%
  \BibitemOpen
  \bibfield  {author} {\bibinfo {author} {\bibfnamefont {I.}~\bibnamefont
  {Kukuljan}}, \bibinfo {author} {\bibfnamefont {S.}~\bibnamefont
  {Sotiriadis}}, \ and\ \bibinfo {author} {\bibfnamefont {G.}~\bibnamefont
  {Takacs}},\ }\bibfield  {title} {\enquote {\bibinfo {title} {{Correlation
  functions of the quantum sine-Gordon model in and out of equilibrium}},}\
  }\href {\doibase 10.1103/PhysRevLett.121.110402} {\bibfield  {journal}
  {\bibinfo  {journal} {Phys. Rev. Lett.}\ }\textbf {\bibinfo {volume} {121}},\
  \bibinfo {pages} {110402} (\bibinfo {year} {2018})}\BibitemShut {NoStop}%
\bibitem [{\citenamefont {Narasimhachar}\ \emph {et~al.}(2019)\citenamefont
  {Narasimhachar}, \citenamefont {Assad}, \citenamefont {Binder}, \citenamefont
  {Thompson}, \citenamefont {Yadin},\ and\ \citenamefont
  {Gu}}]{narasimhachar2019thermodynamic}%
  \BibitemOpen
  \bibfield  {author} {\bibinfo {author} {\bibfnamefont {V.}~\bibnamefont
  {Narasimhachar}}, \bibinfo {author} {\bibfnamefont {S.}~\bibnamefont
  {Assad}}, \bibinfo {author} {\bibfnamefont {F.~C.}\ \bibnamefont {Binder}},
  \bibinfo {author} {\bibfnamefont {J.}~\bibnamefont {Thompson}}, \bibinfo
  {author} {\bibfnamefont {B.}~\bibnamefont {Yadin}}, \ and\ \bibinfo {author}
  {\bibfnamefont {M.}~\bibnamefont {Gu}},\ }\bibfield  {title} {\enquote
  {\bibinfo {title} {Thermodynamic resources in continuous-variable quantum
  systems},}\ }\href@noop {} {\bibfield  {journal} {\bibinfo  {journal}
  {arXiv:1909.07364}\ } (\bibinfo {year} {2019})}\BibitemShut {NoStop}%
\bibitem [{\citenamefont {Coleman}(1975)}]{Coleman75}%
  \BibitemOpen
  \bibfield  {author} {\bibinfo {author} {\bibfnamefont {S.}~\bibnamefont
  {Coleman}},\ }\bibfield  {title} {\enquote {\bibinfo {title} {Quantum
  sine-gordon equation as the massive thirring model},}\ }\href {\doibase
  10.1103/PhysRevD.11.2088} {\bibfield  {journal} {\bibinfo  {journal} {Phys.
  Rev. D}\ }\textbf {\bibinfo {volume} {11}},\ \bibinfo {pages} {2088--2097}
  (\bibinfo {year} {1975})}\BibitemShut {NoStop}%
\bibitem [{\citenamefont {{Mandelstam}}(1975)}]{Mandelstam}%
  \BibitemOpen
  \bibfield  {author} {\bibinfo {author} {\bibfnamefont {S.}~\bibnamefont
  {{Mandelstam}}},\ }\bibfield  {title} {\enquote {\bibinfo {title} {{Soliton
  operators for the quantized sine-Gordon equation}},}\ }\href {\doibase
  10.1103/PhysRevD.11.3026} {\bibfield  {journal} {\bibinfo  {journal} {\prd}\
  }\textbf {\bibinfo {volume} {11}},\ \bibinfo {pages} {3026} (\bibinfo {year}
  {1975})}\BibitemShut {NoStop}%
\bibitem [{\citenamefont {Thirring}(1958)}]{Thirring195891}%
  \BibitemOpen
  \bibfield  {author} {\bibinfo {author} {\bibfnamefont {W.~E.}\ \bibnamefont
  {Thirring}},\ }\bibfield  {title} {\enquote {\bibinfo {title} {A soluble
  relativistic field theory},}\ }\href {\doibase 10.1016/0003-4916(58)90015-0}
  {\bibfield  {journal} {\bibinfo  {journal} {Ann. Phys.}\ }\textbf {\bibinfo
  {volume} {3}},\ \bibinfo {pages} {91--112} (\bibinfo {year}
  {1958})}\BibitemShut {NoStop}%
\bibitem [{\citenamefont {Faddeev}\ and\ \citenamefont
  {Korepin}(1978)}]{Faddeev19781}%
  \BibitemOpen
  \bibfield  {author} {\bibinfo {author} {\bibfnamefont {L.~D.}\ \bibnamefont
  {Faddeev}}\ and\ \bibinfo {author} {\bibfnamefont {V.~E.}\ \bibnamefont
  {Korepin}},\ }\bibfield  {title} {\enquote {\bibinfo {title} {Quantum theory
  of solitons},}\ }\href {\doibase 10.1016/0370-1573(78)90058-3} {\bibfield
  {journal} {\bibinfo  {journal} {Phys. Rep.}\ }\textbf {\bibinfo {volume}
  {42}},\ \bibinfo {pages} {1--87} (\bibinfo {year} {1978})}\BibitemShut
  {NoStop}%
\bibitem [{\citenamefont {Zache}\ \emph {et~al.}(2020)\citenamefont {Zache},
  \citenamefont {Schweigler}, \citenamefont {Erne}, \citenamefont
  {Schmiedmayer},\ and\ \citenamefont {Berges}}]{Zache2020}%
  \BibitemOpen
  \bibfield  {author} {\bibinfo {author} {\bibfnamefont {T.~V.}\ \bibnamefont
  {Zache}}, \bibinfo {author} {\bibfnamefont {T.}~\bibnamefont {Schweigler}},
  \bibinfo {author} {\bibfnamefont {S.}~\bibnamefont {Erne}}, \bibinfo {author}
  {\bibfnamefont {J.}~\bibnamefont {Schmiedmayer}}, \ and\ \bibinfo {author}
  {\bibfnamefont {J.}~\bibnamefont {Berges}},\ }\bibfield  {title} {\enquote
  {\bibinfo {title} {{Extracting the field theory description of a quantum
  many-body system from experimental data}},}\ }\href {\doibase
  10.1103/physrevx.10.011020} {\bibfield  {journal} {\bibinfo  {journal} {Phys.
  Rev. X}\ }\textbf {\bibinfo {volume} {10}},\ \bibinfo {pages} {11020}
  (\bibinfo {year} {2020})}\BibitemShut {NoStop}%
\bibitem [{\citenamefont {Cronin}\ \emph {et~al.}(2009)\citenamefont {Cronin},
  \citenamefont {Schmiedmayer},\ and\ \citenamefont
  {Pritchard}}]{cronin2009optics}%
  \BibitemOpen
  \bibfield  {author} {\bibinfo {author} {\bibfnamefont {A.~D.}\ \bibnamefont
  {Cronin}}, \bibinfo {author} {\bibfnamefont {J.}~\bibnamefont
  {Schmiedmayer}}, \ and\ \bibinfo {author} {\bibfnamefont {D.~E.}\
  \bibnamefont {Pritchard}},\ }\bibfield  {title} {\enquote {\bibinfo {title}
  {Optics and interferometry with atoms and molecules},}\ }\href@noop {}
  {\bibfield  {journal} {\bibinfo  {journal} {Rev. Mod. Phys.}\ }\textbf
  {\bibinfo {volume} {81}},\ \bibinfo {pages} {1051} (\bibinfo {year}
  {2009})}\BibitemShut {NoStop}%
\bibitem [{\citenamefont {Grond}\ \emph {et~al.}(2009)\citenamefont {Grond},
  \citenamefont {Schmiedmayer},\ and\ \citenamefont {Hohenester}}]{Grond09}%
  \BibitemOpen
  \bibfield  {author} {\bibinfo {author} {\bibfnamefont {J.}~\bibnamefont
  {Grond}}, \bibinfo {author} {\bibfnamefont {J.}~\bibnamefont {Schmiedmayer}},
  \ and\ \bibinfo {author} {\bibfnamefont {U.}~\bibnamefont {Hohenester}},\
  }\bibfield  {title} {\enquote {\bibinfo {title} {Optimizing number squeezing
  when splitting a mesoscopic condensate},}\ }\href {\doibase
  10.1103/PhysRevA.79.021603} {\bibfield  {journal} {\bibinfo  {journal} {Phys.
  Rev. A}\ }\textbf {\bibinfo {volume} {79}},\ \bibinfo {pages} {021603}
  (\bibinfo {year} {2009})}\BibitemShut {NoStop}%
\bibitem [{\citenamefont {Werschnik}\ and\ \citenamefont
  {Gross}(2007)}]{werschnik2007quantum}%
  \BibitemOpen
  \bibfield  {author} {\bibinfo {author} {\bibfnamefont {J.}~\bibnamefont
  {Werschnik}}\ and\ \bibinfo {author} {\bibfnamefont {E.}~\bibnamefont
  {Gross}},\ }\bibfield  {title} {\enquote {\bibinfo {title} {Quantum optimal
  control theory},}\ }\href@noop {} {\bibfield  {journal} {\bibinfo  {journal}
  {J. Phys. B}\ }\textbf {\bibinfo {volume} {40}},\ \bibinfo {pages} {R175}
  (\bibinfo {year} {2007})}\BibitemShut {NoStop}%
\bibitem [{\citenamefont {Caneva}\ \emph {et~al.}(2009)\citenamefont {Caneva},
  \citenamefont {Murphy}, \citenamefont {Calarco}, \citenamefont {Fazio},
  \citenamefont {Montangero}, \citenamefont {Giovannetti},\ and\ \citenamefont
  {Santoro}}]{PhysRevLett.103.240501}%
  \BibitemOpen
  \bibfield  {author} {\bibinfo {author} {\bibfnamefont {T.}~\bibnamefont
  {Caneva}}, \bibinfo {author} {\bibfnamefont {M.}~\bibnamefont {Murphy}},
  \bibinfo {author} {\bibfnamefont {T.}~\bibnamefont {Calarco}}, \bibinfo
  {author} {\bibfnamefont {R.}~\bibnamefont {Fazio}}, \bibinfo {author}
  {\bibfnamefont {S.}~\bibnamefont {Montangero}}, \bibinfo {author}
  {\bibfnamefont {V.}~\bibnamefont {Giovannetti}}, \ and\ \bibinfo {author}
  {\bibfnamefont {G.~E.}\ \bibnamefont {Santoro}},\ }\bibfield  {title}
  {\enquote {\bibinfo {title} {Optimal control at the quantum speed limit},}\
  }\href {\doibase 10.1103/PhysRevLett.103.240501} {\bibfield  {journal}
  {\bibinfo  {journal} {Phys. Rev. Lett.}\ }\textbf {\bibinfo {volume} {103}},\
  \bibinfo {pages} {240501} (\bibinfo {year} {2009})}\BibitemShut {NoStop}%
\bibitem [{\citenamefont {Doria}\ \emph {et~al.}(2011)\citenamefont {Doria},
  \citenamefont {Calarco},\ and\ \citenamefont
  {Montangero}}]{PhysRevLett.106.190501}%
  \BibitemOpen
  \bibfield  {author} {\bibinfo {author} {\bibfnamefont {P.}~\bibnamefont
  {Doria}}, \bibinfo {author} {\bibfnamefont {T.}~\bibnamefont {Calarco}}, \
  and\ \bibinfo {author} {\bibfnamefont {S.}~\bibnamefont {Montangero}},\
  }\bibfield  {title} {\enquote {\bibinfo {title} {Optimal control technique
  for many-body quantum dynamics},}\ }\href {\doibase
  10.1103/PhysRevLett.106.190501} {\bibfield  {journal} {\bibinfo  {journal}
  {Phys. Rev. Lett.}\ }\textbf {\bibinfo {volume} {106}},\ \bibinfo {pages}
  {190501} (\bibinfo {year} {2011})}\BibitemShut {NoStop}%
\bibitem [{\citenamefont {Koch}(2016)}]{Koch_2016}%
  \BibitemOpen
  \bibfield  {author} {\bibinfo {author} {\bibfnamefont {C.~P.}\ \bibnamefont
  {Koch}},\ }\bibfield  {title} {\enquote {\bibinfo {title} {Controlling open
  quantum systems: tools, achievements, and limitations},}\ }\href {\doibase
  10.1088/0953-8984/28/21/213001} {\bibfield  {journal} {\bibinfo  {journal}
  {J. Phys.}\ }\textbf {\bibinfo {volume} {28}},\ \bibinfo {pages} {213001}
  (\bibinfo {year} {2016})}\BibitemShut {NoStop}%
\bibitem [{\citenamefont {van Frank}\ \emph {et~al.}(2016)\citenamefont {van
  Frank}, \citenamefont {Bonneau}, \citenamefont {Schmiedmayer}, \citenamefont
  {Hild}, \citenamefont {Gross}, \citenamefont {Cheneau}, \citenamefont
  {Bloch}, \citenamefont {Pichler}, \citenamefont {Negretti}, \citenamefont
  {Calarco},\ and\ \citenamefont {Montangero}}]{VanFrank2015a}%
  \BibitemOpen
  \bibfield  {author} {\bibinfo {author} {\bibfnamefont {S.}~\bibnamefont {van
  Frank}}, \bibinfo {author} {\bibfnamefont {M.}~\bibnamefont {Bonneau}},
  \bibinfo {author} {\bibfnamefont {J.}~\bibnamefont {Schmiedmayer}}, \bibinfo
  {author} {\bibfnamefont {S.}~\bibnamefont {Hild}}, \bibinfo {author}
  {\bibfnamefont {C.}~\bibnamefont {Gross}}, \bibinfo {author} {\bibfnamefont
  {M.}~\bibnamefont {Cheneau}}, \bibinfo {author} {\bibfnamefont
  {I.}~\bibnamefont {Bloch}}, \bibinfo {author} {\bibfnamefont
  {T.}~\bibnamefont {Pichler}}, \bibinfo {author} {\bibfnamefont
  {A.}~\bibnamefont {Negretti}}, \bibinfo {author} {\bibfnamefont
  {T.}~\bibnamefont {Calarco}}, \ and\ \bibinfo {author} {\bibfnamefont
  {S.}~\bibnamefont {Montangero}},\ }\bibfield  {title} {\enquote {\bibinfo
  {title} {{Optimal control of complex atomic quantum systems}},}\ }\href
  {\doibase 10.1038/srep34187} {\bibfield  {journal} {\bibinfo  {journal}
  {Scientific Rep.}\ }\textbf {\bibinfo {volume} {6}},\ \bibinfo {pages}
  {34187} (\bibinfo {year} {2016})}\BibitemShut {NoStop}%
\bibitem [{\citenamefont {Hauke}\ \emph {et~al.}(2014)\citenamefont {Hauke},
  \citenamefont {Lewenstein},\ and\ \citenamefont
  {Eckardt}}]{PhysRevLett113.045303}%
  \BibitemOpen
  \bibfield  {author} {\bibinfo {author} {\bibfnamefont {P.}~\bibnamefont
  {Hauke}}, \bibinfo {author} {\bibfnamefont {M.}~\bibnamefont {Lewenstein}}, \
  and\ \bibinfo {author} {\bibfnamefont {A.}~\bibnamefont {Eckardt}},\
  }\bibfield  {title} {\enquote {\bibinfo {title} {Tomography of band
  insulators from quench dynamics},}\ }\href {\doibase
  10.1103/PhysRevLett.113.045303} {\bibfield  {journal} {\bibinfo  {journal}
  {Phys. Rev. Lett.}\ }\textbf {\bibinfo {volume} {113}},\ \bibinfo {pages}
  {045303} (\bibinfo {year} {2014})}\BibitemShut {NoStop}%
\bibitem [{\citenamefont {Schaff}\ \emph {et~al.}(2014)\citenamefont {Schaff},
  \citenamefont {Langen},\ and\ \citenamefont
  {Schmiedmayer}}]{schaff2014interferometry}%
  \BibitemOpen
  \bibfield  {author} {\bibinfo {author} {\bibfnamefont {J.-F.}\ \bibnamefont
  {Schaff}}, \bibinfo {author} {\bibfnamefont {T.}~\bibnamefont {Langen}}, \
  and\ \bibinfo {author} {\bibfnamefont {J.}~\bibnamefont {Schmiedmayer}},\
  }\bibfield  {title} {\enquote {\bibinfo {title} {Interferometry with
  atoms},}\ }\href@noop {} {\bibfield  {journal} {\bibinfo  {journal} {Rivista
  del Nuovo Cimento della Societa Italiana di Fisica}\ }\textbf {\bibinfo
  {volume} {37}},\ \bibinfo {pages} {509--589} (\bibinfo {year}
  {2014})}\BibitemShut {NoStop}%
\bibitem [{\citenamefont {Eisert}\ \emph {et~al.}(2020)\citenamefont {Eisert},
  \citenamefont {Hangleiter}, \citenamefont {D.}, \citenamefont {Roth},
  \citenamefont {Markham}, \citenamefont {Parekh}, \citenamefont {Chabaud},\
  and\ \citenamefont {Kashefi}}]{Benchmarking}%
  \BibitemOpen
  \bibfield  {author} {\bibinfo {author} {\bibfnamefont {J.}~\bibnamefont
  {Eisert}}, \bibinfo {author} {\bibnamefont {Hangleiter}}, \bibinfo {author}
  {\bibfnamefont {N.~W.}\ \bibnamefont {D.}}, \bibinfo {author} {\bibfnamefont
  {I.}~\bibnamefont {Roth}}, \bibinfo {author} {\bibfnamefont {D.}~\bibnamefont
  {Markham}}, \bibinfo {author} {\bibfnamefont {R.}~\bibnamefont {Parekh}},
  \bibinfo {author} {\bibfnamefont {U.}~\bibnamefont {Chabaud}}, \ and\
  \bibinfo {author} {\bibfnamefont {E.}~\bibnamefont {Kashefi}},\ }\bibfield
  {title} {\enquote {\bibinfo {title} {Quantum certification and
  benchmarking},}\ }\href@noop {} {\bibfield  {journal} {\bibinfo  {journal}
  {Nature Rev. Phys.}\ }\textbf {\bibinfo {volume} {2}},\ \bibinfo {pages}
  {382--390} (\bibinfo {year} {2020})}\BibitemShut {NoStop}%
\bibitem [{\citenamefont {Yang}\ \emph {et~al.}(2020)\citenamefont {Yang},
  \citenamefont {Sun}, \citenamefont {Huang}, \citenamefont {Wang},
  \citenamefont {Deng}, \citenamefont {Dai}, \citenamefont {Yuan},\ and\
  \citenamefont {Pan}}]{yang2020cooling}%
  \BibitemOpen
  \bibfield  {author} {\bibinfo {author} {\bibfnamefont {B.}~\bibnamefont
  {Yang}}, \bibinfo {author} {\bibfnamefont {H.}~\bibnamefont {Sun}}, \bibinfo
  {author} {\bibfnamefont {C.-J.}\ \bibnamefont {Huang}}, \bibinfo {author}
  {\bibfnamefont {H.-Y.}\ \bibnamefont {Wang}}, \bibinfo {author}
  {\bibfnamefont {Y.}~\bibnamefont {Deng}}, \bibinfo {author} {\bibfnamefont
  {H.-N.}\ \bibnamefont {Dai}}, \bibinfo {author} {\bibfnamefont {Z.-S.}\
  \bibnamefont {Yuan}}, \ and\ \bibinfo {author} {\bibfnamefont {J.-W.}\
  \bibnamefont {Pan}},\ }\bibfield  {title} {\enquote {\bibinfo {title}
  {Cooling and entangling ultracold atoms in optical lattices},}\ }\href@noop
  {} {\bibfield  {journal} {\bibinfo  {journal} {Science}\ } (\bibinfo {year}
  {2020})}\BibitemShut {NoStop}%
\bibitem [{\citenamefont {Wildermuth}\ \emph {et~al.}(2005)\citenamefont
  {Wildermuth}, \citenamefont {Hofferberth}, \citenamefont {Lesanovsky},
  \citenamefont {Haller}, \citenamefont {Andersson}, \citenamefont {Groth},
  \citenamefont {Bar-Joseph}, \citenamefont {Kr{\"u}ger},\ and\ \citenamefont
  {Schmiedmayer}}]{wildermuth2005microscopic}%
  \BibitemOpen
  \bibfield  {author} {\bibinfo {author} {\bibfnamefont {S.}~\bibnamefont
  {Wildermuth}}, \bibinfo {author} {\bibfnamefont {S.}~\bibnamefont
  {Hofferberth}}, \bibinfo {author} {\bibfnamefont {I.}~\bibnamefont
  {Lesanovsky}}, \bibinfo {author} {\bibfnamefont {E.}~\bibnamefont {Haller}},
  \bibinfo {author} {\bibfnamefont {L.~M.}\ \bibnamefont {Andersson}}, \bibinfo
  {author} {\bibfnamefont {S.}~\bibnamefont {Groth}}, \bibinfo {author}
  {\bibfnamefont {I.}~\bibnamefont {Bar-Joseph}}, \bibinfo {author}
  {\bibfnamefont {P.}~\bibnamefont {Kr{\"u}ger}}, \ and\ \bibinfo {author}
  {\bibfnamefont {J.}~\bibnamefont {Schmiedmayer}},\ }\bibfield  {title}
  {\enquote {\bibinfo {title} {Microscopic magnetic-field imaging},}\
  }\href@noop {} {\bibfield  {journal} {\bibinfo  {journal} {Nature}\ }\textbf
  {\bibinfo {volume} {435}},\ \bibinfo {pages} {440--440} (\bibinfo {year}
  {2005})}\BibitemShut {NoStop}%
\bibitem [{\citenamefont {Aigner}\ \emph {et~al.}(2008)\citenamefont {Aigner},
  \citenamefont {Della~Pietra}, \citenamefont {Japha}, \citenamefont
  {Entin-Wohlman}, \citenamefont {David}, \citenamefont {Salem}, \citenamefont
  {Folman},\ and\ \citenamefont {Schmiedmayer}}]{aigner2008long}%
  \BibitemOpen
  \bibfield  {author} {\bibinfo {author} {\bibfnamefont {S.}~\bibnamefont
  {Aigner}}, \bibinfo {author} {\bibfnamefont {L.}~\bibnamefont
  {Della~Pietra}}, \bibinfo {author} {\bibfnamefont {Y.}~\bibnamefont {Japha}},
  \bibinfo {author} {\bibfnamefont {O.}~\bibnamefont {Entin-Wohlman}}, \bibinfo
  {author} {\bibfnamefont {T.}~\bibnamefont {David}}, \bibinfo {author}
  {\bibfnamefont {R.}~\bibnamefont {Salem}}, \bibinfo {author} {\bibfnamefont
  {R.}~\bibnamefont {Folman}}, \ and\ \bibinfo {author} {\bibfnamefont
  {J.}~\bibnamefont {Schmiedmayer}},\ }\bibfield  {title} {\enquote {\bibinfo
  {title} {Long-range order in electronic transport through disordered metal
  films},}\ }\href@noop {} {\bibfield  {journal} {\bibinfo  {journal}
  {Science}\ }\textbf {\bibinfo {volume} {319}},\ \bibinfo {pages} {1226--1229}
  (\bibinfo {year} {2008})}\BibitemShut {NoStop}%
\bibitem [{\citenamefont {Andrews}\ \emph {et~al.}(1996)\citenamefont
  {Andrews}, \citenamefont {Mewes}, \citenamefont {Van~Druten}, \citenamefont
  {Durfee}, \citenamefont {Kurn},\ and\ \citenamefont
  {Ketterle}}]{andrews1996direct}%
  \BibitemOpen
  \bibfield  {author} {\bibinfo {author} {\bibfnamefont {M.}~\bibnamefont
  {Andrews}}, \bibinfo {author} {\bibfnamefont {M.-O.}\ \bibnamefont {Mewes}},
  \bibinfo {author} {\bibfnamefont {N.}~\bibnamefont {Van~Druten}}, \bibinfo
  {author} {\bibfnamefont {D.}~\bibnamefont {Durfee}}, \bibinfo {author}
  {\bibfnamefont {D.}~\bibnamefont {Kurn}}, \ and\ \bibinfo {author}
  {\bibfnamefont {W.}~\bibnamefont {Ketterle}},\ }\bibfield  {title} {\enquote
  {\bibinfo {title} {Direct, nondestructive observation of a bose
  condensate},}\ }\href@noop {} {\bibfield  {journal} {\bibinfo  {journal}
  {Science}\ }\textbf {\bibinfo {volume} {273}},\ \bibinfo {pages} {84--87}
  (\bibinfo {year} {1996})}\BibitemShut {NoStop}%
\bibitem [{\citenamefont {Saba}\ \emph {et~al.}(2005)\citenamefont {Saba},
  \citenamefont {Pasquini}, \citenamefont {Sanner}, \citenamefont {Shin},
  \citenamefont {Ketterle},\ and\ \citenamefont {Pritchard}}]{saba2005light}%
  \BibitemOpen
  \bibfield  {author} {\bibinfo {author} {\bibfnamefont {M.}~\bibnamefont
  {Saba}}, \bibinfo {author} {\bibfnamefont {T.}~\bibnamefont {Pasquini}},
  \bibinfo {author} {\bibfnamefont {C.}~\bibnamefont {Sanner}}, \bibinfo
  {author} {\bibfnamefont {Y.}~\bibnamefont {Shin}}, \bibinfo {author}
  {\bibfnamefont {W.}~\bibnamefont {Ketterle}}, \ and\ \bibinfo {author}
  {\bibfnamefont {D.}~\bibnamefont {Pritchard}},\ }\bibfield  {title} {\enquote
  {\bibinfo {title} {Light scattering to determine the relative phase of two
  bose-einstein condensates},}\ }\href@noop {} {\bibfield  {journal} {\bibinfo
  {journal} {Science}\ }\textbf {\bibinfo {volume} {307}},\ \bibinfo {pages}
  {1945--1948} (\bibinfo {year} {2005})}\BibitemShut {NoStop}%
\bibitem [{\citenamefont {Freilich}\ \emph {et~al.}(2010)\citenamefont
  {Freilich}, \citenamefont {Bianchi}, \citenamefont {Kaufman}, \citenamefont
  {Langin},\ and\ \citenamefont {Hall}}]{freilich2010real}%
  \BibitemOpen
  \bibfield  {author} {\bibinfo {author} {\bibfnamefont {D.~V.}\ \bibnamefont
  {Freilich}}, \bibinfo {author} {\bibfnamefont {D.~M.}\ \bibnamefont
  {Bianchi}}, \bibinfo {author} {\bibfnamefont {A.~M.}\ \bibnamefont
  {Kaufman}}, \bibinfo {author} {\bibfnamefont {T.~K.}\ \bibnamefont {Langin}},
  \ and\ \bibinfo {author} {\bibfnamefont {D.~S.}\ \bibnamefont {Hall}},\
  }\bibfield  {title} {\enquote {\bibinfo {title} {{Real-time dynamics of
  single vortex lines and vortex dipoles in a Bose-Einstein condensate}},}\
  }\href@noop {} {\bibfield  {journal} {\bibinfo  {journal} {Science}\ }\textbf
  {\bibinfo {volume} {329}},\ \bibinfo {pages} {1182--1185} (\bibinfo {year}
  {2010})}\BibitemShut {NoStop}%
\bibitem [{\citenamefont {Ku}\ \emph {et~al.}(2016)\citenamefont {Ku},
  \citenamefont {Mukherjee}, \citenamefont {Yefsah},\ and\ \citenamefont
  {Zwierlein}}]{KuZwierlein_Vortex}%
  \BibitemOpen
  \bibfield  {author} {\bibinfo {author} {\bibfnamefont {M.~J.~H.}\
  \bibnamefont {Ku}}, \bibinfo {author} {\bibfnamefont {B.}~\bibnamefont
  {Mukherjee}}, \bibinfo {author} {\bibfnamefont {T.}~\bibnamefont {Yefsah}}, \
  and\ \bibinfo {author} {\bibfnamefont {M.~W.}\ \bibnamefont {Zwierlein}},\
  }\bibfield  {title} {\enquote {\bibinfo {title} {Cascade of solitonic
  excitations in a superfluid fermi gas: From planar solitons to vortex rings
  and lines},}\ }\href {\doibase 10.1103/PhysRevLett.116.045304} {\bibfield
  {journal} {\bibinfo  {journal} {Phys. Rev. Lett.}\ }\textbf {\bibinfo
  {volume} {116}},\ \bibinfo {pages} {045304} (\bibinfo {year}
  {2016})}\BibitemShut {NoStop}%
\bibitem [{\citenamefont {Serafini}\ \emph {et~al.}(2017)\citenamefont
  {Serafini}, \citenamefont {Galantucci}, \citenamefont {Iseni}, \citenamefont
  {Bienaim\'e}, \citenamefont {Bisset}, \citenamefont {Barenghi}, \citenamefont
  {Dalfovo}, \citenamefont {Lamporesi},\ and\ \citenamefont
  {Ferrari}}]{VortexTrento}%
  \BibitemOpen
  \bibfield  {author} {\bibinfo {author} {\bibfnamefont {S.}~\bibnamefont
  {Serafini}}, \bibinfo {author} {\bibfnamefont {L.}~\bibnamefont
  {Galantucci}}, \bibinfo {author} {\bibfnamefont {E.}~\bibnamefont {Iseni}},
  \bibinfo {author} {\bibfnamefont {T.}~\bibnamefont {Bienaim\'e}}, \bibinfo
  {author} {\bibfnamefont {R.~N.}\ \bibnamefont {Bisset}}, \bibinfo {author}
  {\bibfnamefont {C.~F.}\ \bibnamefont {Barenghi}}, \bibinfo {author}
  {\bibfnamefont {F.}~\bibnamefont {Dalfovo}}, \bibinfo {author} {\bibfnamefont
  {G.}~\bibnamefont {Lamporesi}}, \ and\ \bibinfo {author} {\bibfnamefont
  {G.}~\bibnamefont {Ferrari}},\ }\bibfield  {title} {\enquote {\bibinfo
  {title} {{Vortex reconnections and rebounds in trapped atomic Bose-Einstein
  condensates}},}\ }\href {\doibase 10.1103/PhysRevX.7.021031} {\bibfield
  {journal} {\bibinfo  {journal} {Phys. Rev. X}\ }\textbf {\bibinfo {volume}
  {7}},\ \bibinfo {pages} {021031} (\bibinfo {year} {2017})}\BibitemShut
  {NoStop}%
\bibitem [{\citenamefont {Seroka}\ \emph {et~al.}(2019)\citenamefont {Seroka},
  \citenamefont {Curiel}, \citenamefont {Trypogeorgos}, \citenamefont
  {Lundblad},\ and\ \citenamefont {Spielman}}]{Seroka:s}%
  \BibitemOpen
  \bibfield  {author} {\bibinfo {author} {\bibfnamefont {E.~M.}\ \bibnamefont
  {Seroka}}, \bibinfo {author} {\bibfnamefont {A.~V.}\ \bibnamefont {Curiel}},
  \bibinfo {author} {\bibfnamefont {D.}~\bibnamefont {Trypogeorgos}}, \bibinfo
  {author} {\bibfnamefont {N.}~\bibnamefont {Lundblad}}, \ and\ \bibinfo
  {author} {\bibfnamefont {I.~B.}\ \bibnamefont {Spielman}},\ }\bibfield
  {title} {\enquote {\bibinfo {title} {Repeated measurements with minimally
  destructive partial-transfer absorption imaging},}\ }\href {\doibase
  10.1364/OE.27.036611} {\bibfield  {journal} {\bibinfo  {journal} {Opt.
  Express}\ }\textbf {\bibinfo {volume} {27}},\ \bibinfo {pages} {36611--36624}
  (\bibinfo {year} {2019})}\BibitemShut {NoStop}%
\bibitem [{\citenamefont {B{\"u}cker}\ \emph {et~al.}(2009)\citenamefont
  {B{\"u}cker}, \citenamefont {Perrin}, \citenamefont {Manz}, \citenamefont
  {Betz}, \citenamefont {Koller}, \citenamefont {Plisson}, \citenamefont
  {Rottmann}, \citenamefont {Schumm},\ and\ \citenamefont
  {Schmiedmayer}}]{bucker2009single}%
  \BibitemOpen
  \bibfield  {author} {\bibinfo {author} {\bibfnamefont {R.}~\bibnamefont
  {B{\"u}cker}}, \bibinfo {author} {\bibfnamefont {A.}~\bibnamefont {Perrin}},
  \bibinfo {author} {\bibfnamefont {S.}~\bibnamefont {Manz}}, \bibinfo {author}
  {\bibfnamefont {T.}~\bibnamefont {Betz}}, \bibinfo {author} {\bibfnamefont
  {C.}~\bibnamefont {Koller}}, \bibinfo {author} {\bibfnamefont
  {T.}~\bibnamefont {Plisson}}, \bibinfo {author} {\bibfnamefont
  {J.}~\bibnamefont {Rottmann}}, \bibinfo {author} {\bibfnamefont
  {T.}~\bibnamefont {Schumm}}, \ and\ \bibinfo {author} {\bibfnamefont
  {J.}~\bibnamefont {Schmiedmayer}},\ }\bibfield  {title} {\enquote {\bibinfo
  {title} {Single-particle-sensitive imaging of freely propagating ultracold
  atoms},}\ }\href@noop {} {\bibfield  {journal} {\bibinfo  {journal} {New J.
  Phys.}\ }\textbf {\bibinfo {volume} {11}},\ \bibinfo {pages} {103039}
  (\bibinfo {year} {2009})}\BibitemShut {NoStop}%
\bibitem [{\citenamefont {Bergschneider}\ \emph {et~al.}(2018)\citenamefont
  {Bergschneider}, \citenamefont {Klinkhamer}, \citenamefont {Becher},
  \citenamefont {Klemt}, \citenamefont {Z\"urn}, \citenamefont {Preiss},\ and\
  \citenamefont {Jochim}}]{bergschneider_Li}%
  \BibitemOpen
  \bibfield  {author} {\bibinfo {author} {\bibfnamefont {A.}~\bibnamefont
  {Bergschneider}}, \bibinfo {author} {\bibfnamefont {V.~M.}\ \bibnamefont
  {Klinkhamer}}, \bibinfo {author} {\bibfnamefont {J.~H.}\ \bibnamefont
  {Becher}}, \bibinfo {author} {\bibfnamefont {R.}~\bibnamefont {Klemt}},
  \bibinfo {author} {\bibfnamefont {G.}~\bibnamefont {Z\"urn}}, \bibinfo
  {author} {\bibfnamefont {P.~M.}\ \bibnamefont {Preiss}}, \ and\ \bibinfo
  {author} {\bibfnamefont {S.}~\bibnamefont {Jochim}},\ }\bibfield  {title}
  {\enquote {\bibinfo {title} {Spin-resolved single-atom imaging of
  $^{6}\mathrm{Li}$ in free space},}\ }\href {\doibase
  10.1103/PhysRevA.97.063613} {\bibfield  {journal} {\bibinfo  {journal} {Phys.
  Rev. A}\ }\textbf {\bibinfo {volume} {97}},\ \bibinfo {pages} {063613}
  (\bibinfo {year} {2018})}\BibitemShut {NoStop}%
\bibitem [{\citenamefont {Choi}\ \emph {et~al.}(2016)\citenamefont {Choi},
  \citenamefont {Hild}, \citenamefont {Zeiher}, \citenamefont {Schau{\ss}},
  \citenamefont {Rubio-Abadal}, \citenamefont {Yefsah}, \citenamefont
  {Khemani}, \citenamefont {Huse}, \citenamefont {A.}, \citenamefont {Bloch},\
  and\ \citenamefont {Gross}}]{2dMBLScience}%
  \BibitemOpen
  \bibfield  {author} {\bibinfo {author} {\bibfnamefont {J.-Y.}\ \bibnamefont
  {Choi}}, \bibinfo {author} {\bibfnamefont {S.}~\bibnamefont {Hild}}, \bibinfo
  {author} {\bibfnamefont {J.}~\bibnamefont {Zeiher}}, \bibinfo {author}
  {\bibfnamefont {P.}~\bibnamefont {Schau{\ss}}}, \bibinfo {author}
  {\bibfnamefont {A.}~\bibnamefont {Rubio-Abadal}}, \bibinfo {author}
  {\bibfnamefont {T.}~\bibnamefont {Yefsah}}, \bibinfo {author} {\bibfnamefont
  {V.}~\bibnamefont {Khemani}}, \bibinfo {author} {\bibfnamefont
  {D.}~\bibnamefont {Huse}}, \bibinfo {author} {\bibnamefont {A.}}, \bibinfo
  {author} {\bibfnamefont {I.}~\bibnamefont {Bloch}}, \ and\ \bibinfo {author}
  {\bibfnamefont {C.}~\bibnamefont {Gross}},\ }\bibfield  {title} {\enquote
  {\bibinfo {title} {Exploring the many-body localization transition in two
  dimensions},}\ }\href {\doibase 10.1126/science.aaf8834} {\bibfield
  {journal} {\bibinfo  {journal} {Science}\ }\textbf {\bibinfo {volume}
  {352}},\ \bibinfo {pages} {1547} (\bibinfo {year} {2016})}\BibitemShut
  {NoStop}%
\bibitem [{\citenamefont {Fukuhara}\ \emph {et~al.}(2013)\citenamefont
  {Fukuhara}, \citenamefont {Kantian}, \citenamefont {Endres}, \citenamefont
  {Cheneau}, \citenamefont {Schauss}, \citenamefont {Hild}, \citenamefont
  {Bellem}, \citenamefont {Schollw{\"o}ck}, \citenamefont {Giamarchi},
  \citenamefont {Gross}, \citenamefont {Bloch},\ and\ \citenamefont
  {Kuhr}}]{SingleSite}%
  \BibitemOpen
  \bibfield  {author} {\bibinfo {author} {\bibfnamefont {T.}~\bibnamefont
  {Fukuhara}}, \bibinfo {author} {\bibfnamefont {A.}~\bibnamefont {Kantian}},
  \bibinfo {author} {\bibfnamefont {M.}~\bibnamefont {Endres}}, \bibinfo
  {author} {\bibfnamefont {M.}~\bibnamefont {Cheneau}}, \bibinfo {author}
  {\bibfnamefont {P.}~\bibnamefont {Schauss}}, \bibinfo {author} {\bibfnamefont
  {S.}~\bibnamefont {Hild}}, \bibinfo {author} {\bibnamefont {Bellem}},
  \bibinfo {author} {\bibfnamefont {U.}~\bibnamefont {Schollw{\"o}ck}},
  \bibinfo {author} {\bibfnamefont {T.}~\bibnamefont {Giamarchi}}, \bibinfo
  {author} {\bibfnamefont {C.}~\bibnamefont {Gross}}, \bibinfo {author}
  {\bibfnamefont {I.}~\bibnamefont {Bloch}}, \ and\ \bibinfo {author}
  {\bibfnamefont {S.}~\bibnamefont {Kuhr}},\ }\bibfield  {title} {\enquote
  {\bibinfo {title} {Quantum dynamics of a single, mobile spin impurity},}\
  }\href@noop {} {\bibfield  {journal} {\bibinfo  {journal} {Nature Phys.}\
  }\textbf {\bibinfo {volume} {9}},\ \bibinfo {pages} {235} (\bibinfo {year}
  {2013})}\BibitemShut {NoStop}%
\bibitem [{\citenamefont {Bouton}\ \emph {et~al.}(2021)\citenamefont {Bouton},
  \citenamefont {Nettersheim}, \citenamefont {Burgardt}, \citenamefont {Adam},
  \citenamefont {Lutz}, ,\ and\ \citenamefont {Widera}}]{Widera}%
  \BibitemOpen
  \bibfield  {author} {\bibinfo {author} {\bibfnamefont {Q.}~\bibnamefont
  {Bouton}}, \bibinfo {author} {\bibfnamefont {J.}~\bibnamefont {Nettersheim}},
  \bibinfo {author} {\bibfnamefont {S.}~\bibnamefont {Burgardt}}, \bibinfo
  {author} {\bibfnamefont {D.}~\bibnamefont {Adam}}, \bibinfo {author}
  {\bibfnamefont {E.}~\bibnamefont {Lutz}}, , \ and\ \bibinfo {author}
  {\bibfnamefont {A.}~\bibnamefont {Widera}},\ }\bibfield  {title} {\enquote
  {\bibinfo {title} {A quantum heat engine driven by atomic collisions},}\
  }\href {\doibase 10.1038/s41467-021-22222-z} {\bibfield  {journal} {\bibinfo
  {journal} {Nature Comm.}\ }\textbf {\bibinfo {volume} {12}},\ \bibinfo
  {pages} {2063} (\bibinfo {year} {2021})}\BibitemShut {NoStop}%
\bibitem [{\citenamefont {Arrachea}\ \emph {et~al.}(2012)\citenamefont
  {Arrachea}, \citenamefont {Mucciolo}, \citenamefont {Chamon},\ and\
  \citenamefont {Capaz}}]{Nano1}%
  \BibitemOpen
  \bibfield  {author} {\bibinfo {author} {\bibfnamefont {L.}~\bibnamefont
  {Arrachea}}, \bibinfo {author} {\bibfnamefont {E.~R.}\ \bibnamefont
  {Mucciolo}}, \bibinfo {author} {\bibfnamefont {C.}~\bibnamefont {Chamon}}, \
  and\ \bibinfo {author} {\bibfnamefont {R.~B.}\ \bibnamefont {Capaz}},\
  }\bibfield  {title} {\enquote {\bibinfo {title} {Microscopic model of a
  phononic refrigerator},}\ }\href@noop {} {\bibfield  {journal} {\bibinfo
  {journal} {Phys. Rev. B}\ }\textbf {\bibinfo {volume} {86}},\ \bibinfo
  {pages} {125424} (\bibinfo {year} {2012})}\BibitemShut {NoStop}%
\bibitem [{\citenamefont {Chamon}\ \emph {et~al.}(2011)\citenamefont {Chamon},
  \citenamefont {Mucciolo}, \citenamefont {Arrachea},\ and\ \citenamefont
  {Capaz}}]{Nano2}%
  \BibitemOpen
  \bibfield  {author} {\bibinfo {author} {\bibfnamefont {C.}~\bibnamefont
  {Chamon}}, \bibinfo {author} {\bibfnamefont {E.~R.}\ \bibnamefont
  {Mucciolo}}, \bibinfo {author} {\bibfnamefont {L.}~\bibnamefont {Arrachea}},
  \ and\ \bibinfo {author} {\bibfnamefont {R.~B.}\ \bibnamefont {Capaz}},\
  }\bibfield  {title} {\enquote {\bibinfo {title} {Heat pumping in
  nanomechanical systems},}\ }\href@noop {} {\bibfield  {journal} {\bibinfo
  {journal} {Phys. Rev. Lett.}\ }\textbf {\bibinfo {volume} {106}},\ \bibinfo
  {pages} {135504} (\bibinfo {year} {2011})}\BibitemShut {NoStop}%
\bibitem [{\citenamefont {Pigneur}\ \emph {et~al.}(2018)\citenamefont
  {Pigneur}, \citenamefont {Berrada}, \citenamefont {Bonneau}, \citenamefont
  {Schumm}, \citenamefont {Demler},\ and\ \citenamefont
  {Schmiedmayer}}]{Pigneur2018}%
  \BibitemOpen
  \bibfield  {author} {\bibinfo {author} {\bibfnamefont {M.}~\bibnamefont
  {Pigneur}}, \bibinfo {author} {\bibfnamefont {T.}~\bibnamefont {Berrada}},
  \bibinfo {author} {\bibfnamefont {M.}~\bibnamefont {Bonneau}}, \bibinfo
  {author} {\bibfnamefont {T.}~\bibnamefont {Schumm}}, \bibinfo {author}
  {\bibfnamefont {E.}~\bibnamefont {Demler}}, \ and\ \bibinfo {author}
  {\bibfnamefont {J.}~\bibnamefont {Schmiedmayer}},\ }\bibfield  {title}
  {\enquote {\bibinfo {title} {{Relaxation to a phase-locked equilibrium state
  in a one-dimensional bosonic Josephson junction}},}\ }\href {\doibase
  10.1103/PhysRevLett.120.173601} {\bibfield  {journal} {\bibinfo  {journal}
  {Phys. Rev. Lett.}\ }\textbf {\bibinfo {volume} {120}},\ \bibinfo {pages}
  {173601} (\bibinfo {year} {2018})}\BibitemShut {NoStop}%
\bibitem [{\citenamefont {Gelbwaser-Klimovsky}\ \emph
  {et~al.}(2018)\citenamefont {Gelbwaser-Klimovsky}, \citenamefont {Bylinskii},
  \citenamefont {Gangloff}, \citenamefont {Islam}, \citenamefont
  {Aspuru-Guzik},\ and\ \citenamefont {Vuletic}}]{PhysRevLett.120.170601}%
  \BibitemOpen
  \bibfield  {author} {\bibinfo {author} {\bibfnamefont {D.}~\bibnamefont
  {Gelbwaser-Klimovsky}}, \bibinfo {author} {\bibfnamefont {A.}~\bibnamefont
  {Bylinskii}}, \bibinfo {author} {\bibfnamefont {D.}~\bibnamefont {Gangloff}},
  \bibinfo {author} {\bibfnamefont {R.}~\bibnamefont {Islam}}, \bibinfo
  {author} {\bibfnamefont {A.}~\bibnamefont {Aspuru-Guzik}}, \ and\ \bibinfo
  {author} {\bibfnamefont {V.}~\bibnamefont {Vuletic}},\ }\bibfield  {title}
  {\enquote {\bibinfo {title} {Single-atom heat machines enabled by energy
  quantization},}\ }\href {\doibase 10.1103/PhysRevLett.120.170601} {\bibfield
  {journal} {\bibinfo  {journal} {Phys. Rev. Lett.}\ }\textbf {\bibinfo
  {volume} {120}},\ \bibinfo {pages} {170601} (\bibinfo {year}
  {2018})}\BibitemShut {NoStop}%
\bibitem [{\citenamefont {Weedbrook}\ \emph {et~al.}(2012)\citenamefont
  {Weedbrook}, \citenamefont {Pirandola}, \citenamefont {Garcia-Patron},
  \citenamefont {Cerf}, \citenamefont {Ralph}, \citenamefont {Shapiro},\ and\
  \citenamefont {Lloyd}}]{GaussianQuantumInfo}%
  \BibitemOpen
  \bibfield  {author} {\bibinfo {author} {\bibfnamefont {C.}~\bibnamefont
  {Weedbrook}}, \bibinfo {author} {\bibfnamefont {S.}~\bibnamefont
  {Pirandola}}, \bibinfo {author} {\bibfnamefont {R.}~\bibnamefont
  {Garcia-Patron}}, \bibinfo {author} {\bibfnamefont {N.~J.}\ \bibnamefont
  {Cerf}}, \bibinfo {author} {\bibfnamefont {T.~C.}\ \bibnamefont {Ralph}},
  \bibinfo {author} {\bibfnamefont {J.~H.}\ \bibnamefont {Shapiro}}, \ and\
  \bibinfo {author} {\bibfnamefont {S.}~\bibnamefont {Lloyd}},\ }\bibfield
  {title} {\enquote {\bibinfo {title} {Gaussian quantum information},}\ }\href
  {\doibase 10.1103/RevModPhys.84.621} {\bibfield  {journal} {\bibinfo
  {journal} {Rev. Mod. Phys.}\ }\textbf {\bibinfo {volume} {84}},\ \bibinfo
  {pages} {621} (\bibinfo {year} {2012})}\BibitemShut {NoStop}%
\bibitem [{\citenamefont {Eisert}\ and\ \citenamefont
  {Plenio}(2003)}]{Continuous}%
  \BibitemOpen
  \bibfield  {author} {\bibinfo {author} {\bibfnamefont {J.}~\bibnamefont
  {Eisert}}\ and\ \bibinfo {author} {\bibfnamefont {M.~B.}\ \bibnamefont
  {Plenio}},\ }\bibfield  {title} {\enquote {\bibinfo {title} {Introduction to
  the basics of entanglement theory in continuous-variable systems},}\ }\href
  {\doibase 10.1142/S0219749903000371} {\bibfield  {journal} {\bibinfo
  {journal} {Int. J. Quant. Inf.}\ }\textbf {\bibinfo {volume} {1}},\ \bibinfo
  {pages} {479} (\bibinfo {year} {2003})}\BibitemShut {NoStop}%
\bibitem [{Note2()}]{Note2}%
  \BibitemOpen
  \bibinfo {note} {This means that the density matrix has no zero
  eigenvalue.}\BibitemShut {Stop}%
\bibitem [{\citenamefont {Javanainen}(1999)}]{JavanainenPhononApproach}%
  \BibitemOpen
  \bibfield  {author} {\bibinfo {author} {\bibfnamefont {J.}~\bibnamefont
  {Javanainen}},\ }\bibfield  {title} {\enquote {\bibinfo {title} {Phonon
  approach to an array of traps containing bose-einstein condensates},}\ }\href
  {\doibase 10.1103/PhysRevA.60.4902} {\bibfield  {journal} {\bibinfo
  {journal} {Phys. Rev. A}\ }\textbf {\bibinfo {volume} {60}},\ \bibinfo
  {pages} {4902--4909} (\bibinfo {year} {1999})}\BibitemShut {NoStop}%
\bibitem [{\citenamefont {Salasnich}\ \emph {et~al.}(2002)\citenamefont
  {Salasnich}, \citenamefont {Parola},\ and\ \citenamefont
  {Reatto}}]{Salasnich2002}%
  \BibitemOpen
  \bibfield  {author} {\bibinfo {author} {\bibfnamefont {L.}~\bibnamefont
  {Salasnich}}, \bibinfo {author} {\bibfnamefont {A.}~\bibnamefont {Parola}}, \
  and\ \bibinfo {author} {\bibfnamefont {L.}~\bibnamefont {Reatto}},\
  }\bibfield  {title} {\enquote {\bibinfo {title} {Effective wave equations for
  the dynamics of cigar-shaped and disk-shaped bose condensates},}\ }\href@noop
  {} {\bibfield  {journal} {\bibinfo  {journal} {Phys. Rev. A}\ }\textbf
  {\bibinfo {volume} {65}},\ \bibinfo {pages} {043614} (\bibinfo {year}
  {2002})}\BibitemShut {NoStop}%
\bibitem [{\citenamefont {Gluza}\ \emph {et~al.}(2019)\citenamefont {Gluza},
  \citenamefont {Eisert},\ and\ \citenamefont {Farrelly}}]{gge}%
  \BibitemOpen
  \bibfield  {author} {\bibinfo {author} {\bibfnamefont {M.}~\bibnamefont
  {Gluza}}, \bibinfo {author} {\bibfnamefont {J.}~\bibnamefont {Eisert}}, \
  and\ \bibinfo {author} {\bibfnamefont {T.}~\bibnamefont {Farrelly}},\
  }\bibfield  {title} {\enquote {\bibinfo {title} {{Equilibration towards
  generalized Gibbs ensembles in non-interacting theories}},}\ }\href {\doibase
  10.21468/SciPostPhys.7.3.038} {\bibfield  {journal} {\bibinfo  {journal}
  {SciPost Phys.}\ }\textbf {\bibinfo {volume} {7}},\ \bibinfo {pages} {38}
  (\bibinfo {year} {2019})}\BibitemShut {NoStop}%
\bibitem [{\citenamefont {Pigneur}\ and\ \citenamefont
  {Schmiedmayer}(2018)}]{Pigneur2018PRA}%
  \BibitemOpen
  \bibfield  {author} {\bibinfo {author} {\bibfnamefont {M.}~\bibnamefont
  {Pigneur}}\ and\ \bibinfo {author} {\bibfnamefont {J.}~\bibnamefont
  {Schmiedmayer}},\ }\bibfield  {title} {\enquote {\bibinfo {title}
  {{Analytical pendulum model for a bosonic Josephson junction}},}\ }\href
  {\doibase 10.1103/PhysRevA.98.063632} {\bibfield  {journal} {\bibinfo
  {journal} {Phys. Rev. A}\ }\textbf {\bibinfo {volume} {98}},\ \bibinfo
  {pages} {063632} (\bibinfo {year} {2018})}\BibitemShut {NoStop}%
\bibitem [{\citenamefont {Whitlock}\ and\ \citenamefont
  {Bouchoule}(2003)}]{Whitlock03}%
  \BibitemOpen
  \bibfield  {author} {\bibinfo {author} {\bibfnamefont {N.~K.}\ \bibnamefont
  {Whitlock}}\ and\ \bibinfo {author} {\bibfnamefont {I.}~\bibnamefont
  {Bouchoule}},\ }\bibfield  {title} {\enquote {\bibinfo {title} {Relative
  phase fluctuations of two coupled one-dimensional condensates},}\ }\href@noop
  {} {\bibfield  {journal} {\bibinfo  {journal} {Phys. Rev. A}\ }\textbf
  {\bibinfo {volume} {68}},\ \bibinfo {pages} {053609} (\bibinfo {year}
  {2003})}\BibitemShut {NoStop}%
\bibitem [{\citenamefont {Dalfovo}\ \emph {et~al.}(1999)\citenamefont
  {Dalfovo}, \citenamefont {Giorgini}, \citenamefont {Pitaevskii},\ and\
  \citenamefont {Stringari}}]{RMP_Stringari}%
  \BibitemOpen
  \bibfield  {author} {\bibinfo {author} {\bibfnamefont {F.}~\bibnamefont
  {Dalfovo}}, \bibinfo {author} {\bibfnamefont {S.}~\bibnamefont {Giorgini}},
  \bibinfo {author} {\bibfnamefont {L.~P.}\ \bibnamefont {Pitaevskii}}, \ and\
  \bibinfo {author} {\bibfnamefont {S.}~\bibnamefont {Stringari}},\ }\bibfield
  {title} {\enquote {\bibinfo {title} {{Theory of Bose-Einstein condensation in
  trapped gases}},}\ }\href {\doibase 10.1103/RevModPhys.71.463} {\bibfield
  {journal} {\bibinfo  {journal} {Rev. Mod. Phys.}\ }\textbf {\bibinfo {volume}
  {71}},\ \bibinfo {pages} {463--512} (\bibinfo {year} {1999})}\BibitemShut
  {NoStop}%
\end{thebibliography}
%

\end{document}